%% file: stronglyCorrelatedReview_May_22_final.tex
\documentclass[12pt]{iopart}

\usepackage{graphicx}        % standard LaTeX graphics tool
\usepackage{epsfig}
\usepackage{amsmath}    %For more commands for lay-outing formulae
\usepackage{amssymb}
\usepackage{amsfonts}   %For more commands for lay-outing formulae
\usepackage{iopams}    %Must be included for iopart to work
\usepackage[dvips]{color}
\usepackage{latexsym}   %For extra symbols
\usepackage{slashed}
\usepackage{yfonts}
\usepackage{verbatim}
\usepackage{multirow}

\newcommand{\be}{\begin{equation}}
\newcommand{\ee}{\end{equation}}
\newcommand{\bea}{\begin{eqnarray}}
\newcommand{\eea}{\end{eqnarray}}
\newcommand{\ba}{\begin{array}}
\newcommand{\ea}{\end{array}}
\newcommand{\bc}{\begin{center}}
\newcommand{\ec}{\end{center}}

\newcommand{\bm}[1]{\mbox{\boldmath${#1}$}}
\def\lsim{\mathrel {\vcenter {\baselineskip 0pt \kern 0pt
    \hbox{$<$} \kern 0pt \hbox{$\sim$} }}}
\def\gsim{\mathrel {\vcenter {\baselineskip 0pt \kern 0pt
    \hbox{$>$} \kern 0pt \hbox{$\sim$} }}}
\definecolor{Red}{rgb}{1.00, 0.00, 0.00}

\definecolor{Green}{rgb}{0.00, 0.65, 0.00}

\definecolor{Blue}{rgb}{0.00, 0.00, 1.00}

\def\lsim{\mbox{~{\protect\raisebox{0.4ex}{$<$}}\hspace{-1.1em}
	{\protect\raisebox{-0.6ex}{$\sim$}}~}}

\include{allan_macros}

\begin{document}
%\linenumbers
\title[Strongly Correlated Quantum Fluids]{Strongly Correlated Quantum Fluids: Ultracold Quantum Gases, Quantum Chromodynamic Plasmas, and Holographic Duality}

\author{Allan Adams,$^1$ Lincoln D. Carr,$^{2,3}$  Thomas Sch\"afer,$^4$ Peter Steinberg$^5$
       and John E. Thomas$^4$}

\address{$^1$Center for Theoretical Physics, MIT, Cambridge, MA 02139, U.S.A.}

\address{$^2$Physics Institute, University of Heidelberg, D-69120 Heidelberg, Germany}

\address{$^3$Department of Physics, Colorado School of Mines, Golden, CO 80401, U.S.A.}

\address{$^4$Department of Physics, North Carolina State University, Raleigh, NC 27695}

\address{$^5$Brookhaven National Laboratory, Upton, NY 11973, U.S.A.}

\begin{abstract}
Strongly correlated quantum fluids are phases of matter that
are intrinsically quantum mechanical, and that do not have a
simple description in terms of weakly interacting quasi-particles.
Two systems that have recently attracted a great deal of
interest are the quark-gluon plasma, a plasma of strongly
interacting quarks and gluons produced in relativistic
heavy ion collisions, and ultracold atomic Fermi gases,
very dilute clouds of atomic gases confined in optical
or magnetic traps. These systems differ by more than 20
orders of magnitude in temperature, but they were shown
to exhibit very similar hydrodynamic flow. In particular,
both fluids exhibit a robustly low shear viscosity to entropy
density ratio which is characteristic of quantum fluids
described by holographic duality, a mapping from strongly correlated quantum field
theories to weakly curved higher dimensional classical gravity.
This review explores the connection between these fields,
and it also serves as an introduction to the Focus Issue of
\emph{New Journal of Physics} on Strongly Correlated
Quantum Fluids: from Ultracold Quantum Gases to QCD Plasmas.
The presentation is made accessible to the general physics
reader and includes discussions of the latest research
developments in all three areas.
\end{abstract}

\maketitle

\tableofcontents
%%%%%%%%%%%%%%%%%%%%%%%%%%%%%%%%%%%%%%%%%%%%%%%%%%%%%%%%%%%%%%%%%%%%
\section{Introduction}
\label{sec:intro}
%%%%%%%%%%%%%%%%%%%%%%%%%%%%%%%%%%%%%%%%%%%%%%%%%%%%%%%%%%%%%%%%%%%%

 This review covers the convergence between three at first sight
disparate fields: ultracold quantum gases, quantum chromodynamic
(QCD) plasmas, and holographic duality. \textit{Ultracold quantum
gases} have opened up new vistas in many-body physics, from novel
quantum states of matter to quantum computing applications.  There
are over one hundred experiments on ultracold quantum gases around
the world on every continent but Antarctica.
%NB: Most recently, the first African ultracold quantum gas experiment
%is taking place at KwaZulu-Natal University in South Africa.  They have
%their MOT and are close to BEC.
\textit{The QCD plasma}, also called the \emph{quark-gluon plasma}
(QGP), has been the subject of intense experimental investigation
over more than two decades, continuing now at the Relativistic Heavy Ion
Collider (RHIC) at Brookhaven National Laboratory and the Large Hadron
Collider (LHC) at the European Organization for Nuclear Research (CERN).
A QGP is predicted to have occurred in the first microsecond after the
Big Bang, and re-creation of the QGP on earth at present times allows
us to probe the physics of the early universe. \textit{Holographic duality}
is a powerful mapping from strongly interacting quantum field theories,
where the very concept of a quasiparticle can lose meaning, to weakly
curved higher dimensional classical gravitational theories.  %aa--
This duality provides a new approach for modeling strongly interacting
quantum systems,
yielding fresh insights into previously intractable quantum many-body problems
key to understanding experiments such as ultracold quantum gases and the QGP.

What do these three fields have in common?  All treat \emph{strongly
correlated quantum fluids}.  Generically, strong interactions give rise
to strong correlations.  By strongly correlated we mean that we cannot
describe a system by working perturbatively from non-interacting particles
or quasiparticles.  In the case of electrons in condensed matter
systems this means that theories
constructed from single-particle properties, like the Hartree-Fock
approximation, cannot describe a material. In the case of fluids, we
mean that kinetic theories based on quasiparticle degrees of freedom,
in particular the Boltzmann equation, fail.\footnote{Fluids are materials
that obey the equations of hydrodynamics. The word liquid refers to
a phase of matter that cannot be distinguished from a gas in terms
of symmetry, but exhibits short range correlations similar to those
in a solid, and is separated from the gas phase by a line of first
order transitions that terminates at a critical endpoint. A plasma
is a gas of charged particles. Gases, liquids, and plasmas behave
as fluids if probed on very long length scales. Weakly coupled
systems exhibit single particle behavior if probed on microscopic
scales, but strongly coupled systems behave as fluids also on short
scales. Liquids are typically more strongly correlated than gases,
and more likely to behave as a fluid.}
The natural candidates for building quasiparticles are quark and gluons in the
case of the QGP, neutrons and protons in the case of
nuclear matter, and atoms in the case of ultracold atomic gases. In
strongly interacting systems the mean free path for these excitations
is comparable to the interparticle spacing, and quasiparticles lose
their identity. Even though kinetic theory fails, nearly ideal, low
viscosity hydrodynamics is a very good description of these systems.
This is a central prediction of holographic duality, and it has been
verified experimentally for both ultracold quantum gases and the
QGP, as we will explore in this Review.

%%%%%%%%%%%%%%%%%%%%%%%%%%%%%%%%%%%%%%%%%%%%%%%%%%%%%%%%%%%%%%%%%%%%
\begin{figure}[t]
\begin{center}
\vspace*{0.0in}
\includegraphics[width=0.8\textwidth]{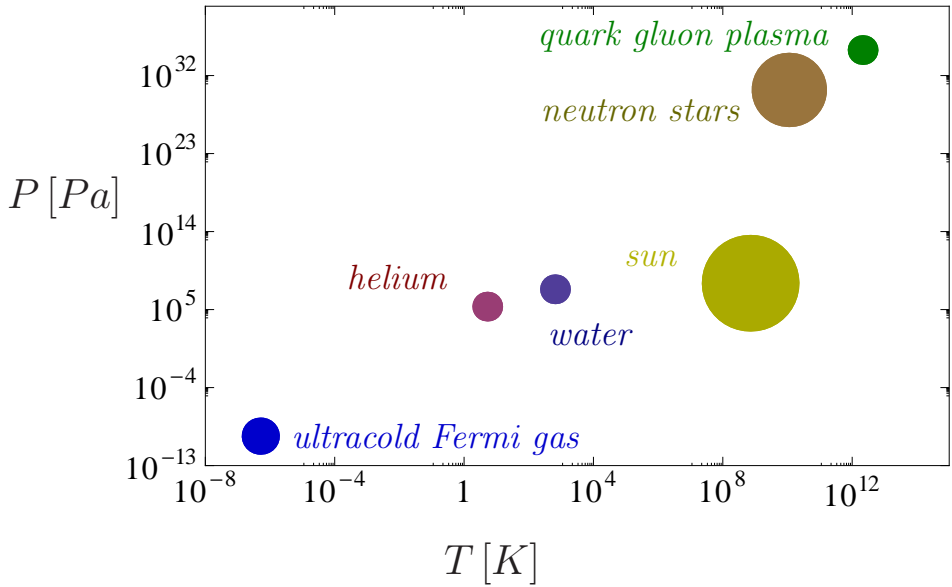}
\caption{\textit{Temperature and pressure scales of extreme quantum
matter.} Ultracold quantum gases are the coldest matter produced to
date, while the quark-gluon plasma is the hottest, together spanning
about 18 orders of magnitude in temperature and more than 40 orders
of magnitude in pressure. Yet these systems exhibit very similar
hydrodynamic behavior, as characterized by the shear viscosity to
entropy density ratio shown in Fig.~\ref{fig:ratio}). We also include
two other well known quantum fluids, liquid helium and hot proto-neutron
star matter, as well as a classical fluid, water, and a classical
plasma, the Coulomb plasma in the sun.}
\label{fig:scales}
\end{center}
\end{figure}
%%%%%%%%%%%%%%%%%%%%%%%%%%%%%%%%%%%%%%%%%%%%%%%%%%%%%%%%%%%%%%%%%%%%

\begin{figure}[t]
\begin{center}
\vspace*{0.0in}
\includegraphics[width=0.8\textwidth]{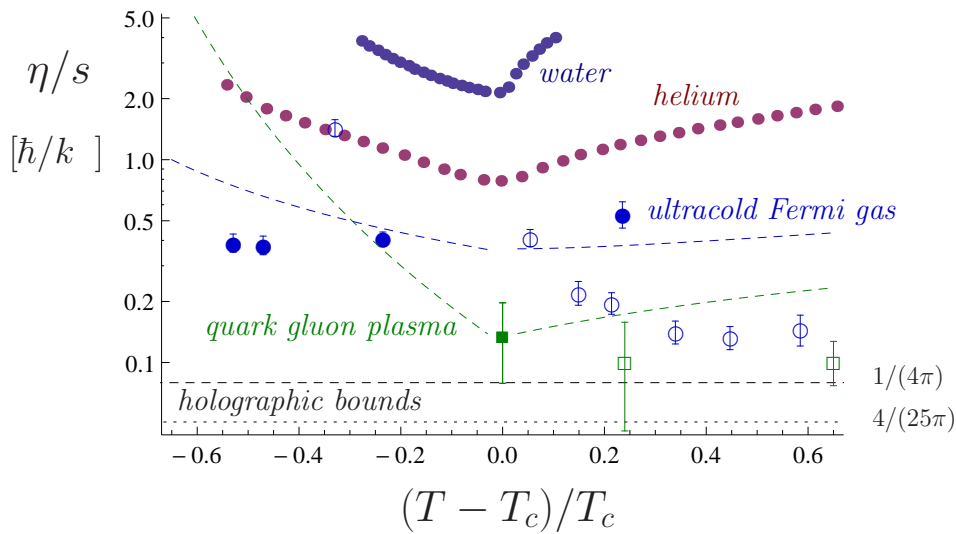}
\caption{\textit{Transport properties of strongly correlated fluids.}
Ratio of shear viscosity $\eta$ to entropy
density $s$ as a function of $(T-T_c)/T_c$, where $T_c$ is the
superfluid transition temperature in the case of ultracold Fermi
gases, the deconfinement temperature in the case of QCD, and the
critical temperature at the endpoint of the liquid gas transition
in the case of water and helium. The data for water and helium are
from~\cite{nist}, the ultracold Fermi gas data are from~\cite{NJPfocusissue4_thomas},
the quark-gluon plasma point (square) is taken from the analysis of
\cite{Song:2011hk}, the lattice QCD data (open squares) from
\cite{Meyer:2007ic}, and the lattice data for the ultracold Fermi
gas (open circles) are the $8^3$ data from \cite{Wlazlowski:2012jb}. The dashed
curves are theory curves from
\cite{Arnold:2000dr,Prakash:1993bt,Massignan:2004,Mannarelli:2012su}.
The theories are scaled by overall factors to match the data near
$T_c$. The lines labeled ``holographic bounds'' correspond to the
KSS bound $\hbar/(4\pi k_B)$~\cite{Kovtun:2004de} and the
Gauss-Bonnet bound $(16/25)\hbar/(4\pi k_B)$~\cite{Brigante:2008gz}.
Similar compilations can be found in \cite{Kovtun:2004de,Csernai:2006zz,Lacey:2006bc}.}
\label{fig:ratio}
\end{center}
\end{figure}

As shown in Fig.~\ref{fig:scales}, strongly correlated quantum fluids
cover an enormous range of scales in temperature and pressure.\footnote{
The points in Fig.~\ref{fig:scales} correspond to the range of
temperatures for which the transport measurements shown in
Fig.~\ref{fig:ratio} have been performed. For the ultracold atomic
Fermi gas experiments described in Sec.~\ref{ssec:experiments} the
critical temperature is roughly 500 nK (the exact value depends
on the trap geometry and the number of particles; Bose gases have
been cooled to temperatures below 1 nK). The data points for helium
and water are centered around the critical endpoint of the liquid
gas transition. The point for the solar plasma corresponds to the
geometric mean of the temperatures in the core and the corona. The
neutron matter point is at $T=1\,{\rm MeV}/k_B=1.2\cdot 10^{10}$ K
and at a density $n=0.1\,n_0$, where $n_0=0.14/{\rm fm}^3$ is nuclear
matter saturation density. Neutron stars are born at $T\simeq 10$
MeV/$k_B$, and they can cool to temperatures below 1 keV/$k_B$. The critical 
temperature of the QGP is $T_c\simeq 150\,
{\rm MeV}/k_B= 1.75\cdot 10^{12}$ K. Experiments with heavy ions explore temperatures
up to $\sim 3T_c$.} We
remind the reader that temperature $T$ and energy $E$ are equivalent
up to a factor of Boltzmann's constant, $k_B =1.3806503 \times 10^{-23}$
J/K, with $E=k_B T$. We focus on fluids that can be studied in bulk,
as opposed to quantum liquids that exist on lattices. We show ultracold
Fermi gases, liquid helium, neutron matter in proto-neutron stars, and
the QGP. For comparison we also show a classical fluid,
water, and a classical plasma, the Coulomb plasma in the sun.

%%%%%%%%%%%%%%%%%%%%%%%%%%%%%%%%%%%%%%%%%%%%%%%%%%%%%%%%%%%%%%%%%%%%
%%%%%%%%%%%%%%%%%%%%%%%%%%%%%%%%%%%%%%%%%%%%%%%%%%%%%%%%%%%%%%%%%%%%

Figure~\ref{fig:ratio} shows that despite the large
range in scale there is a remarkable universality in the transport
behavior of strongly correlated quantum fluids. Transport properties
of the fluid can be characterized in terms of its shear viscosity
$\eta$, which governs dissipation due to internal friction. A
dimensionless measure of dissipative effects is the ratio $\eta/s$
of shear viscosity to entropy density in units of $\hbar/k_B$. Near
the critical point, where the role of correlations is expected to be
strongest, the ratio $\eta/s$ has a minimum. For classical fluids the
minimum value is much bigger than $\hbar/k_B$, but for strongly
correlated quantum fluids $\eta/s$ is of order $\hbar/k_B$, indicating
that dissipation is governed by quantum effects. We observe, in particular,
that $\eta/s$ for the QGP and ultracold Fermi gases is quite similar,
even though the absolute values of $\eta$ and $s$ differ by many orders
of magnitude.\footnote{The theoretical
curves, as well as the data for helium and water corresponds to
systems in the thermodynamic limit. The lattice data for the
QGP and the ultracold Fermi gas have finite volume corrections
that have not been fully quantified. The experimental data point
for the QGP is based on an analysis that assumes
$\eta/s$ to be temperature independent. The data points for the
ultracold Fermi gas show the ratio of trap averages of $\eta$ and
$s$. The local value of $\eta/s$ at the center of the trap is
likely to be smaller than the ratio of the averages; see
Sec.~\ref{ssec:ultracold}.}

Remarkably, these values of $\eta/s$ lie near a lower bound,
$\eta/s \ge \hbar/(4\pi k_B)$, which arises in the study of 4+1
dimensional black holes in classical Einstein gravity. These gravitational
theories are conjectured to be dual to certain 3+1 dimensional quantum
field theories; see Sec.~\ref{sec:hd}. This lower bound is known to be
non-universal; it can be violated in a more general class of theories
dual to a gravitational theory known as Gauss-Bonnet gravity. Imposing
basic physical requirements such as causality and positivity leads to
a slightly smaller bound\footnote{Whether
this value represents a true lower bound, or whether more general
classes of fluids with even smaller values of $\eta/s$ can exist, is
an active area of research; see Sec.~\ref{sec_app_hol}. }
$\eta/s \ge (16/25)\hbar/(4\pi k_B)$.
The main feature of the results obtained using holographic dualities
is that, at strong coupling, $\eta/s$ is both unusually small and
relatively insensitive to the precise strength of the interactions,
so long as they are strong. This is in sharp contrast to the
predictions of kinetic theory for a weakly interacting gas.  As
a result, $\eta/s$ serves as a probe of the strength of correlations
in a quantum fluid.

We have chosen to focus on the fields of ultracold quantum gases
and the QGP not only for their range of energy and density scales,
but also because of their broad intrinsic interest.  Ultracold
fermions are connected to a wide variety of exotic, strongly interacting
systems in nature, ranging from high-temperature superconductors to
nuclear matter.  They are incredibly flexible many-body systems
that allow nearly arbitrary tuning of interactions, symmetries,
spin structure, effective mass, and imposed lattice structures.
The QGP on the other hand explores a very different regime from other
particle physics experiments: thousands of particles are produced,
and these particles form a hot and dense environment recreating the
conditions of the early universe.

Our review is outlined as follows. A key introductory or heuristic
plot is presented in each of our three fields for the general physics
reader.  In Sec.~\ref{sec:ultracold}, Fig.~\ref{fig:crossoverPhaseDiagram}
shows the phase diagram for the Bardeen-Cooper-Schrieffer (BCS) to Bose-Einstein condensate (BEC)
crossover, midway through which the \textit{unitary quantum gas}, a
strongly correlated quantum fluid, is obtained.  In Sec.~\ref{sec_qcd}
the QCD phase diagram is shown in Fig.~\ref{fig_qcd_phase}, including
the strongly and weakly interacting QGP being explored
presently at RHIC and the LHC.  In Sec.~\ref{sec:hd}, not only is an
extensive heuristic description of holographic duality provided in
Sec.~\ref{ssec:heuristics}, but a \textit{holographic dictionary} is
presented in Table~\ref{table:holdict}, mapping quantities in strongly
interacting quantum fields similar to the unitary quantum gas or the
QGP to their gravitational duals.  The three main
sections also cover all 39 papers from three fields comprising this
Focus Issue.  Finally, in Sec.~\ref{sec:conclusions}, rather than a
summary, we instead conclude with open questions in each field.

%%%%%%%%%%%%%%%%%%%%%%%%%%%%%%%%%%%%%%%%%%%%%%%%%%%%%%%%%%%%%%%%%%%%
\section{Ultracold Quantum Gases}
\label{sec:ultracold}
%%%%%%%%%%%%%%%%%%%%%%%%%%%%%%%%%%%%%%%%%%%%%%%%%%%%%%%%%%%%%%%%%%%%

Ultracold quantum gases provide a unique table-top paradigm for exploring
the properties of quantum many-body systems in nature~\cite{dalfovo1999,leggett2001,lewensteinM2007,bloch2008,giorgini2008,carr2009b,Schafer:2009dj},
from the thermodynamics of high temperature superconductors to the
hydrodynamics of QGPs.  These gases are made of mainly
alkali metal atoms but more recently other atoms as well as diatomic
molecules.  They can be fermionic or bosonic with a wide variety of
internal hyperfine spin structures.  They can be made strongly or
weakly interacting and both attractive and repulsive.  They are
contained in a variety of magnetic and optical traps in one, two, and
three dimensions, including optical lattices.  The latter give rise
to arbitrary lattice structures.  Because these gases are dilute and
very cold, they are described by first principles theories built up
from low-energy binary scattering between atoms, and well-known
interactions with magnetic and optical fields.  The collection of atoms
can be probed and manipulated by external laser beams and pulses, as
well as external magnetic fields.

In this review we focus on atomic Fermi gases~\cite{giorgini2008,bloch2008,ZwierleinFermiReview,zwerger2011},
in particular strongly interacting Fermi gases.  Several have been
cooled to degeneracy using evaporative cooling methods. The most
widely studied atoms are $^{40}$K~\cite{jin1999} and
$^6$Li~\cite{truscott2001,schreck2001,Granade2001,hadzibabic2003}.\footnote{Degeneracy has also been achieved
in metastable $^3$He$^*$~\cite{Vassen3He*}, in $^{171}$Yb and
$^{173}$Yb~\cite{Takahashi171173Yb}, and recently in $^{87}$Sr~\cite{Killian87Sr} and $^{161}$Dy~\cite{luMW2012}.}
Experiments are carried out at temperatures in the nanokelvin to
microkelvin range, with typical densities from $10^{11}$ to $10^{14}$
atoms/cc. For a $^6$Li atom at nanokelvin temperatures, the thermal
de Broglie wavelength
\be
\lambda_{\mathrm{dB}} = \hbar\sqrt{\frac{2\pi}{m k_B T}}
\label{eqn:deBroglie}
\ee
is on the order of several $\mu$m. Quantum degeneracy occurs when
the de Broglie wavelength is greater than or on the order of the particle
spacing, $\lambda_{\mathrm{dB}}n^{-1/3}\gtrsim 1$, where $n$ is the density;
this condition is equivalent to $T/T_F \geq 1$, where $T_F$ is the Fermi
temperature.\footnote{In this review, $T_F$ is always defined with
respect to the non-interacting Fermi gas, $T_F=\hbar^2k_F^2/(2mk_B)$ with
$k_F=(3\pi^2 n)^{1/3}$ for a two-component gas.} Our interest is in
ultracold Fermi gases that are quantum degenerate: $T/T_F \leq 1$.

 A cloud of
trapped dilute fermions is typically about 100 $\mu$m in size, and
is deformed by harmonic trapping fields into prolate or oblate forms,
commonly called a cigar or a pancake. In the degenerate regime
the cloud is stabilized against collapse by Pauli pressure
\cite{jin1999,Granade2001,truscott2001}. The size of the cloud
depends on the interplay between the trapping potential, the
Pauli pressure, and interactions between the atoms. Because of
the low density and  ultracold temperatures these interactions
are dominated by an effective $s$-wave contact interaction.
The scattering amplitude is of the form
\be
f(k) =   \frac{1}{-1/a+r_0k^2/2-ik}\, ,
\ee
where $a$ is the $s$-wave scattering length and $r_0$ is the
effective range. Higher partial waves as well as short range corrections are
suppressed by powers of $r_0/\lambda_{\mathrm{dB}}$ and $r_0n^{1/3}$.\footnote{The range
of the atomic potential
is on the order of the van der Waals length $l=(mC_6/\hbar^2)^{1/4}$,
where $C_6$ controls the van der Waals tail of the atomic potential,
$V\sim C_6/r^6$.  We assume that the $p$-wave
scattering length is natural, meaning $a_p\sim r_0$.} The
scattering length is widely tunable by a \emph{Feshbach resonance}
\cite{chinCheng2010}, an external magnetic field that brings a
weakly bound excited molecular state into resonance with the
unbound atomic scattering state.

%%%%%%%%%%%%%%%%%%%%%%%%%%%%%%%%%%%%%%%%%%%%%%%%%%%%%%%%%%%%%%%%%%%%%%%%%
\begin{figure}[t]
\begin{center}
\vspace*{0.0in}\includegraphics[width=0.8\textwidth]{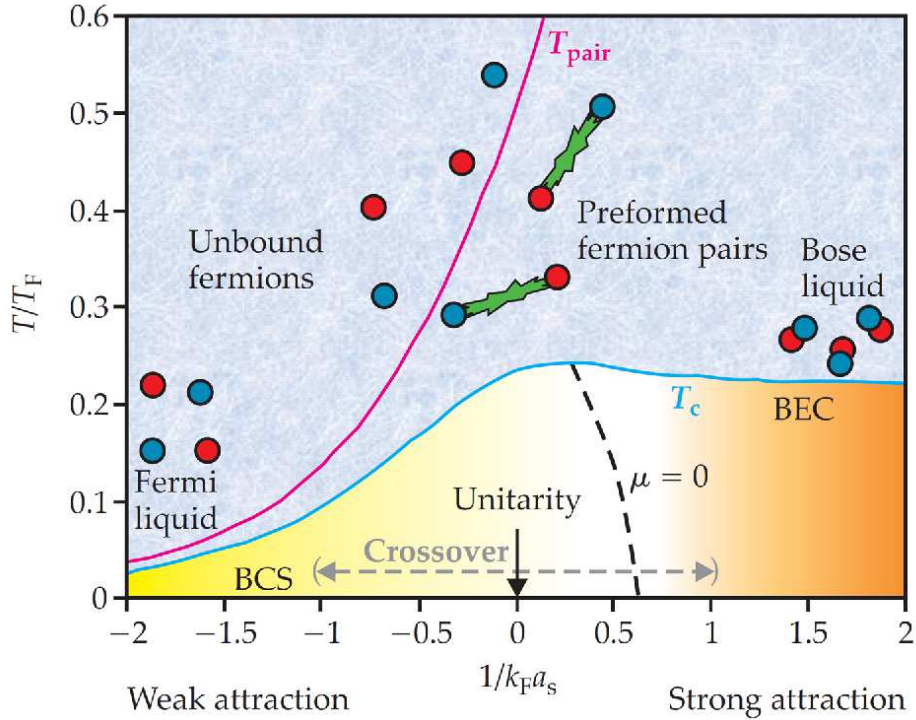}
\caption{\textit{Ultracold Fermi gas phase diagram.} Sketch of the
Bardeen-Cooper-Schrieffer (BCS) to Bose-Einstein condensate (BEC)
crossover for ultracold Fermi gases.  When the scattering length
$a_s$ passes through a pole, so that $1/(k_F a_s) \to 0$, one
obtains a strongly correlated fluid, the \emph{unitary gas}.  The
critical temperature $T_c$ for the phase transition only approaches
the pairing temperature $T_{\mathrm{pair}}$ in the limit $1/(k_F a)
\to -\infty$.  The \emph{crossover region} is the strongly interacting
regime, loosely defined by $|1/(k_F a_s)| < 1$. Note that we denote
the scattering length by $a$ in the text. Used with permission
from Ref.~\cite{sademelo2008}.}
\label{fig:crossoverPhaseDiagram}
\end{center}
\end{figure}
%%%%%%%%%%%%%%%%%%%%%%%%%%%%%%%%%%%%%%%%%%%%%%%%%%%%%%%%%%%%%%%%%%%%%%%%%

Each of the different trapped atomic elements used in ultracold quantum
gas experiments has an internal spin structure due to hyperfine structure
of the atom, that is, the combination of the nuclear spin and, in the
case of the alkali metals, the electron outside the closed shell.  For instance, $^6$Li
has a nuclear spin of 1 and one unpaired electron. The two lowest hyperfine states have a total spin of
1/2, and the remaining four have a total spin of 3/2.
By selecting out hyperfine states
through appropriate laser-induced transitions and trapping and cooling
methods, experiments can thus work with a variety of spin structures.
The case in which two hyperfine states are trapped is effectively
equivalent to a spin 1/2 atom. Tuning the scattering length by using a
Feshbach resonance, one obtains three distinct regimes, shown in
Fig.~\ref{fig:crossoverPhaseDiagram}.  The first is for weak attractive
scattering, $- k_Fa \ll 1$, where $k_F$ is the
Fermi momentum.  Then for
temperatures well below the Fermi temperature $T_F$
one obtains
a BCS state~\cite{leggett1975}, or $s$-wave superconductivity.  We call
this an \emph{atomic Fermi superfluid}, since our systems are in fact
neutral. In such a state fermions of opposite spin join to make Cooper
pairs, but their average pair size $\xi_c$ is greater than the
interparticle spacing $n^{-1/3}$,  so that
they are overlapping: $\xi_c n^{1/3} \gg 1$.  Tuning $a$ as in the
phase diagram, we observe that the scattering length passes through a
pole; note that the figure shows temperature as a function of the
inverse scattering length, $1/(k_F a)$.  Thus as $a \to \pm\infty$,
$1/(k_F a) \to 0$, and one obtains a second regime, called the
\emph{unitary gas}.  This middle regime is a strongly correlated
fluid, and one finds $\xi_c n^{1/3} \simeq 1$, i.e., the pair
size is about equal to the interparticle spacing.  Finally, for
large positive scattering lengths, the paired fermions make much
more tightly bound molecules, and one obtains a molecular BEC,
similar to the well-known atomic BECs.  This regime is depicted
on the far right of Fig.~\ref{fig:crossoverPhaseDiagram}.  In practice
these molecules are still quite large, thousands of Bohr or more, but
still much smaller than the interparticle spacing, so that
$\xi_c n^{1/3} \ll 1$.

The upper curve on the figure depicts the pair formation temperature
$T_{\mathrm{pair}}$, which in general is distinct from the critical temperature
for superfluidity, $T_c$ \cite{Chen:2005}.
Note that superfluidity is associated with the
spontaneous breakdown of a global symmetry, the $U(1)$ phase symmetry
of the wave function, and $T_c$ is therefore always well defined.
$T_{\mathrm{pair}}$, on the other hand, is not associated with a symmetry
or a local order parameter and may not be well defined. This remark
is particularly relevant in the BCS regime, where the size of the
pairs is large compared to the inter-particle spacing. Physically we
expect that in the BCS regime there are no pre-formed pairs, and $T_c$
and $T_{\mathrm{pair}}$ are very close together.

Although we refer to these systems as ultracold, in terms of the
dimensionless ratio $T/T_F$, and in comparison to solid state
systems, they are quite hot. In the unitary regime the phase
transition occurs at $T_c/T_F =0.167(23)$~\cite{kuJHMark2011},
compared to typical solid state superconductors in which $T_c/T_F
\lsim 0.01$. The unitary Fermi gas is a very high $T_c$ superfluid.
As indicated in Fig.~\ref{fig:crossoverPhaseDiagram}, in the
BCS regime the temperature required to achieve a phase transition
to an atomic Fermi superfluid is quite low. In this regime,
the critical temperature is given by the weak coupling
expression~\cite{Gorkov:1961}
\begin{equation}
\label{Tc_BCS}
\frac{T_c}{T_F} \simeq
 \frac{4\cdot 2^{1/3} e^\gamma}{\pi e^{7/3}}
   \exp\left(-\frac{\pi}{k_F |a|}\right) \,,
\end{equation}
where $\gamma\simeq 0.577$ is the Euler constant. The
numerical value of the factor in front of the exponent is
$0.277$. We observe that even though Eq.~(\ref{Tc_BCS})
is formally valid only in the limit $ k_F |a| \ll 1$
it also provides a reasonable estimate for $T_c$ at
unitarity. This is somewhat of an accident, because
higher order corrections in $k_F|a|$ are divergent
in the unitary limit.

 In the following, we explore the unitary regime of the BCS-BEC
crossover for ultracold Fermi gases. In Sec.~\ref{ssec:experiments}
we present an overview of experiments on these systems.  In
Sec.~\ref{ssec:uni} we focus on universal aspects of unitary
gases. In the strongly correlated regime the scattering length
diverges, and the remaining length scales in the problem are
the Fermi length $1/k_F$ and the de Broglie wave length
$\lambda_{\mathrm{dB}}$, given in Eq.~(\ref{eqn:deBroglie}).  Thus many theoretical statements can be made
despite the lack of a small parameter or perturbative calculations.
In Sec.~\ref{ssec:exp-eos} and \ref{ssec:phase} we treat the
thermodynamics and the structure of the phase diagram for unitary gases, and
in Sec.~\ref{ssec:transport} we focus on transport properties.
Section~\ref{ssec:lattices} presents an overview of ultracold
Fermi gases in optical lattices. Finally, in Sec.~\ref{ssec:newDirections}
we treat new directions in unitary gases as presented in this Focus Issue,
including novel experimental probes,
solitons, imbalanced systems and polarons, disorder, quantum phase transitions, Efimov physics, and the use of
three hyperfine states to explore SU(3) physics and connections
to the QGP.

%%%%%%%%%%%%%%%%%%%%%%%%%%%%%%%%%%%%%%%%%%%%%%%%%%%%%%%%%%%%%%%%%%%%%%
\subsection{Ultracold Fermi Gas Experiments}
\label{ssec:experiments}
%%%%%%%%%%%%%%%%%%%%%%%%%%%%%%%%%%%%%%%%%%%%%%%%%%%%%%%%%%%%%%%%%%%%%%

Historically, atomic Fermi gases were first brought to degeneracy at JILA in 1999~\cite{jin1999}, using a mixture of two hyperfine states in $^{40}$K to enable $s$-wave scattering in a magnetic trap. Dual species radio-frequency-induced  evaporation  produced a degenerate, weakly interacting sample, with $T/T_F\simeq 0.5$.  Later, degeneracy was achieved in fermionic $^6$Li by direct evaporation from a MOT-loaded optical trap~\cite{Granade2001} and by sympathetic cooling with another species~\cite{truscott2001,schreck2001,hadzibabic2003}, producing a lower $T/T_F$. However, the minimum reduced temperature was initially limited to $T/T_F \simeq 0.2$, which may have been a consequence of trap-noise-induced heating~\cite{Savard1997} or, at the lowest temperatures, Fermi hole heating~\cite{timmermans2001} in combination with evaporative cooling~\cite{carr2004a}.

Optical traps enabled a dramatic improvement in the efficiency of evaporation and the creation of strongly interacting Fermi gases, through the use of magnetically-tunable collisional resonances, or Feshbach resonances~\cite{Tiesinga1993}. Feshbach resonances in fermionic atoms were initially characterized in 2002  by several groups~\cite{OHaraZeroCrossing2002,JochimFeshbach2002,JinFeshbach2002}. For a recent review of Feshbach resonances see Ref.~\cite{chinCheng2010}. In a Feshbach resonance, a bias magnetic field tunes the total energy of a pair of colliding atoms in the incoming open (triplet) channel into resonance with a molecular bound state in an energetically closed (singlet) channel. At resonance, the zero-energy $s$-wave scattering length $a$ diverges and the collision cross section attains the  {\it unitary} limit, proportional to the square of the de Broglie wavelength, i.e., $\sigma=4\pi/k^2$, where $\hbar k$ is the relative momentum. The collision cross section therefore {\it increases} with decreasing temperature, enabling highly efficient evaporative cooling in optical traps and much lower reduced temperatures $T/T_F\simeq 0.05$.

An optical trap is formed by a focused laser beam. Atoms are polarized by the field and attracted to the high
intensity region, when the laser is detuned below resonance with the resonant optical transition, so that the induced dipole moment is in phase with the field. For large detunings, obtained using infrared lasers, the trapping potential is independent of the atomic hyperfine state, enabling several species to be trapped, which is ideal for Fermi gases~\cite{OHaraFermionOptTrap1999}.  Forced evaporation is accomplished by slowly lowering the intensity of the optical trap laser beam. Near a Feshbach resonance, a highly degenerate sample can be produced in a few seconds~\cite{OHara:2002}.

%%%%%%%%%%%%%%%%%%%%%%%%%%%%%%%%%%%%%%%%%%%%%%%%%%%%%%%%%%%%%%%%%%%%%%%%%%%%%
\begin{figure}[t]
\begin{center}
\includegraphics[width=0.45\textwidth]{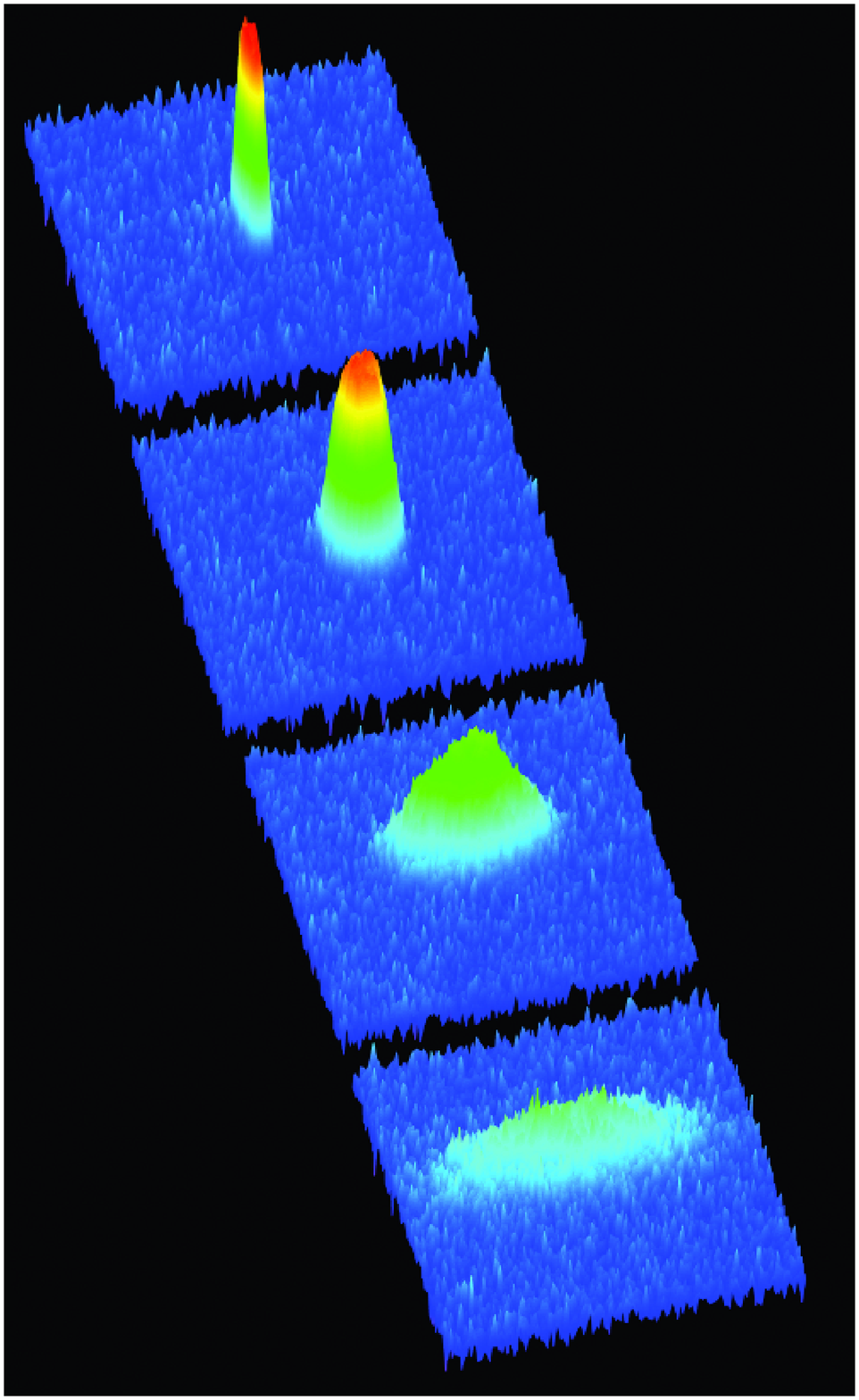}
\hspace*{0.17\textwidth}
\includegraphics[width=0.31\textwidth]{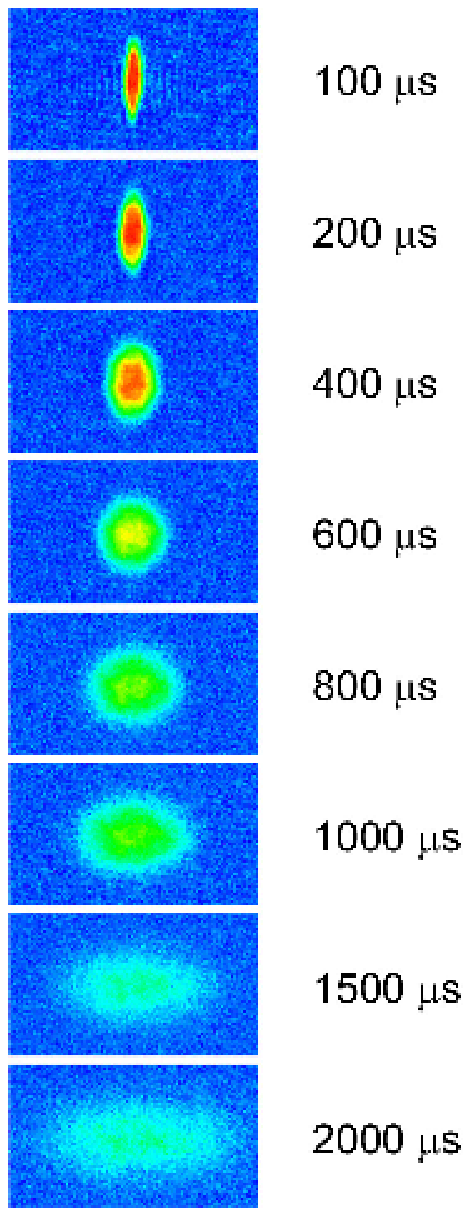}
\caption{\textit{Experimental images.}   Elliptic flow of a strongly-interacting Fermi gas as a function of time after release from a cigar-shaped optical trap: from top to bottom, $100\,\mu$s to 2 ms after release. The pressure gradient is much larger in the initially narrow directions of the cloud than in the long direction, causing the gas to expand much more rapidly along the initially narrow directions, inverting the aspect ratio. Achieving nearly perfect elliptic flow requires extremely low shear viscosity, as is the case also for a quark-gluon plasma.  The sequence of images is created by recreating similar initial conditions and destructively imaging the cloud at different times after release~\cite{OHara:2002}.}
\label{fig:exptImages}
\end{center}
\end{figure}
%%%%%%%%%%%%%%%%%%%%%%%%%%%%%%%%%%%%%%%%%%%%%%%%%%%%%%%%%%%%%%%%%%%%%%%%%%%%%

A milestone in the Fermi gas field was the observation in 2002 of a  strongly interacting, degenerate Fermi gas~\cite{OHara:2002}, in the so-called BEC-BCS crossover regime, using this method. In contrast to Bose gases, which are unstable and undergo  three-body loss on millisecond time scales near Feshbach resonances, the two-component Fermi gas was found to be stable, as the Pauli principle suppresses three-body $s$-wave scattering~\cite{petrov2004}.  Released from the cigar-shaped trapping potential of the focused beam, the cloud was observed to expand much more rapidly in the initially narrow direction, compared to the long direction, as a consequence of the much larger pressure gradient along the narrow axis. Consequently, the aspect ratio inverts from a cigar to an ellipse, as shown in Fig.~\ref{fig:exptImages}. Remarkably, the same type of \textit{elliptic flow} is also observed in the momentum distribution of an expanding quark-gluon-plasma produced in an off-center collision of two heavy ions; see Sec.~\ref{sec_flow}. There, the temperature is nineteen orders of magnitude hotter and the particle density is twenty-five orders of magnitude greater than that of the Fermi gas. In both systems, however, this nearly perfect ``elliptic" flow, Fig.~\ref{fig:exptImages},  is a consequence of extremely low viscosity hydrodynamics, which persists in the normal, non-superfluid unitary gas and deeply connects these two apparently disparate fields.

The creation of a degenerate Fermi gas near a Feshbach resonance was followed in 2003 by the first measurements of the interaction energy~\cite{GehmMechStab2003,SalomonExpInt2003} and the creation of Bose-condensed dimer molecules~\cite{Salomonmolec2003,GrimmBEC2003,JinBEC2003,KetterleBEC2003}. In 2004, condensed fermionic atom pairs were observed using a fast magnetic field sweep to project the pairs onto stable molecular dimers~\cite{regal2004,zwierlein2004}. Using this method, the first phase diagram in the crossover region was obtained (see Fig.~\ref{fig:crossoverPhaseDiagram} below) as a function of magnetic field and temperature, albeit using the temperature of the ideal gas obtained by an adiabatic sweep to a non-interacting regime~\cite{regal2004}. Evidence of superfluidity in a Fermi gas was provided by measurements of collective mode frequencies and damping rates versus temperature~\cite{kinast2004} and magnetic field~\cite{bartenstein2004}, and measurement of the pairing gap by radio-frequency spectroscopy~\cite{GrimmGap2004}. The observation of a vortex lattice in 2005 provided direct verification of Fermi superfluid behavior~\cite{zwierlein2005a}.

Also in 2005, initial thermodynamic measurements were done by adding a controlled amount of energy to the cloud and measuring an  empirical temperature from the corresponding cloud spatial profile~\cite{Kinast2005}. However, the results were model-dependent, as calibration of the empirical temperature  relied on comparing the measured cloud profiles with theoretical predictions. Model-independent measurements were soon to follow, based on universal behavior in the unitary regime, where the local thermodynamic quantities, such as the pressure, are functions only of the density $n$ and temperature $T$~\cite{HoUniversal2004}.

Model-independent measurements of the total energy $E$ of a resonantly interacting Fermi gas are based on the Virial theorem, which holds for a unitary  gas as a consequence of universality and yields the energy directly from the cloud profile~\cite{ThomasUniversal2005}.  Using entropic cooling~\cite{carr2004d,carr2004g}, a model-independent measurement of the total entropy $S$ was accomplished by an adiabatic sweep of the bias magnetic field from resonance to the weakly interacting regime, where the entropy can be calculated from the cloud profile~\cite{luoL2007}. As $T=\partial E/\partial S$, these measurements enabled the first model-independent temperature calibration and estimates of the critical parameters in the strongly interacting regime~\cite{ThermoLuo2009}. A refined temperature calibration is used in the measurement of universal quantum viscosity, as  described in this Focus Issue~\cite{NJPfocusissue4_thomas}.

Measurements of global thermodynamic quantities from the cloud profiles in the strongly-interacting regime are now superseded by model-independent measurements of local quantities~\cite{ThermoUeda2010,navonN2010}. Using the Gibbs-Duhem relation
\be
dP=n\,d\mu
\label{eqn:gibbsDuhem}
\ee at constant temperature, the local pressure is obtainable from the integrated column density, where the local chemical potential $\mu$ is determined by the known trap potential~\cite{HoPressure2010,hazzard2011}. Combined with a temperature measurement, the local equation of state $P(\mu,T)$ or $P(n,T)$ is determined.  The most precise local measurements avoid temperature measurement, which introduces the most uncertainty, by determining the pressure, density and compressibility from the cloud profiles. The resulting equation of state  reveals clearly a lambda transition, and provides the best measurement of the critical parameters for a unitary Fermi gas~\cite{kuJHMark2011}, as described in detail in Sec.~\ref{ssec:exp-eos}. Measurements of equilibrium thermodynamic quantities are now used as stringent tests of predictions, as described in this Focus Issue by Hu~\cite{NJPfocusissue36_drummond}. These thermodynamic measurements are connected to universal hydrodynamics and transport measurements, as described in~\cite{NJPfocusissue4_thomas}.

%%%%%%%%%%%%%%%%%%%%%%%%%%%%%%%%%%%%%%%%%%%%%%%%%%%%%%%%%%%%%%%%%%%%%%%%%%%
\begin{figure}[t]
\begin{center}
\includegraphics[width=0.45\textwidth]{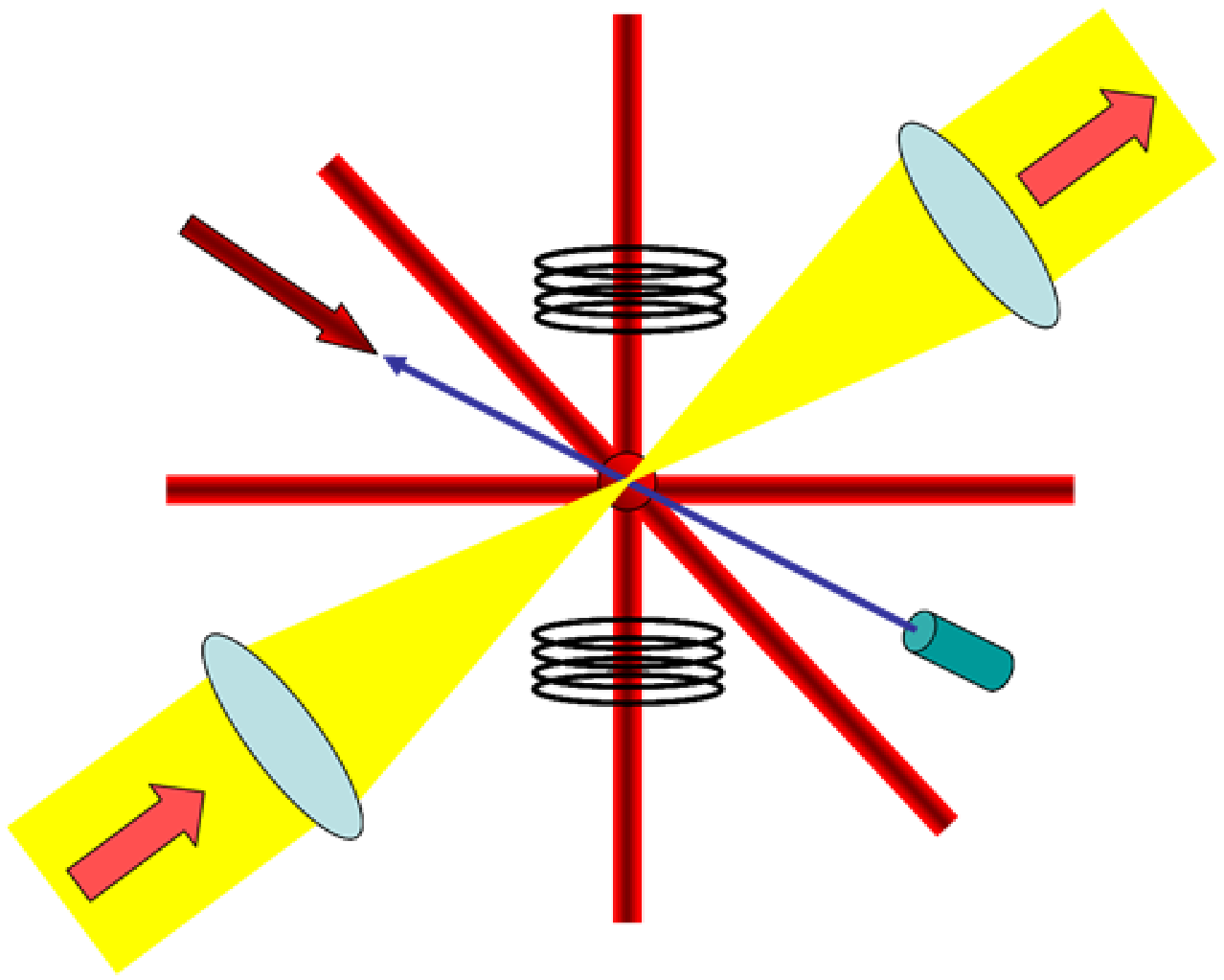}
\hspace*{0.03\textwidth}
\includegraphics[width=0.45\textwidth]{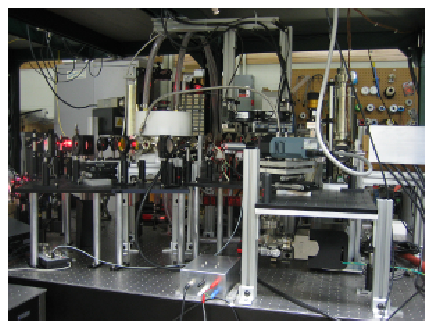}
\caption{\textit{Ultracold quantum gas experimental apparatus.}  Left: sketch of the experimental apparatus for ultracold fermions.  Right: Apparatus for the Duke  experiments (currently at North Carolina State University).  Compare to sketch of QGP experiment at the LHC in Fig.~\ref{fig_atlas}: the quantum gas experiment is about ten times smaller (2.5 meters vs. 26 meters), but the size of the trapped ultracold gas is 11 orders of magnitude larger (a few hundred micrometers vs. a few femtometers).  The ultracold quantum gas is at nanokelvin temperatures, or  pico-eV, compared to the deconfinement temperature of $\simeq 2\times 10^{12}$K in the QGP, or 200 MeV, created by colliding gold nuclei at energies of 100 GeV/nucleon.}
\label{fig:MOTOpticalTrap}
\end{center}
\end{figure}
%%%%%%%%%%%%%%%%%%%%%%%%%%%%%%%%%%%%%%%%%%%%%%%%%%%%%%%%%%%%%%%%%%%%%%%%%%%%

We proceed to describe the all-optical methods developed at Duke in 2002~\cite{Granade2001,OHara:2002}, as one specific example of experimental techniques, which are closely tied to the theme of this Focus Issue,  viscosity and transport measurements on Fermi gases in the universal regime~\cite{NJPfocusissue4_thomas}.   A degenerate, strongly interacting Fermi gas is readily made by all-optical methods~\cite{OHara:2002}. As sketched in the left panel of Fig.~\ref{fig:MOTOpticalTrap}, a magneto-optical trap (MOT) is used to pre-cool a 50-50 mixture of spin-up and spin-down $^6$Li atoms, which is loaded into a CO$_2$ laser optical trap and magnetically tuned to an $s$-wave Feshbach resonance. Atoms from the source, (lower right, green cylinder), form an atomic beam (blue arrow) that is slowed by radiation pressure forces from a resonant laser beam (top left, opposing red arrow). For $^6$Li, the deceleration is $2 \times 10^{6}$ m/s$^2$, slowing the atoms from oven thermal speeds of a km/s to tens of m/s sec over a distance of a fraction of a meter.  Six laser beams (three thick red lines) then propagate toward the center of the MOT (point of intersection of three thick red lines), creating inward damping forces that cool the atoms. Opposing magnetic fields, created by two coils (stacked black circles, top and bottom,) spatially tune the atomic resonance frequency, causing the six beams to produce  a harmonic restoring force at the MOT center. Typical MOT clouds are a few millimeters across and contain several hundred million atoms. A  trapping laser beam (shown in yellow) is focused (lenses indicated by two light blue ovals) at the MOT center and loaded. After turning off the MOT beams and the MOT magnetic field, an additional bias magnetic field tunes the atoms to a collisional (Feshbach) resonance.  Forced evaporation near resonance, by lowering the trap depth, rapidly cools the cloud to quantum degeneracy, i.e., $T/T_F \ll 1$, producing a cigar-shaped cloud that is typically a few microns in diameter and several hundred microns long, containing several hundred thousand atoms. In the right panel of Fig.~\ref{fig:MOTOpticalTrap} is shown the experimental apparatus from the Duke laboratory, currently located at North Carolina State University.  From right to left in the photo: the oven assembly where hot fermions are produced (aluminum housing); camera to produce density images (blue device in foreground); Zeeman slower to bring atoms into MOT (middle, behind camera); bias field magnets containing MOT in ultrahigh vacuum (white plastic housings); ZnSe lens for the CO$_2$ laser trapping beam and optical table (left).

In the simplest case, the optical trap consists of a single laser beam, focused into the center of the MOT and detuned well below the atomic resonance frequency to suppress spontaneous scattering, which would otherwise heat the atoms. For an optical trapping laser detuned below the atomic resonance, the induced atomic dipole moment is in-phase with the trapping laser field, so that the atoms are attracted to the high field region at the trap focus, i.e., the effective trapping potential is $U=-\alpha \langle E^2\rangle/2$, where the polarizability $\alpha >0$ and $\langle E^2\rangle$ is proportional to the trap laser intensity, time-averaged over a few optical cycles. The effective potential then has the same spatial profile as the intensity of the focused trap laser. For ultracold atoms, the energy per particle is typically quite small compared to the depth of the  optical trap, so that the atoms vibrate in a nearly harmonic confining potential. The vibration frequencies of the atoms in each direction $\omega_i$, $i=x,y,z$ of the trap are readily determined by parametric resonance: the trap laser intensity is modulated and the size of the cloud is measured as a function of modulation frequency. When the modulation frequency is twice the harmonic oscillation frequency, the energy of the atoms increases, causing the density profile to increase in size. This method permits precise characterization of the trap parameters.

All information about the cloud is obtained by absorption imaging: a spatially uniform short (several $\mu$s) low intensity laser pulse is transmitted through the atom cloud, which partially absorbs the light. The shadow cast by the absorption is imaged onto a CCD (charge coupled device) array to record the image, which is analyzed to extract the column density, integrated along the line of sight. This ``laser flash photography" method provides real space images with a resolution of a few microns, on a time scale short compared to the time scale over which the atoms move significantly compared to the spatial resolution.  Both non-destructive and destructive imaging techniques are possible.  In the latter case the entire cloud is destroyed by the laser pulse in order to make the most complete possible image.  One then runs the same experiment multiple times, with nearly identical initial conditions, to obtain an average picture of temporal evolution. While the CCD measures just the density or $g^{(1)}$ correlations, in fact it is possible to extract density-density or $g^{(2)}$ correlations from the noise on an image~\cite{altman2004,greiner2005}.

What do experimental measurements actually look like? In Fig.~\ref{fig:exptImages} are shown absorption images from the 2002 Duke experiment on elliptic flow~\cite{OHara:2002}. In the experiments, $N=7.5\times 10^4$ atoms in each of the two lowest hyperfine states were cooled to degeneracy in a CO$_2$ laser trap, with a reduced temperature $T/T_{FI}$ between $0.08$ and $0.2$. Here,  $T_{FI}=\hbar\bar{\omega}(6N)^{1/3}/k_B$ is the Fermi temperature for an ideal gas at the center of a harmonic trap,\footnote{$T_{FI}$ is defined by $T_{FI}=T_F(n_0(0))$, where $T_F(n)=\hbar^2\,k_F(n)^2/(2mk_B)$ is the local Fermi temperature of a non-interacting gas evaluated at the center of the trap. An equivalent definition is that $k_BT_{FI}=E_{FI}$, where $E_{FI}=\hbar\bar{\omega}(3N)^{1/3}$ is the Fermi energy of $N$ non-interacting fermions in a harmonic trap. The advantage of $T_{FI}$ is that it only depends on $N$ and $\bar\omega$.} where $\bar{\omega}=2\pi\times 2160 (65)$ Hz is the geometric mean of the trap oscillation frequencies, measured by parametric resonance as described above. For these parameters, the Fermi temperature is $T_{FI}=7.9\,\mu$K.  For an ideal gas in the trap, the  Fermi radii are $\sigma_x=3.6\,\mu$m in the narrow directions and $\sigma_z = 103\,\mu$m in the long direction.

%%%%%%%%%%%%%%%%%%%%%%%%%%%%%%%%%%%%%%%%%%%%%%%%%%%%%%%%%%%%%%%%%%%%%%
\subsection{Scale Invariance and Universality}
\label{ssec:uni}
%%%%%%%%%%%%%%%%%%%%%%%%%%%%%%%%%%%%%%%%%%%%%%%%%%%%%%%%%%%%%%%%%%%%%

Studies of trapped ultracold Fermi gases have provided important information
about the phase diagram, the equation of state, transport properties,
and quasiparticle properties of strongly correlated Fermi gases.
This is possible because under the conditions typically encountered
in the experiments local properties of the trapped gas directly correspond
to equilibrium properties of the homogeneous Fermi gas. Consider the
ground state of $N$ harmonically trapped fermions. Hohenberg and
Kohn showed that the solution of the $N$-body Schr\"odinger equation
is equivalent to the minimum of the energy functional
\cite{Hohenberg:1964}
\be
 E[n(\mathbf{x})] = \int d\mathbf{x} \left( {\cal E}(n(\mathbf{x}))
 + n(\mathbf{x})U(\mathbf{x})\right)\, ,
\ee
where $n(\mathbf{x})$ is the density, subject to the condition
$\int d\mathbf{x}\, n(\mathbf{x})=N$, ${\cal E}(n)$ is the energy density
functional, and $U(\mathbf{x})$ is the trap potential. If the density
is sufficiently smooth we can write ${\cal E}(n)$ as a function
of the local density and its gradients. On dimensional grounds
we have
\be
{\cal E}(n(\mathbf{x})) = \frac{c_0\hbar^2}{m}\, n(\mathbf{x})^{5/3} +
  \frac{c_1\hbar^2}{m}\frac{(\vec\nabla n(\mathbf{x}))^2}{n(\mathbf{x})}
 + O\left(\nabla^4n(\mathbf{x})\right)\, ,
\ee
where $c_0,c_1,\ldots$ are dimensionless constants. At unitarity
the coefficients $c_i$ are pure numbers, but for a finite
scattering length they become functions of $na^3$.  To first
approximation we can neglect the gradient terms. Then the
density is given by $n(\mathbf{x})=n_{\mathrm{eq}}(\mu-U(\mathbf{x}))$,
where $n_{\mathrm{eq}}(\mu)$ is the equilibrium density at the chemical potential
$\mu$ and zero temperature. This approximation is known as the
local density approximation. Gradient terms are suppressed by
$(\omega_\perp/\mu)^2\sim 1/(\lambda_z N)^{2/3}$, where $\lambda_z=
\omega_z/\omega_\perp$ is the trap deformation. Typical experiments
involve $\lambda_z\simeq 0.025$ to 0.1 and $N\geq 10^5$, so
corrections beyond the local density approximation are quite
small. These arguments generalize to systems at non-zero
temperature. In this case the density of the trapped
system is $n(\mathbf{x})=n_{\mathrm{eq}}(\mu-U(\mathbf{x}),T)$.

 The equilibrium density can be determined from the
equation of state, $P=P(\mu,T)$, through the thermodynamic
relation\footnote{Here and in the remainder of this review
we have dropped the subscript ``eq.''} $n=\partial P/ \partial \mu$.
In the following we also frequently refer
to the relation $P=P(n,T)$ as the equation of state. At
unitarity the interaction is scale invariant and the only
scales in the many-body system are the inter-particle distance
$n^{-1/3}$ and the de Broglie wave length, given in Eq.~(\ref{eqn:deBroglie}). Dimensional analysis implies that the
equation of state must be of the form
\be
\label{P_uni}
 P(n,T) = \frac{\hbar^2 n^{5/3}}{m}f(n\lambda^3_{\mathrm{dB}})\, ,
\ee
where $f(x)$ is a universal function. At zero temperature
the pressure is proportional to $n^{5/3}/m$. This implies, in
particular, that the pressure is given by a numerical constant
times the pressure of a free Fermi gas. The same is true for
the energy per particle and the chemical potential. It has
become standard to denote the ratio of the energy per particle
of the unitary gas and the free Fermi gas as the Bertsch
parameter $\xi$,
\be
 \frac{E}{N} = \xi\left(\frac{E}{N}\right)_0\, .
\ee
Bertsch posed the calculation of the parameter $\xi$
as a challenge problem to the many-body physics community
in 1999~\cite{Bertsch:1999}. At the time, the problem
was stated in the context of a toy model for dilute
neutron matter; see Sec.~\ref{sec_phases}.

Using thermodynamic identities we can show that Eq.~(\ref{P_uni})
implies $P=\frac{2}{3}\epsilon$, where $\epsilon$ is the
energy density. This relation is analogous to
the equation of state of a scale-invariant relativistic gas,
$P=\frac{1}{3}\epsilon$, as discussed in Sec.~\ref{sec_sQGP}. For a
trapped gas the relation between pressure and energy density
implies a Virial theorem: in a harmonic trap, the internal
energy of the system is equal to the potential energy of the
trapping potential~\cite{Thomas:2005,Werner:2008},
\be
\label{virial_th}
\int d\mathbf{x} \, \epsilon(\mathbf{x}) =
  \int d\mathbf{x}\, n(\mathbf{x})U(\mathbf{x})\, .
\ee
These universal relations have been extended in many
ways; see~\cite{Castin:2011} for a review. An important
class of relations, discovered by Tan, connects the
derivative of thermodynamic quantities with respect to
$1/a$ to short range correlation in the gas. Tan defined
the contact density ${\cal C}$ via
\cite{Tan:2008a,Tan:2008b,Tan:2008c}
\be
\frac{d\epsilon}{d(a^{-1})} = -
 \frac{\hbar^2}{4\pi m}\, {\cal C}\, ,
 \label{eqn:Tan1}
\ee
where the derivative is taken at constant entropy density. The
contact density appears in a number of thermodynamic relations.
The universal equation of state, for example, is given by
\be
P =\frac{2}{3}\, \epsilon + \frac{\hbar^2}{12\pi ma}
 \, {\cal C}\, .
\ee
More remarkable is the fact that ${\cal C}$ controls short
distance correlations in the dilute Fermi gas. The tail of
the momentum distribution is given by
\be
 n_\sigma(k)\to \frac{C}{k^4}  \hspace{0.5cm}
  \left( |a|^{-1} \ll k \ll r_0^{-1} \right)\,  ,
\ee
where $C=\int d^3x\, {\cal C}(x)$ is the integrated contact,
$n_\sigma(k)$ is the momentum distribution\footnote{The momentum
distribution is normalized as $\int d\mathbf{k}/(2\pi)^3 n_\sigma(k)=
N_\sigma$, where $N_\sigma$ is the total number of atoms in state
$\sigma$.} in the spin state $\sigma$, and $r_0$ is the range of
the interaction. There are similar expression for the asymptotic
behavior of other correlation observables like the static and
dynamic structure factors, and the dynamic shear viscosity; see
\cite{Braaten:2010} for a review. In this Focus Issue, Kuhnle \textit{et
al.} present a comprehensive set of measurements of the contact
as a function of interaction strength and reduced temperature
\cite{NJPfocusissue16_vale}. These results can be compared to
new theoretical predictions discussed by Hu \textit{et al.}~\cite{NJPfocusissue36_drummond}.

 Below the critical temperature for superfluidity the
superfluid flow velocity $v_s$ can be viewed as an additional
thermodynamic variable. The response of the pressure to
the superfluid velocity defines the superfluid mass
density
\be
\rho_s =mn_s = -\left. \frac{\partial^2 P}{\partial v_s^2}
  \right|_{v_n=0}\, ,
\ee
where the derivative is taken in the rest frame of the
normal fluid, meaning $v_n=0$. The superfluid mass
density can be measured using rotating clouds~\cite{Ho:2010}.
The second moment of the trap integrated value of the
superfluid mass density determines the quenching of the
moment of inertia. New measurements of the moment of
inertia can be found in~\cite{NJPfocusissue40_grimm}.

 For small values of $n|a|^3$ the equation of state $P(n,T)$
can be computed in perturbation theory. This program was
initiated by Lee, Yang, and Huang~\cite{Lee:1957,Huang:1957}.
At unitarity weak coupling methods can be used in the
limit of high temperature. This is based on the observation
that the binary cross section at unitarity is $\sigma=4\pi/k^2$.
At high temperature the mean momentum is large and the average
thermal cross section is small. The equation of state can be
written as an expansion in $n\lambda_{\mathrm{dB}}^3$, which is
the well-known Virial expansion. We have
\be
\label{Virial_exp}
 P = n k_B T \left\{ 1+b_2 (n\lambda_{\mathrm{dB}}^3)
   +O((n\lambda_{\mathrm{dB}}^3)^2) \right\}\, ,
\ee
with $b_2=-1/(2\sqrt{2})$ at unitarity~\cite{Beth:1936,Ho:2004}.
Analytic approaches in the non-perturbative regime $n\lambda_{\mathrm{dB}}^3
\sim 1$ are based on extrapolating to the unitary limit from different
regimes in the phase diagram. For this purpose the phase diagram has
been studied as a function of the strength of the interaction, the
number of species, and the number of spatial dimensions. The oldest
theory of this type is Nozi\'eres-Schmitt-Rink (NSR) theory
\cite{nozieres1985,perali2004,fukushimaN2007}, which is based on a
set of many-body diagrams that correctly describe both the BCS and BEC
limits. NSR theory works surprisingly well, despite the formal lack
of a small parameter at unitarity.  For example, the basic form
of the critical temperature sketched in Fig.~\ref{fig:crossoverPhaseDiagram}
is correctly reproduced. Modern theories of this type are typically
based on self-consistent T-matrix approximations; see
\cite{chen:2005,hausmann2007}. Another idea is to generalize
the unitary Fermi gas to $2N$ spin states~\cite{Nikolic:2006}. Mean
field theory is reliable in the limit $N\to \infty$, and the
interesting case $N=1$ can be studied by expanding in $1/N$. This
method is of interest in connection with holographic dualities,
because the gravitational dual is expected to be classical in the limit that
the number of degrees of freedom is large. Finally, it was proposed
to use the number of dimensions as a control parameter. The
unitary limit is perturbative in both $D=2$ and $D=4$ spatial
dimensions~\cite{Nussinov:2004}. The interesting case $D=3$ can be
studied as an expansion around $D=2+\epsilon$ or $D=4-\epsilon$
dimensions~\cite{Nishida:2006br}.

 These methods are promising, but currently the only techniques
that provide reliable and systematically improvable results in
the regime $n\lambda_{\mathrm{dB}}^3\simeq 1$ are quantum Monte Carlo
calculations. At zero temperature the standard technique is Green
function Monte Carlo (GFMC)~\cite{Carlson:2003zz,Astrakharchik:2004zz}.
This method relies on a variational initial wave function, which
is used as the initial condition for an imaginary time diffusion
process. The Monte Carlo method suffers from a fermion sign problem
which is addressed using the fixed node approximation.
At finite temperature a number of groups have performed imaginary
time path integral Monte Carlo (PIMC) calculations
\cite{Bulgac:2005pj,Lee:2005it,goulko2010,Drut:2011tf}. These calculations
do not rely on variational input, and they do not suffer from a sign
problem, but they are formulated on a space-time lattice and require
an extrapolation to zero lattice spacing. A new technique is bold
diagrammatic Monte Carlo (BDMC), which is based on sampling the sum
of all Feynman diagrams. The method suffers from a sign problem, but
convergence in the regime above the critical temperature for
superfluidity was found to be very good~\cite{VanHoucke:2011ux}.

%%%%%%%%%%%%%%%%%%%%%%%%%%%%%%%%%%%%%%%%%%%%%%%%%%%%%%%%%%%%%%%%%%%%%%
\subsection{Experimental Determination of the Equation of State}
\label{ssec:exp-eos}
%%%%%%%%%%%%%%%%%%%%%%%%%%%%%%%%%%%%%%%%%%%%%%%%%%%%%%%%%%%%%%%%%%%%%%

The equation of state describes a functional relation between key thermodynamic variables, such as the pressure $P(\mu,T)$ as a function of chemical potential $\mu$ and temperature $T$.  In Fermi gases, as stated in Sec.~\ref{ssec:experiments}, what we can actually measure is the density profile of a trapped cloud.  There are various techniques for translating density measurements into thermodynamic quantities,  all relying on the local density approximation.

The first thermodynamic measurements took place at Duke in 2005, where global thermodynamic variables were measured. In the initial experiments~\cite{Kinast2005}, a controlled amount of energy was added to the trapped cloud by abruptly releasing it from the optical trap, allowing it to expand hydrodynamically by a known factor, and then recapturing the cloud after a selected expansion time, thereby increasing the potential energy. After allowing  the gas to equilibrate, the cloud profile was measured to determine an empirical temperature, which was later calibrated by comparing to theoretical cloud profiles,  predicted as a function of reduced temperature. The resulting energy versus temperature curve was compared to that measured for an ideal Fermi gas in the same trap, and showed a departure from ideal gas behavior at a certain temperature, which yielded an estimate of the superfluid to normal fluid transition temperature.  However, the results suffered from being model-dependent, as calibration of the empirical temperature  relied on comparison of the measured cloud profiles with theoretical predictions. To avoid this  model dependence, in 2006 the JILA group measured the potential energy of the strongly interacting cloud of $^{40}$K as a function of the ideal Fermi gas temperature that was obtained after an adiabatic sweep of the bias field to the noninteracting regime above resonance~\cite{Stewart2006}.

\begin{figure}[t]
\begin{center}
\includegraphics[width=0.6\textwidth]{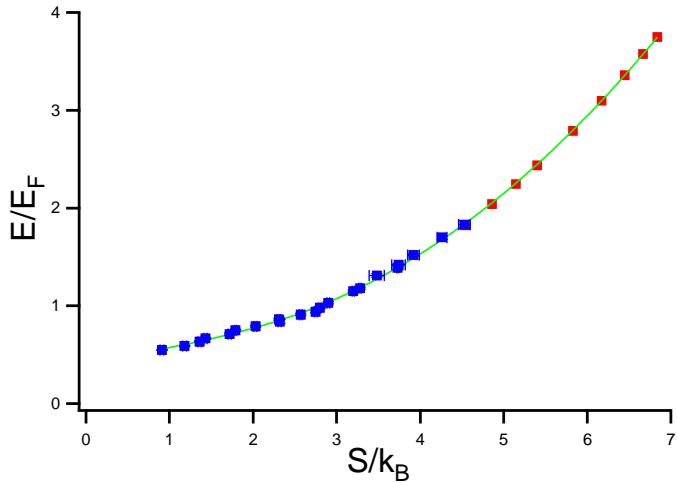}
\end{center}
\caption{Total energy per particle of a strongly interacting Fermi
gas in the universal regime versus the entropy per particle. The blue dots show
the entropy obtained by adiabatically sweeping the magnetic field from $840G$ to $1200$ G, where the gas is weakly interacting.  The red dots are the theoretical calculations using a second Virial coefficient approximation. The green curve is a fit using two power laws, which determines the temperature $T=\partial E/\partial S$. From~\cite{NJPfocusissue4_thomas}.  \label{fig:energyvsentropy}}
\end{figure}

Model-independent determination of thermodynamic quantities was done  by the Duke group in 2007~\cite{luoL2007}, where the total energy $E$ and total entropy $S$ of a trapped cloud were measured from cloud profiles,  by exploiting the universal behavior of a unitary Fermi gas at a Feshbach resonance. From Eq.~(\ref{virial_th}) we know that for a harmonic trapping potential the total energy is twice the average potential energy,  $E=2\langle U\rangle=3\,m\omega_z^2\langle z^2\rangle$. Hence, by measuring the harmonic oscillator frequency and mean square cloud size, the total energy is readily determined from cloud images. This method was demonstrated experimentally in~\cite{ThomasUniversal2005}. With the energy in the strongly interacting regime measured from cloud images, the entropy is determined by adiabatically sweeping the magnetic field to a weakly interacting regime. There, the entropy and mean square cloud size are readily calculated as a function of the temperature, yielding the entropy as a function of the mean square cloud size of the weakly interacting gas. Adiabatic behavior is verified by a round-trip sweep. Fig.~\ref{fig:energyvsentropy} shows the energy per particle as a function of the entropy per particle, in the universal regime. The temperature is determined by fitting a smooth curve~\cite{ThermoLuo2009,NJPfocusissue4_thomas}. Hu, Liu, and Drummond combined the demonstrated universal behavior of the global thermodynamic quantities by reanalyzing the measurements  in $^{40}$K and showing that the $^{40}$K and $^6$Li data fit on a single thermodynamic curve~\cite{HuLiuDrummond2007}.

%%%%%%%%%%%%%%%%%%%%%%%%%%%%%%%%%%%%%%%%%%%%%%%%%%%%%%%%%%%%%%%%%%%%%%%%%%%%%
%%%%%%%%%%%%%%%%%%%%%%%%%%%%%%%%%%%%%%%%%%%%%%%%%%%%%%%%%%%%%%%%%%%%%%%%%%%%%

As already mentioned briefly in Sec.~\ref{ssec:experiments}, model-independent measurement of global thermodynamic variables was superseded in 2010 by model-independent measurement of local thermodynamic quantities, which can be directly compared to predictions, within the local density approximation.  Equation~\ref{eqn:gibbsDuhem}, determines the local pressure $P$  from the local density $n$ and local chemical potential, $\mu=\mu_g-U$, where $\mu_g$ is the global chemical potential and $U$ is the known trapping potential. Absorption imaging directly yields the column density $\tilde{n}(x,z)=\int\, dy\,n(x,y,z)$, for an imaging beam propagating along $y$. In a cylindrically symmetric harmonic trap $U=m\omega_\perp^2(x^2+y^2)/2+m\omega_z^2\,z^2/2$, where $\omega_{\perp}$ and $\omega_z$ are the radial and axial trapping frequencies, respectively, we can write $-d\mu=dU=m\omega_\perp^2\rho\,d\rho=m\omega_\perp^2\,dx\,dy/(2\pi)$ from which we see that the pressure is determined from the doubly integrated density, i.e., the integrated column density $\bar{n}(z)=\int\,dx\, dy\,n(x,y,z)$. For a 50-50 mixture of spin-up and spin-down fermions with $n$ the total density, the pressure is determined as a function of $z$, $P(z,T)=m\omega_\perp^2\,\bar{n}(z)/(2\pi)$. For $x=y=0$, the corresponding chemical potential is $\mu(z)=\mu_g-m\omega_z^2\,z^2/2$, so that $P(\mu,T)$ is determined if the temperature and global chemical potential can be determined.

\begin{figure}[t]
\begin{center}
\includegraphics[width=0.6\textwidth]{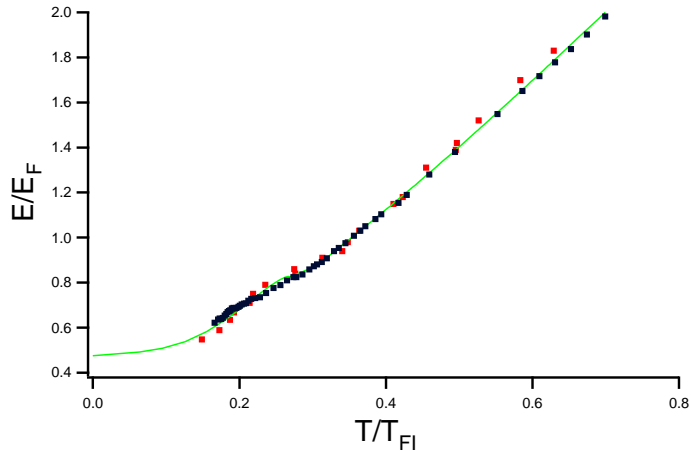}
\end{center}
\caption{Measured energy versus the temperature obtained from the
calibration of Ref.~\cite{NJPfocusissue4_thomas} (red dots); For comparison, we
show the data obtained by the ENS group~\cite{navonN2010} (black dots)
and the theory of Hu \textit{et al.}~\cite{NJPfocusissue36_drummond} (green curve).
From~\cite{NJPfocusissue4_thomas}. \label{fig:energyvsT}}
\end{figure}

To determine the temperature, the Tokyo group~\cite{ThermoUeda2010}   used the temperature calibration  by the Duke group to obtain $T$ from the total energy and hence from the mean square cloud size~\cite{ThermoLuo2009}. An improved version of this calibration is described in Ref.~\cite{NJPfocusissue4_thomas} of this Focus Issue.  The ENS group~\cite{navonN2010}  directly determined the temperature by using a weakly interacting $^7$Li impurity to measure the temperature of a strongly interacting $^6$Li gas. The global chemical potential was determined from the wings of the density profile using a fit based on a Virial expansion, yielding $P(\mu,T)$, from which all other local thermodynamic quantities, such as the energy density and entropy density, were determined. These results enable a direct comparison with predictions. Further, the total entropy $S$ and total energy $E$ of the trapped cloud are readily determined by integration, and can be compared with the corresponding global quantities measured by the Duke group. The total energy versus temperature is shown in Fig~\ref{fig:energyvsT}, where the improved temperature calibration of Ref.~\cite{NJPfocusissue4_thomas} from this Focus Issue is displayed for the Duke data.  As quite different methods were employed to make the measurements, their close agreement indicates the correctness of the data.

%%%%%%%%%%%%%%%%%%%%%%%%%%%%%%%%%%%%%%%%%%%%%%%%%%%%%%%%%%%%%%%%%%%%%%%%%%%%%
%%%%%%%%%%%%%%%%%%%%%%%%%%%%%%%%%%%%%%%%%%%%%%%%%%%%%%%%%%%%%%%%%%%%%%%%%%%%%

\begin{figure}[t]
\begin{center}
\vspace*{0.0in}\includegraphics[width=0.75\textwidth]{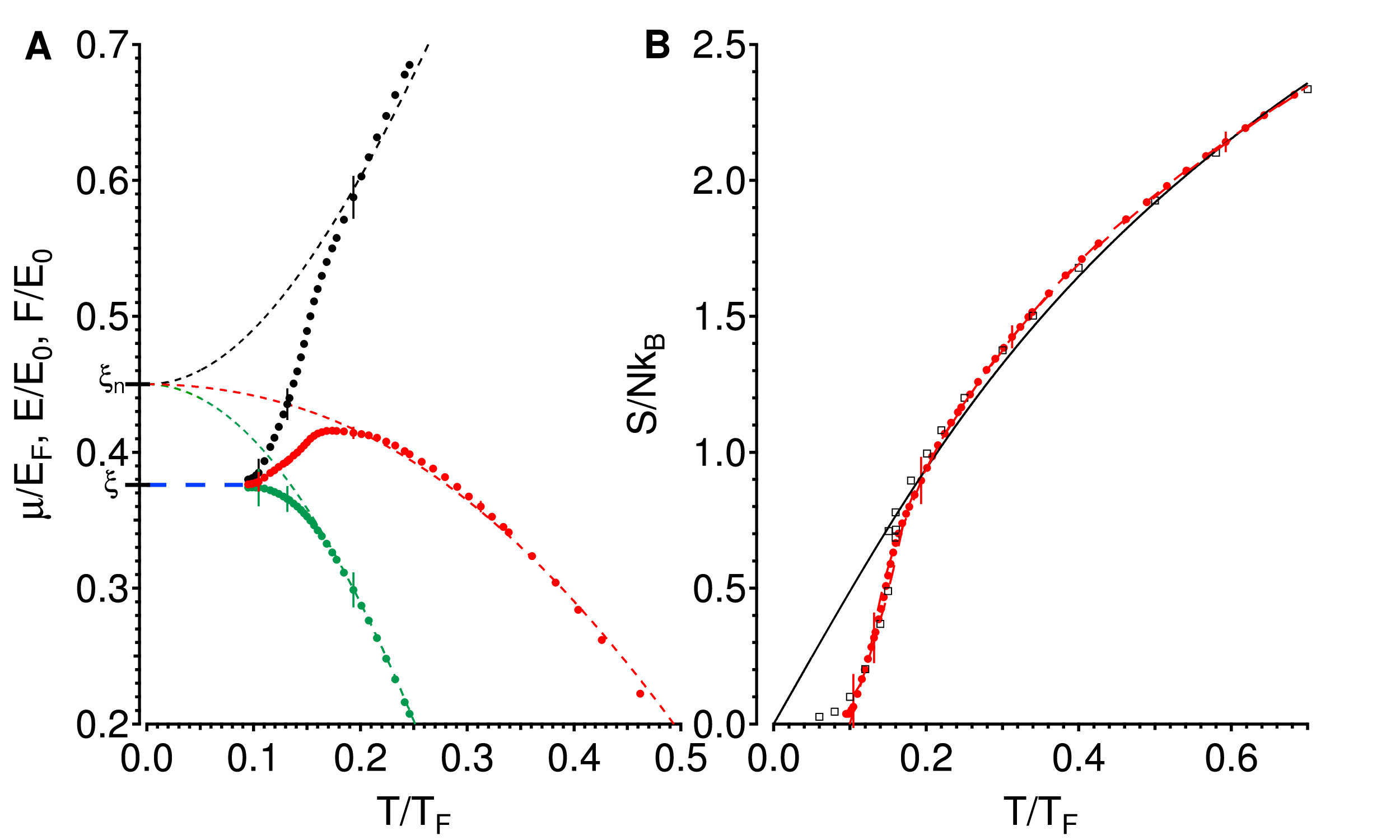}\\
\caption{\textit{Experimentally measured thermodynamics for a unitary
Fermi gas at MIT.} Left panel: energy, free energy, chemical potential,
as a function of temperature. Right panel: entropy, as a function of
temperature. See the text for an explanation of the curves.
Used with permission from Ref.~\cite{kuJHMark2011}.}
%Bottom panels: equation of state expressed in terms of density and pressure.
\label{fig:equationOfState}
\end{center}
\end{figure}

The latest studies of local thermodynamics by the MIT group~\cite{kuJHMark2011} use refined methods, where measurements of the isothermal compressibility $\kappa$ directly from the density profiles replaces temperature measurements, yielding an equation of state $n(P,\kappa)$.  This  eliminates the determination of the temperature and local chemical potential from fits  at the edges of the cloud, which produced the most uncertainty in previous work. For this method, the three dimensional density $n(x,y,z)$ is determined by tomographic imaging, using an inverse Abel transform \cite{InverseAbel} to determine $n$ from the measured column density. The trap potential is carefully characterized by determining the surfaces of constant density, hence constant chemical potential, in a very shallow trap where the axial $z$ trap potential is almost perfectly harmonic, and therefore known. Equation~\ref{eqn:gibbsDuhem} yields the pressure
\begin{equation}
P(U,T)=\int_{-\infty}^\mu d\mu\,n(\mu',T)=\int_U^\infty dU\,n(U,T),
\label{eq:pressure}
\end{equation}
 where the unknown global potential $\mu_g$ is not needed, since the integral is over the known trap potential~\cite{kuJHMark2011}. The compressibility  is the change in density with respect to a change in the local trapping potential $U$, and is determined from the density profile by
 \begin{equation}
 \kappa=\left.\frac{1}{n}\frac{\partial n}{\partial P}\right|_T =-\left.\frac{1}{n^2}\frac{\partial n}{\partial U}\right|_T,
 \label{eq:compressibility}
 \end{equation}
 since $dP=n\,d\mu=-n\,dU$ at constant $T$.  Thus, the actual observed equation of state   is the functional relation $n(\kappa,P)$, measured directly from the density distribution, the clear and direct experimental observable for ultracold quantum gases as discussed in Sec.~\ref{ssec:experiments}. From $n(P,\kappa)$, all other local thermodynamic quantities are determined~\cite{kuJHMark2011}, as shown in Fig.~\ref{fig:equationOfState}.

%%%%%%%%%%%%%%%%%%%%%%%%%%%%%%%%%%%%%%%%%%%%%%%%%%%%%%%%%%%%%%%%%%%%%%%%%%%%%
%%%%%%%%%%%%%%%%%%%%%%%%%%%%%%%%%%%%%%%%%%%%%%%%%%%%%%%%%%%%%%%%%%%%%%%%%%%%%

 These experiments provide the best current value for the Bertsch
parameter of $\xi = 0.376(5)$, which is consistent with the value
obtained in measurements of global quantities, $\xi=1+\beta=0.39(2)$
\cite{ThermoLuo2009}. These measurements can also be compared to
theory predictions. The two most recent quantum Monte Carlo
calculations give $\xi \leq 0.383(1)$~\cite{Forbes:2010gt} and
$\xi=0.3968^{+0.0076}_{-0.0077}$~\cite{Endres:2012cw}. See
\cite{Endres:2012cw} for an extensive compilation of analytic
results and earlier Monte Carlo calculations.

Figure~\ref{fig:equationOfState} shows several representations
of the equation of state, and provides a glimpse of the level
of precision that can be achieved in present experiments.  In
the left panel are shown the chemical potential $\mu$, energy $E$,
and free energy $F$. The right panel shows the entropy per particle
$S/(Nk_B)$ versus $T/T_F$. The chemical potential (red solid circles)
is normalized by the Fermi energy; energy (black solid circles) and
free energy (green solid circles) are normalized by $E_0 = \frac{3}{5}
N E_F$, which is the energy per particle in a uniform Fermi gas. At
high temperatures all quantities approximately track those for a
non-interacting Fermi gas, shifted by $\xi_n -1$ with $\xi_n\simeq
0.45$ (dashed curves). The peak in the chemical potential roughly
coincides with the onset of superfluidity. In the very low temperature
regime, $\mu/E_F$, $E/E_0$ and $F/F_0$ all approach $\xi$ (blue dashed
line). At high temperatures, the entropy closely tracks that
of a non-interacting Fermi gas (black solid curve). The open squares
are from the self-consistent T-matrix calculation~\cite{hausmann2007}.
A few representative error bars are shown, representing mean $\pm$
standard deviation.

%%%%%%%%%%%%%%%%%%%%%%%%%%%%%%%%%%%%%%%%%%%%%%%%%%%%%%%%%%%%%%%%%%%%%%%%%%%
\subsection{Experimental Studies of the Phase Transition}
\label{ssec:phase}
%%%%%%%%%%%%%%%%%%%%%%%%%%%%%%%%%%%%%%%%%%%%%%%%%%%%%%%%%%%%%%%%%%%%%%%%%%%

%%%%%%%%%%%%%%%%%%%%%%%%%%%%%%%%%%%%%%%%%%%%%%%%%%%%%%%%%%%%%%%%%%%%%%%%%%%%%
\begin{figure}[t]
\begin{center}
\vspace*{0.0in}\includegraphics[width=0.55\textwidth]{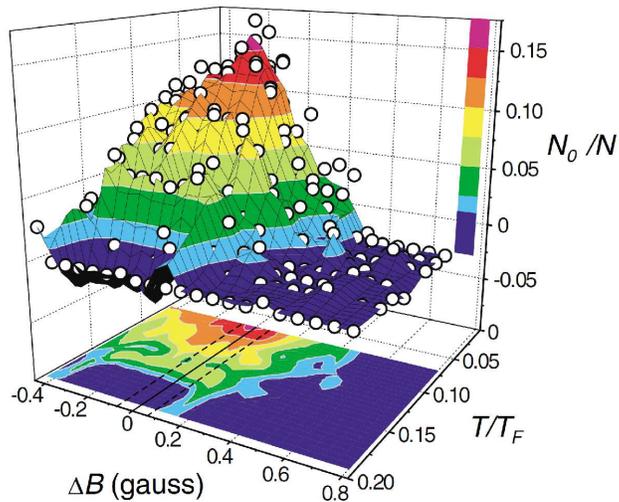}
\caption{\textit{Experimental Fermi gas phase diagram.} Experimental measurements of the Bardeen-Cooper-Schrieffer (BCS) to Bose-Einstein condensate (BEC) crossover for ultracold Fermi gases (compare to sketch in Fig.~\ref{fig:crossoverPhaseDiagram}).  Since a magnetic Feshbach resonance is used to tune the interaction strength, one axis is magnetic field.  In this measurement, the BCS side is on the $B>0$ side, so Fig.~\ref{fig:crossoverPhaseDiagram} should be reversed for comparison; it is not possible to go deep into the BEC side because the lifetimes become too short in this experiment.  $\Delta B < 0.6$ contains the strongly interacting region $k_F |a| > 1$, and the dashed lines indicate uncertainty in the precise position of the Feshbach resonance.  Used with permission from Ref.~\cite{regal2004}.}
\label{fig:realCrossoverPhaseDiagram}
\end{center}
\end{figure}
%%%%%%%%%%%%%%%%%%%%%%%%%%%%%%%%%%%%%%%%%%%%%%%%%%%%%%%%%%%%%%%%%%%%%%%%%%%%%

The first measurements of the phase transition at unitarity were made
at JILA, shown in Fig.~\ref{fig:realCrossoverPhaseDiagram}, by a pair
projection technique.  After the pairs were created in the BCS and
unitary regimes, a rapid magnetic field sweep was used to pairwise
project the fermions onto molecules, to protect them during subsequent
expansion measurements, where the molecular momentum distribution
was measured to determine the condensed pair fraction. The fraction
of near-zero momentum molecules is interpreted as the percentage of
condensed Fermi unitary or BCS pairs.  In this early experiment
$^{40}$K was used, and the measured temperature was that of an
ideal Fermi gas, rather than the temperature of the interacting gas.
This ideal Fermi gas temperature was obtained by ballistic expansion
after an adiabatic sweep to the weakly interacting regime above
resonance, yielding the condensed pair fraction versus ideal gas
temperature and magnetic field. Despite the uncertainties regarding
the temperature and the pair conversion efficiency we observe that
the shape of the transition line is qualitatively consistent with theoretical
expectations, as summarized in Fig.~\ref{fig:crossoverPhaseDiagram}.

Initial estimates of the critical parameters of the trapped gas
were done by the Duke group, first based on model-dependent measurements
of the energy versus temperature~\cite{Kinast2005} and later based on
model-independent measurements of the total energy $E$ and entropy $S$,
obtained as described above. Assuming that a phase-transition would
be manifested in a change in the scaling of $E$ with $S$, two power-laws
were used to fit the $E(S)$ data,  one to fit the high temperature data
and one to fit the low temperature data, which joined at a point $S_c$.
The continuity of the energy and temperature were used as a constraint,
and the critical entropy $S_c$ was estimated from the joining point that
minimized the $\chi^2$ for the fit~\cite{ThermoLuo2009}. While the
results obtained by this method were  consistent with predictions, they
suffered from a dependence on the form of the fit function, making the
uncertainty difficult to quantify. Further, the global observables suffer
from the averaging that masks an abrupt local phase transition, initially
near the trap center.

 Using the refined local measurements described in Sec.~\ref{ssec:exp-eos}, the MIT group
has  traced out the phase transition in the unitary regime,
without any need for rapid magnetic sweeps, fitting parameters,
thermometry, or indeed any kind of theory besides elementary
thermodynamic considerations.  The main goal of these experiments was
to obtain a cusp-like signal of the phase transition to superfluidity
in a unitary Fermi gas, by focusing in on second order derivatives of
the pressure, where a cusp appears.  Although superfluidity was
established by creation of vortex lattices~\cite{zwierlein2005a}, the
actual phase transition itself had only been indirectly observed.

%%%%%%%%%%%%%%%%%%%%%%%%%%%%%%%%%%%%%%%%%%%%%%%%%%%%%%%%%%%%%%%%%%%%%%
\begin{figure}[t]
\begin{center}
\vspace*{0.0in}\includegraphics[width=4.0in]{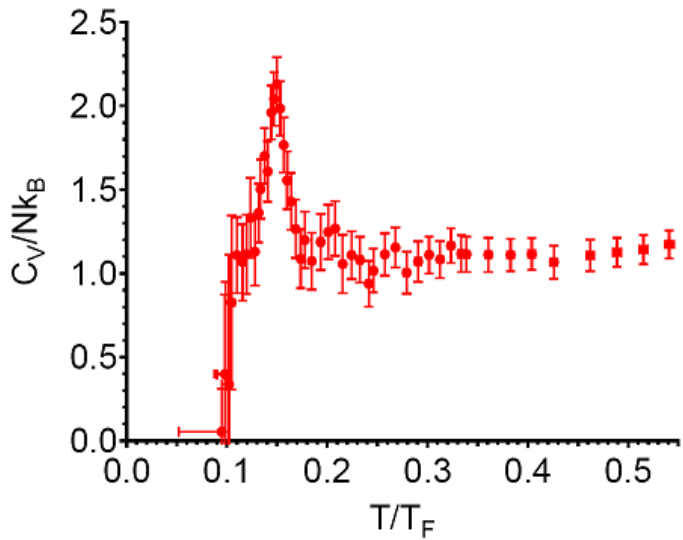}
\caption{\textit{Superfluid phase transition of a unitary Fermi gas.}
Specific heat per particle, $C_V/(Nk_B)$, as a function of quantum
degeneracy, $T/T_F$.  The phase transition is clearly evident at
$T_c/T_F = 0.167 (13)$.  Adapted with permission from
Ref.~\cite{kuJHMark2011}.}
\label{fig:SpecificHeat}
\end{center}
\end{figure}
%%%%%%%%%%%%%%%%%%%%%%%%%%%%%%%%%%%%%%%%%%%%%%%%%%%%%%%%%%%%%%%%%%%%%%

 In Fig.~\ref{fig:SpecificHeat} is shown the specific heat per
particle, clearly displaying the superfluid phase transition.
Experimentally, the specific heat is derived from the compressibility
$\kappa$ and the pressure $P$~\cite{kuJHMark2011},
\begin{equation}
\frac{C_V}{k_B N} = \frac{5}{2}\frac{T_F}{T}
   \left(\tilde{p}-\frac{1}{\tilde{\kappa}}\right)\, ,
\label{eqn:specificHeat}
\end{equation}
where $\tilde{\kappa} = \kappa/\kappa_0$ and $\tilde{p}\equiv P/P_0$
are normalized to the non-interacting Fermi gas compressibility
$\kappa_0=\frac{3}{2}\frac{1}{nE_F}$ and pressure $P_0=\frac{2}{5}
n E_F$, respectively. Using $n=(\partial P/\partial\mu)_T$,  the
compressibility can be written $\kappa=(1/n^2)(\partial n/\partial\mu)_T
=(1/n^2)(\partial^2 P/\partial^2\mu)_T$. As $\kappa$ is a second derivative
of the pressure,  the specific heat shows a clear cusp-like signature,
Fig.~\ref{fig:SpecificHeat}. Qualitatively, the behavior of $C_V$
can be understood as follows: As one approaches the phase transition
from above, $T/T_c > 1$, the compressibility increases due to the
attraction between fermions; below the phase transition, $T/T_c < 1$,
the compressibility  decreases because fermions are bound into
pairs, and it becomes more difficult to squeeze the gas, i.e., to
change the single particle density.

%%%%%%%%%%%%%%%%%%%%%%%%%%%%%%%%%%%%%%%%%%%%%%%%%%%%%%%%%%%%%%%%%%%%%%
\subsection{Universal Hydrodynamics and Transport}
\label{ssec:transport}
%%%%%%%%%%%%%%%%%%%%%%%%%%%%%%%%%%%%%%%%%%%%%%%%%%%%%%%%%%%%%%%%%%%%%%

Transport properties of the unitary Fermi gas are of interest for several
reasons. The first reason is related to the main theme of this review:
holographic dualities suggest a new kind of universality in the transport
properties of strongly interacting quantum fluids. We expect, in particular,
that the shear viscosity to entropy density ratio is close to the value
$\eta/s\sim \hbar/(4\pi k_B)$ originally discovered in the QGP,
and first obtained theoretically using the $AdS/CFT$ correspondence
\cite{Kovtun:2004de}, where \ads\ is a special maximally-symmetric spacetime described in detail in Sec.~\ref{SEC:EssentialHolography}, and CFT stands for {\em conformal field theory}.
The second reason is that transport properties
are very sensitive to the strength of the interaction, and the types
of quasiparticles present in the system. The Bertsch parameter, which
characterizes the effect of interactions on the energy per particle,
varies by about a factor of two between the weak coupling (BCS) and
strong coupling (unitarity) limits. The shear viscosity, on the other
hand, changes by many orders of magnitude. Finally, quantum limited
transport has also been observed in systems that are of great
practical significance, in particular in the strange metal phase
of high $T_c$ compounds; see the contribution to this Focus Issue by Guo \textit{et al.}~\cite{NJPfocusissue2_levin}.

 Transport properties have been studied experimentally by exciting
hydrodynamic modes, such as  collective oscillations
\cite{kinast2004,Kinast:2004b,Bartenstein:2004,Kinast:2005,Altmeyer:2006},
collective flow~\cite{OHara:2002,Cao:2010wa}, sound~\cite{Joseph:2006},
and rotational modes~\cite{Clancy:2007}. In a system that can
be described in terms of quasiparticles the hydrodynamic
description is valid if the Knudsen number ${\it Kn}=l_{\mathrm{mfp}}/
L$, the ratio of the mean free path $l_{\mathrm{mfp}}$ to the system
size $L$, is small.\footnote{Criteria for the validity of hydrodynamics
can also be formulated if there is no underlying quasiparticle
description, a situation that is of great interest in connection with
holographic dualities. In this case, hydrodynamics is based on
a gradient expansion of the conserved currents. The ratio of
the $O(v)$ to $O(\partial v)$ terms in the stress tensor is
known as the Reynolds number, ${\it Re}=vLmn/\eta$. Validity of
the gradient expansion requires that the Reynolds number is large.}
In the unitary gas the mean free path is $l_{\mathrm{mfp}}=1/(n\sigma)$,
where $n$ is the density and $\sigma=4\pi/k^2$ is the universal cross
section. In the high temperature limit the thermal average cross section
is $\sigma=4\lambda_{\mathrm{dB}}^2$.
%% I averaged 1/k^2 over exp[-k^2/(m*T)]. this is a bit crude
%% but not that far off. -- Thomas
The Knudsen number of a unitary gas confined in a cigar-shaped
harmonic trap is
\be
{\it Kn} = \frac{3\pi^{1/2}}{4(3\lambda_z N)^{1/3}}
 \left(\frac{T}{T_{FI}}\right)^2\, ,
\ee
where we have taken $L$ as the radius in the narrow, or $z$ direction. Here,
$N$ is the number of particles, $\lambda_z$ was defined previously as
the aspect ratio of the trap, and $T_{FI}$ is the global Fermi temperature
for a harmonically trapped ideal gas; see Sec.~\ref{ssec:experiments}.
Using $N=2\cdot 10^5$ and $\lambda_z=0.045$ as in~\cite{Kinast:2005}
we conclude that hydrodynamics is expected to be valid for $T\lsim 5
T_{FI}$. For the Fermi gas viscosity measurements described in
this Focus Issue~\cite{Cao:2010wa,NJPfocusissue4_thomas} the maximum
temperature is $T\simeq 1.5\,T_{FI}$ and ${\it Kn}\leq 0.09$.

Nearly ideal hydrodynamic behavior was first observed in
the expansion of a unitary Fermi gas after release from a deformed
trap~\cite{OHara:2002}; see Fig.~\ref{fig:exptImages}. For a
ballistic gas the expansion reflects the isotropic local momentum
distribution in the trap. As a result the gas expands in all
directions and the cloud slowly becomes spherical. For a hydrodynamic
system the expansion is driven by gradients in the pressure.
In the case of a deformed cloud the gradients are largest
in the short direction of the trap, and the expansion
takes place mostly in the transverse direction. As a result
the cloud eventually becomes elongated along what was originally
the short direction. This phenomenon is analogous to the elliptic
flow observed in heavy ion collisions, as described in Sec.~\ref{sec_flow}.
What is also remarkable is the fact that even though the gas
becomes more dilute as it expands this effect is compensated
by the growth in the mean cross section. As a consequence,
the gas remains hydrodynamic throughout the expansion.
Ballistic behavior sets in eventually only because of
imperfections, such as the fact that the scattering length
is not truly infinite.

 The role of dissipative effects, in particular shear viscosity,
was first studied in collective modes. The radial breathing mode
can be excited by removing the confining potential, letting the
gas expand for a short period of time, and then restoring the
potential. Hydrodynamic behavior can be established by measuring
the frequency of the breathing mode. For an ideal fluid $\omega
=\sqrt{10/3}\,\omega_0$ whereas in a ballistic system $\omega=2
\omega_0$~\cite{Stringari:2004,Bulgac:2004}. The transition from
ballistic behavior in the BCS limit to hydrodynamics in the unitary
limit was observed experimentally in~\cite{Kinast:2004b,Bartenstein:2004}.
In the hydrodynamic regime damping is expected to be dominated
by dissipative terms in the equations of fluid dynamics. The
energy dissipation is given by
\bea
\dot{E} &=& - \int d\mathbf{x}\, \Bigg\{\frac{\eta(\mathbf{x})}{2}\,
  \left(\nabla_iv_j+\nabla_jv_i-\frac{2}{3}\delta_{ij}
       (\nabla\cdot \mathbf{v}) \right)^2  \\
  & & \hspace{2.5cm}\mbox{}
   + \zeta(\mathbf{x})\, \big( \nabla\cdot \mathbf{v}\big)^2
   + \frac{\kappa(\mathbf{x})}{T} (\nabla T)^2\Bigg\}
  \nonumber ,
\eea
where $v_i$ is the fluid velocity, $\eta$ is the shear viscosity,
$\zeta$ is the bulk viscosity, $\kappa$ is the thermal
conductivity, and all derivatives, divergences, and gradients are spatial. At unitarity the system is scale invariant and the
bulk viscosity is expected to vanish~\cite{Son:2005tj,Castin:2011}.
This prediction was experimentally checked by Cao \textit{et al.}~\cite{NJPfocusissue4_thomas,Dusling:2011dq}. Thermal conductivity is not
important because the system remains isothermal in ideal
hydrodynamics. Temperature gradients only appear due to shear
viscosity, and their contribution to dissipation is higher order
in the gradient expansion. This means that damping is dominated by shear
viscosity.

 Collective modes in ideal fluid dynamics are described by scaling
solutions of Euler's equation. This means that shape of the density
profile does not change during the evolution, and that the velocity
is linear in the coordinates. The solution is analogous to Hubble
flows in cosmology, and the Bjorken expansion of a QGP,
as discussed in Sec.~\ref{sec_flow}. In the case of a scaling solution
the shear stresses $\partial_i v_j$ are spatially constant, and the
energy dissipated only depends on the spatial integral of $\eta$.
On dimensional grounds we can write $\eta= \hbar n\alpha_n$. For
a scale invariant system $\alpha_n$ is only a function of the
dimensionless variable $n^{2/3}\hbar^2/(mk_BT)$, i.e., a function
of the reduced temperature $T/T_F$. The reduced temperature
varies across the trap, but for a given fluid element it remains
approximately constant during the hydrodynamic evolution of the
system. This implies that the damping constant of a collective
mode is related to the spatial average of the shear viscosity in
the initial equilibrium state,
\be
\label{alpha_n}
\langle \alpha_n \rangle =\frac{1}{N} \int d\mathbf{x} \, \eta(x)\, .
\ee
The extracted values of $\langle \alpha_n\rangle$ can be converted
to the trap averaged shear viscosity to entropy density ratio by using
the measured entropy per particle. This type of analysis was originally
carried out in~\cite{Schafer:2007pr,Turlapov:2007}, where it was observed that
$\eta/s\lsim 0.5\hbar/k_B$ in the vicinity of $T_c$. More recently,
Cao \textit{et al.} showed that in the high temperature regime $\alpha_n$
exhibits the scaling behavior expected from the solution of the
linearized Boltzmann equation~\cite{Cao:2010wa}. The shear viscosity
due to elastic 2-body scattering has the form $\eta\simeq npl_{\mathrm{mfp}}$,
where $p\sim \lambda^{-1}_{\mathrm{dB}}$ is the mean quasiparticle momentum.
As we saw above, $l_{\mathrm{mfp}}\sim 1/(n\sigma)\sim 1/(n\lambda_{\mathrm{dB}}^2)$. This
implies that at high temperature $\eta\sim \lambda_{\mathrm{dB}}^{-3}$. The
coefficient of proportionality was determined by
Bruun \textit{et al.}~\cite{Massignan:2004,Bruun:2005}. They find
\be
\label{eta_unit}
 \eta =  \frac{15}{32\sqrt{\pi}}\frac{(mk_BT)^{3/2}}{\hbar^2}\, .
\ee

There is an important problem related to Eq.~(\ref{eta_unit}) that affects
the extraction of the shear viscosity from experiments with
scaling flows. Equation~(\ref{eta_unit}), which is reliable in
high temperature or low density part of the cloud, is independent
of the density.\footnote{This is a general property of the viscosity
of dilute gases, and was first noticed by Maxwell. The result was
experimentally confirmed by Maxwell himself, who measured the
damping of oscillating discs in a partially evacuated container
\cite{Maxwell:1866}}.
As a consequence the integral in Eq.~(\ref{alpha_n}) diverges
in the low density region. A solution to this problem was
proposed in~\cite{Bruun:2007,Schaefer:2009px}: In the low
density regime the viscous relaxation time $\tau_R\simeq
\eta/(nk_BT)$, which is the time is takes for the dissipative
stresses to relax to the Navier Stokes form $\eta(\nabla_i
v_j +\nabla_j v_i-\frac{2}{3}\delta_{ij}\nabla_k v_k)$, becomes
very large. Since
the dissipative stresses are initially zero, taking relaxation
into account suppresses the contribution from the dilute
corona. A simplified version of this approach was used in
Cao \textit{et al.}~\cite{Cao:2010wa,NJPfocusissue4_thomas}. The relaxation time
and its relation to the spectral function of the shear tensor
is discussed in the contribution by Braby \textit{et al.}~\cite{NJPfocusissue29_schaefer}.

 The shear viscosity drops with temperature and is expected to
reach a minimum near $T_c$. In this regime quantum effects are
important. T-matrix calculations can be found in~\cite{Enss:2010qh}
and in the contributions by LeClair and Guo \textit{et al.} in this Focus Issue~\cite{NJPfocusissue11_leclair,NJPfocusissue2_levin}. It is also possible that
in this regime quasiparticle descriptions break down completely,
and the most efficient description of the unitary gas near
$T_c$ is in terms of a suitable weakly coupled holographic
dual. Progress towards constructing holographic duals of
non-relativistic quantum fluids is summarized in
Sec.~\ref{ssec:coldatoms}.

 In the low temperature superfluid regime the appropriate description
is superfluid (two-fluid) hydrodynamics~\cite{Khalatnikov:1965}.
Superfluid hydrodynamics predicts the existence of additional
hydrodynamic modes, in particular second sound, and contains
additional transport coefficients that come into play if there is
relative motion between the superfluid and normal components
of the fluid. There are proposals for exciting second sound
modes in the literature~\cite{Taylor:2009}, but these ideas
have not been confirmed yet. At very low temperature the
shear viscosity is expected to be dominated by phonons, similar
to liquid helium or dilute Bose gases. Elastic phonon scattering
gives $\eta\sim 1/T^5$~\cite{Rupak:2007vp}, whereas inelastic
processes can give a slower increase at low temperature, $\eta
\sim 1/T$~\cite{Mannarelli:2012su}. These predictions are difficult
to verify experimentally because the phonon free path is quite
large.

 The short mean free path in the strongly interacting normal
fluid suggests that not only the viscosity, but also other
transport coefficients may exhibit universal behavior. The
spin diffusion constant was recently studied by Sommer \textit{et al.}~\cite{Sommer:2011,NJPfocusissue17_zwierlein}. In the first paper Sommer
\textit{et al.} observed collisions between polarized Fermi gas clouds.
The colliding clouds are initially very far from equilibrium,
but at late times the system relaxes diffusively. The
corresponding relaxation time can be used to measure the
spin drag and the spin diffusion constant $D_s$. The spin
diffusion constant in the homogeneous system is defined
by Fick's law,
\be
  \mathbf{\jmath}_{s} = -D_s\nabla M \, ,
\ee
where $\mathbf{\jmath}_s$ is the spin current, and $M=n_\uparrow-
n_\downarrow$ is the polarization. In this Focus Issue Sommer
\textit{et al.} follow up on these studies by measuring the damping
of the spin dipole mode in strongly polarized gases
\cite{NJPfocusissue17_zwierlein}. A theoretical study of the spin drag
relaxation rate for a repulsive gas is presented by
Duine \textit{et al.} in this Focus Issue~\cite{NJPfocusissue25_stoof}. They show
that spin fluctuations enhance the spin drag in the
vicinity of the Stoner ferromagnetic transition.  Similarities
in the spin transport in the unitary gas and graphene
are studied in this Focus Issue by M\"uller and Nguyen~\cite{NJPfocusissue34_mueller}.

 A calculation of the diffusion constant based on the two-body
Boltzmann equation can be found in~\cite{NJPfocusissue38_bruun,Sommer:2011}.
They find
\be
 D_s = \frac{9\pi^{3/2}}{32\sqrt{2}}\frac{\hbar}{m}
   \left(\frac{T}{T_F}\right)^{3/2}\, .
\ee
Similar to the shear viscosity, the spin diffusion constant
drops with decreasing temperature. Near the critical temperature $D_s$
is expected to approach the universal value $D_s\sim \hbar/m$.
This behavior is indeed observed in the experiment~\cite{Sommer:2011};
see also the recent analysis of Bruun and Pethick~\cite{Bruun:2011}.
It is interesting to compare this result to the observed minimum
of the shear viscosity. Shear viscosity governs the rate of
momentum diffusion. The associated diffusion constant is
$D_\eta = \eta/(mn)$. Near $T_c$ we have $\eta/s\simeq 0.5
\hbar/k_B$ and $s/n\simeq k_B$. This implies $D_\eta\simeq 0.5
\hbar/m$, comparable to what is seen in spin diffusion. A
similar correlation between shear viscosity and diffusion was
observed in the QGP, as discussed in Sec.~\ref{sec_heavyq}.

\subsection{The BCS-BEC Crossover in Lattices}
\label{ssec:lattices}

A question quite distinct from that of the BCS-BEC crossover in ultracold Fermi gases in the continuum is how fermions pair, and reach the unitary regime, in a discrete context.  Optical lattices, or crystals of light, are created from interfering laser beams.  They make a sinusoidal standing wave that traps fermions via the AC Stark effect.  The strongly discretized regime, i.e., in the lowest band(s) and tight binding approximations, is quite distinct from that explored by both the QGP and continuum ultracold quantum gas experiments.  Essential features of weakly interacting fermions in this context have already been explored and understood~\cite{kohlM2005}; however, the strongly interacting regime remains in question.  Moreover, the lattice regime presents an additional challenge for holographic duality, as it too has a unitary point where $1/(k_F a)\to 0$.  The question of how to correctly model this problem in the context of a lattice is subtle, as simply attaching a band index to fermionic fields leads to hundreds of bands deep in the BEC regime, and is therefore numerically and practically intractable.

In the continuum, a single channel model treating only fermions qualitatively reproduces the phase diagram of Figs.~\ref{fig:crossoverPhaseDiagram} and~\ref{fig:realCrossoverPhaseDiagram}. This qualitative approach can be taken for instance by using the generalized BCS ansatz, technically valid only for weak interaction and high density, at the mean-field level for arbitrary interaction and finding the chemical potential self-consistently by fixing the average number of particles~\cite{Leggett_06}.  This method was first used by Eagles in the context of superconductivity in low carrier concentration systems~\cite{Eagles_69} and later by Leggett~\cite{Leggett_80} and Nozi\'eres and Schmitt-Rink~\cite{nozieres1985} explicitly for the BCS-BEC crossover at zero temperature and finite temperature, respectively, together called NSR theory.  In NSR one begins with a lattice, but then quickly takes a continuum or low momentum limit.  Interestingly, if this limit is not taken one obtains a completely incorrect prediction: instead of the BEC critical temperature tending to a constant non-zero value as $1/(k_F a) \to \infty$, as in Fig.~\ref{fig:crossoverPhaseDiagram}, the critical temperature tends to zero algebraically.  This unphysical tendency is because in order to hop or tunnel between lattice sites, a pair of fermions must be broken and tunnel one by one.  Then for very strong pair binding energy the process is prohibitively expensive.  In fact, even on the BCS side such models utilizing a single band have been shown to fail quantitatively for quite small values of $k_F a < 0$~\cite{buchler2010,vonStecher11}.  This model is called the \emph{Hubbard Hamiltonian}~\cite{hubbard1963} and is also a proposed model for high-temperature superconductivity~\cite{andersonPW2002}.

A solution to this quandry is to introduce a \emph{two-channel model} incorporating both fermions and bosons~\cite{timmermans1999,holland2001,kokkelmans2002,milstein2002}.  The fermions then represent the unbound atoms scattering at threshold (open channel) while the bosons represent the weakly bound molecule brought into resonance (closed channel).  Such two-channel models are mentioned briefly in the condensed matter context as far back as 1985~\cite{ranninger1985} and explored seriously in partial form starting in 1995~\cite{ranninger1995}; they continue to be explored as the \emph{Cooperon model}, in current research on high temperature superconductivity~\cite{yangKY2011}.  However, only in ultracold quantum gases have all hopping, interaction, and interconversion terms been included to make a \emph{Fermi-Bose Hubbard Hamiltonian} in the lattice~\cite{carr2005b}.  A series of papers attempted this approach with steady improvement over time, starting from simply attaching a band index to each channel~\cite{carr2005b,zhou2005a} up to performing a renormalization procedure to produce effective molecules in the lattice and minimize the number of bands that must be included~\cite{dickerscheid2005,diener06,duanLM2005b,kestner2010,carr2012h}.  The latter method has led to a Hamiltonian that is so far from the original Hubbard Hamiltonian that it has been given a new name, the \emph{Fermi Resonance Hamiltonian}.

The Fermi Resonance Hamiltonian is
\begin{eqnarray}
\label{eq:effHami}
\hat{H}_{\mathrm{eff}}=&-t_f\sum_{\sigma\in\left\{\uparrow,\downarrow\right\}}
  \sum_{\langle i,j\rangle}\hat{a}_{i\sigma}^{\dagger}\hat{a}_{j\sigma}
  +E_0\sum_{\sigma\in\left\{\uparrow,\downarrow\right\}}\sum_i \hat{n}_{i\sigma}^{\left(f\right)}
  -\sum_{\alpha\in\mathcal{M}}\sum_{i,j}t^{\alpha}_{i,j}\hat{d}_{i,\alpha}^{\dagger}\hat{d}_{j,\alpha}
\\
 &\nonumber\mbox{}
 +\sum_{\alpha\in\mathcal{M}}\bar{\nu}_{\alpha}\sum_i\hat{n}_{i\alpha}^{\left(b\right)}
 +\sum_{\alpha\in\mathcal{M}}\sum_{ijk}g_{i-j,i-k}^{\alpha}
\left[\hat{d}_{i,\alpha}^{\dagger}\hat{a}_{j,\uparrow}\hat{a}_{k,\downarrow}+\mathrm{h.c.}\right]\, ,
\end{eqnarray}
where $\hat{a}_{i\sigma}^{\dagger}$ creates a particle with spin $\sigma$ in the lowest open channel band Wannier state centered at lattice site $i$; $\hat{d}_{i,\alpha}^{\dagger}$ creates a particle in the $\alpha^{\mathrm{th}}$ dressed molecule Wannier state centered at site $i$; $\hat{n}_{i\sigma}^{\left(f\right)}$ is the number operator for fermions in the lowest Bloch band; and $\hat{n}_{i\alpha}^{\left(b\right)}$ is the number operator for the $\alpha^{\mathrm{th}}$ dressed molecule state.  The set of dressed molecules $\mathcal{M}$ which are included dynamically can be determined on energetic and symmetry grounds from the two-particle solution.  In order, the terms in Eq.~(\ref{eq:effHami}) represent tunneling of atoms in the lowest Bloch band between neighboring lattice sites $i$ and $j$; the energy $E_0=\sum_{\mathbf{q}}E_{\mathbf{1},\mathbf{q}}/N^3$ of a fermion in the lowest band with respect to the zero of energy; tunneling of the dressed molecular center of mass between two lattice sites $i$ and $j$, not necessarily nearest neighbors; detunings of the dressed molecules from the lowest band two-particle scattering continuum; and resonant coupling between the lowest band fermions at sites $j$ and $k$ in different internal states and a dressed molecule at site $i$.  The Fermi Resonance Hamiltonian is a two-channel resonance model, between unpaired fermions in the lowest band, and dressed molecules nearby in energy~\cite{carr2012h}.  Among other unusual predictions it makes, in contrast to usual Hubbard physics, is significant diagonal hopping,  pairing between atoms which do not lie along a principal axis of the lattice to form a dressed molecule, and multiple molecular bound states induced by the lattice.  Thus the lattice problem is indeed quite different both from the continuum crossover problem and well-known Hubbard physics.

In practice, the solution of the crossover problem in the lattice must occur in three steps.  First, for a given lattice strength and interacting strength the two-body problem must be solved exactly, including a renormalization procedure to get effective Wannier states for the molecules.  Then the coefficients in Eq.~(\ref{eq:effHami}) are calculated.  Then the Hamiltonian itself must be solved.  Potential solution methods range from matrix product state methods in 1D to mean field, quantum Monte Carlo, and dynamical mean field theory methods in higher dimensions.  Holographic duality may offer new approaches to this Hamiltonian in the future.  The crossover problem on the lattice remains very much an open problem and therefore we return to it in Sec.~\ref{sec:conclusions}.

\subsection{Recent and New Directions in Crossover Physics}
\label{ssec:newDirections}

In this subsection, we cover some of the new directions in unitary Fermi gases and related systems not discussed in previous sections, as explored in this Focus Issue.

%These two treated previously, and I have trouble fitting them in to the discussion below, so I don't cover them here:
%[Levin dissipative transport]
%[le Clair new calculations for $\eta/s$]
%Anyone is welcome to add them in if they see a good way to do it!

\subsubsection{New Experimental Probes}

We require new experimental probes into unitary gases in order to better understand their behavior.  Thus nearly all papers on ultracold fermions in this Focus Issue suggest an experimentally measurable effect.  However, three papers realize new probes directly in experiments.  First, Riedl \emph{et al.}~\cite{NJPfocusissue40_grimm} measure the quenching of the moment of inertia in the unitary regime.  This quenching is a well-known signal of superfluidity, as starting with liquid $^4$He it has been demonstrated that the moment of inertia drops below its classical value as one decreases a system through its critical temperature for the superfluid transition.  Unlike other methods of measurement in ultracold Fermi gases such as frequency and damping rate of collective excitations in response to small perturbations, quenching of the moment of inertia zeroes in specifically on superfluid effects.  Thus one can distinguish between nearly ideal fluid dynamics in the normal phase, which is chracterized by very low viscosity but allows rotational flow, and irrotational superfluid hydrodynamics.  A second new experimental probe is spin diffusion measurement, covered by three sets of contributors, one experimental~\cite{NJPfocusissue17_zwierlein} and two theoretical~\cite{NJPfocusissue25_stoof,NJPfocusissue38_bruun}, as we already touched on in Sec.~\ref{ssec:transport}.  The spin diffusion constant $D_s$ and the momentum diffusion constant $D_\eta$ exhibit a similar temperature dependence, and an analogous relation between heavy quark and momentum diffusion is expected in the QGP. Overall, diffusion constants offer another point of comparison for bounds predicted by holographic duality.  Finally, a third new experimental probe is Tan's contact density, given in its original form in Eq.~(\ref{eqn:Tan1}), and covered in this Focus Issue in both an experimental~\cite{NJPfocusissue16_vale} and a theoretical contribution~\cite{NJPfocusissue36_drummond}.  Tan's contact density connects microscopic scattering properties to macroscopic thermodynamic observables throughout the BCS-BEC crossover.  Finally, as shown in Fig.~\ref{fig:energyvsT}, an improved temperature calibration is realized in this Focus Issue to better characterize the key ratio of the viscosity to the entropy~\cite{NJPfocusissue4_thomas}.

\subsubsection{Solitons to Polarons}

We turn now to new directions in theory.  The simplest and oldest approach to describing the BCS-BEC crossover is BCS mean field theory, in the form of the Boguliubov-de Gennes (BdG) equations describing quasiparticles in the BCS phase.  Despite the fact that these equations are technically only valid for weak interactions $ - k_F a \ll 1$, and are based on a quasiparticle picture, one can attempt an extrapolation through the crossover by self-consistently solving the gap equation, BdG equations, and density and normalization of quasiparticles.  Both Spuntarelli \emph{et al.}~\cite{NJPfocusissue33_carr} and Baksmaty \emph{et al.}~\cite{NJPfocusissue12_pu} describe this approach in detail in this Focus Issue; the method is numerically challenging as it requires solution of simultaneous nonlinear partial differential equations.  Spuntarelli predicts features of dark solitons through the crossover.  Solitons appear as robust local minima in the gap with a strong indicator in the density as well, even at unitarity.  Solitons are also described by holographic duality in this Focus Issue by Ker\"anen \emph{et al.}~\cite{NJPfocusissue10_keranen}, providing a point of comparison to mean field theory.  Further recent work on solitons through the BCS-BEC crossover can be found in Refs.~\cite{scottRG2010,liaoR2011,scottRG2012}.  Baksmaty {\it et al.} use the BdG approach to shed light on experimental data in imbalanced Fermi gases, i.e., those with more spin up than spin down fermions.  Imbalanced Fermi gases in elongated traps are shown to display strong violation of the local density approximation in the form of phase separation, a quite different perspective from the thermodynamic one.

Imbalanced Fermi gases, sometimes also called spin-polarized Fermi gases, open up new territory for observing quantum phase transitions and have been used recently to explore a Fermi liquid description of the strongly interacting normal phase~\cite{nascimbene2010} and to observe a Fulde-Ferrel-Larkin-Ovchinnikov (FFLO) state in one-dimensional Fermi gases~\cite{HuletFFLO}.  The FFLO state is a pairing between two different Fermi surfaces, resulting in a finite center-of-mass momentum of the pairs.  Spin-dependent scattering is suggested as a novel experimental probe in a theoretical contribution by Sheehy~\cite{NJPfocusissue28_sheehy}, to measure the pairing spin gap in locally imbalanced Fermi gases, where the imbalance is induced by coupling to a spin-dependent potential.  Imbalanced gases are also used in spin diffusion measurements, as described in Sec~\ref{ssec:transport}. In the limit in which a Fermi gas is strongly imbalanced, a polaron model can be used, where the minority species is dressed by the Fermi sea of the majority species and the resulting quasiparticle is called a polaron; in the extreme case a single atom of the minority species is considered, leading to a single polaron.  Exactly how such concepts apply at unitarity where quasiparticle concepts tend to breakdown is an open question.  In this Focus Issue Sadeghzadeh {\it et al.}~\cite{NJPfocusissue15_recati} predict the existence of a new metastable state consisting of a Fermi sea of polarons.

\subsubsection{Disorder and Quantum Phase Transitions}

As discussed in Sec.~\ref{ssec:lattices}, lattice physics is only barely beginning to be treated in Fermi gases through the BCS-BEC crossover.  Two new directions are treated for ultracold Fermions in optical lattices in this Focus Issue.  First, Han and S\'a de Melo~\cite{NJPfocusissue14_sademelo} treat the BCS-BEC crossover in the presence of disorder.  They find the superfluid near unitary to be much more robust against disorder than in the BCS and BEC regimes, and give some useful observations on how the different practical realizations of disorder in optical lattices can lead to different considerations for crossover physics.  Second, lattices give one access to strongly correlated systems which have nothing to do with the BCS-BEC crossover.  That is, near quantum phase transitions one generally obtains scale-invariant phenomena that can potentially be studied with holographic duality.  Zhang {\it et al.}~\cite{NJPfocusissue24_chin} cover experiments on one such system in this Focus Issue, ultracold bosons near the Mott insulator to superfluid transition in the Hubbard model in two dimensions.  Then the same kinds of issues as we have discussed throughout Sec.~\ref{sec:ultracold} reappear, from a universal equation of state to transport.  The first half of this paper forms a very readable overview for the general reader, while the second half gives a status report on experiments.

The massive ground state degeneracy induced by lattices is the principle underlying the achievement of strongly correlated states in bosons.  One can obtain similar degeneracy either by rotating a system to achieve such degeneracy in the lowest Landau level and get quantum Hall physics, or by taking advantage of the spin degree of freedom.  The latter strategy is pursued by Shlyapnikov and Tsvelik~\cite{NJPfocusissue7_shlyapnikov}.  Chromium, a Bose-condensed atom with spin $S=3$, provides a practical working example, and they find a commensurate-incommensurate quantum phase transition, among other possibilities in a rich phase diagram.

\subsubsection{Trimers: From Efimov States to SU(3)}

Although we very briefly mentioned three-body losses in Sec.~\ref{ssec:experiments}, in fact three-body losses are the dominant loss mechanism at unitarity, whether for fermions or for bosons.  Three body physics shows some real surprises in ultracold fermions: even when there is no two body bound state at all, one finds a denumerably infinite and universal\footnote{By universal we mean that the details of the scattering potential don't matter for Efimov states.} set of three-body bound states in the strongly interacting regime, called \textit{Efimov} states.  Efimov states can occur for various combinations of bosons and fermions of same or different masses.  The evidence for such series of states is found in three-body losses.  Braaten and Hammer wrote a review back in 2006~\cite{braatenE2006}.  However, this area continues to develop rapidly, motivated strongly by ultracold experiments.  In this Focus Issue, Wang and Esry cover this important area~\cite{NJPfocusissue27_esry}, extending our knowledge of Efimov physics beyond the low-energy threshold collisions typical of ultracold gases to the higher energy regime relevant to colliding clouds.  They treat both positive and negative scattering length and both broad and narrow Feshbach resonances~\cite{chinCheng2010}: in narrow Feshbach resonances Efimov effects are less universal in nature, requiring a second scattering potential parameter beyond the scattering length, namely, the effective range.  Losses in fermions also show a puzzling feature, a maximum at a magnetic field below the pole of the Feshbach resonance where the scattering length diverges.  Zhang and Ho suggest an explanation for this anomaly in terms of an interplay between atom-atom, atom-dimer, and dimer-dimer interactions.  Based on these considerations, and incorporating temperature and trapping parameters, they develop rate equations which match all experimental loss features, covering data in four different labs taken over seven years.

Finally, so far we have discussed solely SU(2) physics.  However, the hyperfine spin manifold of ultracold fermions allows for a variety of spin structures to be created.  In particular, the preparation of degenerate three-state mixtures with three-fold ($SU(3)$) symmetric attractive interactions opens up new territory in the connection between cold atoms and nuclear matter.  Such a three-state gas  can be used to explore pairwise superfluidity and spontaneous symmetry breaking, color superconductivity, and a superfluid to trimer gas transition that mimics the deconfinement to hadronization transition in quark matter~\cite{NJPfocusissue8_ohara}. In symmetric three-component gases, it is possible that more than one pairing field can simultaneously become nonzero~\cite{NJPfocusissue13_torma}. Magnetism and domain formation also can be studied~\cite{NJPfocusissue30_hofstetter}. A key issue is to perform experiments at low  density in order to reduce the 3-body decay rates $\propto n^3$, which are not Pauli-suppressed in three-state systems. Three-body decay arises from recombination, which is enhanced and modified near a  Feshbach resonance, where Efimov states exist with binding energies well beyond the threshold region~\cite{NJPfocusissue27_esry}. By tuning away from resonance  and using radio-frequency dressing to precisely symmetrize the pair-wise interactions, and confining the atoms in a lattice to increase the critical reduced temperature, the goal of creating cold-atom analogs of quark matter appears to be within reach~\cite{NJPfocusissue8_ohara}.

%%%%%%%%%%%%%%%%%%%%%%%%%%%%%%%%%%%%%%%%%%%%%%%%%%%%%%%%%%%%%%%%%%%%
\section{Quantum Chromodynamics, the Quark-Gluon Plasma, and Heavy Ion Collisions}
\label{sec_qcd}
%%%%%%%%%%%%%%%%%%%%%%%%%%%%%%%%%%%%%%%%%%%%%%%%%%%%%%%%%%%%%%%%%%%%

In this section we will give an introduction to the physics of
the quark-gluon plasma (QGP). This is a vast subject and we will not
attempt to provide a comprehensive review here. Recent reviews
of various issues related to quark matter and the QGP can be found in~\cite{Alford:2007xm,Shuryak:2008eq,Heinz:2009xj,BraunMunzinger:2009zz,Schafer:2009dj,CasalderreySolana:2011us}
and the results from the experimental program at the Relativistic Heavy Ion Collider (RHIC) at Brookhaven National Laboratory are summarized
in~\cite{rhic:2005}. Some results from the Large Hadron Collider (LHC) program at the European Organization for Nuclear Research (CERN) can be found
in the contributions~\cite{Steinberg,Snellings:2011sz,Nagle:2011uz}
and in the recent review \cite{Muller:2012zq}.

We will focus on observables and experimental results that
are important in establishing the presence of strong correlations
in the QGP, and that provide direct connections
to the physics of ultracold atomic gases and holographic dualities.
In Sec.~\ref{sec_phases} we introduce quantum chromodynamics (QCD), the theory of the
strong interaction, and discuss the QCD phase diagram. The
differences between weakly coupled QCD plasmas and strongly
coupled QCD fluids are discussed in Sec.~\ref{sec_sQGP}.
The experimental study of high temperature QCD matter is based
on heavy ion collisions. We provide an overview of the experimental
program in Sec.~\ref{sec_hic} and discuss specific observables
(multiplicities, flow, energy loss and heavy quarks) in
Sec.~\ref{sec_bulk}-\ref{sec_heavyq}. Finally, a discussion of
open issues can be found in Sec.~\ref{sec_qgp_open}.

%%%%%%%%%%%%%%%%%%%%%%%%%%%%%%%%%%%%%%%%%%%%%%%%%%%%%%%%%%%%%%%%%%%%
\subsection{Quantum Chromodynamics and the Phase Diagram}
\label{sec_phases}
%%%%%%%%%%%%%%%%%%%%%%%%%%%%%%%%%%%%%%%%%%%%%%%%%%%%%%%%%%%%%%%%%%%%

QCD describes the behavior of strongly
interacting matter, such as protons, neutrons, and nuclei, as well
as the hot and dense matter created in heavy ion collisions, in
terms of quarks and gluons and their interactions. The complicated
phenomenology of the strong interaction is encoded in a deceptively
simple Lagrangian. The Lagrangian is formulated in terms of quark
fields $q_{\alpha\, f}^c$ and gluon fields $A_\mu^a$. Here, $\alpha=1,
\ldots,4$ is a Dirac spinor index (corresponding, in the Dirac
representation, to quarks and anti-quarks with spin up and down),
$c=1,\ldots,N_c$ is a color index, and $f={\it up}, {\it down},
{\it strange}, {\it charm},{\it bottom}, {\it top}$ is a flavor
index. In our world the number of colors is $N_c=3$, but as a
theoretical laboratory it is useful to consider theories with
$N_c\neq 3$.  The dynamics of the theory is governed by the color
degrees of freedom. The gluon field $A_\mu^a$ is a vector field (like
the photon) labeled by an adjoint color index $a=1,\ldots,8$.
The octet of gluon fields can be used to construct a matrix valued
field $A_\mu=A_\mu^a \frac{\lambda^a}{2}$, where $\lambda^a$ is a set
of traceless, Hermitian, $3\times 3$ matrices.
Repeated indices are assumed to be summed throughout our treatment. The QCD Lagrangian is
\be
\label{l_qcd}
 {\cal L } =  - \frac{1}{4} G_{\mu\nu}^a G_{\mu\nu}^a
  + \sum_f^{N_f} \bar{q}_f ( i\gamma^\mu D_\mu - m_f) q_f\, ,
\ee
where $G^a_{\mu\nu}$ is the QCD field strength tensor
\be
 G_{\mu\nu}^a = \partial_\mu A_\nu^a - \partial_\nu A_\mu^a
  + gf^{abc} A_\mu^b A_\nu^c\, ,
\ee
and $f^{abc}=4i\,{\rm Tr}([\lambda^a,\lambda^b]\lambda^c)$ is a
set of numbers called the $SU(3)$ structure constants. The
covariant derivative acting on the quark fields is
\be
 i D_\mu q =  \left(
 i\partial_\mu + g A_\mu^a \frac{\lambda^a}{2}\right) q\, ,
\ee
and $m_f$ is the mass of the quarks. The different terms in
Eq.~(\ref{l_qcd}) describe the interaction between quarks and
gluons, as well as nonlinear three and four-gluon interactions.
It is important to note that, except for the number of flavors
and their masses, the structure of the QCD Lagrangian is completely
fixed by a local $SU(3)$ color symmetry.

 For many purposes we can consider the light flavors (up, down,
and strange) to be approximately massless, and the heavy flavors
(charm, bottom, top) to be infinitely massive. In this limit the QCD
Lagrangian contains a single dimensionless parameter, the coupling
constant $g$. If quantum effects are taken into account the coupling
becomes scale dependent. At leading order the running coupling
constant is
\be
\label{g_1l}
 g^2(q^2) = \frac{16\pi^2}
  {b_0\log(q^2/\Lambda_{\it QCD}^2)}\, , \hspace{1cm}
 b_0=\frac{11}{3}N_c-\frac{2}{3}N_f\, ,
\ee
where $q$ is a characteristic momentum and
$N_f$ is the number of active flavors ($N_f=3$ if we consider charm,
bottom and top to be infinitely heavy). The result in Eq.~(\ref{g_1l}) implies
that, as a quantum theory, QCD is not characterized by a dimensionless coupling
but by a dimensionful scale, the QCD scale parameter $\Lambda_{\it QCD}$. This
effect is called dimensional transmutation~\cite{Coleman:1973jx}. We also
observe that the coupling decreases with increasing momentum. This is the
phenomenon of asymptotic freedom~\cite{Gross:1973id,Politzer:1973fx}. The
flip side of asymptotic freedom is anti-screening, or confinement:
the effective interaction between quarks increases with distance.

 In massless QCD  the scale parameter is an arbitrary parameter (a QCD
``standard kilogram''), but once QCD is embedded into the standard model
and quarks acquire masses by electroweak symmetry breaking the QCD scale
is fixed by the choice of units in the standard model. The numerical
value of $\Lambda_{QCD}$ depends on the renormalization scheme used to
derive Eq.~(\ref{g_1l}). Physical masses, as well as the value of
$b_0$, are independent of this choice. In the \textit{modified minimal
subtraction} ($\overline{MS}$) one finds $\Lambda_{QCD}\simeq 200$
MeV; see the QCD section in~\cite{Nakamura:2010zzi}.

%%%%%%%%%%%%%%%%%%%%%%%%%%%%%%%%%%%%%%%%%%%%%%%%%%%%%%%%%%%%%%%%%%%%%%%%%
\begin{figure}[t]
\bc\includegraphics[width=10cm]{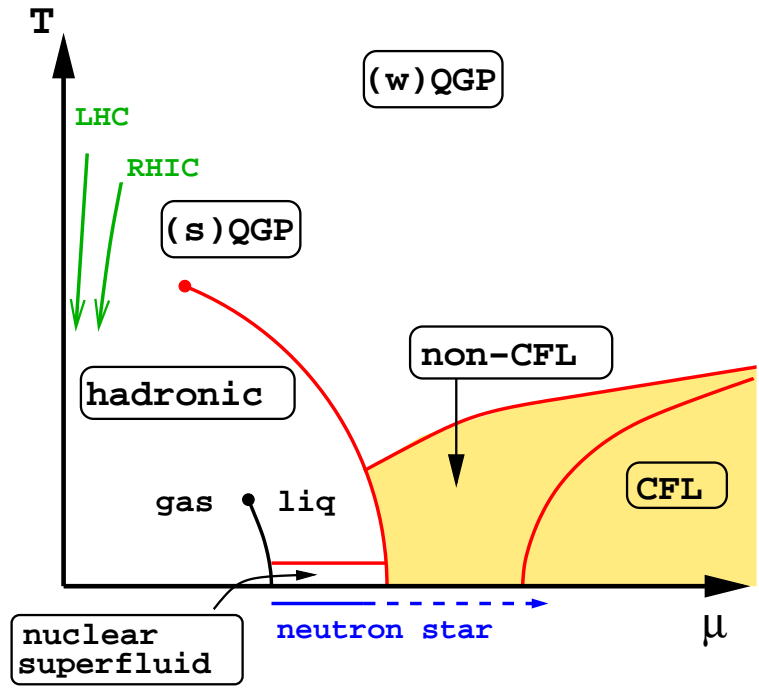}\ec
\caption{\label{fig_qcd_phase}
Schematic phase diagram of QCD as a function of temperature $T$ and
baryon chemical potential $\mu$. QGP refers to the QGP,
and sQGP/wQGP denote the regions of the phase diagram where the plasma
is strongly/weakly coupled. The CFL (color-flavor locked) phase is the
color superconducting phase that occurs at asymptotically large chemical
potential. LHC/RHIC denote the regions of the phase diagram that are
being explored by the experimental heavy ion programs at the LHC and
RHIC. The red and black points denote the critical endpoints
of the chiral and nuclear liquid-gas phase transitions, respectively. }
\end{figure}
%%%%%%%%%%%%%%%%%%%%%%%%%%%%%%%%%%%%%%%%%%%%%%%%%%%%%%%%%%%%%%%%%%%%%%%%%

 Asymptotic freedom and the symmetries of QCD determine the main
phases of strongly interacting matter that appear in the QCD phase
diagram shown in Fig.~\ref{fig_qcd_phase}. In this figure we show the
phases of QCD as a function of the temperature $T$ and the baryon
chemical potential $\mu$. The chemical potential $\mu$ controls the
baryon density $\rho$, defined as 1/3 times the number density of
quarks minus the number density of anti-quarks.

 At zero temperature and chemical potential the interaction between
quarks is dominated by large distances and the effective coupling is
large. As a consequence, quarks and gluons are permanently confined in
color singlet hadrons, with masses of order $\Lambda_{QCD}$. For example, the proton
has a mass of $m_p=935$ MeV.\footnote{Note that we are quoting the
mass in units of energy, setting the speed of light equal to unity}.

If we think of the proton as composed of three \textit{constituent quarks}
this implies that quarks have effective masses $m_Q\simeq m_P/3\simeq
\Lambda_{QCD}$. This should be compared to the bare up and down
quark masses which are of the order 10 MeV.

Strong interactions between quarks and anti-quarks lead to a vacuum
condensate of $\bar{q}q$ pairs, $\langle\bar{q}q\rangle\simeq
-\Lambda^3_{QCD}$~\cite{GellMann:1968rz,Coleman:1980mx,tHooft:1982}.
This vacuum condensate spontaneously breaks the approximate chiral
$SU(3)_L\times SU(3)_R$ flavor symmetry of the QCD Lagrangian. Chiral
symmetry breaking implies the existence of Goldstone bosons, particles
with masses smaller than $\Lambda_{QCD}$. These particles are known as
pions, kaons, and etas.\footnote{The $SU(3)_L\times SU(3)_R$ is explicitly
broken by quark masses and the mass of the charged pion is $m_\pi=139$
MeV, which is not much smaller than $\Lambda_{QCD}$. The lightest
non-Goldstone particle is the rho meson, which has a mass $m_\rho=770$
MeV.}

 The quark-anti-quark condensate $\langle\bar{q}q\rangle$ is
analogous to the di-fermion condensate $\langle qq\rangle$ that
characterizes BCS superconductors~\cite{leggett1975}.
BCS pairing involves particles
with opposite momenta near the Fermi surface. The chiral condensate
is an order parameter for pairing of fermions and anti-fermions
near the surface of the \textit{Dirac sea}. The most important difference
between BCS pairing and chiral condensation is that because of
the finite density of states near the Fermi surface BCS pairing
can take place at weak coupling. Chiral condensation, on the other
hand, only happens at strong coupling.\footnote{In
ultracold atomic gases in the BEC regime pair condensation involves
pre-formed pairs. Whether or not pre-formed $\bar{q}q$ pairs exist in
in the QGP above $T_c$ is still being investigated; see for example
\cite{Asakawa:2000tr}. We should note that there is no completely
rigorous criterion for the existence of a bound state embedded in
a plasma. In practice, researchers have looked for peaks in the
spectral function associated with the correlator $\langle
\bar{q}q(x)\bar{q}q(0)\rangle$.}  We will see below that in
QCD at large baryon density there is a transition between $\bar{q}q$
and $qq$ pairing. This transition is driven by the competition
between the stronger coupling in the $\bar{q}q$ channel and the
growing density of states for $qq$ pairing.

At very high temperature quarks and gluons have thermal momenta
$p\sim T\gg\Lambda_{QCD}$ (Note that we are writing momenta in units
of energy, again setting $c=1$). Asymptotic freedom implies that
these particles are weakly interacting, and that they form a
plasma of mobile color charges, the QGP~\cite{Shuryak:1977ut,Shuryak:1978ij}. We note that the argument
that the QGP at asymptotically high temperature is weakly coupled
is somewhat more subtle than it might appear at first sight. If two
particles in the plasma interact via large angle scattering then the
momentum transfer is large, and the effective coupling is weak
because of asymptotic freedom. However, the color Coulomb interaction
is dominated by small angle scattering, and it is not immediately clear
why the effective interaction that governs small angle scattering is
weak. The important point is that in a high temperature plasma there
is a large thermal population ($n\sim T^3$) of mobile charges that
screen the interaction at distances beyond the Debye length $r_D\sim
1/(gT)$. We also note that even in the limit $T\gg\Lambda_{QCD}$ the
QGP contains a non-perturbative sector of static magnetic color
fields~\cite{Linde:1980ts}. This sector is strongly coupled, but
it does not contribute to thermodynamic or transport properties
of the plasma in the limit $T\to\infty$.

 The plasma phase exhibits neither color confinement nor chiral
symmetry breaking. This means that the high temperature QGP phase must
be separated from the low temperature hadronic phase by a phase transition.
The nature of this transition is very sensitive to the values of the quark
masses. In QCD with massless $u,d$ and infinitely massive $s,c,b,t$ quarks
the transition is second order~\cite{Pisarski:1983ms}. In the case of
massless (or sufficiently light) $u,d,s$ quarks the transition is first
order. Lattice simulations show that for realistic quark masses, $m_u
\simeq m_d\simeq 10$ MeV and $m_s\simeq 120$ MeV, the phase transition
is a rapid crossover~\cite{Aoki:2006we}. The transition temperature,
defined in terms of the chiral susceptibility, is $T_c\simeq 151\pm 3
\pm 3$ MeV~\cite{Aoki:2006br,Aoki:2009sc}.

 The transition is believed to strengthen as a function of chemical
potential, so that there is a critical $\mu$ at which the crossover
turns into a first order phase transition~\cite{Stephanov:2004wx}.
This point is the critical endpoint of the chiral phase transition.
Due to the fermion sign problem it is very difficult to locate the
critical endpoint using simulations on the lattice. A number
of exploratory calculations have been performed
\cite{Fodor:2001pe,Allton:2002zi,Karsch:2003va,Fodor:2004nz},
but at this point it is not even clear whether the idea that the
transition strengthens as the baryon chemical potential increases
is correct~\cite{deForcrand:2010he}. The critical endpoint is
interesting because it is the only point on the phase transition
line at which the correlation length diverges (there is a similar
endpoint on the nuclear liquid-gas transition line). This means
that the critical point may manifest itself in heavy ion collisions
in terms of enhanced fluctuations~\cite{Stephanov:1998dy}.
The idea is that one can tune the baryon chemical potential
by changing the collision energy, since lower beam energy allows
for more stopping of the baryons in the initial state and therefore
leads to higher chemical potential. As the collision energy is
varied one looks for non-monotonic behavior of fluctuation and
correlation observables. A typical observable
is the variance (divided by the mean) of the net number of protons
(protons minus anti-protons) in a finite sub-volume, $\langle
(\Delta N_p)^2\rangle/\langle N_p\rangle$ with $\Delta N_p=N_p
-\langle N_p\rangle$. A beam energy scan is now under way at RHIC,
and similar scans have been performed at lower energy as part of
the CERN fixed target program~\cite{Gazdzicki:2011fx,Gupta:2011wh}.

At low temperature the first phase one encounters as the chemical
potential is increased from zero is nuclear matter, a strongly
correlated superfluid composed of approximately non-relativistic
neutrons and protons. It is interesting to note that nuclear matter in
weak $n\leftrightarrow p+e+\bar\nu_e$ equilibrium is neutron
rich,\footnote{Isolated neutrons are unstable with regard to the decay
into protons, electrons, and anti-neutrinos. If we ignore the role
of electrons then dense matter is composed of equal numbers of protons
and neutrons because this configuration optimizes the total Fermi
energy. Taking electrons into account we observe that for a given
density of protons and electrons the Fermi energy of the electrons
is much bigger than that of protons, and the lowest energy state is
neutron rich.} and that the neutron-neutron scattering
length\footnote{1 fm (Fermi)=$10^{-15}$ m is the typical unit of
length in nuclear and particle physics. 1 fm is approximately equal
to the radius of the proton and neutron.} $a_{nn}\simeq 20$ fm
is much larger than the average inter-particle spacing, $r_{nn}
\simeq 1.9$ fm at nuclear matter saturation density. As a consequence,
dilute nuclear matter is closely related to the unitary atomic Fermi
gases discussed in Sec.~\ref{sec:ultracold}.

 At very large chemical potential we can use arguments similar
to those in the high temperature limit to establish that
quarks and gluons are weakly coupled. The main difference
between cold quark matter and the hot QGP is
that because of the large density of states near the quark Fermi
surface even weak interactions can cause qualitative changes
in the ground state of dense matter. In particular, attractive
interactions between quark pairs lead to color superconductivity
and the formation of a $\langle qq\rangle$ condensate. Since
quarks carry color, flavor, and spin labels, many superconducting
phases are possible. The most symmetric of these, known as the
color-flavor locked (CFL) phase, is predicted to exist at very
high density~\cite{Alford:1998mk,Schafer:1999fe}. In the CFL phase
the diquark order parameter is $\langle q^A_{\alpha f} q^B_{\beta g}
\rangle \sim \epsilon_{\alpha\beta} \epsilon^{ABC}\epsilon_{fgC}$.
This order parameter has a number of interesting properties. It
breaks the $U(1)$ symmetry associated with baryon number, leading
to superfluidity, and it breaks the chiral $SU(3)_L \times SU(3)_R$
symmetry. Except for Goldstone modes the spectrum is fully gapped;
fermions acquire a BCS-pairing gap, and gauge fields are screened
by the Meissner effect. This implies that the CFL phase, even
though it arises from a superdense liquid of quarks, shares many
properties of superfluid nuclear matter.

 The CFL phase involves equal pair-condensates $\langle ud\rangle
=\langle us\rangle = \langle ds\rangle$  of all three light quark
flavors. As the density is lowered effects of the non-zero strange quark
mass become more important, and less symmetric phases are likely to
appear~\cite{Alford:2007xm}. Calculations based on weak coupling suggest
that the first non-CFL phase is a CFL-like phase with a Bose condensate
of kaons, and the second phase involves a standing meson wave
superimposed on the kaon condensate~\cite{Bedaque:2001je,Schafer:2005ym}.
Other possibilities which may appear in strong coupling include a phase
with up-down pairing only (2SC), or separate spin-one condensates of
up, down and strange quarks.

 Some guidance for analyzing the competition between these phases
may come from studying analog models of the quark-hadron phase
transition based on ultracold atomic gases. A number of authors have
pointed out to the possibility of a BCS-BEC crossover in dense
quark matter~\cite{Abuki:2010jq}. The basic idea is that as the
density is lowered, quark Cooper pairs become more strongly bound
and form a diquark Bose condensate. At lower densities one may
also encounter a mixture of condensed diquarks and unpaired quarks.
The physics of this system can be studied using ultracold boson-fermion
mixtures~\cite{truscott2001,schreck2001,Maeda:2009ev,Maeda2011}. Finally, it may be possible to study
the transition from nucleons, bound states of three quarks, to
color superconducting diquarks using three-species systems of
ultracold fermions~\cite{Rapp:2006rx,NJPfocusissue8_ohara,NJPfocusissue13_torma,NJPfocusissue30_hofstetter}.

Color superconductivity affects both the equation of state and
transport properties of dense quark matter. These effects may
manifest themselves in the structure and evolution of neutron
stars. Possible signatures of pairing appear in the mass-radius
relation, the cooling curve, and the relation between the spin-down
rate and the magnetic field. There are two contributions in this
Focus Issue that touch on these issues. Shovkovy and Wang study the
bulk viscosity of normal quark matter in the large amplitude
regime~\cite{Shovkovy:2010xk}. Anglani, Mannarelli and Ruggieri
investigate the interaction of collective modes in the CFL
phase~\cite{Anglani:2011cw}. This calculation is a first step
towards more accurate calculations of transport properties of the
CFL phase~\cite{Mannarelli:2010vd,Schafer:2010cs}.

%%%%%%%%%%%%%%%%%%%%%%%%%%%%%%%%%%%%%%%%%%%%%%%%%%%%%%%%%%%%%%%%%%%%
\subsection{Weakly vs. Strongly Coupled Plasmas}
\label{sec_sQGP}
%%%%%%%%%%%%%%%%%%%%%%%%%%%%%%%%%%%%%%%%%%%%%%%%%%%%%%%%%%%%%%%%%%%%

 At asymptotically high temperature the coupling is weak and
properties of the QGP can be systematically
computed. One of the most basic properties is the equation
of state. In the following we will consider a related quantity,
the entropy density. The perturbative expansion of the entropy
density is~\cite{Kajantie:2002wa,Kraemmer:2003gd}
\be
\label{s_pqcd}
 s = T^3 \left\{ c_0 + c_2 g^2 + c_3 g^3 + \ldots \right\}\, ,
\ee
where $g$ is the QCD coupling constant from Eq.~(\ref{g_1l}) and we have chosen
units such that $\hbar=k_B=1$. The expansion is performed with
the coupling $g=g(\bar\mu)$ evaluated at a fixed scale $\bar\mu$.
Higher order terms contain logarithms of the form $\log(2\pi T/
\bar\mu)$, which combine with lower order terms to eliminate
the dependence on the arbitrary scale $\bar\mu$, and lead to
an expansion in terms of the running coupling $g(q)$ evaluated
at a scale $q\simeq 2\pi T$. The first term in Eq.~(\ref{s_pqcd})
corresponds to the Stefan-Boltzmann law, and $c_0$ is proportional
to the number of degrees of freedom,
\be
c_0 = \frac{2\pi^2}{45}\left(
  2\left( N_c^2-1 \right) + 4N_cN_f\frac{7}{8} \right)\,,
\ee
where $2(N_c^2-1)$ is the number of degrees of freedom from the gluons and
$4N_cN_f$ from the quarks.  The naive perturbative expansion is an expansion in powers of
$g^2$. Odd powers of $g$ appear because of infrared divergences.
As we approach the phase transition the coupling becomes large
and higher order terms are no longer small. Indeed, the convergence
properties of the expansion in Eq.~(\ref{s_pqcd}) are extremely poor:
the series shows no signs of converging unless the coupling is taken
to be much smaller than one, $g\ll 1$, corresponding to completely
unrealistic temperatures on the order of 1 TeV. The convergence can
be improved significantly by using a self-consistent quasiparticle
expansion~\cite{Blaizot:2003tw}. This means that the perturbative
expansion is formulated not in terms of free quarks and gluons, but
in terms of quasi-quarks and quasi-gluons which have effective masses
and effective interactions.

 Quasi-particle expansions fit lattice data quantitatively down
to temperatures $T\sim 2T_c$. In this regime $s/s_0\simeq 0.85$, where
$s_0$ is the entropy density of a non-interacting plasma. In the past
this was frequently taken as evidence that quasiparticles are not
strongly coupled at temperatures relevant to the early stages of
heavy ion collisions at RHIC or the LHC. A new perspective on this
questions is provided by holographic duality; see
Sec.~\ref{sec:hd}.  Holographic duality maps strongly interacting quantum field theories
onto one higher dimensional classical gravity, in particular anti-de Sitter space.
In the original version of the mapping
the quantum field theory is a QCD-like theory known as $N=4$ supersymmetric (SUSY) Yang-Mills
theory~\cite{D'Hoker:2002aw}.  This theory is conformal.   That means
its coupling constant does not run,  $g^2(q^2)={\rm const}$, and there
is no analog of $\Lambda_{QCD}$. It also does not have a phase transition,
and so the theory is in the plasma phase for all values of the coupling.
The particle content of $N=4$ SUSY Yang-Mills theory differs from that
of QCD. The theory has no quarks, and in addition to gluons it contains
supersymmetric fermionic partners of gluons called gluinos as well as additional
colored scalar fields. Nevertheless, the plasma phase shares many
features of the QGP, and by focusing on a ratio like $s/s_0$ we can
remove the difference in the number of degrees of freedom.  Later versions
of holographic duality relax conformal and supersymmetric requirements, but still do not map
precisely onto QCD.

Using holographic duality it was found that in the limit
of strong coupling  and a large number of colors the entropy density
of SUSY Yang-Mills theory is $s/s_0=0.75$~\cite{Gubser:1998nz}. This
value is remarkably close to the result observed in lattice QCD
calculations near the phase transition, casting doubt on an
interpretation of the data in terms of weakly coupled quasiparticles.
More generally, holographic duality demonstrates that
thermodynamic properties of the plasma need not be very sensitive
to the strength of the interaction.

Holographic duality also shows that, in contrast to
equilibrium properties, transport properties of the plasma are
very sensitive probes of the strength of the interaction. In
perturbative QCD the shear viscosity of three flavor QCD is
\cite{Baym:1990uj,Arnold:2000dr,Arnold:2003zc}
\be
\label{eta_qcd}
 \eta = \frac{kT^3}{g^4\log(\mu^*/m_D)},
\ee
where $k=106.67$, $\mu^*=2.96T$ and $m_D\sim gT$ is the screening mass.
Shear viscosity is related to the rate of momentum diffusion. A simple
estimate of the shear viscosity is~\cite{Danielewicz:1984ww}
\be
\label{eta_mfp}
 \eta\simeq \frac{1}{3}np\,l_{\mathrm{mfp}}
      \simeq \frac{1}{3}\frac{p}{\sigma_T},
\ee
where $n$ is the density, $p$ is the mean momentum, $l_{\mathrm{mfp}}$ is
the mean free path, and $\sigma_T$ is the transport cross section. In
a perturbative QCD plasma the typical momentum is $p\sim T$, and the
transport cross section is $\sigma_T\sim g^4\log(g)T^{-2}$. The time
scale for momentum diffusion is $\eta/(sT)\sim 1/(g^4\log(g)T)$. We
note that in the weak coupling limit this number is parametrically
large. In a QGP at $T=200$ MeV,  just above the phase
transition, we have $T^{-1}\simeq 1$ fm/c and $\eta/s\simeq 9.2/g^4$.
A typical value of the coupling is $g\simeq 2$ (corresponding to
$\alpha_s\simeq 0.3$), which implies $\eta/(sT)\simeq 0.6$ fm/c.

 The strong coupling limit of $\eta/s$ in the SUSY Yang-Mills plasma
was studied by Policastro, Son, and Starinets~\cite{Policastro:2001yc}.
They find
\be
\label{eta_s_ads}
 \frac{\eta}{s} = \frac{1}{4\pi}
   \left( 1 + O\left( \lambda^{-3/2} \right) \right) ,
\ee
where $\lambda=g^2N_c$ is the 't Hooft coupling. This result implies
that at strong coupling momentum equilibration is almost an order of
magnitude more rapid than it is in weak coupling. The ratio $\eta/s$
is also close to a bound that was argued to arise from the uncertainty
relation~\cite{Danielewicz:1984ww}. The idea is that, because of quantum
mechanics, the product $p\,l_{\mathrm{mfp}}$ in Eq.~(\ref{eta_mfp}) cannot
become smaller than Planck's constant $\hbar$ (note that we are using
units in which $\hbar=1$). This implies $\eta/s \lsim \frac{1}{3}
(\frac{n}{s})\simeq 0.09$, where we have used the entropy per particle
of a free gas~\cite{Danielewicz:1984ww}. The uncertainty argument has
never been made precise, since the mean free path estimate is not valid at
strong coupling.  However, the holographic duality calculation was discovered to be
quite general. It was shown that the strong coupling limit of $\eta/s$
is universal in a large class of theories that have gravitational duals,
and that the $O(\lambda^{-3/2})$ corrections are positive
\cite{Buchel:2003tz,Iqbal:2008by}. These observations led Kovtun, Son
and Starinets to make the conjecture that
\be
\label{kss_bound}
\frac{\eta}{s} \geq \frac{1}{4\pi}
\ee
is a universal bound~\cite{Kovtun:2004de} that applies to all fluids.
The status of this conjecture is discussed in more detail in
Sec.~\ref{sec:hd}, and sketched in Fig.~\ref{fig:ratio}.
There are some known counterexamples involving
theories with gravitational duals described by higher derivative
gravity. However, it seems clear that Eq.~(\ref{kss_bound}) applies
to a large class of theories that include generalizations of QCD.

Experiments on collective flow in heavy ion collisions at RHIC
and the LHC (discussed in Sec.~\ref{sec_bulk}) indicate that the
viscosity of the QGP near the critical temperature is indeed close
to the proposed bound, and that equilibration must be very rapid.
Together with the observation of a large energy loss of highly
energetic probes of the plasma these results led to the conclusion
that the QGP produced at RHIC must be strongly coupled
\cite{Gyulassy:2004zy,rhic:2005}.

 In the case of the SUSY Yang-Mills plasma one can show that strong
coupling implies the absence of well defined quasiparticles. This
can be seen most clearly by studying the spectral function $\rho(
\omega)$ of the stress tensor correlation function. Kubo's formula
relates the value of the spectral function at zero energy to the
shear viscosity. The spectral function at finite energy carries
information about the physical excitations that contribute to momentum
relaxation. In weak coupling the spectral function has a peak at zero
energy. The width of the peak, $\Gamma\sim g^4\log(g)T$, is related to
the quasiparticle life time. The spectral function in the strong
coupling limit was computed in~\cite{Teaney:2006nc,Kovtun:2006pf}.
It was found that the spectral density is completely featureless:
the intercept at zero energy smoothly connects to the continuum
contribution $\rho(\omega)\sim \omega^4$.

 In the case of the QGP the evidence regarding the existence of
quasiparticles is ambiguous. The stress tensor correlation
function in lattice QCD was studied by H.~Meyer~\cite{Meyer:2007ic}.
The reconstructed spectral function is smooth, but the resolution was
not sufficient to exclude the presence of quasiparticles. There
are a number of results that have been interpreted as favoring
the existence of quasiparticles. One is the fact that lattice
calculations of fluctuations of conserved charges,  like baryon
number and strangeness, are compatible with the behavior of a free
gas, even near $T_c$~\cite{Koch:2005vg}. An experimental observation
that has been cited as evidence for the presence of quasiparticles
is the approximate quark-number scaling of the elliptic flow
parameter $v_2$~\cite{Voloshin:2002wa}. Another experimental
observable that may shed some light on the quasiparticle structure
is the heavy quark diffusion constant $D$, which we will discuss in
Sec.~\ref{sec_heavyq}. The main observation is that kinetic theory
and holographic duality make very different predictions about
the relation between the momentum relaxation time $\eta/(sT)$ and the
heavy quark relaxation time $m_QD/T$~\cite{Shuryak:2008eq}.

%%%%%%%%%%%%%%%%%%%%%%%%%%%%%%%%%%%%%%%%%%%%%%%%%%%%%%%%%%%%%%%%%%%%%%%%%%%%
\begin{figure}
\begin{center}
\includegraphics[width=12cm]{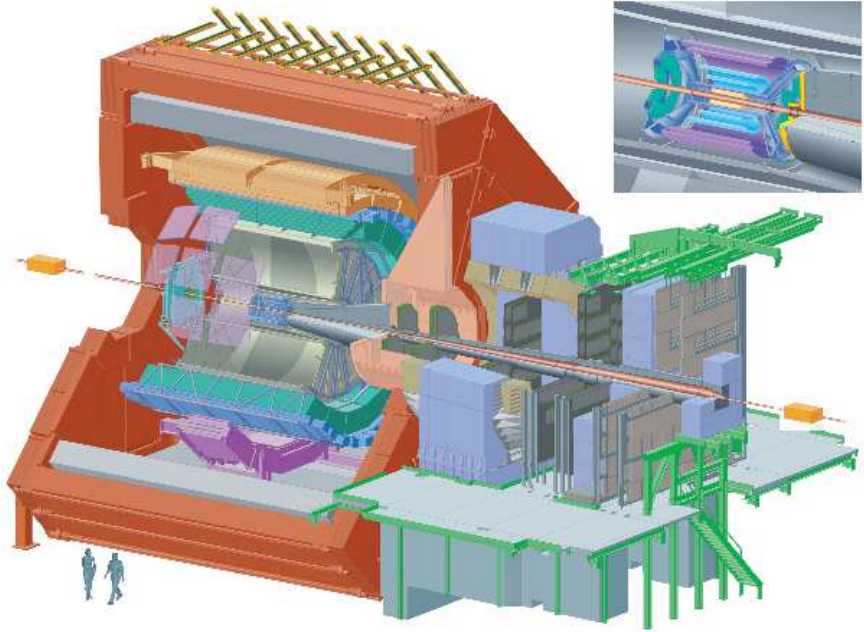}
\end{center}
\caption{\textit{Schematic layout of the ALICE detector at the Large Hadron
Collider (LHC).} The beam lines are the thin yellow tubes entering
from both sides. The outermost (red) layer is a magnet. The magnetic
field helps to discriminate between particles of different charge-to-mass
ratios. The central gray barrel is a time projection chamber and the
outer layers inside the magnet contain calorimeters, transition radiation
and time-of-flight detectors. The innermost layers shown in the blow-up
contain silicon detectors for precise determination of the initial
production vertices.   }
\label{fig_atlas}
\end{figure}
%%%%%%%%%%%%%%%%%%%%%%%%%%%%%%%%%%%%%%%%%%%%%%%%%%%%%%%%%%%%%%%%%%%%%%%%%%%

%%%%%%%%%%%%%%%%%%%%%%%%%%%%%%%%%%%%%%%%%%%%%%%%%%%%%%%%%%%%%%%%%%%%
\subsection{Nuclear Collisions: Initial Conditions}
\label{sec_hic}
%%%%%%%%%%%%%%%%%%%%%%%%%%%%%%%%%%%%%%%%%%%%%%%%%%%%%%%%%%%%%%%%%%%%

 The experimental study of heavy ion collisions at relativistic energies,
which were long seen as the best environments for the production of the
QGP, began in the late 1980's with fixed target programs both at CERN near
Geneva, and Brookhaven National Laboratory (BNL), outside of New York City;
see~\cite{Stock:2004cf,Satz:2008zzb,Baym:2001in} for historical
overviews.
These two programs were essential in building up a vital experimental
community, formed of larger and larger collaborations as the experiments
became more and more sophisticated. The programs at CERN and BNL demonstrated
that heavy ion collisions produce strongly interacting, approximately
equilibrated matter. The next step was taken with the turn-on of the
Relativistic Heavy Ion Collider (RHIC) at Brookhaven in 2000 which
brought heavy ion physics into the collider era, with accessibility
to processes calculable with perturbative QCD.
Colliders are accelerators with counter-circulating beams in which
the entire collision energy is available for particle production. RHIC
collides gold ions with a beam energy such that every nucleon in each
of the beams has an energy of 100 GeV. Since the rest mass energy of a
nucleon is about 1 GeV, this implies a relativistic $\gamma$ factor of
about 100. When two bunches of counter rotating beams overlap in each
of the RHIC experimental halls, at a rate of thousands of times per
second, the one billion ions in each bunch typically induce less than
one collision.  In a fraction of these collisions, particularly the
ones where the nuclei are head-on, thousands of particles are produced
by the conversion of kinetic energy into mass energy.

The particles are recorded in detectors which were originally located
in four experimental areas located around the ring.  The early RHIC
program had two large multipurpose detectors, STAR and PHENIX, with
roughly 500 collaborators each, and two smaller detectors, BRAHMS
and PHOBOS, with about 50 collaborators each~\cite{rhic:2005}. STAR
and PHENIX are both large spectrometers.  STAR focuses primarily
on charged hadrons using a \textit{time projection chamber} of 4m diameter that
creates a 3D image of collision events at a relatively low rate.
 PHENIX is a combination of drift chambers, particle identification
counters and calorimeters that make precise measurements of both charged
particles and photons. Both of these detectors were inspired by the collider
detectors of the previous generation and make measurements mainly at large
angles with respect to the beam directions. To make sure physics was not
missed in other regions of phase space, BRAHMS and PHOBOS were both
designed to make more limited measurements near the beam axis, BRAHMS
with a narrow-band spectrometer and PHOBOS with a large-coverage
single-layer silicon detector, sensitive to roughly 75\% of the total
number of charged particles produced per event.  A range of the data
from these experiments will be shown in later sections, and in several
of the contributions to this Focus Issue
\cite{Steinberg,Snellings:2011sz,Nagle:2011uz}. As of 2011, RHIC
has completed
its eleventh experimental run, after colliding ions with a wide range
of energies (from 7.7 GeV to 200 GeV) and nuclear species (protons,
deuterons, copper and gold).

%%%%%%%%%%%%%%%%%%%%%%%%%%%%%%%%%%%%%%%%%%%%%%%%%%%%%%%%%%%%%%%%%%%%%%%%%%%%
%\begin{figure}
%\begin{center}
%\includegraphics[width=8cm]{PHOBOSevent.eps}
%\end{center}
%\caption{Event display taken by the PHOBOS detector.}
%\label{fig_atlas}
%\end{figure}
%%%%%%%%%%%%%%%%%%%%%%%%%%%%%%%%%%%%%%%%%%%%%%%%%%%%%%%%%%%%%%%%%%%%%%%%%%%

As RHIC was taking data, the Large Hadron Collider (LHC) at CERN was
being built, along with its three large detector systems ALICE, ATLAS
and CMS.  The LHC provides
heavy ion collisions  of primarily lead, slightly larger than gold,  at
center of mass energies up to 28 times that reached by RHIC. This
translates into higher particle densities and temperatures and, even
more significantly, into much higher rates for large momentum-transfer
processes,  such as jets, photons, and heavy flavor, discussed below.
The ALICE detector was designed around another large time projection
chamber, similar to but larger than the one in STAR at RHIC, with a set
of additional detectors to identify different hadron species as well
as photons and electrons in limited angular regions. The layout of
the ALICE detector is shown in Fig.~\ref{fig_atlas}. The two larger
experiments, ATLAS and CMS, were general purpose detectors deploying
charged particle tracking and hermetic calorimetry over a large solid
angle.  However, the stringent design requirements to search for new high
mass particles led to detectors that were quite capable for the higher
multiplicities expected in heavy ion collisions at the LHC, and which have
excellent capabilities for high energy processes, similar to that needed
for Higgs and searches beyond the Standard Model. As of late 2011, the
LHC has completed its second lead ion run, as well as a first feasibility
study for future proton lead collisions.

%%%%%%%%%%%%%%%%%%%%%%%%%%%%%%%%%%%%%%%%%%%%%%%%%%%%%%%%%%%%%%%%%%%%%%%%%%%%
\begin{figure}
\begin{center}
\includegraphics[width=6.5cm]{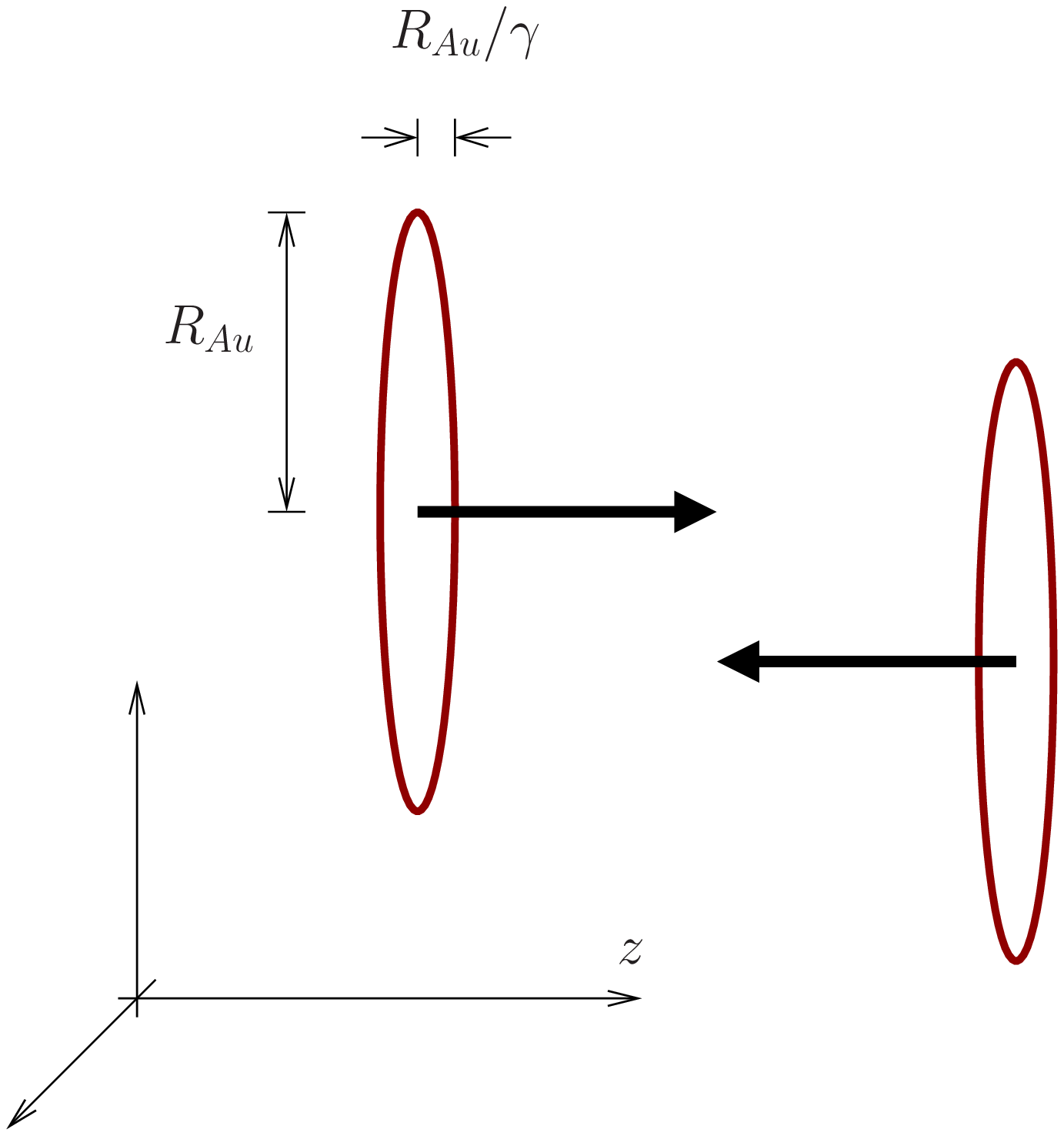}
\hspace{1cm}
\includegraphics[width=6.5cm]{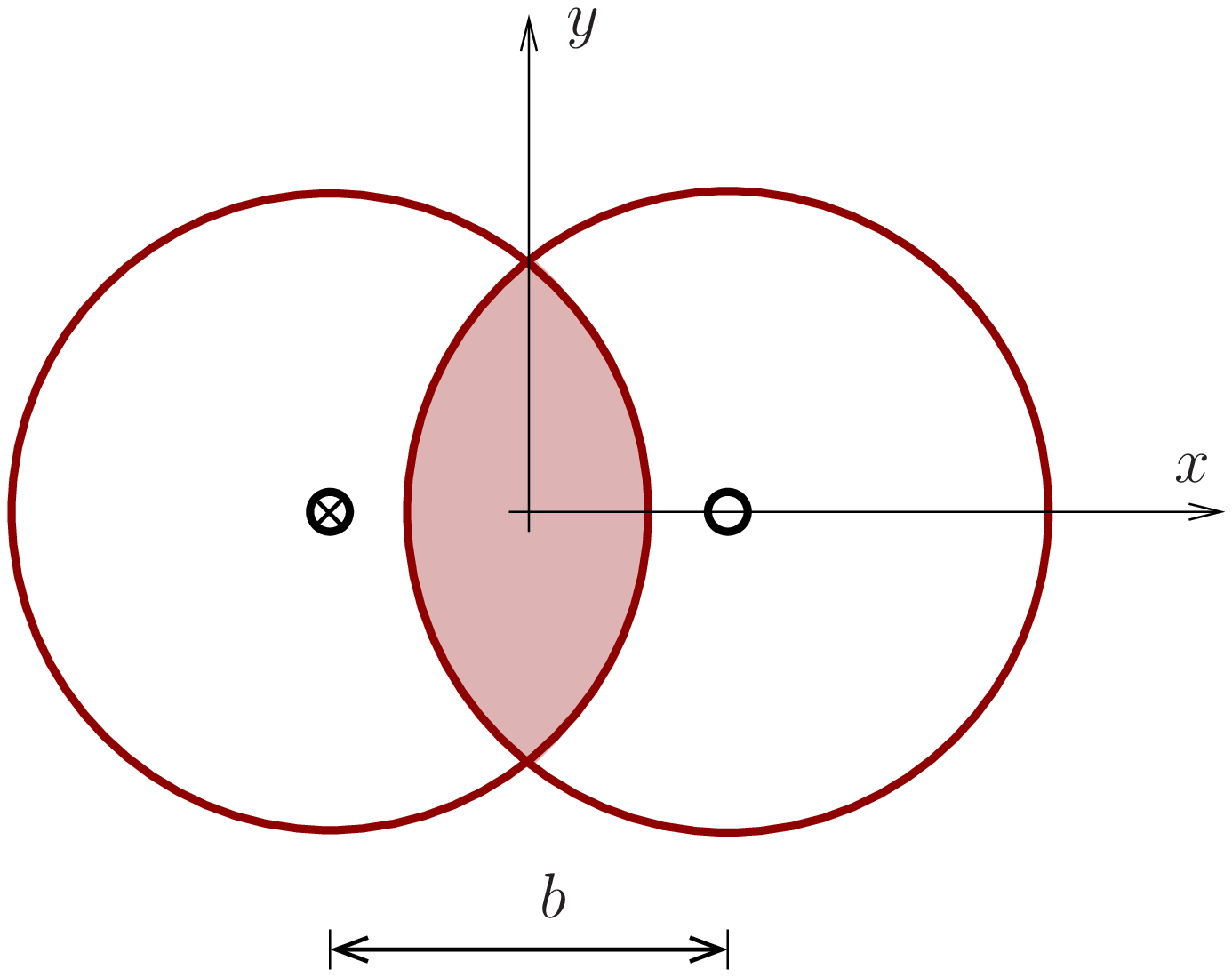}
\end{center}
\caption{\textit{Geometry of a high energy heavy ion collision.} Left:
collision of two Lorentz contracted gold nuclei.
The beam direction is the $z$-axis. Right:
same collision in the transverse plane. The impact parameter is
along the $x$-axis, and the remaining transverse direction is
the $y$-axis.}
\label{fig_coll_geom}
\end{figure}
%%%%%%%%%%%%%%%%%%%%%%%%%%%%%%%%%%%%%%%%%%%%%%%%%%%%%%%%%%%%%%%%%%%%%%%%%%%

In the following we will concentrate on results from Au+Au at the top
RHIC energy of $100$ GeV per nucleon. The transverse radius of a Au
nucleus is approximately 6 fm, and the duration of a heavy ion event
is $\tau \sim (6-10)$ fm/c. The estimate for the lifetime comes from
hydrodynamic simulations which we will describe in Sec.~\ref{sec_flow}.
The simplest observable in a heavy ion experiment, typically published
very soon after a new machine becomes operational, is the total
multiplicity of produced particles. In Au+Au collisions at 100 GeV per
nucleon the total multiplicity is about 7000;  see Sec.~\ref{sec_bulk}.
Somewhat more detailed information is provided by the spectra $dN/d^3p$
of produced particles (e.g. Ref.~\cite{Adams:2003xp,Adler:2003cb}).
The momenta can be decomposed into a transverse momentum $p_T^2=p_x^2
+p_y^2$ and a longitudinal momentum $p_z$;  see Fig.~\ref{fig_coll_geom}.
In the relativistic regime the natural variable to describe the motion
in the $z$ direction is the rapidity,
\be
 y = \frac{1}{2}\log\left(\frac{E+p_z}{E-p_z}\right).
\ee
At RHIC the energy of the colliding nuclei is 100+100 GeV per
nucleon, and the separation in rapidity is $\Delta y=10.6$.

%%%%%%%%%%%%%%%%%%%%%%%%%%%%%%%%%%%%%%%%%%%%%%%%%%%%%%%%%%%%%%%%%%%%
\begin{figure}
\begin{center}
\includegraphics[width=14cm]{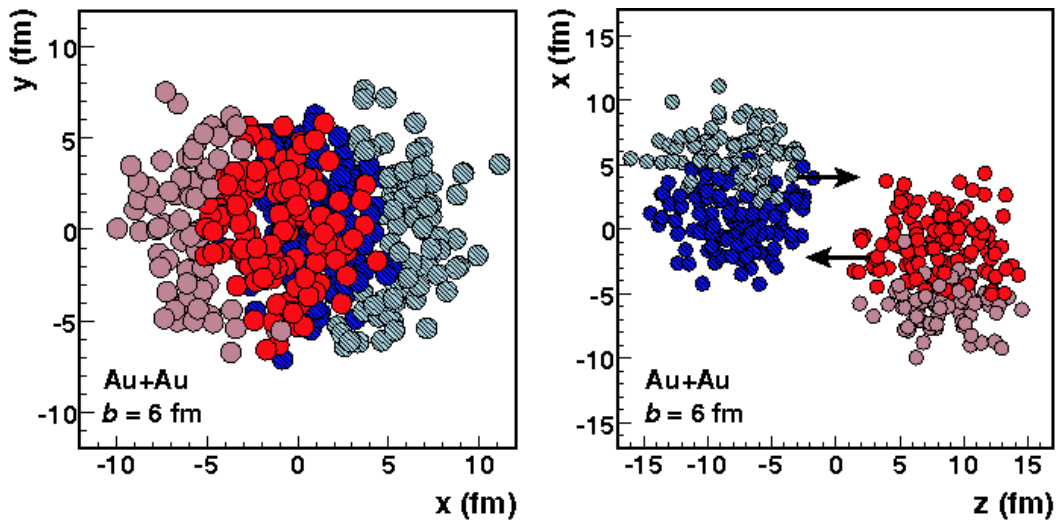}
\end{center}
\caption{
Glauber Monte Carlo calculation showing the collision of two gold
nuclei at an impact parameter $b=6$ fm   head-on (left panel)
and from a side view (right panel). The Glauber model is a geometric
model for the high energy scattering of composite quantum mechanical
states. Spectator nucleons from the two nuclei are shown in pale
red and blue, participants are shown as dark red and blue circles.
Figure from~\cite{Miller:2007ri}.}
\label{glauber}
\end{figure}
%%%%%%%%%%%%%%%%%%%%%%%%%%%%%%%%%%%%%%%%%%%%%%%%%%%%%%%%%%%%%%%%%%%%

 A simple picture of the initial state of the fireball created
in the collision was suggested by Bjorken~\cite{Bjorken:1983}.
He proposed that the two highly Lorentz contracted nuclei
pass through each other and create a longitudinally expanding
fireball in which particles are produced. In the original model
the number of produced particles is independent of rapidity,
and the subsequent evolution is invariant under boosts along the
$z$ axis. A simple model for the initial energy density in the
transverse plane is the Glauber model~\cite{Glauber:1959,Miller:2007ri}.
The Glauber model is based on the observation that high energy
scattering can be described in the eikonal (geometric optics)
approximation. The initial entropy density in the transverse
plane is
\bea
\label{Glauber}
 s({\bf x}_\perp,b) &\propto&
   \,   T_A({\bf x}_\perp+{\bf b}/2) \Big[1 - \exp\left(-\sigma_{NN}
   \,   T_A({\bf x}_\perp-{\bf b}/2) \right)\Big] \nonumber  \\
 & & +
   \,  T_A({\bf x}_\perp-{\bf b}/2) \Big[1  - \exp\left(-\sigma_{NN}
   \,  T_A({\bf x}_\perp+{\bf b}/2) \right) \Big] \, ,
\eea
where ${\bf b}$ is the impact parameter,
\be
\label{T_A}
T_A({\bf x}_\perp) = \int dz \, \rho_A({\bf x})
\ee
is the thickness function, ${\bf x}\equiv(x,y,z)$ and ${\bf x}_\perp\equiv(x,y)$, and $\sigma_{NN}(\sqrt{s})$ is the
nucleon-nucleon cross section. We also define the nuclear density as
$\rho_A({\bf x})$. The idea behind the Glauber model is that the
initial entropy density is proportional to the number of nucleons
per unit area which experience an inelastic collision, which we
call the \textit{number of participants}, $N_{\mathrm{part}}$.  Other variants
exist. For instance, one can distribute the energy density according
to the number of binary nucleon-nucleon collisions, $N_{\mathrm{coll}}$;
see Ref.~\cite{Kolb:2001qz} for a comparison. A more sophisticated
theory of the initial energy density is provided by the \textit{color glass
condensate} (CGC)~\cite{McLerran:1993ni,Hirano:2005xf}. This model
leads to somewhat steeper initial transverse energy density
distributions.

It should also be noted that the simple Glauber model is typically not
used by modern calculations, since it does not account for the large
fluctuations which are currently attributed to the event-wise variations
in the nucleon configuration coming from each nucleus and the collision
process itself.  Figure~\ref{glauber} illustrates this by means of a single
event simulated by a  \textit{Glauber Monte Carlo} code, which counts participants
and collisions by means of a simple prescription, that collisions occur
when two nucleons are within $d<\sqrt{\sigma_{NN}/\pi}$, where $\sigma_{NN}$
is the total nucleon-nucleon inelastic cross section~\cite{Miller:2007ri}.

%%%%%%%%%%%%%%%%%%%%%%%%%%%%%%%%%%%%%%%%%%%%%%%%%%%%%%%%%%%%%%%%%%%%
\subsection{Particle Multiplicities}
\label{sec_bulk}
%%%%%%%%%%%%%%%%%%%%%%%%%%%%%%%%%%%%%%%%%%%%%%%%%%%%%%%%%%%%%%%%%%%%

%%%%%%%%%%%%%%%%%%%%%%%%%%%%%%%%%%%%%%%%%%%%%%%%%%%%%%%%%%%%%%%%%%%%
\begin{figure}
\begin{center}
\includegraphics[width=7cm]{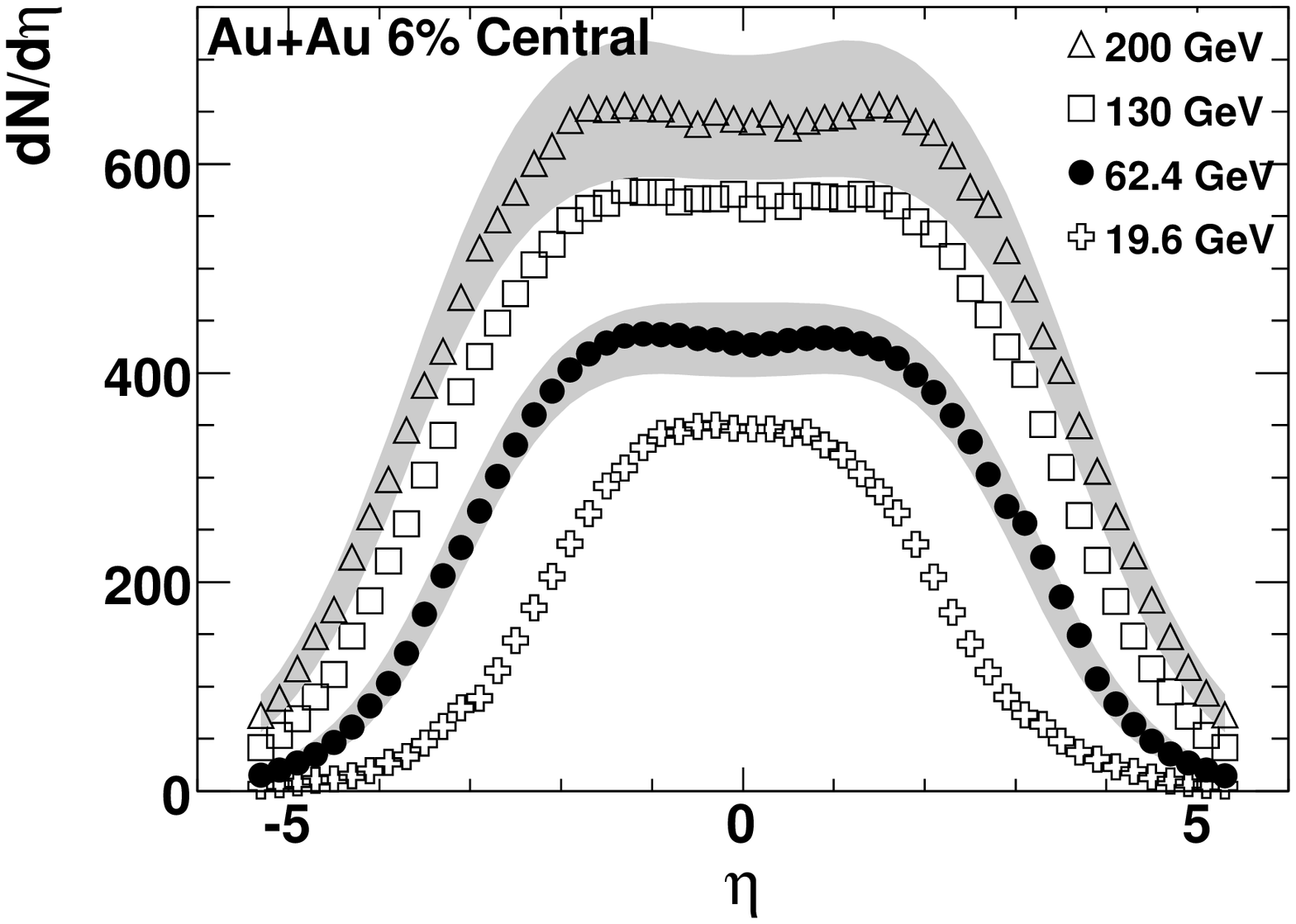}
\includegraphics[width=7cm]{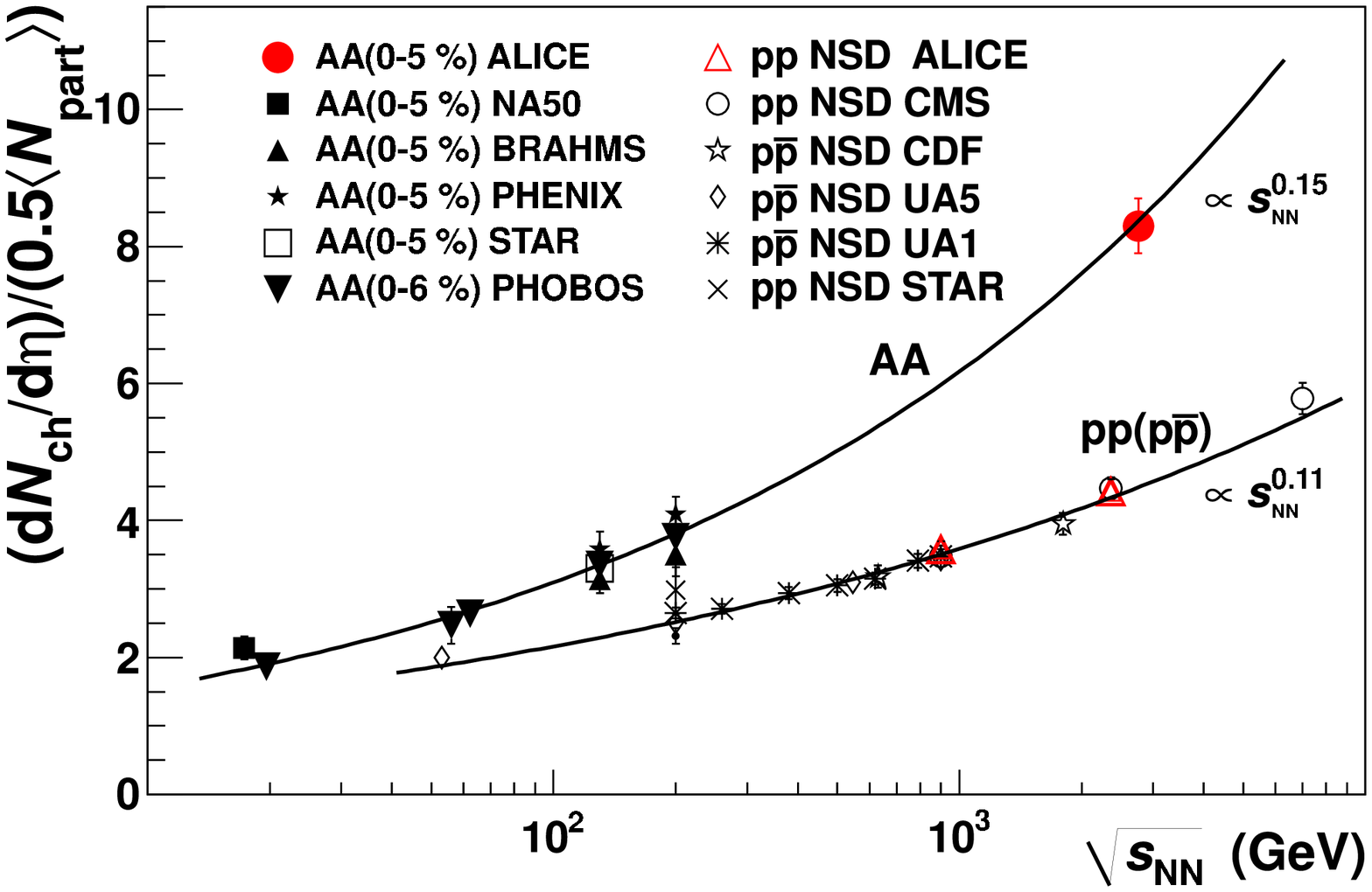}
\end{center}
\caption{
Left: number of charged particles per unit
pseudorapidity $\eta$ as as a function of $\eta$ for $Au+Au$ collisions
at several different energies. Pseudorapidity is defined like
rapidity, but using the approximation $E \simeq |\vec{p}|$, and is
more easily measured experimentally than the rapidity $y$. For light
particles, such as pions, $\eta\simeq y$;  note, however, that
corrections to this relation are biggest at $\eta=0$. Data from
the PHOBOS collaboration at RHIC~\cite{rhic:2005}.
Right: multiplicity $dN/d\eta$ per participant
pair at midrapidity for both nucleon-nucleon and nucleus-nucleus
collisions as a function of the collision energy. Data and
compilation of earlier results from the ALICE collaboration at
the LHC; see~\cite{Aamodt:2010pb} for original references.}
\label{fig_mult}
\end{figure}
%%%%%%%%%%%%%%%%%%%%%%%%%%%%%%%%%%%%%%%%%%%%%%%%%%%%%%%%%%%%%%%%%%%%

Particle multiplicities are important both as first measurements at
new machines, but also as a means of testing the current theoretical
approaches in their ability to predict the degrees of freedom available
at higher energies.

 From the theoretical perspective, multiplicities are argued to be a
good proxy for the initial gluon density.  This relies on one essential
non-trivial argument, that subsequent scatterings in the QGP, after
the initial phase where hard collisions occur and energy is deposited
in the interaction region, do not increase the overall entropy, and
thus do not increase the overall multiplicity.  This is certainly not
exactly true, as both thermalization and viscous effects during the
hydrodynamic evolution will produce some amount of entropy,  but there
is evidence that the total entropy is dominated by initial particle
production~\cite{Fries:2009wh}. Assuming that entropy is conserved
one can derive the  \textit{Bjorken estimate} for the initial entropy density,
\be
 s_0 = \frac{3.6}{\pi R^2\tau_0}\left(\frac{dN}{dy}\right)\, ,
\ee
where $\tau_0$ is the thermalization time, $R$ is the nuclear radius
and $dN/dy$ is the total number of particles per unit rapidity. This
estimate is based on a model of the hydrodynamic evolution which we
will discuss in more detail in Sec.~\ref{sec_flow}. The left panel in
Fig.~\ref{fig_mult} shows experimental results for $dN/dy$ of charged
particles at RHIC. Assuming that $N({\rm all})=1.5N({\rm charged})$
(most particles are pions) we get $dN/dy|_{y=0}\simeq 975$ for Au+Au
collisions as 100 GeV/nucleon. A conservative estimate for the
equilibration time is $\tau_0=1$ fm/c; see Sec.~\ref{sec_qgp_open}.
This corresponds to an initial entropy density $s_0\simeq 30\,
{\rm fm}^{-3}$. For a weakly interacting QGP this implies an initial
temperature $T_0 \simeq 230$ MeV, and an initial energy density
$\epsilon_0=5\,{\rm GeV}/{\rm fm}^3$. This number is significantly
larger than the critical energy density for forming a QGP, $\epsilon_{\mathrm{crit}}\simeq 1\,{\rm GeV}/{\rm fm}^3$
\cite{Bazavov:2009zn}.

 Experiments typically focus on studying how multiplicities scale with
beam energy and collision geometry. Empirically the multiplicity per
unit rapidity scales as a  small power of the beam energy;  see
the right panel of Fig.~\ref{fig_mult}. In nucleus-nucleus collisions
data ranging from the fixed target SPS program all the way to the
LHC are described by $dN/dy|_{y=0}\sim s^\alpha$, where $s$ is the
square of the center of mass energy in the nucleon-nucleon system
and $\alpha\simeq 0.15$. In nucleon-nucleon collisions the exponent
is somewhat smaller, $\alpha\simeq 0.11$.

 As a first approximation, it is
natural to expect that heavy ion collisions can be built by a linear
superposition of proton-proton or,  more accurately, nucleon-nucleon
collisions.  This is the basis of the  \textit{wounded nucleon model} of the
1970's which was invoked to explain how multiplicities in proton-nucleus
collisions tended to scale as $N_{ch}^{p+A} = N_{ch}^{p+p} \times N_{\mathrm{part}}
/2$~\cite{Bialas:1977pd,Bialas:1976ed}. In this expression, $N_{\mathrm{part}}$ is
the number of participating nucleons estimated either using a Glauber
model, or by counting the number of slow proton tracks, e.g. in emulsion.
In this picture, soft particle production was expected to result from
the excitation and subsequent independent decay of individual nucleons.
Hard processes, in which single  \textit{partons} (quarks and gluons) from one
nucleon scattered off partons of another nucleon, were also discovered
in the 1970's.  As these were not simply decay products of a soft excitation
but were short-range elastic scatterings, they were expected to scale
linearly with the number of binary collisions, also predictable using a
Glauber model.

In order to compare to experimental results,  calculations typically scale the
produced multiplicity by the number of participants, since the multiplicity
is dominated by soft particles,  with transverse momenta under 2 GeV.
The experiments use the observed distribution of multiplicities at
a fixed beam energy to estimate the  \textit{centrality} of each collision,
usually expressed as a percentage of the total inelastic cross section.
The 10\% highest multiplicity events are the  0-10\% centrality bin,
the next highest are the  10-20\% and so on.  These bins are then
compared to the events with the 10\% smallest impact parameter, or
10\% highest multiplicity events, in either a simple Glauber model or
one which includes some model of the experimental
fluctuations~\cite{Miller:2007ri}.

The first comparisons of theory and experiment at RHIC using the multiplicity
in the most central events were surprising.  While many models predicted
a combination of soft and semi-hard physics processes, it was the models
that were predominantly composed of soft processes that came closest to
the experimental data.  This included particularly models based on parton
saturation, which were preferred by the early data.  The situation is a bit
different at the LHC, as described briefly in the contribution by
Steinberg~\cite{Steinberg}. At the higher energies, models which include a
combination of hard and soft processes seem to do better at predicting the
multiplicities in the most central collisions. However, the  \textit{centrality
dependence}, i.e.,  the relationship between the number of measured charged
particles and the number of participating nucleons (or effective
nucleon-nucleon collisions), is essentially the same at RHIC and the
LHC~\cite{Aamodt:2010cz}. At present it is not clear how to reconcile
this observation with the apparent increase in the fraction of bulk
particle production stemming from hard processes.

%%%%%%%%%%%%%%%%%%%%%%%%%%%%%%%%%%%%%%%%%%%%%%%%%%%%%%%%%%%%%%%%%%%%
\subsection{Hydrodynamic Flow}
\label{sec_flow}
%%%%%%%%%%%%%%%%%%%%%%%%%%%%%%%%%%%%%%%%%%%%%%%%%%%%%%%%%%%%%%%%%%%%

 The particle density measured by the early experiments is not only
valuable to compare with different production models. It is also an
essential empirical input that can be used to determine the initial
conditions for hydrodynamical models of the subsequent evolution. The
hydrodynamic description is valid if the system is in local thermal
equilibrium, and if the thermodynamic variables are varying smoothly.
If the system has a kinetic description in terms of quasiparticles this
implies that the ratio of the mean free path over the characteristic
length scale of the system,  known as the  \textit{Knudsen number}, must be
small. There is evidence that this condition is satisfied during
the early stages of the evolution, but it must eventually break
down during the late, dilute, stage of the evolution,
since the mean free path increases with diluteness.  The time
when the mean free path becomes larger than the system
size,  more precisely, the expansion time multiplied by the mean
velocity, is called  \textit{freezeout}.

There are several experimental observations that provide evidence
for the assumption that heavy ion collisions create a thermalized
state.  The first observation is that the overall abundances of produced
particles, including very rare ones, is described by a simple thermal
model that contains only two parameters, the freezeout temperature $T$
and baryon chemical potential $\mu$~\cite{BraunMunzinger:2009zz}. The
second observations is that for momenta less than about 2 GeV  the spectra
$dN/d^3p$ of produced particles follow a Boltzmann distribution
characterized by the freezeout temperature and a collective radial
expansion velocity~\cite{Heinz:2009xj}. This  \textit{radial flow} is clearly
seen from the fact that the spectra of heavy particles, which receive
a larger momentum boost from the collective velocity field, have a
larger apparent temperature than the spectra of light particles.

The third piece of evidence in support of not only thermalization,
but early thermalization, that means equilibration significantly
before freezeout, is the observation of strong azimuthal
anisotropies, typically called  \textit{elliptic flow}, in
non-central heavy ion collisions. Elliptic flow represents the
collective response of the system to pressure gradients in the
initial state. The analogous phenomenon in ultracold atomic gases
is discussed in Sec.~\ref{ssec:experiments}.
At finite impact parameter the initial state has
the shape of an ellipse, with the short axis along the $x$-direction,
and the long axis along the $y$-direction;  see Fig.~\ref{fig_coll_geom}.
This implies that pressure gradients along the $x$-axis are larger than
along the $y$-axis. Hydrodynamic evolution converts the initial pressure
gradients to velocity gradients in the final state. Elliptic flow is
sensitive to early thermalization because the initial anisotropy that
drives elliptic flow disappears with time. Elliptic flow is quantified
in terms of the second Fourier coefficient $v_2$ of the particle
distribution in the transverse plane,
\be
\label{v_2}
 \left. p_0\frac{dN}{d^3p}\right|_{p_z=0} = v_0(p_T)
 \Big( 1 + 2v_1(p_T)\cos(\phi)
         + 2v_2(p_T)\cos(2\phi) +\ldots \Big) ,
\ee
where $\phi$ is the angle between the momentum vector and the $x$-axis.
Odd Fourier moments like $v_1$ and $v_3$ arise mainly from fluctuations
in the initial state~\cite{Alver:2010gr,Teaney:2010vd}. Techniques for
measuring $v_2$ and systematic trends in the results are described in
some detail in the contribution by Snellings~\cite{Snellings:2011sz}.
Snellings emphasizes that measurements of $v_2$ constrain the
equation of state and the transport properties of the QGP.
A similar conclusion is reached in the contribution of Lisa
et al.~\cite{NJPfocusissue9_lisa} to this Focus Issue. These authors focus on
direct measurements of the shape of the final state using
HBT (Hanbury-Brown-Twiss) interferometry.

 In this section we will explain how elliptic flow can be used
to constrain the ratio $\eta/s$ introduced in Sec.~\ref{sec_sQGP}.
In a relativistic fluid the equations of energy and momentum
conservation can be written as a single equation
\be
\label{rel_hydro}
 \partial_\mu T^{\mu\nu}= 0 \, ,
\ee
where $T^{\mu\nu}$ is the energy momentum tensor. In ideal fluid
dynamics the form of $T_{\mu\nu}$ is completely fixed by
Lorentz invariance,
\be
\label{T_ideal}
 T^{\mu\nu} = (\epsilon+P)u^\mu u^\nu + P \eta^{\mu\nu}\, ,
\ee
where $u^\mu$ is the fluid velocity ($u^2=-1$) and $\eta^{\mu\nu}=
{\rm diag}(-1,1,1,1)$ is the metric tensor. The hydrodynamic
equations have to be supplemented by an equation of state
$P=P(\epsilon)$. The four equations given in Eq.~(\ref{rel_hydro})
can be split into longitudinal and transverse components relative
to the fluid velocity. The longitudinal equation is equivalent to
entropy conservation,
\be
\label{rel_s_cont}
 \partial_\mu (su^\mu)=0\, ,
\ee
and the transverse equation is the relativistic Euler equation,
\be
\label{rel_euler}
 Du_\mu = -\frac{1}{\epsilon+P}\nabla_\mu^\perp P\, ,
\ee
where $D=u^\mu \partial_\mu$ and $\nabla_\mu^\perp=\partial_\mu-u_\mu D$.
We observe that the inertia of a relativistic fluid is governed by
$\epsilon+P$. In the non-relativistic limit we can have $u^\mu\simeq
(1,\vec{v})$ and $\epsilon+P\simeq \rho$ (energy density is dominated
by rest mass energy). We also find $D\simeq\partial_t+v\cdot\partial$
and $\vec{\nabla}^\perp\simeq \vec{\partial}$. These approximations
lead to the usual Euler equation. We also note that $\nabla_\mu^\perp
P = c_s^2\nabla_\mu^\perp \epsilon$, where $c_s$ is the speed of sound.
This implies that for a given initial energy density gradient the
resulting acceleration is determined by the speed of sound. Ideal
fluid dynamics corresponds to the limit that variations in the
hydrodynamic variables occur on scales much larger than the mean
free path. Dissipation arises from the leading gradient terms in
the energy momentum tensor. In the rest frame of the fluid these
corrections have the same form as in non-relativistic fluids. We
have
\be
\label{del_T_ij}
\delta T^{ij}= -\eta  \left(\partial^i u^j+\partial^j u^i
  - \frac{2}{3}\delta^{ij}\partial\cdot u \right)
 - \zeta \delta^{ij} \partial \cdot u \, ,
\ee
where $\eta$ and $\zeta$ are the shear and bulk viscosity.

The application of hydrodynamics to relativistic heavy ion
collisions goes back to the work of Landau~\cite{Landau:1953}
and Bjorken~\cite{Bjorken:1983}. Bjorken discussed a simple
scaling solution of the equations of fluid dynamics that corresponds
to the space-time picture discussed in Sec.~\ref{sec_hic}. This
solution provides a natural starting point for more detailed
studies in the ultra-relativistic domain. In the Bjorken
solution the initial entropy density is independent of rapidity,
and the subsequent evolution is invariant under boosts along
the $z$ axis. The evolution in proper time is the same for
all comoving observers. The flow velocity is
\be
\label{u_bj}
 u_\mu = \gamma(1,0,0,v_z)= (t/\tau,0,0,z/\tau),
\ee
where $\gamma=\sqrt{1-v_z^2}$ is the boost factor and $\tau=
\sqrt{t^2-z^2}$ is the proper time. The velocity field in
Eq.~(\ref{u_bj}) solves the relativistic Euler equation (\ref{rel_euler}).
In particular, there is no longitudinal acceleration. The remaining
hydrodynamic variables are determined by entropy conservation.
Eq.~(\ref{rel_s_cont}) gives
\be
\frac{d}{d\tau}\left[\tau s(\tau) \right] = 0
\ee
and $s(\tau)=s_0\tau_0/\tau$. For an ideal relativistic gas $s\sim T^3$
and $T\sim 1/\tau^{1/3}$. We saw in the previous section that the initial
entropy density is constrained by the final state multiplicity. Typical
parameters at RHIC are $\tau_0\simeq (0.6-1.6)$ fm and $T_0\simeq (300-425)$
MeV. We note that the initial temperature is significantly larger than
the critical temperature for the QCD phase transition. The temperature
drops as a function of $\tau$ and eventually the system becomes too dilute
for the hydrodynamic evolution to make sense. At this point, the
hydrodynamic description is matched to kinetic theory, generally
using the formalism described by Cooper and Frye in the early
1970's~\cite{Cooper:1974mv}, and the spectra of produced particles
are computed.

%%%%%%%%%%%%%%%%%%%%%%%%%%%%%%%%%%%%%%%%%%%%%%%%%%%%%%%%%%%%%%%%%%%%%
\begin{figure}[t]
\bc\includegraphics[width=10cm]{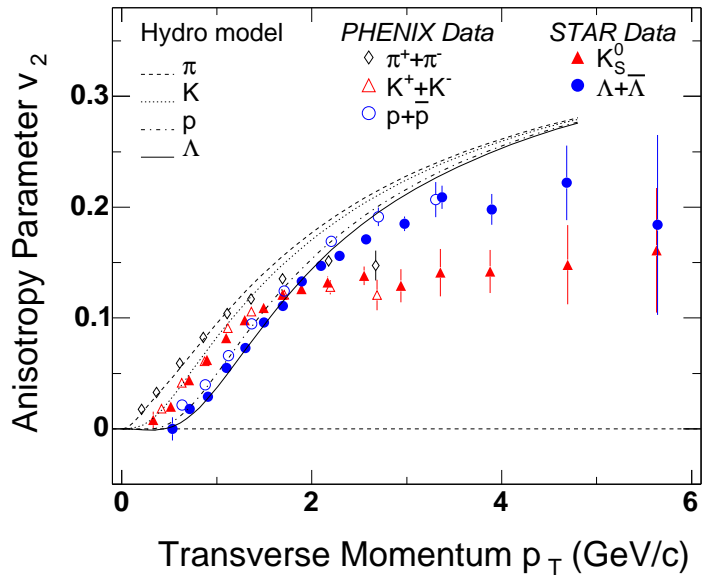}\ec
\caption{\label{fig_v2}
Differential elliptic flow $v_2(p_T)$ of identified
hadrons from minimum bias Au+Au collisions at 200 GeV
per nucleon compared to predictions from ideal (non-dissipative)
hydrodynamics, from~\cite{Heinz:2005zg}.}
\end{figure}
%%%%%%%%%%%%%%%%%%%%%%%%%%%%%%%%%%%%%%%%%%%%%%%%%%%%%%%%%%%%%%%%%%%%%

 In order to quantitatively describe the observed particle
distributions several improvements of the simple Bjorken model are
necessary. First, one has to include the transverse expansion of
the system~\cite{Baym:1984sr}. Second, one has to include deviations
from boost invariance in the longitudinal directions. Rapidity
distributions at RHIC most likely result from physics somewhere in
between the Bjorken scenario, which
assumes boost invariance, and the Landau picture, which assumes
complete stopping of the initial nuclei~\cite{Steinberg:2004vy}. One
also has to include realistic equations of state and take into account
the geometry of the initial state. Results for $v_2(p_T)$ obtained from
a calculation in ideal hydrodynamics are shown in Fig.~\ref{fig_v2}.
This calculation focuses on the central rapidity regime and maintains
the assumption of boost invariance. The figure shows the result for a
given centrality class, corresponding to a specific range of impact
parameters. Hydrodynamic calculations show that the elliptic flow
response $v_2$ is approximately linear in the spatial anisotropy
$\langle y^2-x^2\rangle/\langle x^2+y^2 \rangle$ of the initial
state. We observe that ideal hydrodynamics provides an excellent
fit to the RHIC data for $p_T\lsim (1.0-1.5)$ GeV, depending
on the particle species.  This includes the observed
hierarchy in the $p_T$ dependence of $v_2$ for different species
at low $p_T$. The mass splitting of $v_2(p_T)$ reflects an approximate
transverse energy $E_T=(p_T^2+m^2)$ scaling of the particle spectra
in hydrodynamics.

 Having established a baseline description of the spectra using
ideal hydrodynamics we can now discuss the role of dissipative
effects. We begin with the effect of shear and bulk viscosity
on the Bjorken solution. The scaling flow given in Eq.~(\ref{u_bj})
is a solution of the relativistic Navier-Stokes equation. If the
transverse expansion of the system is neglected viscosity does
not affect the flow profile but it does generates entropy. We find
\be
\label{bj_ns}
\frac{1}{s}\frac{ds}{d\tau} = -\frac{1}{\tau}
 \left( 1 - \frac{\frac{4}{3}\eta+\zeta}{sT\tau}\right)\, .
\ee
The applicability of the Navier-Stokes equation requires that
the viscous correction is small~\cite{Danielewicz:1984ww},
\be
\label{DG}
\frac{\eta}{s}+\frac{3}{4}\frac{\zeta}{s}
\ll \frac{3}{4}(T\tau)\, .
\ee
For the Bjorken solution $T\tau\sim \tau^{2/3}$ grows with time,
and this condition is most restrictive during the early stages of
the evolution. Using $\tau_0=1$ fm and $T_0=300$ MeV gives $\eta/s<0.6$.
This result implies that hydrodynamics cannot be used in relativistic
heavy ions collisions unless the QGP is strongly coupled and the
shear viscosity is small.

It is instructive to study the viscous contribution to the stress
tensor in more detail. Neglecting bulk viscosity the stresses in
the central rapidity slice are given by
\be
T_{zz} = P -\frac{4}{3}\frac{\eta}{\tau},
\hspace{0.5cm}
T_{xx}=T_{yy} =  P + \frac{2}{3}\frac{\eta}{\tau} \, .
\ee
This means that shear viscosity decreases the longitudinal
pressure and increases the transverse one. In the Bjorken scenario
there is no acceleration, but if pressure gradients are taken into
account shear viscosity will tend to increase radial flow. At finite
impact parameter shear viscosity reduces the pressure along the
$x$-direction, and increases the pressure in the $y$-direction. As a
consequence there is less acceleration in the $x$-direction, and
elliptic flow is suppressed.

Viscosity modifies the stress tensor, and via the matching
to kinetic theory at freezeout,  this modification changes the
distribution functions $f_p$ of produced particles. In
Ref.~\cite{Teaney:2003kp} a simple quadratic ansatz for the
leading correction $\delta f$ to the distribution function
was proposed,
\be
\label{del_f}
\delta f_p = \frac{1}{2T^3} \frac{\eta}{s}f_0(1\pm f_0)
   p_\alpha p_\beta  \partial^{\langle \alpha} u^{\beta\rangle} \, ,
\ee
where $f_0$ is the Bose-Einstein/Fermi-Dirac distribution
and $\partial^{\langle \alpha} u^{\beta\rangle}$ is the symmetric
traceless tensor that appears in Eq.~(\ref{del_T_ij}). This
ansatz is a very good approximation to the result of a more
involved calculation using kinetic theory~\cite{Arnold:2000dr}.
The modified distribution function leads to a modification of the
single particle spectrum. For a simple Bjorken expansion and at
large $p_T$ we find
\be
\label{del_N}
 \frac{\delta (dN)}{dN_0} = \frac{1}{3\tau_fT_f}\frac{\eta}{s}
    \left( \frac{p_T}{T_f} \right)^2 \,,
\ee
where $dN_0$ is the number of particles produced in ideal
hydrodynamics, $\delta (dN)$ is the dissipative correction,
and $\tau_f$ is the freezeout time. There is an analogous formula
for the second Fourier moment of the spectrum, related to $v_2$
\cite{Teaney:2003kp}. We observe that the dissipative correction
to the spectrum is controlled by the same parameter $\eta/(s\tau T)$
that appeared in the entropy equation. We also note that the viscous
term grows with $p_T$. These results are in agreement with experiment:
Deviations from ideal hydrodynamics grow with $p_T$, and they are
larger in smaller systems (which freeze out earlier). More detailed
analyses can be found in the contributions by Snellings and
Nagle, Bearden and Zajc~\cite{Snellings:2011sz,Nagle:2011uz}. A
conservative bound for
$\eta/s$ at RHIC is $\eta/s<0.4$, but the best fits tend to give values
that are even smaller $\eta/s\simeq (0.1-0.2)$.

%%%%%%%%%%%%%%%%%%%%%%%%%%%%%%%%%%%%%%%%%%%%%%%%%%%%%%%%%%%%%%%%%%%%
\subsection{Jet Quenching}
\label{sec_jets}
%%%%%%%%%%%%%%%%%%%%%%%%%%%%%%%%%%%%%%%%%%%%%%%%%%%%%%%%%%%%%%%%%%%%

%%%%%%%%%%%%%%%%%%%%%%%%%%%%%%%%%%%%%%%%%%%%%%%%%%%%%%%%%%%%%%%%%%%%
\begin{figure}
\begin{center}
\includegraphics[width=12cm]{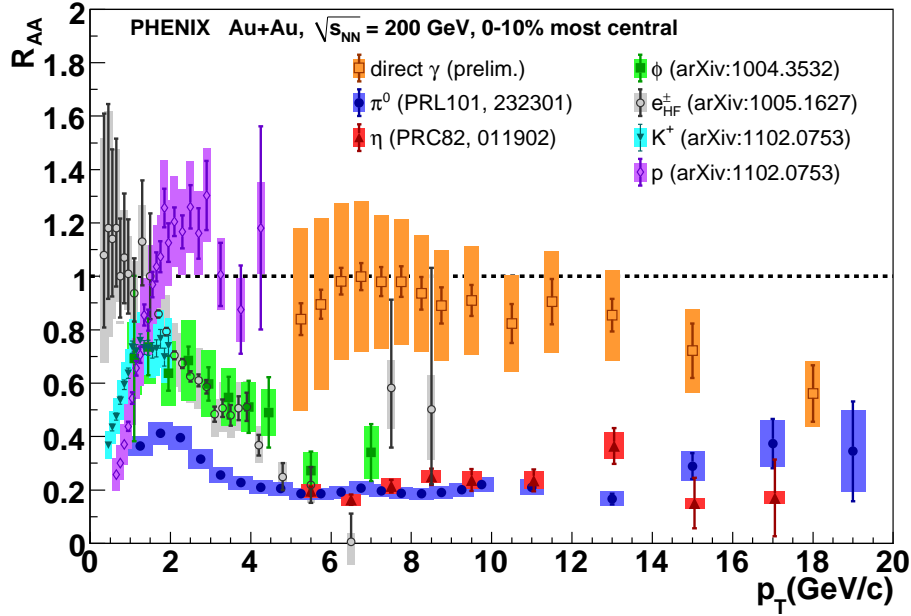}
\end{center}
\caption{
Nuclear suppression factor $R_{AA}$ for a variety of hadron species,
photons and electrons measured in PHENIX. Note that the dominant
source of electrons is the decay of heavy quarks (bottom and charm).
References to data are indicated on the figure itself. The definition
of $R_{AA}$ is discussed in the text. The value $R_{AA}=1$ implies that
for this particular probe a nucleus-nucleus collision behaves
like a simple superposition of nucleon-nucleon collisions. Figure
provided courtesy of Carla Vale, BNL.}
\label{raa_rhic}
\end{figure}
%%%%%%%%%%%%%%%%%%%%%%%%%%%%%%%%%%%%%%%%%%%%%%%%%%%%%%%%%%%%%%%%%%%%

Another way to directly probe the density of gluons present in the initial
state is to scatter fast partons, formed in the initial collisions, which
can be detected in the final state as hadronic jets. This idea was proposed
as far back as 1982 by Bjorken, placed on firmer theoretical footing in
the early 1990's, and finally discovered experimentally by RHIC experiments
in the early part of the 2000's~\cite{Adcox:2001jp}. Figure~\ref{raa_rhic}
shows the  \textit{nuclear suppression factor}, defined as
\begin{equation}
R_{AA} = \frac{1}{N_{\mathrm{coll}}} \frac{dN/dp_{T}^{AA}}{dN/dp_{T}^{pp}}\, ,
\end{equation}
i.e. the yield of a particle species in heavy ion collisions,  typically
measured as a function of its transverse momentum, divided by the similar
yield in proton-proton collisions, divided by the number of binary collisions
calculated from a Glauber model.  This ratio essentially tests the hypothesis
that each individual binary collision has an equal probability to induce a
hard process identical to that found in proton-proton collisions.
The latter acts as a reference system in which hard processes
can usually be calculated perturbatively~\cite{Ellis:1990ek}. As can
be seen in Fig.~\ref{raa_rhic}, only the production of direct photons
appears to be unmodified, and even then only below 12-13 GeV, above which
isospin effects are
expected~\cite{Arleo:2006xb}.  Conversely, light hadrons
like $\pi^0$ and $\eta$ are suppressed by about a factor of five
above 5 GeV and perhaps slowly rise at high $p_T$.  Electrons from
charmed hadron decays are unsuppressed at low $p_T$ but quickly fall
to a level similar to that found for light hadrons, as
discussed in more detail in the next section.

%%%%%%%%%%%%%%%%%%%%%%%%%%%%%%%%%%%%%%%%%%%%%%%%%%%%%%%%%%%%%%%%%%%%
\begin{figure}[t]
\begin{center}
\includegraphics[width=10cm]{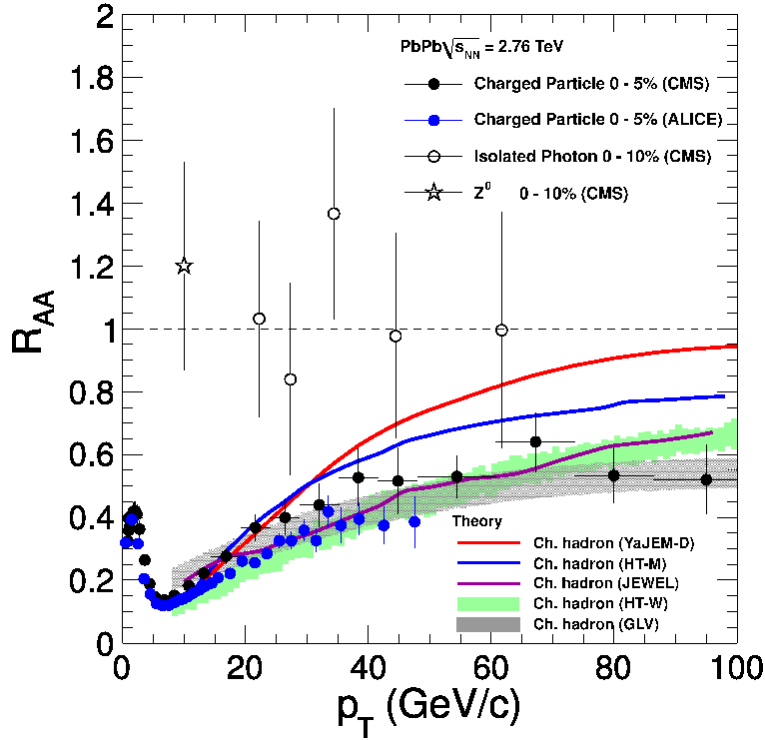}
\end{center}
\caption{
Nuclear suppression factor $R_{AA}$ for a variety of hadron species,
photons and $Z^0$ bosons measured in $Pb$-$Pb$ collisions at the LHC,
from \cite{Muller:2012zq}. The plot shows results from the CMS and
ALICE collaboration compared to various theoretical models. See
\cite{Aamodt:2010jd,CMS:2012aa} for the original data and
\cite{Muller:2012zq} for references to the theory predictions.}
\label{raa_lhc}
\end{figure}
%%%%%%%%%%%%%%%%%%%%%%%%%%%%%%%%%%%%%%%%%%%%%%%%%%%%%%%%%%%%%%%%%%%%

Significant effort has been devoted to the development of a perturbative
QCD-based formalism that describes the energy loss of fast partons moving
through a hot, dense medium.  The main transport parameter that appears
in the theory is the transverse momentum diffusion constant $\hat{q}$,
\cite{Baier:1996kr}
\be
\label{q-hat}
 \hat{q} = \rho \int dq_T^2 \, q_T^2 \, \frac{d\sigma}{dq_T^2}\, ,
\ee
which determines the mean transverse momentum  kick of the fast
parton per unit length traveled. Here, $d\sigma/dq_T^2$ is the differential
cross section for scattering of the parton off the constituents of the
dense medium, and $\rho$ is the density of the medium. Recent work has
focused on non-perturbative definitions of $\hat{q}$ that make no reference
to the structure of the dense medium;  see~\cite{CasalderreySolana:2011us}
for a review. The energy loss per unit length scales as $dE/dz\sim \hat{q}L$
for short path lengths ($L<L_{c}$), and as $dE/dz\sim \sqrt{\hat{q}E}$
for long path lengths. The characteristic length scale is
$L_{c}\sim \sqrt{E/\hat{q}}$.

Extracting $\hat{q}$ from experimental data on jet quenching at RHIC
has proven to be difficult. Based on an analysis of $R_{AA}$ using a
Monte-Carlo implementation of the theory of perturbative energy loss
\cite{Baier:1996kr,Salgado:2003gb} the PHENIX collaboration has reported
$\hat{q}=13.2^{+2.1}_{-3.2}{\rm GeV}^2/{\rm fm}$ (the $2\sigma$ errors are
$\hat{q}=13.2^{+6.3}_{-5.2}{\rm GeV}^2/{\rm fm}$)~\cite{Adare:2008cg}.
This number is large compared to expectations for a perturbative QCD
plasma, $\hat{q}\simeq (1-2){\rm GeV}^2/{\rm fm}$~\cite{Baier:2002tc}.
However, there are significant uncertainties associated with different
implementations of energy loss. Bass  \textit{et al.}~consider three different
methods and find that equally good descriptions for $R_{AA}$ can be
obtained for values of $\hat{q}$ ranging from $\hat{q}=2.5\,{\rm GeV}^2
/{\rm fm}$ to $\hat{q}=10\,{\rm GeV}^2/{\rm fm}$~\cite{Bass:2008rv}.

%%%%%%%%%%%%%%%%%%%%%%%%%%%%%%%%%%%%%%%%%%%%%%%%%%%%%%%%%%%%%%%%%%%%
\begin{figure}[t]
\begin{center}
\includegraphics[width=10cm]{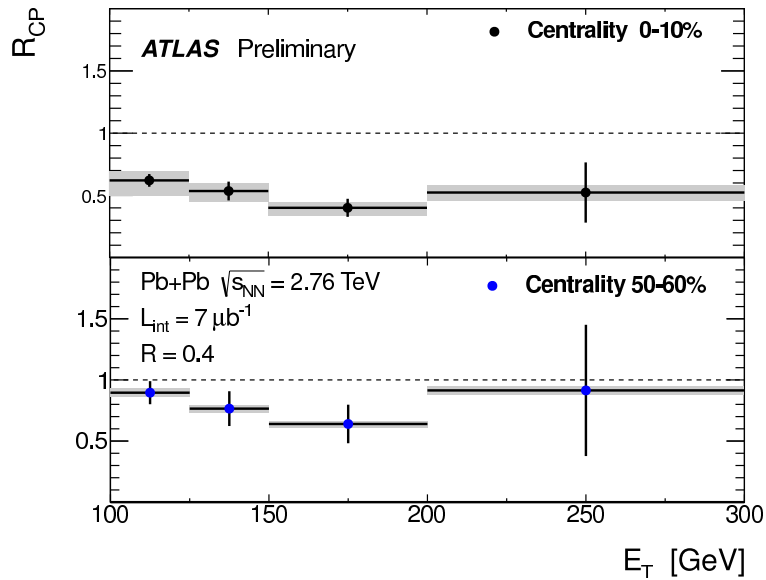}
\end{center}
\caption{
Jet central-to-peripheral ratio $R_{CP}$ for different centralities
in $Pb$-$Pb$ collisions at the LHC, measured by the ATLAS collaboration
\cite{Cole:2011zz}. The plots show the ratio of the jet yield in
central $(0-10)\%$ and semi-peripheral $(50-60)\%$ events relative
to the yield in very peripheral collisions as a function of the
transverse energy $E_T$ of the jet.}
\label{rcp_lhc}
\end{figure}
%%%%%%%%%%%%%%%%%%%%%%%%%%%%%%%%%%%%%%%%%%%%%%%%%%%%%%%%%%%%%%%%%%%%

The LHC offers significantly larger jet production cross sections,
a much increased $p_T$ range, and the capability of detailed studies of
not only leading particles but identified jets and jet shapes. First
results from the LHC are shown in Figs.~\ref{raa_lhc} and \ref{rcp_lhc}.
Figure~\ref{raa_lhc} shows the observable $R_{AA}$ introduced above. The
low $p_T$ behavior agrees with results at RHIC, despite the much larger
collision energy. At high $p_T$ the suppression is smaller, $R_{AA}\sim
0.5$, but there is no hint of $R_{AA}$ approaching 1. This is seen even
more dramatically in Fig.~\ref{rcp_lhc}, which shows the suppression
factor $R_{CP}$ for identified jets, not just high $p_T$ hadrons. The
ratio $R_{CP}$ is defined relative to the yield in very peripheral
nucleus-nucleus collisions. We observe that $R_{CP}$ is approximately
equal to 0.5 for jets with transverse energies as high as 300 GeV.
These results have important implications for the dependence of energy
loss on the energy density of the medium, but the theoretical analysis
of the LHC data is still very much in progress. A discussion of the
first jet data from the LHC can be found in the contribution in this
Focus Issue by Steinberg~\cite{Steinberg}.

 There are some important connections between $\hat{q}$ and other
transport parameters. In a quasiparticle description the cross
section in Eq.~(\ref{q-hat}) is the same cross section that governs
momentum diffusion of approximately thermal particles, i.e., shear
viscosity. This implies that $\eta/s\propto T^3/\hat{q}$, where the
constant of proportionality is roughly 1. In the context of kinetic
theory a small shear viscosity therefore implies strong jet quenching
\cite{Majumder:2007zh}. In detail the situation is more complicated.
One can show that the viscous correction to $v_2(p_T)$ at moderate
$p_T$ is proportional to $1/\sqrt{\hat{q}}$, not $1/\hat{q}$
\cite{Dusling:2009df}. Also, elliptic flow at large $p_T$ is mainly
a probe of the path length dependence of energy loss~\cite{Adare:2010sp}.
In the next section we will see that there are important connections
between energy loss of light quarks, and energy loss of heavy
quarks.

%%%%%%%%%%%%%%%%%%%%%%%%%%%%%%%%%%%%%%%%%%%%%%%%%%%%%%%%%%%%%%%%%%%%
\subsection{Heavy Quarks}
\label{sec_heavyq}
%%%%%%%%%%%%%%%%%%%%%%%%%%%%%%%%%%%%%%%%%%%%%%%%%%%%%%%%%%%%%%%%%%%%

%%%%%%%%%%%%%%%%%%%%%%%%%%%%%%%%%%%%%%%%%%%%%%%%%%%%%%%%%%%%%%%%%%%%
\begin{figure}
\begin{center}
\includegraphics[width=8.5cm]{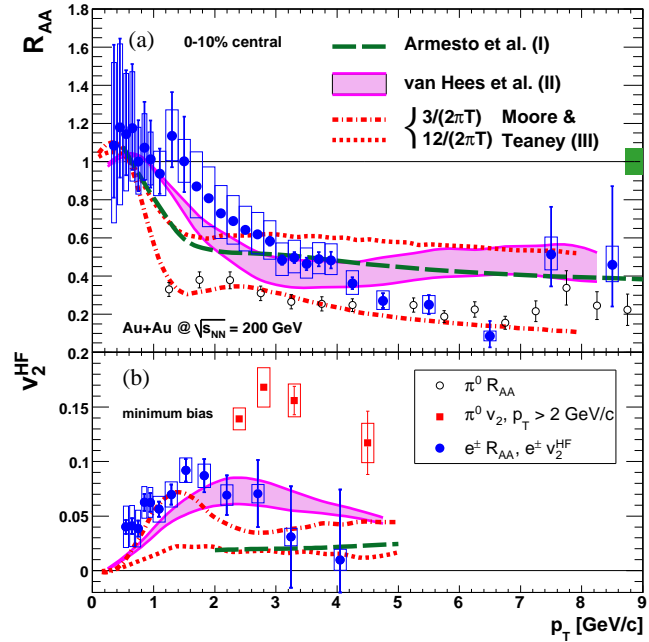}
\end{center}
\caption{
(a) $R_{\rm AA}$ of heavy-flavor electrons in 0-10\% central collisions
\cite{Adare:2006nq} compared with $\pi^0$ data~\cite{Adler:2003qi} and
model calculations (curves I~\cite{Armesto:2005mz}, II~\cite{vanHees:2007me},
and III~\cite{Moore:2004tg}). Model III is the Langevin calculation
discussed in the text for two different values of the diffusion constant.
Model II contains some extra processes, in particular final state
coalescence. Model I is based on energy loss. The box to the far right
at $R_{\rm AA} = 1$ shows the uncertainty in $T_{AA}$. (b) $v_2^{\rm HF}$
of heavy-flavor electrons in minimum bias collisions compared with $\pi^0$
data~\cite{Adler:2005rg} and the same models. }
\label{fig_hq_v2}
\end{figure}
%%%%%%%%%%%%%%%%%%%%%%%%%%%%%%%%%%%%%%%%%%%%%%%%%%%%%%%%%%%%%%%%%%%%%

 An important diagnostic of the properties of the QGP is the drag
force on a heavy quark initially produced in a hard collision during
the pre-thermal-equilibrium stage of the collision. We will see
below that the drag force can be related to the heavy quark diffusion
constant. Combining experimental constraints on heavy quark diffusion and
shear viscosity, which is related to momentum diffusion, provides
additional information about transport properties of the nearly
perfect fluid created in heavy ion collisions.

 The diffusion of heavy quarks in a QCD plasma was first studied
by Svetitsky~\cite{Svetitsky:1987gq}. In the following we will follow
the arguments presented by Teaney and Moore~\cite{Moore:2004tg}.
Consider a small density $n_Q$ of heavy quarks inside a QGP.
In a heavy ion collision, heavy charm and bottom
quarks are produced in hard collisions between quarks and gluons
during the pre-equilibrium stage of the reaction. If there is
enough time, and if the interaction is sufficiently strong, then
the heavy quarks can reach thermal equilibrium.\footnote{The charm
and bottom quark masses are $m_c\simeq 1.3$ GeV and $m_b\simeq 4.2$
GeV, respectively. This implies that even if thermal equilibrium is
reached, the density of heavy quarks is suppressed by large Boltzmann
factors $\exp(m_Q/T)$. We will see below that the expected thermalization
times are larger than those of light quarks and gluons by a factor
$m_Q/T$.} In this case the
the time evolution of the density can be described by a diffusion
equation
\be
\label{diff_equ}
\frac{\partial n_Q}{\partial t}= D{\bm \nabla}^2 n_Q\, ,
\ee
where $D$ is the diffusion constant. We can relate $D$ to the drag
force on the quark by using a stochastic (Langevin) equation.
\be
\label{Langevin}
 \frac{d{\bf p}}{dt} = -\eta_D {\bf p} +{\bf \xi}(t),
 \hspace{1cm}
 \langle \xi_i(t)\xi_j(t')\rangle = \kappa\delta_{ij}\delta(t-t').
\ee
Here, ${\bf p}$ is the momentum of the particle, $\eta_D$ is the
drag coefficient, and ${\bf \xi}(t)$ is a stochastic force. In
a kinetic theory picture the stochastic force models collisions
with quarks and gluons in the plasma. The coefficient $\kappa$ is
related to the mean square momentum change per unit time, $3\kappa
=\langle (\Delta{\bf p})^2\rangle/(\Delta t)$. The Langevin equation
can be integrated to determine the mean squared momentum. In the
long time limit ($t\gg \eta_D^{-1}$) the particle thermalizes and
we expect that $\langle {\bf p}^{\, 2}\rangle = 3mT$. This requirement
leads to the Einstein relation
\be
\label{Einstein}
 \eta_D=\frac{\kappa}{2mT} .
\ee
The relation between $\eta_D$ and the diffusion constant can be
determined from the mean square displacement. At late times
$\langle [\Delta {\bf x}(t)]^2\rangle = 6D|t|$ and
\be
\label{Einstein:2}
 D= \frac{T}{m\eta_D}=\frac{2T^2}{\kappa}.
\ee
The diffusion constant for heavy quarks in a QGP
can be determined by computing the mean square momentum transfer
per unit time. At weak coupling and for approximately thermal heavy
quarks the diffusion constant is dominated by heavy quark scattering
on light quarks and gluons, $qQ\to qQ$ and $gQ\to gQ$. These processes
are similar to the processes that determine the shear viscosity,
except that heavy quarks move slowly and the interaction only
involves the color Coulomb interactions, whereas shear viscosity
is sensitive to both electric and magnetic interactions. The leading
order result in QCD with three light flavors is
\cite{Svetitsky:1987gq,Moore:2004tg}
\be
\label{D_QCD}
 D=\frac{6\pi}{g^4T\log(2T/m_D)}\, .
\ee
Comparing this result with Eq.~(\ref{eta_qcd}) we observe
that heavy quark and momentum diffusion are related. In
the relevant range of coupling constants, and keeping
terms beyond the leading logarithm, one finds $DT\simeq
6(\eta/s)$. We note
that the heavy quark relaxation time $\eta_D^{-1}$ contains an
extra factor $m_Q/T$ compared to the hydrodynamic relaxation
time $\eta/(sT)$. For charm quarks at $T=200$ MeV this factor
is $m_c/T\simeq 7$. This implies that even if hydrodynamic
behavior is reached very quickly, $\eta/(sT)\sim 0.2$ fm, charms
quarks have barely enough time to equilibrate, $m_cD/T\sim 8$ fm.

 This simple relation between $DT$ and $\eta/s$ is broken
in the strong coupling limit of the SUSY Yang-Mills plasma. Using
holographic duality one finds
\cite{Herzog:2006gh,CasalderreySolana:2006rq,Gubser:2006bz}
\be
\label{D_AdS}
 D = \frac{2}{\pi T}\frac{1}{\sqrt{\lambda}}\, .
\ee
This result implies that in the strong coupling limit there is
no bound on the diffusion constant, and that even very heavy
quarks can possibly equilibrate. It is not clear how $\lambda$
should be chosen in order to compare Eq.~(\ref{D_AdS}) to
experiments in QCD. Gubser has advocated a value $\lambda\simeq
6\pi$ and concludes that $D\simeq 1/(2\pi T)$~\cite{Gubser:2006qh}.

 Experimental information on the diffusion constant comes from
the observation of single electrons from charm and bottom decays,
as shown in Fig.~\ref{fig_hq_v2}~\cite{Adare:2006nq}. There is
data on both the nuclear modification factor $R_{AA} = dN_{Au+Au}
/(\langle T_{AA} \rangle d\sigma_{p+p})$, where $dN_{Au+Au}$ is
the differential yield in Au+Au, $\langle T_{AA} \rangle$ is the
nuclear overlap, and $d\sigma_{p+p}$ is the differential cross
section in p+p, as well as on the elliptic flow parameter $v_2$.
The nuclear modification factor is sensitive to energy loss,
which in turn is governed by the drag force. Drag also implies
that the elliptic flow of light quarks and gluons induces a
non-vanishing $v_2$ parameter for heavy quarks. Different models
for the drag force are shown as the dashed and dash-dotted lines
in Fig.~\ref{fig_hq_v2}.
We observe that these models can account qualitatively for the behavior
of $R_{AA}$ but fail to reproduce
the observed flow. This indicates that charm quarks show some
degree of equilibration. A Langevin simulation with $(2\pi T)D
=(4-6)$ provides a qualitative description of both $R_{AA}$ and
$v_2$. It is interesting to note that this result, combined with
the estimate $DT\simeq 6 (\eta/s)$, gives $(4\pi)\eta/s\simeq
(1.3-2.0)$,  which is close to the result obtained from the elliptic
flow of light particles~\cite{Adare:2006nq}.

There are a number of uncertainties in this analysis that will
be addressed in the future. Experiments have not yet been able to
determine the flavor of the heavy quark, and there are significant
theoretical uncertainties in predictions of the charm and bottom
spectra. Future experiments will be able to identify the flavor of
the heavy quark.  Two contributions in this Focus Issue address transport
properties of heavy quarks. Meyer~\cite{NJPfocusissue35_meyer} presents
a lattice study of the Euclidean chromo-electric field correlation
function. This correlation function is related by a Kubo formula
to the diffusion constant. Rapp and Riek~\cite{Riek:2010py}
perform a non-perturbative T-matrix analysis of the charm quark
diffusion constant and the charmonium spectral function. These
and other studies, together with ongoing efforts to measure the
spectra of identified heavy quarks, will shed light on the
relation between the different transport properties of the
plasma such as diffusion, energy loss, and shear viscosity, and help to
characterize the degree to which the initial state in a heavy
ion collision thermalizes.

%%%%%%%%%%%%%%%%%%%%%%%%%%%%%%%%%%%%%%%%%%%%%%%%%%%%%%%%%%%%%%%%%%%%
\section{Holographic Duality}
\label{sec:hd}
%%%%%%%%%%%%%%%%%%%%%%%%%%%%%%%%%%%%%%%%%%%%%%%%%%%%%%%%%%%%%%%%%%%%

Holography is a duality relating quantum field theory {(QFT)} and gravity.
Roughly speaking, holographic duality maps the quantum physics of strongly correlated many-body systems to the
classical dynamics of black hole horizons in one higher dimension, replacing quasiparticles with geometry as the salient degrees of freedom.
As such, holography is very much in the tradition of emergent critical phenomena: when the system is strongly coupled, new weakly-coupled degrees of freedom dynamically emerge.
The novelty is that the emergent fields live in a dynamical spacetime with an extra spatial dimension. This extra dimension plays the role of an energy scale in the QFT, with motion along the extra dimension representing a change of scale, or renormalization group (RG) flow, in the QFT.
Holography thus translates problems in quantum many-body physics, such as thermodynamics and transport physics, into equivalent problems in classical gravity.

The power of this reorganization is that various phenomena which are easy to see, or perhaps even universal, in one presentation may be very surprising in the dual presentation, but just as universal.
For example, it is a classic result that the physics of black hole horizons in general relativity (GR) is largely independent of the details of the {black hole}~\cite{Israel:1967wq,Israel:1967za,Carter:1971zc,Hawking:1971tu,Hawking:1971vc}, with small fluctuations of the horizon obeying the equations of viscous hydrodynamics~\cite{Price:1986yy}.
The dissipation of waves in this fluid encodes the absorption of energy by the black hole, with the shear viscosity taking a universal value determined by basic properties of the
Einstein-Hilbert action~\cite{Price:1986yy,Kovtun:2003wp}.
Holographically, this is a remarkable fact: it tells us that any generic strongly-interacting quantum many-body system at finite temperature and density and with a sufficiently large number of degrees of freedom per unit volume (necessary for the application of holographic duality) can be expected to behave, at low energies, like a nearly-perfect liquid, and not like a gas of long lived quasiparticles as one might naively imagine.

More generally, holography gives us an entirely new way to define quantum field theories.
Explicitly, holography provides a recipe~\cite{Maldacena:1997re,Gubser:1998bc,Witten:1998qj} for using classical gravity in $(d+1)$-dimensions to compute quantum amplitudes that manifestly satisfy the consistency conditions of a $d$-dimensional QFT (locality, causality, etc.).\footnote{This recipe is colloquially referred to as the {\em holographic dictionary}, a sketch of which is presented in Table (\ref{table:holdict}).}  In general we do not know how to identify the corresponding QFT in terms of more familiar tools, for example by specifying a Hamiltonian governing the interactions of a set of well-defined quasiparticles.  Moreover, as we shall explore in some detail below, the holographic description is generally reliable precisely in situations when most traditional techniques are not: when $\l$, the typical coupling in the QFT, is strong, $\l\gg1$, and when $N$, the number of degrees of freedom per unit volume, is large, $N\gg1$.\footnote{More generally, holographic duality relates QFT in $d$-dimensions to quantum gravity in $(d+1)$-dimensions, with the gravitational description becoming classical when the  QFT is strongly-coupled, as discussed in Sec. (\ref{sssec:strong}).  We will focus on regimes where the QFT is strongly coupled and the gravitational dual classical, but we emphasize that holography remains true even when quantum gravitational (string theoretic) effects become important.}
In such cases we can simply take the gravitational description as a constructive definition of the QFT. Such {\em holographic QFTs} thus define a special subset of the space of well-defined QFTs which does not depend on any quasiparticle picture or conventional perturbation theory.
Importantly, some of the most theoretically and experimentally interesting real-world systems manifestly do not have any well-defined quasiparticles upon which to base a standard QFT.  Holography thus provides an entirely new way to construct consistent models of these strongly correlated quantum many-body systems, replacing quasiparticles with geometry as the central organizing principle.

An important corollary is that it is typically more fruitful to use holographic QFTs as windows on a general class of phenomena, or to study general concepts in the space of QFTs, than to try to exactly reproduce or solve a specific QFT of previous interest.   Indeed, building ``the holographic dual'' of one's favorite QFT is generally quixotic, as the regimes of validity of holographic QFTs typically exclude the theories of previous interest.  The classic example is $SU(N)$ gauge theory: QCD is given by $N=3$, while the holographic dual is classical only when $N\gg1$ and weakly-coupled only when the 't\,Hooft coupling is also large, $\l\gg1$.  It is thus futile to try to reproduce QCD exactly.  Where holography has proven useful, rather, is in studying general properties of $SU(N)$ gauge theories at high temperature and densities where many interesting phenomena arise which appear to be relatively insensitive to the precise value of $N$.  For example, holographic models suggest that any strongly-coupled QGP should behave as a liquid whose viscosity is very low and relatively insensitive to the precise value of the coupling, in sharp contrast to the large viscosity and strong coupling-dependence predicted by weakly-coupled QCD.  The low-viscosity of the QGP observed at RHIC (see Sec. \ref{sec_flow}) thus suggests that the RHIC fireball is indeed an extremely strongly-coupled quantum liquid.

To be sure, the simplest holographic models are in many ways very different from QCD -- in the most well-understood example there are no quarks and no mass gap!
What is remarkable from this point of view is that many of the striking features of the simplest holographic models persist even after the inclusion of quarks, a mass gap, and other phenomenologically important ingredients.  For example, in more realistic holographic models of QCD which contain fundamental quarks and display confinement, the physics at low energy is again governed by hydrodynamics with an exceptionally low and coupling-insensitive viscosity.
Such holographic models may thus be treated of as computationally tractable toy models which exhibit a rich set of behaviors analogous to those observed in the lab, the study of which can reveal qualitative, and sometimes even quantitative, general properties of the larger class of strongly-coupled QFTs.

It is important to stress that while the original discovery of holographic duality~\cite{Maldacena:1997re}\ required esoteric tools such as supersymmetry and conformal invariance,\footnote{Indeed, holographic duality is often referred to as the {\em \ads-CFT correspondence}  and also as {\em gauge-gravity duality}, among many other names.} the duality itself does not depend on supersymmetry or conformality.\footnote{The role of all this structure is simple: given enough symmetry, it becomes possible to compute quantities in an interacting QFT at both weak and strong coupling.  Holographic duality was discovered when Maldacena pointed out~\cite{Maldacena:1997re}\ that scattering amplitudes in the {\em maximally} supersymmetric 4d QFT were in fact identical to amplitudes for the low-energy scattering of closed strings off a maximally symmetric 5d black hole in string theory.  Conformal symmetry still plays an important role in holographic QFTs, but no more so than it does in any interacting QFT: conformal symmetry arises dynamically at fixed points of the RG.}  In fact, most current work on holographic duality involves non-supersymmetric, non-conformal QFTs.  Indeed, we now have a long list of examples in which each of these constraints is weakened or removed (see e.g.~\cite{Aharony:2002up,Son:2008ye,Balasubramanian:2008dm,Kachru:2008yh} and references therein).  To date there is no example of a violation of a sharp holographic duality.  From the current point of view, holographic duality is simply a true, if as yet unproven, fact about quantum field theories and quantum gravity.

While our understanding of the duality remains in various ways incomplete, holographic model building has already generated novel insights, including predictions for the anomalously low viscosity of cold atom gasses, which follow from the same universal horizon physics as that of strongly-coupled $SU(N)$ plasmas, and lessons about jet quenching and rapid thermalization in the RHIC fireball\cite{CasalderreySolana:2011us}.
Beyond providing new tools with which to model specific phenomena,
holography has sometimes suggested new organizing principles for strongly quantum dynamics, as seen for example  in semi-holographic models of non-Fermi liquids~\cite{Faulkner:2010tq,Iqbal:2011ae,Nickel:2010pr}, or in the application of classical numerical relativity to study far-from-equilibrium dynamics in extreme quantum liquids~\cite{Chesler:2010bi}.

The goal of this section is to provide a brief introduction to the basic structure of holography, first heuristically and then more precisely, as well as its origins and a few of its applications.  We will also review a few major themes in the holographic study of extreme quantum matter and give a broad, if selective view of some of the key open questions in the field.  Of necessity, we omit many important topics and references.  For a more detailed introduction, several excellent reviews are available, including~\cite{CasalderreySolana:2011us,Iqbal:2011ae,Hartnoll:2009sz,2009arXiv0909.3553H,McGreevy:2009xe,Aharony:1999ti,Son:2007vk,Schafer:2009dj,Erdmenger:2007cm,Herzog:2009xv,Gubser:2009md,Lee:2010fy}, from which we have drawn heavily in preparing various parts of this section.

%%%%%%%%%%%%%%%%%%%%%%%%%%%%%%%%%%%%%%%%%%%%%%%%%%%%%%%%%%%%%%%%
%%%%%%%%%%%%%%%%%%%%%%%%%%%%%%%%%%%%%%%%%%%%%%%%%%%%%%%%%%%%%%%%
%%%%%%%%%%%%%%%%%%%%%%%%%%%%%%%%%%%%%%%%%%%%%%%%%%%%%%%%%%%%%%%%
%
%
%	HEURISTICS
%
%
%%%%%%%%%%%%%%%%
\subsection{Why Should Holography Be True?  Two Heuristic Pictures.}\label{ssec:heuristics}

Why should a quantum field theory (QFT) in $d$-dimensions have anything to do with gravity in $d+1$, or vice versa?  To give some intuition we describe two heuristic arguments that motivate such a holographic correspondence. We first consider a system that includes gravity, {i.e.,} objects falling into a black hole, and argue that it should admit an equivalent description without gravity in one lower dimension. We then examine a strictly non-gravitational system, a QFT on a lattice, and argue that it should be related to an equivalent theory with gravity in one higher dimension.

\subsubsection{Starting with Gravity: Falling into a Black Hole}\label{ss:GR2Hol}

Imagine standing far from a black hole\footnote{For example, in a spaceship following a distant circular orbit of the black hole.  We emphasize that everything in Sec. (\ref{ss:GR2Hol}) follows from standard results about black holes in GR.  For a clear and concise review of the physics of black holes in GR; see~\cite{Hughes:2005wj}\ and references therein.  More complete references can be found in~\cite{Gallo:2008bh}.} in $(d+1)$-dimensions and sending a robotic probe on a straight line trajectory directly toward the black hole.  To follow the trajectory of the robot, we have attached to it a strobe light which flashes once a second.  According to the robot, it will itself reach the horizon in finite proper time, {as measured for example by a clock onboard the robot,} and will thus have flashed its strobe only a finite number of times before crossing the horizon into the black hole interior.

However, according to you, the distant observer, the story looks very different.  As the robot falls into the gravitational well of the black hole, the light emitted by the flashing strobe must do work to escape the gravitational potential, and is thus red-shifted.  By the constancy of the speed of light, this means that the flashes arrive to you at ever greater intervals.   As the robot approaches the horizon, the observed red-shift in fact diverges, so that the final flash emitted at the moment of crossing the horizon takes an infinite time to arrive at the distant observer; {light} from flashes after the robot has crossed the horizon will never make it out to you.  Thus, what you actually see is not the robot falling into the black hole, but rather the robot approaching the black hole and gradually slowing down and compressing into a thin membrane on the surface of the black hole.  Indeed, to you, it appears as if the robot has stopped falling toward the massive black hole entirely, as if it had stopped responding to gravity at all.

Now imagine sending in a swarm of robots in an isotropic shell around the black hole, so as to increase the mass of the black hole by a small amount corresponding to the matter and energy in the collapsing shell.  Again, according to you, the distant observer, the shell will never actually appear to fall into the black hole.  Rather, you will see it approach the black hole, slow down, and effectively stop, forming a non-gravitating membrane which stretches over the true horizon.  As each new bit of matter reaches this \textit{stretched horizon}, it generates a disturbance in the membrane which spreads in waves through the stretched horizon as if through a fluid. Remarkably, all of the observable behavior of this system can be precisely captured by a non-gravitational viscous hydrodynamical model for a fluid on the stretched horizon which lives in only $d$, rather than $(d+1)$ spacetime dimensions.

% figure 1: Lattice
\begin{figure}[!t]
\centerline{\epsfig{figure=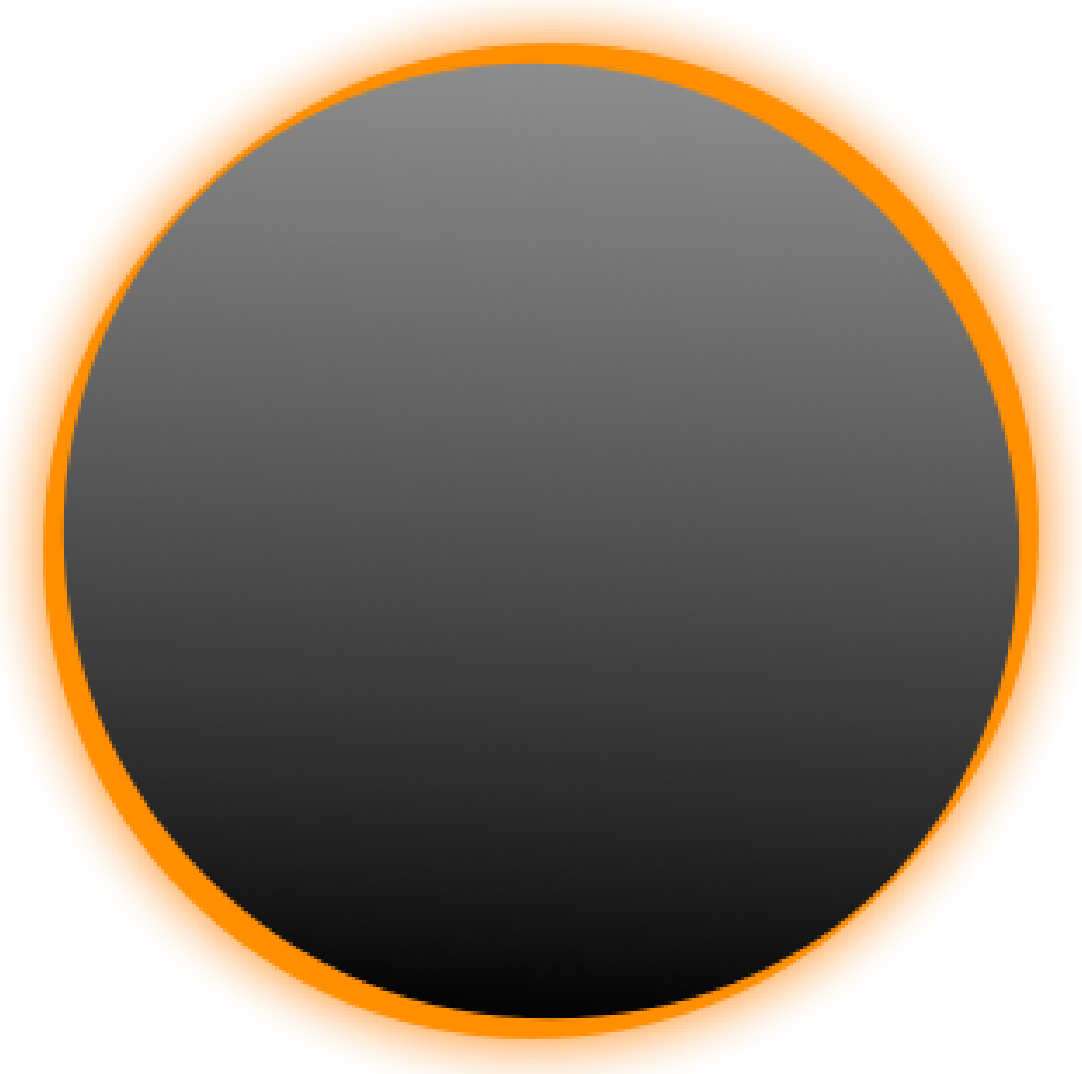, height=2.4 in}~~~~~~~~~~~~~~~~\epsfig{figure=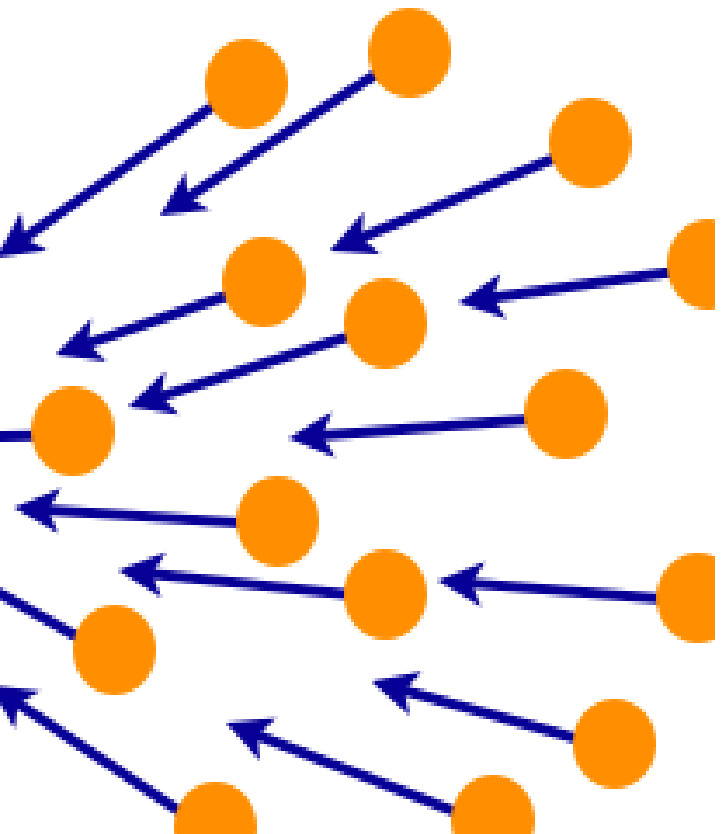, height=2.4 in}}
\caption{\em Two views of {a swarm of robots falling toward a black hole in an isotropic shell formation.}  [Left] A distant observer sees the robots slow down and spread over the horizon.  According to her, the proper description of the robots is as a non-gravitating fluid stretched over the horizon. [Right] An observer falling along with the robots would see them continue steadily towards and beyond the horizon, responding as one would expect to the black hole's gravitational pull. According to the in-falling observer, the appropriate description is general relativity in the bulk.}
%\label{FIG:Lattice}
\end{figure}

If we describe this same process from the perspective of one of the robotic probes, the physics looks very different. In this frame, nothing special happens as the robots approach and pass the horizon.  On the contrary, the swarm of robots continues to fall towards the black hole, interacting with each other and with the gravity of the black hole in the full volume of $(d+1)$-dimensions. These dynamics can be modeled in a straighforward manner with GR in $(d+1)$-dimensions coupled to the matter and energy of the robots and the black hole.

This leaves us with two very different descriptions of our probes as they approach the horizon, one involving the gravitational dynamics of probes falling into a $(d+1)$-dimensional black hole, the other involving the hydrodynamic response to our probes of a $d$-dimensional fluid.  Both descriptions accurately capture the behavior of our probes as viewed by two different observers.  But there cannot be two facts of the matter about the physics of our probes before they cross the horizon: two observers cannot accurately observe contradictory events unfold.\footnote{Explicitly, so long as both observers remain outside the black hole, it remains physically possible for them to exchange information and compare their observations.} The gravitational dynamics of a $(d+1)$-dimensional black hole must thus be, in some deeply non-local way, equivalent to the hydrodynamics of a $d$-dimensional fluid.

\subsubsection{Starting without Gravity: Taking the Renormalization Group Literally.}

Rather than start with gravity, let's start with a familiar non-gravitational field theory.  Consider a
system on a lattice with lattice spacing $a$ and Hamiltonian,
\be\label{EQ:LatHam}
H = \sum_{x,i}J_{i}(x)\CO^{i}(x)\,.
\ee
Here, $x$ labels the sites in the lattice, $i$ labels the various operators $\CO^{i}(x)$ defined at each site,
and the $J_{i}(x)$ are coupling constants / sources for the operators\footnote{We write sources with index down and operators with index up, for later notational convenience.} $\CO^{i}$.
Note that the sources will in general depend on both space and time; we take $x$ to stand in for both as is typical spacetime notation in relativity.
Given this setup, what we generally want to compute is the physics of the ground state and the {low energy} excitations over this ground state as a function of the microscopic coupling constants at the lattice scale.
In general, however, exactly diagonalizing the Hamiltonian is intractably difficult.

Kadanoff and Wilson taught us a beautiful approach to this problem: the Renormalization Group (RG)~\cite{Kadanoff:1966wm,Wilson:1973jj,Wilson:1974mb}.
%
% figure 1: Lattice
\begin{figure}[!b]
\centerline{\epsfig{figure=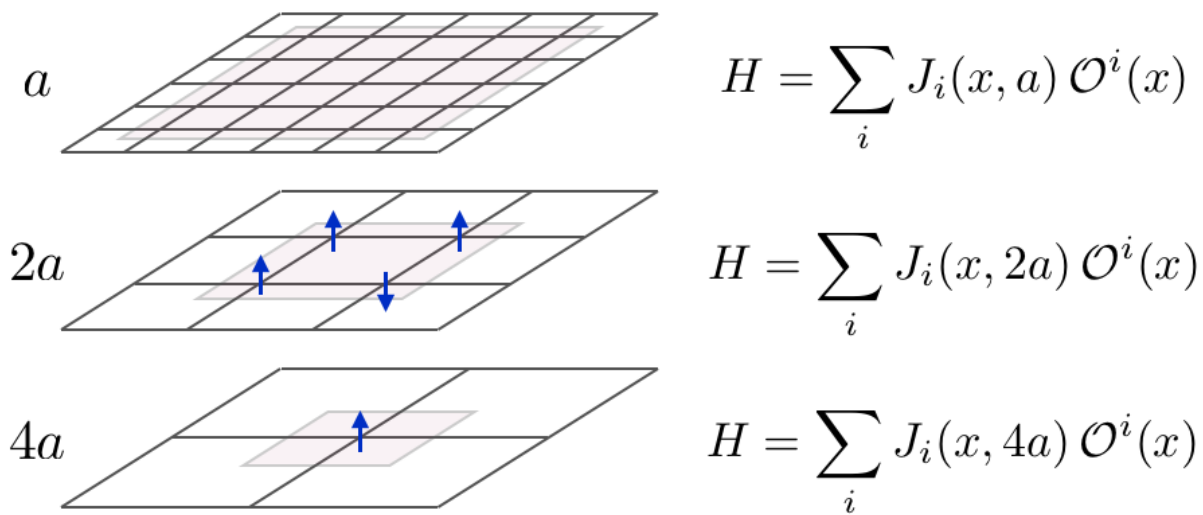, height=2.7 in}}
\caption{\em Coarse-graining spins on a lattice a-l\'a Kadanoff and Wilson. At each step we coarse-grain by replacing the degrees of freedom on a block of sites by an average value on a single site, rescaling the lattice spacing $a$ and tuning the couplings $J_{i}(x,a)$ in the Hamiltonian $H$ so that the physics of the ground state and low-lying excitations remain invariant under the scaling operation.}
%\label{FIG:Lattice}
\end{figure}
The basic move is to iteratively coarse grain the lattice, making the lattice spacing larger with each step, so that at each step a single site represents the average of multiple sites in the previous lattice.
We then tune the couplings so as to preserve the physics of the ground state and the low-energy excitations over it, replacing the fixed coupling $J_{i}\!\(x\)$ with a scale-dependent coupling $J_{i}\!\(x,\scale\)$, where $\scale$  denotes the length scale at which we probe the system.
By iterating this procedure, we eventually arrive at an effective description of long wavelength modes of the system.
As explained by Kadanoff and Wilson, the resulting flow of the couplings with scale $\scale$ can be encoded in a beta function which is, remarkably, local in energy scale,
\be
\scale {\p\over\p \scale} J_{i}(x,\scale) = \b_{i}(J_{j}(x,\scale),\scale)
\ee
When the beta function can be determined, for example in perturbation theory, this renormalisation group approach is extremely powerful.
Indeed, even finding the fixed points of the RG flow, corresponding to scale-invariant or conformal points governed by conformal field theories (CFTs), can be enormously revealing, so much so that we often {\it define} scale-dependent QFTs with non-trivial RG flows by starting with a well-understood CFT and turning on a relevant deformation to generate the desired RG flow.

However, in many complex, strongly coupled systems, including many strongly correlated systems governed by interesting conformal fixed points, the beta functions cannot be straightforwardly derived.
It is thus tempting to search for a reorganization of the RG which determines the correct RG flow of the couplings without requiring an explicit calculation of the $\b$-functions.

To this end, consider the following recasting of the Kadanoff-Wilson picture.  As before, consider our series of coarse-grained lattices with coarse grained Hamiltonians and suitably tuned couplings, $J_{i}(x,\scale)$.  Now imagine arranging each coarse-graining of the lattice into a stack ordered by scale, so that at each subsequent level of the stack the lattice spacing grows, $\scale$=$\{$$a$, $2a$, $4a$,$\dots$$\}$.  By construction, motion down the stack reproduces RG flow in the original lattice theory.  In this one-higher-dimensional hyper-lattice, however, the couplings at each scale, $J_{i}(x,\scale)$, look a lot like
fluctuating fields in a one-higher-dimensional lattice, varying both in space-time
and in the scale direction.

This raises an interesting question.  We know that the true solution $J_{i}(x,\scale)$ solves the $\b$-function equations for our original lattice theory, but we do not know what the correct $\b$-functions for our lattice model are.
Might there be some simple equation of motion on this one-higher-dimensional lattice whose solution is also $J_{i}(x,\scale)$? More generally, might there be a conventional local field theory\footnote{By a conventional local QFT we mean a QFT with a finite number of local fields governed by a Lagrangian with canonical quadratic kinetic terms and local interactions.  One can certainly imagine other possibilities, but this structure turns out to be particularly useful in what follows.} whose dynamics encode the full, and unknown, $\b$-functions of the original lattice theory?

% figure 2: Duality
\begin{figure}[!t]
\centerline{\epsfig{figure=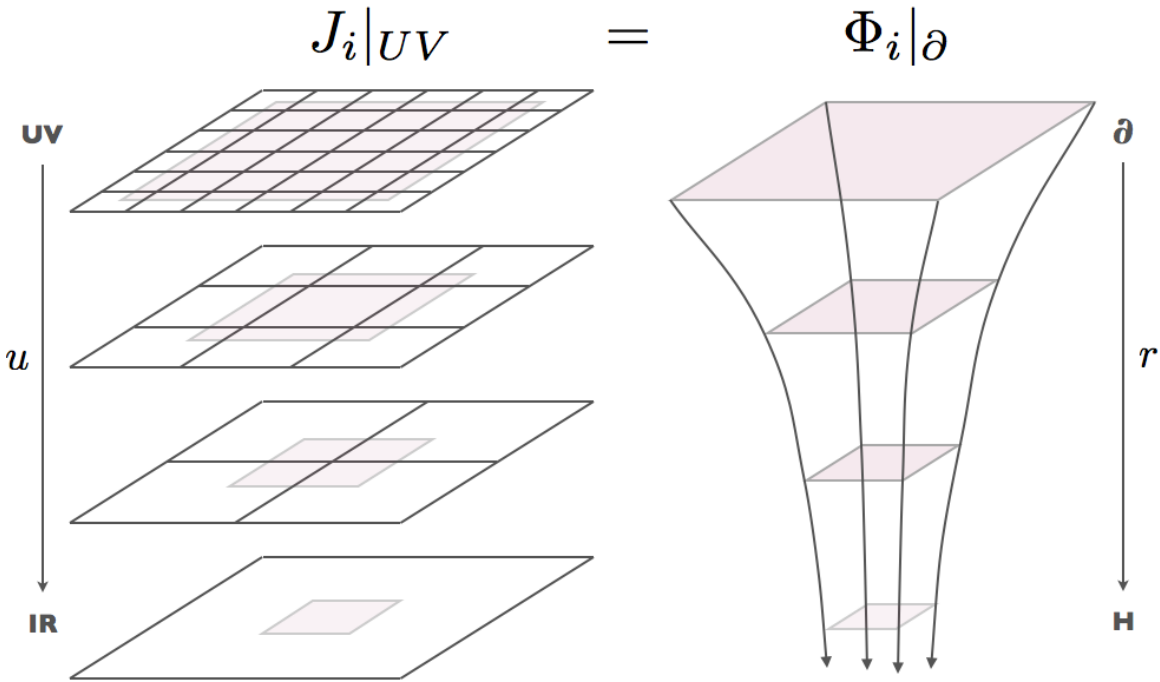, height=3.7 in}}
\caption{\em  Holography promotes the tower of coarse-grained lattices to a one-higher-dimensional lattice, with the RG scale recast as a spatial dimension and the running couplings $J_{i}$ replaced by dynamical fields $\Phi_{i}$ which {asymptotically approach} the UV couplings at the UV top of the stack and which satisfies a gravitational equation of motion in the bulk of the stack.}
\label{FIG:Duality}
\end{figure}

Consider the full RG-stack of lattices labeled by a new RG-coordinate, $r$, which runs from the original ultraviolet (UV) lattice cutoff, $r=a$, to the deep {infrared (IR)}, $r\to\infty$.   We seek a simple QFT defined on this one-higher-dimensional space whose $(d+1)$-dimensional bulk fields, $\Phi_{i}(x,r)$, are in 1-to-1 correspondence with the couplings, $J_{i}(x)$, of the underlying lattice theory, with the values of the bulk fields at the UV end of the stack, $r = a$, determined by the microscopic couplings,
\be
\Phi_{i}(x,a)~=~ J_{i}(x)\,.
\ee
As a consequence, the bulk field $\Phi(x,r)$ must have the same charges, tensor structure and other quantum numbers as the corresponding coupling,\footnote{Recall that our couplings are defined as the coefficients of the associated operators in the Hamiltonian, and so have definite quantum numbers.  Note too that, at fixed points of the $\b$-functions where the system is scale-invariant, the dimension of an operator is also a quantum number. This is also true in perturbations around such conformal fixed points.} $J_{i}(x)$.
Thus to each scalar operator $\CO(x)$ is associated a bulk scalar field $\Phi(x,r)$ such that $\Phi(x,a)\CO(x)$ is a scalar operator we can add to our Hamiltonian, to each current operator $\CJ^{\m}(x)$ is associated a bulk vector field, $A_{\mu}(x,r)$  such that $A_{\m}(x,a)\CJ^{\m}(x)$ is a scalar, and so forth.
In particular, to the canonical stress-tensor $T^{\m\n}(x)$ is associated a canonical spin-two field in the bulk, $g_{\mu\nu}(x,r)$.

What can the Lagrangian of this bulk QFT be?  It is not the original lattice Hamiltonian -- the fields of the bulk QFT encode the dynamics of the {\em couplings} in RG space, not of the operators on the original lattice.  A priori, the natural thing to do is to write down the most general effective field theory allowed by the fields and symmetries of the system.  However, a simple argument greatly constrains the possibilities.
Any QFT defined by perturbation from a fixed point CFT as discussed above
includes among its operators a canonical stress-energy tensor, $T_{\m\n}$, arising as the Noether current corresponding to the space-time symmetry group.
The bulk QFT should thus include a corresponding canonical spin-2 field, $g_{\m\n}$.  But by the Weinberg and Weinberg-Witten theorems,\footnote{Weinberg~\cite{Weinberg:1965nx}\ argued that any {Lorentz-invariant} theory of a spin-2 field must respect the equivalence principle or suffer IR pathologies; the Weinberg-Witten theorem~\cite{Weinberg:1980kq}\ further states that any massless spin-two field must either be the graviton or suffer one of a host of pathologies, for example not having a conserved stress tensor.  For an illuminating discussion of these constraints; see~\cite{Jenkins:2009un} and references therein.} and assuming the existence of a Lorentz-invariant continuum limit, any spin-2 field must either decouple at low-energy, which would imply that sources of momentum or energy in the original lattice theory do not affect the system, or it must couple universally according to the equivalence principle.  In other words, the bulk field theory must be either topological or be a theory of gravity which reduces, at long wavelength and low energy, to GR.

If all this can be done consistently, we would appear to have two distinct descriptions of our system: the original lattice theory with lattice spacing $a$ and sources $J_{i}(x)$ which satisfy {first-order} $\b$-function equations; and a gravitational description living in a one-higher-dimensional bulk whose fields, $\Phi_{i}(x,r)$, are in 1-to-1 correspondence with the sources $J_{i}(x)$ but which satisfy conventional {second-order} local field equations.  In other words, we have two ways to compute correlation functions in our system: we may either work in the lattice theory and solve the QFT problem; or we may work in the gravitational description and solve a GR problem.

In order for these two descriptions to have any chance of being equivalent\footnote{By equivalent we mean that their partition functions are exactly equal, so that both descriptions contain precisely the same information, as we shall spell out in detail in Sec.~\ref{sssec:strong}.  Note that this implies  their free energies and other thermodynamic data must similarly, and identically, match.} the bulk theory must satisfy two among many rather unusual properties, both of which turn out to be basic properties of black holes.   First, the entropy as computed in the two descriptions must be the same. In the original lattice QFT, the entropy will generically be extensive, with the entropy $S_{\mathrm{QFT}}$ of the $d$-dimensional QFT in a region $R_{d}$ scaling linearly with its spatial  volume, $S_{\mathrm{QFT}}(R_{d})\propto V(R_{d})$.  As a result, the entropy as measured in the one-greater-dimensional bulk gravity, $S_{\mathrm{GR}}$ must be {\em sub}-extensive, scaling as the area bounding a region, $S_{\mathrm{GR}}(R_{d+1}) \propto A(R_{d+1}) \propto V(\p R_{d+1})$.   This is a basic feature of any theory of gravity:\footnote{Various other theories, including topological theories, share the property of sub-extensivity.  However, by Weinberg-Witten, to have both an area law and an interacting spin-2 field, gravity appears to be the only option.} the entropy in a volume is bounded by the entropy of a black hole that fits inside that volume~\cite{Bousso:2002ju}.  Since the entropy of a black hole is proportional to the surface area of its horizon, $S_{\mathrm{BH}}={1\over4}A_{H}$, this tells us that the entropy of any region in a theory of gravity is bounded by its surface area, $S_{\mathrm{GR}}(R_{d})\le{1\over4}A(R_{d})$.  Indeed, the fact that the entropy of a black hole can be entirely associated with its surface, and that as a result the entropy in gravity is necessarily sub-extensive, is the origin of the term {\em holography}~\cite{'tHooft:1993gx,Susskind:1994vu}.\footnote{The term holography is used by analogy with familiar holograms, and is meant to convey the surprising fact that the information about the $(d+1)$-dimensional bulk spacetime is being reliably encoded in the data of the QFT living on the $d$-dimensional boundary.  The analogy is in various ways quite apt.  For example, information which is local in the $(d+1)$-dimensional bulk is encoded non-locally in the $d$-dimensional boundary.  However, one should not take the analogy too literally; for example, the encoding here involves quantum-mechanics and gravity.}

Second, the $\b$-function equations in the boundary lattice theory are first-order in the scale $r$, while the gravitational equations of motion in the bulk are second-order in $r$.  Physically this means that specifying the sources in the lattice QFT at one scale completely determines the coupling at all other scales, {i.e.,} completely fixes the RG trajectory.  By contrast, in the gravitational system we have two solutions to our second-order equation, so we must further specify the derivative of the couplings with scale to uniquely fix the RG trajectory.  How can these be equivalent?  Somehow it must be the case that the gravitational system automatically selects one of the two solutions of the bulk equations as physical.

It is again the physics of black holes which explains how this can happen.  Imagine solving a simple wave equation in the neighborhood of a black hole horizon.  In principle there are two solutions to the {second-order} wave equation.  However, since things can fall into a black hole horizon but nothing can escape, it is natural to organize the solutions near the horizon into in-going and out-going waves.
Moreover, when working in Euclidean signature
%(aka imaginary time)
as is done for example in computing thermodynamic data, only the in-going solution is regular at the horizon; the out-going solution is always irregular at the horizon.
Thus, while the gravitational field equations are indeed {second-order,} the presence of a black hole horizon effectively adds a second boundary condition, so that we again need only specify a single boundary condition to completely determine the solution.
Since the location of the horizon encodes the thermodynamic variables (temperature, entropy) of the system, the physical meaning of this second boundary condition is to specify the state of the QFT.

These heuristics suggest a connection between quantum field theory in $d$ spacetime dimensions and quantum gravity in $d+1$ spacetime dimensions, with the fields in the gravitational system  exactly paired with sources in the dual QFT and with the extra spatial coordinate in the gravitational bulk playing the role of an RG scale.  The next sections will describe a precise and computationally effective theory of this connection known as holographic duality, and review some of the lessons gleaned from its study to date.

\subsubsection{Coda: The Surprise of Locality}

It is worth pausing to contemplate just how
remarkable this proposal truly is.
Consider a lattice theory in $d$-dimensions which has, in its {Hamiltonian,} a real parameter $q$.  You might think of $q$ as a global charge, or as a momentum, or something else entirely, which can in principle be varied adiabatically.  All correlation functions in the theory are thus functions of $d$-dimensional spacetime, $(x,t)$, as well as of our parameter, $q$.

We might be tempted to take a stack of such lattices, each with a different value of $q$, call the ``space'' labeled by $(x,t,q)$ a $(d+1)$-dimensional spacetime, and claim that the resulting tensor product of lattice QFTs is well described by a $(d+1)$-dimensional QFT living on this $(d+1)$-dimensional spacetime.

This would be wrong.  Being $(d+1)$-dimensional means that all observables should transform appropriately under translations, rotations, accelerations, and general coordinate transformations amongst $x$, $t$ and $q$.  They should also respect $(d+1)$-dimensional causality and locality.  But that is absurd!   What does it mean to rotate between the $x$ direction and the $q$ parameter?  Meanwhile, there is absolutely no reason for correlators of two operators which live on lattices with very different values of $q$ to vanish outside some $q-t$ light-cone, or for operators from distant $q$-slices to have vanishing equal-time commutators in whatever state you choose to call the ground state.  Locality and causality are almost sure to be violated, and badly. This is not a $(d+1)$-dimensional system.

{To summarize,} saying that the system is well-described by a local, causal, unitary $(d+1)$-dimensional QFT implies a long list of constraints on the $(d+1)$-dimensional correlation functions.  These translate into constraints on the original lattice theory and on the $q$-dependence of its correlation functions.  It is by no means obvious that any such stack of lattices can ever have a causal, local $(d+1)$-dimensional description.  From this point of view, holographic duality, in which the role of $q$ is played by the energy scale and in which the $(d+1)$-dimensional description is in fact local, causal, and even classical when the $d$-dimensional QFT is very strongly interacting, is nothing short of astonishing.  Precisely why, when, and how this locality arises is perhaps the deepest question in holography.\footnote{See for example~\cite{Heemskerk:2009pn} for a discussion of locality in holographic duality, and~\cite{Fitzpatrick:2011hu} for an interesting spin of the problem.}

%%%%%%%%%%%%%%%%%%%%%%%%%%%%%%%%%%%%%%%%%%%%%%%%%%%%%%%%%%%%%%%%
%%%%%%%%%%%%%%%%%%%%%%%%%%%%%%%%%%%%%%%%%%%%%%%%%%%%%%%%%%%%%%%%
%%%%%%%%%%%%%%%%%%%%%%%%%%%%%%%%%%%%%%%%%%%%%%%%%%%%%%%%%%%%%%%%
%
%
%	The Structure of Holography
%
%
%%%%%%%%%%%%%%%%
\subsection{Essential Holography}\label{SEC:EssentialHolography}

Holographic duality~\cite{Maldacena:1997re,Gubser:1998bc,Witten:1998qj}\ is a precise equivalence between certain $d$-dimensional quantum field theories and $(d+1)$-dimensional gravitational theories  which provides a sharp realization of the heuristics described in Sec.~\ref{ssec:heuristics}.
In particular, all the basic features noted above will reappear below: the fields in the bulk correspond to the couplings in the QFT; the RG flow of the QFT is encoded in the radial evolution of the gravitational theory along an extra dimension; and black hole horizons play a key role.
Of particular importance is the precise geometry of the emergent spacetime, as well as the precise relationship between observables in the QFT and those of the gravitational dual.  We will explore a few simple examples along the way.

For clarity of presentation, we will focus our discussion on the simplest possible QFTs from the point of view {of} the RG, i.e., Lorentz-invariant conformal field theories (CFTs) which are fixed-points of the RG.  Since the RG maps to radial evolution in the gravitational dual, QFTs corresponding to RG fixed-points must be dual to geometries which are translationally-invariant along the emergent radial direction up to overall rescalings; adding Lorentz-invariance then entirely fixes the dual geometry to be Anti-de Sitter space (\ads), whose geometry we will now describe.\footnote{The first examples of holographic duality involved such \ads/CFT dual pairs, and hence holographic duality is often referred to as AdS/CFT.  Just as we can always construct a general QFT by perturbing away from a CFT fixed point, we can extend such AdS/CFT dualities to more general QFTs and geometries by perturbing both sides in corresponding ways, so we do not lose too much by focusing on these examples.}

%%%%%%%%%%%%%%%%
\subsubsection{Gravity and Matter in Anti-de Sitter Space}

\ads$_{d+1}$ is a homogeneous, isotropic spacetime with $d+1$ spacetime dimensions and constant negative curvature whose metric can be conveniently written as,\footnote{For the remainder of our discussion of holography we will work in \textit{natural units} in which $\hbar=1$ and $c=1$.}
\be
%ds^{2} = L^{2}{-dt^{2}+d\vec{x}^{2} +dr^{2} \over r^{2}}\,,
ds^{2} = {L^{2}\over r^{2}}\[-dt^{2}+d\vec{x}^{2} +dr^{2} \]\,,
\ee
where $\vec{x}$
are ($d$-1) spatial coordinates, $t$ is a timelike coordinate, and r is a final ``radial'' spatial coordinate.
The constant parameter $L$, known as the {\textit{\ads-radius},} sets the radius of curvature, with the Ricci curvature scalar given by the constant
\be\label{eq:adsricci}
R=-{d(d+1)\over L^{2}}\,.
\ee
The space is thus weakly curved when $L$ is large and strongly curved when $L$ is small.
At fixed values of the radial coordinate $r_{*}$, the metric reduces to the $d$-dimensional Minkowski metric on flat space-time rescaled by {$L^{2}/r_{*}^{2}$,}
\be
ds^{2}_{r_{*}} =  {L^{2} \over r_{*}^{2}}\[-dt^{2}+d\vec{x}^{2}\]\,,
\ee
It is thus useful to think of \ads\ as a stack of slices of flat space, with $r$ controlling the physical scale on that slice.  The slice at $r\to0$ is referred to as the \textit{boundary}.
Note that the volume element of the boundary slice at $r\to0$ diverges, while the proper distance to the boundary also diverges.  Thus the boundary at coordinate $r\to0$ should be understood as the set of points at \textit{spatial infinity}.\footnote{In computations, it will often be useful to regulate this divergence by introducing a cutoff at $r= a$, with $a\ll L$, removing the regulator, $a\to0$, only at the end of all computations.}

Importantly, the full geometry is invariant under not just translations, rotations and Lorentz boosts along each slice, but also under general scaling transformations,
\be\label{eq:scaling}
(r,\vec{x},t) \to (\alpha r,\alpha \vec{x},\alpha t)
\ee
as well as rotations and boosts that mix $r$ with the other dimensions.  The full isometry group of \ads$_{d+1}$ is in fact
identical to the conformal group in $d$ space-time dimensions.\footnote{Note that our regulator at $r=a$ explicitly breaks conformal symmetry but does not break space-time rotational or translational invariance in $x$ and $t$, as a more conventional UV regulator such as a hard momentum cutoff would.  Matching regulators between bulk and boundary turns out to be quite subtle, \cf~\cite{Heemskerk:2010hk}.}

% Harmonic Trap
It is often useful to think of \ads\ as a harmonic trap for gravity.  More precisely, the constant negative curvature of \ads\ acts as a harmonic trap such that if you sit inside \ads\ at an arbitrary point and throw a ball, it will inevitably fall back to you in finite time.  Indeed, if you fire out a photon, that photon will run away to the boundary at $r=0$ and return, again in finite observed time.\footnote{Unlike the electromagnetic harmonic traps used in trapping cold atoms, \ads\ is homogeneous and isotropic: there is no center to which all objects return.  Rather, the curvature ensures that any two initially parallel trajectories always curve towards each other, no matter where they begin, and then oscillate.}  And yet while nothing, not even gravity, gets out of the \ads\ box, the spacetime remains completely homogenous and isotropic.  \ads\ thus acts as a peculiarly graceful IR regulator for gravity which breaks none of the symmetries of flat space.

All of this implies an intimate connection between the radial coordinate in the bulk of \ads\ and spatial scales along the boundary: probing short distances (or high energies) along the boundary corresponds to probing the bulk only near the boundary at $r=0$, while probing long distances (or low energies) along the boundary corresponds to data deep in the interior of \ads.  Roughly, then, we can think of the region near the boundary of \ads\ as associated {with} the UV physics of the boundary, and of the deep interior of \ads\ as associated to the IR {physics} of the boundary.

% figure 3: Lattice
\begin{figure}[!h]
\centerline{\epsfig{figure=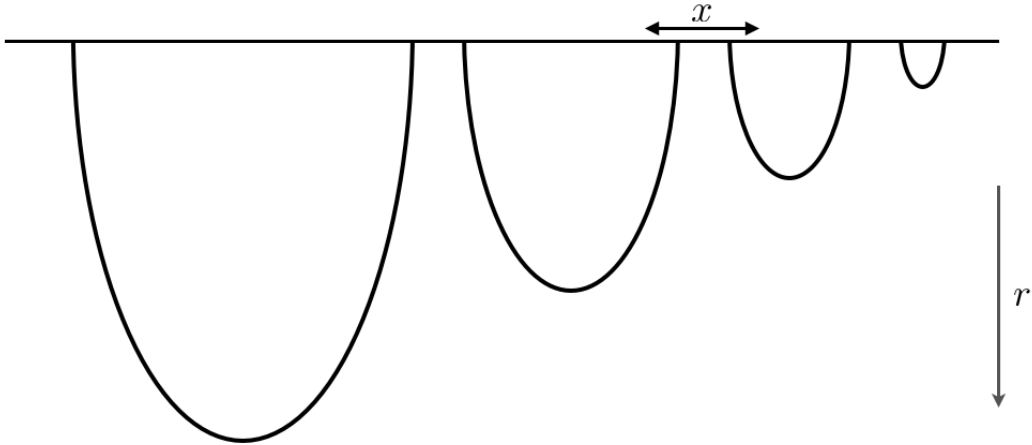, height=2.4 in}}
\caption{\em Hanging rope in \ads.  When the ends are close together in the spatial direction, $x$, so that the rope probes short distances and high energies in the QFT, the stretches only slightly into the bulk of the \ads\ spacetime.  When the ends are far apart, probing long distances and low energies in the QFT, the rope stretches deep into the \ads\ bulk.  Thus is the boundary of \ads\ associated to the UV of the QFT, while the deep bulk is associated with the IR.}
%\label{FIG:Lattice}
\end{figure}

To visualize this association, consider a fixed tension rope whose ends are glued to the boundary at $r=0$.  Due to the constant negative curvature of \ads, the bulk of the rope is drawn into the bulk of the \ads\ spacetime, forming an arc drooping away from the boundary.  If the ends of rope are held close together -- probing short distances or high energy in the 4d boundary -- the rope barely hangs into bulk.  If we instead pull the ends of the rope far apart -- probing long distances or low energy in the 4d boundary -- the rope dips much further into the bulk.  In this way, probing the IR of the boundary corresponds to probing the deep interior of the \ads, $r\to\infty$, while probing the UV of the boundary corresponds to focusing on the near-boundary region of the geometry, $r\to0$.  Note that the near-boundary cutoff at $r=a\ll1$ thus acts as a UV regulator, while the horizon at $r=\rH$ provides a natural IR regulator.

Importantly, \ads\ is a solution to the equations of motion of a generic Wilsonian action for the metric
\be
I_{\mathrm{Gravity}} = {1\over 16\pi G_{N}} \int d^{d+1}\!x\,\sqrt{-g}\(  -2\Lambda \,+ R \,+c_{2} R^{2} \,+c_{3} R^{3} +\dots  \)\,.
\ee
Here, $G_{N}$ is the Newton constant, $g$ is the determinant of the spacetime metric, $g={\rm det}(g_{\m\n})$, $R$ is the Ricci curvature scalar built out of two derivatives of the metric, $R\sim \p\p g$, $\Lambda$ is a cosmological constant (a.k.a. the tension of the vacuum),  and the $\dots$ represent all other scalars one can build out of the metric and its derivatives.  The $R$ term in the action plays the role of a two-derivative kinetic term for the metric, $g_{\m\n}$.  GR is defined by retaining only the lowest dimension kinetic term, $R$, giving the {\em Einstein-Hilbert} action whose equation of motion is the {\em Einstein equation},
\be\label{eq:einstein}
R_{\m\n}-\half g_{\m\n} R = -\Lambda g_{\mu\nu}
\ee
The  $c_{n} R^{n} +\dots$ terms then represent higher-derivative corrections to GR.
The reader can verify that the \ads\ metric solves the Einstein equation (\ref{eq:einstein}) with the \ads-radius $L$ determined by the cosmological constant, $\Lambda$, as $\Lambda=-{d(d-2)\over 2L^{2}}$.  {One can then show}~\cite{Adams:2008zk}\ that \ads\ continues to be a solution even when we include generic higher-curvature corrections: all that changes is the precise relationship between $\Lambda$ and $L$.

Importantly, the Ricci scalar in \ads\ scales as $R\sim {1\over L^{2}}$ (c.f. (\ref{eq:adsricci})).  As a result, we may neglect higher-curvature corrections in \ads\ when $L^{2}$ is large compared to the appropriate powers of the dimensionful couplings $c_{n}$.  For example, if the higher-derivative operators are generated by quantum gravity effects, the dimensional couplings are controlled by the Planck length, $\ell_{p}$.  Quantum corrections to the metric can thus be neglected in \ads\ when $L$ is large compared to the Planck length, ${L\over\ell_{p}}\gg1$.  Similarly, GR receives corrections from string theory even at the classical level.  Such corrections become important at a characteristic length scale known as the {\em string scale}, $\ell_{s}$.  In \ads, these stringy corrections can be neglected so long as ${L\over\ell_{s}}\gg1$.
Our generic Wilsonian action for quantum gravity in \ads\ is thus well-approximated by classical GR when the \ads-radius is large compared to all scales in the problem,
\be
{L\over\ell_{p}}\gg1
\,,~~~~~~~~
{L\over\ell_{s}}\gg1
\,,~~~~~~~~
\dots\,,
\ee
where the $\dots$ denote other sources of corrections to GR which may kick in at other length scales.

\paragraph{Black Holes in \ads}\label{sssec:adsbh}

If pure empty \ads\ is a ground state for gravity, finite-temperature states correspond to black holes inside \ads.  The simplest such asymptotically-\ads\ black hole is the \ads-Schwarzchild black brane,
\be\label{eq:bhmetric}
ds^{2} = {L^{2}\over r^{2}} \[-f(r)\,dt^{2}+d\vec{x}^{2} +{1\over f(r)}dr^{2}\]\,\,,
\ee
with emblackening factor
\be
f(r) = 1-{r^{d}\over r_{H}^{d}} \,.
\ee
Near the asymptotic boundary at $r\to0$, $f\to1$, so this metric is asymptotically \ads$_{d+1}$.  At $r=r_{H}$, however, $f\to0$, signaling the presence of a black hole horizon.
This horizon in turn shields us from a physical singularity at $r\to\infty$ by ensuring that nothing which is inside the horizon, and thus sensitive to the singularity, can ever escape to influence events in the rest of the spacetime.  Since the entire solution is translationally invariant in the $(d-1)$ spatial directions, $\vec{x}$, this black brane is not a compact object, but rather extended in all directions
other than $r$.

Importantly, a classical black hole is a thermodynamic object with a definite temperature, energy and entropy, as shown by Bekenstein and Hawking (see~\cite{Bousso:2002ju} and references therein).  If this were not the case, we could violate the third law of thermodynamics by throwing all of our waste heat into a black hole and thus build perfectly efficient Carnot engines.  Black holes in \ads\ similarly carry definite temperatures, energies and entropies, though the details are a bit different from the flat space case~\cite{Hawking:1982dh}.  In the case at hand, the
Hawking temperature $T$, energy density $\eps$ and entropy density $s$ of the black brane are then,
\be
T = {d\over 4\pi \,\rH}
\,, ~~~~~~
\eps = {d-1\over16\pi\,\rH^{d}} \({L\over\ell_{p}}\)^{\!d-1}
\,, ~~~~~~
s ={1\over 4\, \rH^{\,d-1}} \({L\over\ell_{p}}\)^{\!d-1}
\,,
\ee
where $\ell_{p}$ is the Planck length defined by $G_{N}=\ell_{p}^{d-1}$.  The ratio ${L\over\ell_{p}}$ thus measures the \ads-scale in Planck units.

Physically, these thermodynamic relations are telling us that if we increase $\eps$ by throwing additional mass or energy into the black brane, the horizon swells outward toward the asymptotic boundary ($r_{H}\to0$) and the black brane heats up.  The specific heat of this black brane is thus positive.  This means that black holes in \ads\ can come to equilibrium.  Again, this can be traced to \ads\ playing the role of a harmonic trap for gravity: any radiation that the black hole evaporates away will, in finite time, fall back into the horizon.\footnote{This is very different from black holes in asymptotically flat space, whose specific heat is negative and which never come to equilibrium in the absence of external forces.  Black holes in \ads\ thus behave like familiar systems in the thermodynamic limit, while black holes in flat space behave more like finite volume
sub-systems.}

\paragraph{Charged Black Holes in \ads}
More generally, we can add charge to our black brane by adding a non-trivial Maxwell field,
\be
ds^{2} = L^{2}{-f(r)dt^{2}+d\vec{x}^{2} +{1\over f(r)}dr^{2} \over r^{2}} \,,
~~~~~~~~
A = A_{t}(r) dt\,,
\ee
with emblackening factor
\be\label{eq:emblackening}
f=1-M \,r^{d} + Q^{2} \,r^{2(d-1)}\,,
\ee
and electromagnetic scalar potential\footnote{Note that we are working with units in which $\hbar = 1$ and $c=1$.  For example, $M$ and $Q$ scale as energy and charge densities on the (non-compact) horizon, $M\sim E/V_{d-1}\sim L^{-d}$ and $Q\sim 1/V_{d-1}\sim L^{-(d-1)}$, as required by Eq (\ref{eq:emblackening}).  Similarly, both the $d+1$-dimensional scalar potential, $A_{t}$, and the $d$-dimensional chemical potential, $\mu$, scale as $1\over L$.}
\be
A_{t}(r) = \m \,\(1-\({r\over r_{H}}\)^{d-2}\)\,,
\ee
where $\m= {2Q\over\CC} \,{r_{H}^{~d-2}}$ and ${\cal C}=\sqrt{2(d-2)\over d-1}$.

This metric and gauge field together extremize the Einstein-Maxwell action,\footnote{Here $F_{\m\n}=\p_{[\m}A_{\n]}$ is the totally antisymmetric electromagnetic field strength tensor ($E^{i}=-F^{0,i}$ and $B^{i} =-\eps^{ijk}F_{jk}$) and
$F^{2}=F^{\m\n}F_{\m\n}=(\vec{E}^{2}\!-\!\vec{B}^{2})$ is the Maxwell kinetic term.}
\be
I_{ME} = {1\over 16\pi G_{N}}\int d^{d}x\,\sqrt{-g}\[-2\Lambda +R - {L^{2}\over 4e^{2}} F^{2}\]\,,
\ee
again with $\Lambda={-d(d-2)\over 2 L^{2}}$.
In this geometry, the horizon lies at the radial position $r=r_{H}$ implicitly defined
as the value of $r$ where $f(r)$ vanishes.  $M$ and $Q$ then determine the Hawking temperature of the horizon,
\be
T = {d\over 4\pi \rH }\(1-{d\!-\!2\over d}Q^{2}\rH^{2d-2}\)\,,
\ee
as well as its energy, entropy and charge densities,
\be
\eps = M\, {d-1\over16\pi} \({L\over\ell_{p}}\)^{\!d-1}
\,, ~~~~~~
s ={1\over 4\, \rH^{\,d-1}} \({L\over\ell_{p}}\)^{\!d-1}
\,, ~~~~~~
\rho =Q\,{d-1\over8\pi\CC} \({L\over\ell_{p}}\)^{\!d-1}
\,.
\ee
It is then straightforward to check that these variables satisfy the first law of thermodynamics, $d\eps = T ds + \m\,d\rho$, with $\mu$ playing the role of the chemical potential.

A curious feature of these charged black holes is that the zero-temperature limit has finite entropy density.  To see this, note that we can tune $Q$ to make $T$ vanish, $Q\to Q_{*}=\sqrt{d\over d-2}{1\over \rH^{d-1}}$.  In this limit, $\rH$, and thus the entropy density, remains non-zero.  This finite zero-temperature entropy represents an enormous degeneracy of the ground state of this black hole,\footnote{The precise microcanonical counting of this ground-state entropy was a long-standing puzzle in quantum gravity which was first solved, for a specific set of black holes in string theory, by Strominger and Vafa in 1996~\cite{Strominger:1996sh}.  A similar counting can now be performed for a much larger set of black holes, and in fact played a very important role in the discovery of holographic duality.} and correspondingly a large number of potential instabilities of these black hole solutions; these turn out to play an interesting role in holographic descriptions of quantum phase transitions.  For our present purposes, let it suffice to say that this is an interesting feature of these black branes which will crop up from time to time {in the following}.

\paragraph{Matter in \ads: The Effective Action as a Functional of Boundary Conditions}

The fact that \ads\ behaves like a homogeneous harmonic trap has important consequences when we consider matter fields in \ads.
Consider for simplicity a scalar field $\Phi$ with mass $m^{2}$ in the background of an uncharged \ads\ black brane (\ref{eq:bhmetric}) with classical action,
\be\label{EQ:ScalarAction}
I_{\Phi} \propto \int \!d^{d+1}\!x\, \sqrt{-g}\(-\half\(\p\Phi\)^{2}-{m^{2}\over2}\Phi^{2} +\dots\)\,,
\ee
where the $\dots$ denote interactions with other matter fields which we will for the moment neglect.  Working for a moment with a simple plane wave with momentum $k$ in the boundary dimensions, $\Phi(x,r)=\Phi(r)e^{ik\cdot x}$, the wave equation following from the above action becomes
\be
r^{2}f \,\Phi''(r) -r\[rf'\!-\!(d\!-\!1)f\]\Phi'(r) -\[k^{2}r^{2}\!+\!m^{2}L^{2}\]\Phi(r) = 0\,.
\ee
This equation has two simple degenerate points, one at the boundary where $r=0$ and the other at the horizon where $f=0$.  Let's study a general solution in the neighborhood of each.

Near the boundary at $r\to0$, where $f\to 1$, the general solution of this equation reduces to,
\be\label{EQ:PhiAsymp}
\Phi \sim \phi_{d-\Delta}(k) \,r^{d-\Delta} + \phi_{\Delta}(k)\, r^{\Delta} + \dots\,,
\ee
where the scaling dimension $\Delta$ is determined by\footnote{Similar scaling behavior is obtained for any set of matter fields, with the their resulting scaling dimensions $\Delta$ determined by their precise spins, masses and interactions.}
$\Delta(\Delta-d)=m^{2}L^{2}$.  So long as $m^{2}\ge -d^{2}/4L^{2}$, the scaling dimension is real.  This {\textit{Breitenloner-Freedman} (BF)} bound tells us that a small negative $m^{2}$ does not lead to an instability in \ads\ as it would in flat space -- instead, the would-be instability is lifted by the harmonic potential generated by the \ads\ curvature.  Note that in the flat space limit, $L\to\infty$, this window of ``allowed tachyons'' disappears.  So long as the BF bound is satisfied, specifying a solution reduces to specifying the two radial integration constants, $\phi_{d-\Delta}$ and $\phi_{\Delta}$.

A key fact about this asymptotic behavior is that, so long as $m^{2}L^{2}>1-{d^{2}\over4}$, the mode $\phi_{d-\Delta}$ is {\em non-normalizable} according to the natural norm on a constant-time spatial slice,
$\Sigma_{t}$,
\be
\(\Phi_{1},\Phi_{2}\) = -i\int_{\Sigma_{t}}dz d\vec{x}
\sqrt{-g}g^{tt} \(\Phi_{1}^{*}\p_{t}\Phi_{2}-\Phi_{2}^{*}\p_{t}\Phi_{1}\)\,.
\ee
Varying this non-normalizable mode thus corresponds to a large (divergent) change in the action.  We must thus fix the non-normalizable mode to make the variational problem in the bulk well-posed, for example by specifying a boundary condition for the bulk field $\Phi$ near the boundary, $r\to0$.
This leaves us with a one-parameter family of solutions labeled by the remaining integration constant, $\phi_{\Delta}$, which can be chosen arbitrarily.

We thus draw a general conclusion for studying matter fields in \ads: in order for the least-action principle to be well-defined in \ads, we must fix the values of the non-normalizable modes of all fields.
This is typically done by fixing boundary conditions for the non-normalizable modes of all fields at the \ads\ boundary, $r\to0$.

Near the horizon at $r_{H}$, where $f\sim f_{H}(1-r)\to0$, the equation of motion again degenerates.  Here, however, the physics is rather different.  Physically, this equation has two kinds of solutions: one which is regular, so that the degenerating $f\Phi''$ term is negligible; and one which is irregular at the horizon, so that the $f\Phi''$ term is not negligible.  But why are half the solutions irregular?  The reason was alluded to above: near the horizon, all waves can be expressed as a superposition of ingoing and outgoing waves.  To see that this is precisely what is going on, let's study our scalar equation at non-zero frequency.  Recalling that the magnitude of the $d$-momentum $k$ in the metric (\ref{eq:bhmetric}) is $k^{2} = -{r^{2}\over f} \w^{2}+r^{2}\vec{k}^{2}$, the dominant terms in the scalar equation near the horizon are
\be
f\,\p_{r}(f\,\p_{r} \Phi) +{\w^{2}}\Phi \sim 0 \,.
\ee
It is useful to switch to the near-horizon coordinate  $\r$ defined by $f\p_{r} = \p_{\r}$, i.e., $e^{-\rho}\sim(1-r)^{1/f_{H}}$, in terms of which the horizon lies at $\r\to\infty$.  In terms of this coordinate, the general solution can be expressed as a superposition of ingoing and outgoing waves as
\be
\Phi =     \Phi_{\mathrm{in}}(\rho)e^{-i\w(t-\r)}
            + \Phi_{\mathrm{out}}(\rho) e^{-i\w(t+\r)}
\,.\ee
For future reference, note that the in(out)-going solutions satisfy the first-order constraint at $r_{H}$,
\be\label{EQ:Ingoing}
f\,\p_{r} \Phi_{\mathrm{in}} =i\w\Phi_{\mathrm{in}}\,,
~~~~~~~~
f\,\p_{r} \Phi_{\mathrm{out}} =-i\w\Phi_{\mathrm{out}}\,.
\ee
Note that, in terms of the original radial coordinate, $r$, the phases of both of these time-dependent solutions accumulate near the horizon.   One can check that, in the limit of zero-frequency, the in-going wave is regular at the horizon while the out-going wave is irregular, matching our zero-frequency analysis above.  Note, too, that this suggests a natural prescription for analytic continuation to Euclidean time: we should choose the continuation so that the ingoing wave is regular at the Euclidean horizon, $e^{i\w\r}\to e^{-\w_{E}\r}$, i.e., we should define $\w=i\w_{E}$, where $\w_{E}>0$.

Again, this turns out to be a general result: in the presence of a black hole horizon, the Laplacian in any tensor representation takes the form $\square \sim r^{2}f \p_{r}^{2}+{r^{2}\w^{2}\over f}\dots$, so we must impose a boundary condition on our solutions such that they correspond to regular, in-falling modes at the horizon.

Thus, while our equation of motion is second order, there is only one linearly independent solution which is regular everywhere in the bulk and in-going at the horizon.
Our gravity problem in \ads, in the presence of a black hole horizon, becomes effectively a first-order problem.\footnote{A similar story holds for fermions, though the details differ because the Dirac equation is naturally first-order~\cite{Iqbal:2009fd}.}
Explicitly, fixing the non-normalizable mode $\phi_{d-\Delta}$ at the boundary and imposing in-falling boundary conditions at the horizon completely determines our solution for $\Phi$ throughout the bulk of \ads, and as a result determines $\phi_{\Delta}$.  However, in order to actually compute $\phi_{\Delta}$ given $\phi_{d-\Delta}$ we must solve our problem not just at the boundary but all the way through the bulk to the horizon, where we impose the appropriate boundary conditions.  The relation between $\phi_{d-\Delta}$ and $\phi_{\Delta}$ is thus also determined by IR physics, not just UV physics.

It is illuminating to see this work in detail in an analytically solvable example.  Consider again our scalar field $\Phi$ but now in pure \ads, {i.e.,} with no black hole.  This corresponds to studying our dual QFT at zero temperature and zero density.  Expanding in plane waves, $e^{-i(\w t -kx)}$, the general solution to the bulk equation of motion can be found in closed form,
\be\label{EQ:ExactSol}
\Phi(t,x,r) = \[\phi_{\mathrm{reg}}\, r^{d\over2}\,K_{\nu}\(\kappa r\)+\phi_{\mathrm{irreg}}\, r^{d\over2}\,I_{\nu}\(\kappa r\)\]e^{-i(\w t -kx)}\,\,,
\ee
where $\nu=\Delta-{d\over2}$, $\kappa=\sqrt{\w^{2}+\vec{k}^{2}}$,  $K_{\nu}$ and $I_{\nu}$ are modified Bessel functions, {and $\phi_{\mathrm{reg}}$ and $\phi_{\mathrm{irreg}}$ are the two integration constants.}  Since $I_{\nu}(\kappa r)\sim e^{\kappa r}$ near the ``\ads\ horizon'' at $r\to\infty$, regularity requires that $\phi_{\mathrm{irreg}}=0$.  We thus have just a single integration constant, $\phi_{\mathrm{reg}}$, which fixes the overall normalization of the in-falling mode in $\Phi$.  Near the boundary, the two asymptotic integration constants $\phi_{d-\Delta}$ and $\phi_{\Delta}$ in (\ref{EQ:PhiAsymp}) are not independent.  We can see how they are related by setting $\phi_{\mathrm{irreg}}=0$ in (\ref{EQ:ExactSol}) and expanding the solution near the boundary.  After a little algebra, this gives,
\be\label{EQ:SourceResponse}
\phi_{\Delta} = {\Gamma({d\over2}-\Delta)\over2^{\Delta-{d\over2}}\Gamma(\Delta-{d\over2})}  \(\w^{2}+k^{2}\)^{\Delta-{d\over2}}  \phi_{d-\Delta}\,.
\ee
Thus, once we impose regularity on our bulk solutions, we need only one boundary condition in the UV at $r\to0$ to fully specify a solution of the {$2^{\mathrm{nd}}$} order bulk equations of motion.  Meanwhile, we know we need to fix the non-normalizable mode $\phi_{d-\Delta}$ at the boundary so that our variational problem is well posed.  Once we impose regularity in the bulk, {i.e.,} in-falling at the horizon, the value of $\phi_{\Delta}$ is completely determined.

~

This teaches us an important lesson.  Suppose we want to study the partition function for our gravitational theory in \ads, integrating over all bulk fields, $\Phi(x,r)$.  To make the variational problem well-posed, we must specify a boundary condition for each of the bulk fields at the boundary of \ads, as discussed above,
\be\label{EQ:NNRenorm}
\phi_{d-\Delta}(x) =\lim_{r\to0}r^{\Delta-d}\Phi(x,r) \,\,,
\ee
where the factor of $r^{\Delta-d}$ should be thought of as the appropriate wavefunction renormalization of the bulk field $\Phi(x,r)$.  This means that the partition function in \ads, and thus the effective action {$\Gamma_{\mathrm{\ads}}=-\ln[Z_{\mathrm{\ads}}]$,} is a functional of the boundary conditions for all bulk fields,
\be
Z_{\mathrm{\ads}}[\phi_{d-\Delta}(x)]\equiv Z_{\mathrm{\ads}}[\Phi[\phi_{d-\Delta}(x)]]\,.
%= e^{\Gamma_{\mathrm{\ads}}[\phi_{d-\Delta}(x)]}
\ee

\paragraph{What One Means by ``A Theory of Gravity in \ads''}
\label{sssec:means}

In discussing holography, we will regularly refer to {classical gravity, quantum gravity and perturbative gravity} in \ads, so it's worth taking a moment to define our terms here.

By {quantum gravity in \ads} we mean a complete quantum description of the gravitational system, {i.e.,} string theory in \ads.  In general, string theory is a rich and complicated quantum theory which has no simple perturbative classical description.  {However, in some cases closed strings have a tractable} semiclassical low-energy expansion involving a metric, $g_{\m\n}$, minimally coupled to a host of dynamical matter  fields, $\{\Phi_{i}\}$, governed by an effective action of the form,
\be
I_{\mathrm{GR}} = {1\over 16\pi G_{N}} \int d^{d+1}\!x\,\sqrt{-g}\(  -2\Lambda + R + \dots + \CL_{\mathrm{matter}}(\Phi_{i})  \)\,.
\ee
Here $\CL_{\mathrm{matter}}(\Phi_{i})$ is the Lagrangian density for the matter fields and $R$ is the Ricci curvature scalar, which is roughly two derivatives of the metric.  $\sqrt{g}R$ plays the role of a kinetic term for the metric in {GR}.  The $\dots$ then represent an infinite tower of higher-derivative terms involving the metric, for example $R^{2}$, $R^{4}$, etc, which represent closed string corrections to {GR}.  Since these higher-derivative terms are also higher-dimension irrelevant operators, they must appear with dimensionful couplings.  In general these couplings are controlled by two independent length scales: $\ell_{p}$, the {Planck} length, which controls quantum corrections to the dynamics of classical strings, and $\ell_{s}$, which controls classical ``stringy'' corrections to the dynamics of point particles.

When expanding the theory around \ads\ with \ads-length $L$ (and thus with Ricci curvature $R\sim L^{-2}$), we thus have two dimensionless parameters controlling the various possible corrections to {GR}: ${\ell_{p}\over L}$, which tells us whether quantum corrections are important, and ${\ell_{s}\over L}$, which tells us whether stringy corrections are important.  Note that ensuring that the curvature is weak in Planck units, ${\ell_{p}\over L}\ll1$, does ensure that the system can be treated classically, but does not tell us that the gravity must be exactly GR: the curvature of our \ads\ may still be large compared to the string scale, so higher-curvature corrections to the Lagrangian of GR may not be negligible.  For pure classical GR to be a good approximation, two independent conditions must thus hold,
\be
{\ell_{p}\over L}\ll1 ~~~~~~ {\rm and} ~~~~~~ {\ell_{s}\over L}\ll1\,,
\ee
i.e., the curvature must be small compared both to the {Planck} scale and to the string scale.

%%%%%%%%%%%%%%%%
\subsubsection{Field Theories in Flat Space}

Our next job is to define the $d$-dimensional QFT we want to study.  A canonical way to do so begins by specifying a list of local operators $\CO_{i}$ labeled by their Lorentz structure, their charges $q_{i}$ and their scaling dimensions $\Delta_{i}$,  all defined at some UV fixed point.  To generate an RG flow of interest, we perturb away from this fixed point by turning on appropriate sources $J_{i}$.  The basic observables of the theory are then correlation functions of products of local operators,
\be
%\vev{\CO_{1}(x_{1}) \dots \CO_{n}(x_{n})}_{J^{*}}\,.
\vev{\CO_{1}(x_{1}) \dots \CO_{n}(x_{n})}\,.
\ee
These can be conveniently encoded in terms of the quantum generating functional,
\be
Z_{\mathrm{QFT}}[J_{i}(x)] = \vev{ e^{\int dx^{d}  J_{i}(x)\CO_{i}(x)}}\,,
\ee
where the $J_{i}(x)$ represent a set of sources and couplings for the operators $\CO_{i}$.
Note that the scaling dimensions and tensor structure of the sources $J_{i}$ are completely determined by those of $\CO_{i}$.
This allows us to express correlation functions as derivatives of the partition function,
\be
\vev{\CO_{1}(x_{1}) \dots \CO_{n}(x_{n})}
=
%\left.
{\delta^{n}   \ln Z_{\mathrm{QFT}}[J(x)]  \over \delta J_{1}(x_{1})\dots \delta J_{n}(x_{n})}
{\Big |}_{J_{i}=0} \,,
\ee
where the restriction  $|_{J_{i}=0}$ means evaluate the final result with all remaining sources turned off.
Solving the theory then requires one to compute the partition function, $Z_{\mathrm{QFT}}[J_{i}(x)]$.

\paragraph{Canonical Example: $SU(N)$ Yang-Mills at {Large $N$}}

The classic example of a field theory with a well-understood and controlled holographic dual, to which we will appeal below, is a special version of $SU(N)$ Yang-Mills gauge theory in flat 4d spacetime called the $\CN=4$ theory.  Here $\CN$ refers to the amount of supersymmetry enjoyed by the theory; $\CN=4$ is the most one can have in 4d without including gravity.  For our purposes, the role of supersymmetry is nothing more than a way of turning off quantum mechanics without totally trivializing the theory: with $\CN=4$ supersymmetry, some quantities do not receive quantum corrections beyond one-loop, {and} so can be computed at weak coupling and reliably extrapolated to strong coupling. In particular, the $b$-function of the theory can be computed exactly and is identically zero for all values of the coupling.\footnote{Our ability to compute various quantities at both strong and weak coupling in the $\CN=4$ theory was key to the original discovery of holographic duality, and is the reason this example is {both illuminating and} canonical.  However, that is the extent of the role of supersymmetry, to provide unusually simple and tractable examples.  It is no more a necessary feature of holography than spherical symmetry is a necessary property of hydrogen. The maximally-supersymmetric $\CN=4$ theory is an illuminating example to study, so we turn to it now, but we stress that the general structure of holographic duality thus revealed is more general, and does not depend on supersymmetry.}

The  basic ingredients of the $\CN=4$ theory are a gauge group, $G=SU(N)$, a gauge field, $A_{\mu}(x)$, transforming in the {adjoint} of $G$,  six scalars $\phi^{I}(x)$ also transforming in the {adjoint} of $G$ and further enjoying a global $SO(6)$ symmetry, and large set of fermions.  {We can safely neglect the fermions for now, as their only role} is to ensure supersymmetry.  The Lagrangian is
\be
\CL_{\mathrm{YM}} = -{1\over 4g^{2}}\(\Tr F^{2} + \Tr |D\phi_{I}|^{2} + \dots\)\,,
\ee
where $g^{2}_{\mathrm{YM}}$ is the gauge coupling, $D_{\mu}$ is the gauge-covariant derivative, $\dots$ denotes fermionic terms {we neglect,} and all fields are canonically normalized with an overall factor of $-{1\over 4g^{2}}$ in front of their kinetic terms.

What are the good operators of this theory?  {We may suppose that} any good gauge-invariant combination of the fundamental fields is a perfectly good observable. However, since we have a global symmetry, it is useful to write this list in terms of $SO(6)$ {irreducible representations.} For example, the following three Lorentz scalars,
\be
\CO = \sum_{I}\Tr \phi_{I}^{2}\,,
~~~~~~~~
\CO_{IJ} = \Tr \phi_{(I}\phi_{J)}\,,
~~~~~~~~
\CO^{\mu} = \Tr \CJ^{\mu}\,,
\ee
transform as a scalar, a symmetric {2-tensor} and a vector, respectively, under the global $SO(6)$, with the latter corresponding to a charge current operator.  In the weak-coupling limit, their scaling dimensions $\Delta_{i}$ are just their engineering dimensions which can be read off the Lagrangian.\footnote{Here, \textit{engineering dimension} refers to the naive dimension of an operator as determined by dimensional analysis on the kinetic terms in the classical Lagrangian, while the true scaling dimension is determined by the scale-dependence of correlation functions containing the operator in the full interacting quantum theory.  For a review of engineering and general scaling dimensions see e.g. the text by Sachdev~\cite{SachdevBook}.}

Now, as pointed out by 't Hooft~\cite{G't1974461}\ (see also~\cite{Coleman:1980nk}), anytime you have a gauge theory of $N\times N$ matrices, there is a natural way to organize the {Feynman} diagrams in terms of underlying genus-$g$ surfaces.\footnote{Here the \textit{genus} refers to the number of handles on a 2-d surface: a sphere or plane has genus zero, a donut has genus 1, a donut with 2 holes has genus 2, etc.}  Explicitly, since every field comes with two gauge indices, and since all indices in a gauge invariant observable must be contracted, every {Feynman} diagram can be written as a {\textit{ribbon diagram}} in which each propagator is a ribbon with a gauge running along each side of the ribbon.  One can then show that any given {Feynman} diagram, with any number of underlying loops, can be drawn without the index lines crossing only on a genus-$g$ Riemann surface, where the specific genus $g$ depends on precisely how the index lines are contracted to make the given {Feynman} diagram.  Explicitly, a diagram with genus $g=0$ can be drawn on a piece of paper with no lines crossing, a diagram of genus $g=1$ must be drawn on a torus to have no lines cross, etc.

{A remarkable result} is that if we reorganize the loop expansion in $g^{2}_{\mathrm{YM}}$ as a double expansion in $N$ and $\l=g^{2}_{\mathrm{YM}}N$, then every observable can be expressed as a power series expansion in $N$ where the powers of $N$ for a given diagram is determined only by the genus $g$ on which that diagram can be drawn with no crossings.  For example, the free energy $\CF=\ln Z$ takes the form
\be
\CF = N^{2} f_{0}(\l) + f_{1}(\l) +{1\over N^{2}} f_{2}(\l) + \dots = N^{2}\sum_{g}{f_{g}(\l)\over N^{2g}}\,,
\ee
where $f_{g}(\l)$ is the sum of all diagrams arising at genus $g$ and is a function of only $\l$, independent of $N$.   A similar result {is obtained} for every possible observable, with the only difference being the overall factor of $N$,
\be
\CA = N^{m} \(\CA_{0}(\l) + {1\over N^{2}}\CA_{1}(\l) +{1\over N^{4}} \CA_{2}(\l) + \dots\) = N^{m}\sum_{g}{\CA_{g}(\l)\over N^{2g}}
\ee
This double expansion implies a remarkable simplification at large $N$: if $N\gg1$, the only {term} which matters in the genus expansion is the leading {planar} term; any diagram that cannot be drawn on a piece of paper without crossings, while possibly large in numerical value, is dwarfed by the much larger terms coming from the planar diagrams.

Another way of expressing this simplification at {large $N$} is on terms of {\em {large $N$} factorization}.  Consider a set of single-trace operators\footnote{The meaning of single-trace operators is slightly subtle, but can roughly be understood as the set of operators which have local, extensive classical limits.} of the form $\CO_i =\Tr(\dots)$.  At {large $N$,} correlation functions of single trace operators factorize,
\be
\vev{\CO_1(x_1)\CO_2(x_2)} = \vev{\CO_1(x_1)}\vev{\CO_2(x_2)} + O\({1\over N^2}\)\,.
\ee
This follows because the disconnected parts of the diagrams necessarily involve more closed index loops than connected parts, and thus additional powers of $N$.  This does not imply that the {large $N$} theory is free, for while these correlators do factorize, the individual one-point functions remain non-trivial. {For example, the anomalous dimensions of all operators are} controlled by the 't~Hooft coupling, $\l$.

We thus see that {large $N$} gauge theories have two natural coupling constants: $1\over N$, which controls the genus expansion and factorization; and $\l$, which controls the perturbative corrections to each term in the genus expansion and the anomalous dimensions of {large $N$} factorized operators.
Suggestively, this is precisely the structure of closed string perturbation theory, where every amplitude is expressed as a sum of contributions from various genera, with the {loop-counting} parameter given by the string coupling $g_{s}$ and the amplitude of a given genus determined by an auxiliary quantum field theory whose interactions are determined by the string tension {$\a'$ via the relation}
\be
\CA = g_{s}^{n} \(\CA_{0}(\a') + {g_{s}^{2}}\CA_{1}(\a') +{g_{s}^{4}} \CA_{2}(\a') + \dots\) = g_{s}^{n}\sum_{g} g_{s}^{2g}\CA_{g}(\a')\,.
\ee
This analogy led various people to speculate that {large $N$} gauge theories should be captured by some theory of closed strings.  This turned out to be wrong, but only just {barely:} at {large $N$,} these theories {\em are} dual to closed sting theories, but the closed strings live in one higher dimension!

As an aside, while not every theory is a gauge theory of $N\times N$ matrices, it is useful to keep in mind this example when thinking about more general examples. What $N^{2}$ is really measuring is the number of degrees of freedom per unit volume, {or per} lattice site if we put the theory on a lattice.  Similarly, $\l$ is measuring the strength of the dominant interactions when the number of degrees of freedom grows large.  Many theories which are not simple gauge theories nonetheless have a useful notion of $N$ and $\lambda$.  Exactly when one obtains such a controlled double expansion remains in general a key open question in holography.

\subsubsection{The Holographic Dictionary}

The basic claim of holography is that every $d$-dimensional QFT defined as above
can be exactly reorganized into a $(d+1)$-dimensional quantum theory of gravity and matter propagating in \ads$_{d+1}$.  The precise relationship between these two theories is known as the \textit{holographic dictionary},
a sketch of which is outlined in Table (\ref{table:holdict}).  The rest of this subsection will develop various of the entries in the dictionary, including a few canonical examples of computations performed via the holographic dictionary.

\begin{table}[t!]\label{table:holdict}
\begin{center}
\begin{tabular}{| l r c l r |  }
%  \multicolumn{5}{c}{The Holographic Dictionary} \\
  \hline
  Boundary QFT &~~~~~~~~~~~~~~&~~~~~~~~~~~~~~~&~~~~&  Bulk Gravity \\
%& & & & \\
  \hline
  \hline
%  \hline
\vspace{-0.2cm}& & & & \\
  Operator 				&$\CO(x)$		&$\longleftrightarrow$	& $\Phi(x,r)$		&  Field \\
  Spin 					&$s_{\CO}$		&$\longleftrightarrow$	&$s_{\Phi}$ 		& Spin \\
  Global Charge 			&$q_{\CO}$		&$\longleftrightarrow$	&$q_{\Phi}$ 		& Gauge Charge \\
  Scaling dimension 			&$\Delta_{\CO}$	&$\longleftrightarrow$	&$m_{\Phi}$ 		& Mass \\
  Source  					&$J(x)$	 		&$\longleftrightarrow$	&$\Phi(x,r)|_{\p}$ 	& \!\!\!\! Boundary Value (B.V.) \\
  Expectation Value 			&$\vev{\CO(x)}$ 	&$\longleftrightarrow$	& $\Pi_{\Phi}(x,r)|_{\p}$ 	& \!\!\!\!\!\!\!\!B.V. of Radial Momentum \\
\vspace{-0.2cm}& & & & \\
  Global Symmetry Group		&$G$			&$\longleftrightarrow$	&$G$				&Gauge Symmetry Group\\
  Source for Global Current	&$\CA_{\m}(x)$		&$\longleftrightarrow$	& $A_{\m}(x,r)|_{\p}$ 	&B.V. of Gauge Field\\
  Expectation of Current		&$\vev{\CJ^{\m}(x)}$	&$\longleftrightarrow$	& $\Pi_{A}^{\m}(x,r)|_{\p}$ 	& B.V. of  Momentum\\
\vspace{-0.2cm}& & & & \\
  Stress Tensor 				&$T^{\m\n}(x)$	 	&$\longleftrightarrow$	& $g_{\m\n}(x,r)$		&Spacetime Metric\\
  Source for Stress-Energy 	&$h_{\m\n}(x)$		&$\longleftrightarrow$	& $g_{\m\n}(x,r)|_{\p}$ 	& B.V. of Metric\\
  Expected Stress-Energy	\!\!\!\!			&$\vev{T^{\m\n}(x)}$	&$\longleftrightarrow$	& $\Pi_{g}^{\m\n}(x,r)|_{\p}$ 	& B.V. of Momentum \\
\vspace{-0.2cm}& & & & \\
  \# of Degrees of Freedom 	&\multirow{2}{*}{$N^{2}$}	&\multirow{2}{*}{$\longleftrightarrow$}	&\multirow{2}{*}{$\({L\over \ell_{p}}\)^{d-1}$}	& Radius of Curvature \\
  Per Spacetime Point		&&&&  In Planck Units\\
\vspace{-0.2cm}& & & & \\
  Characteristic Strength	&\multirow{2}{*}{$\l$}	&\multirow{2}{*}{$\longleftrightarrow$}	& \multirow{2}{*}{$\({L\over\ell_{s}}\)^{d}$}	& Radius of Curvature \\
  of Interactions			&&&&  In String Units\\
\vspace{-0.2cm}& & & & \\
  \hline
  \hline
\vspace{-0.2cm}& & & & \\
  QFT Partition Function  	
  &\multirow{2}{*}{$Z_{\mathrm{QFT_{d}}}[J_{i}]$}
  &\multirow{2}{*}{$\longleftrightarrow$}
  &\multirow{2}{*}{$Z_{\mathrm{QG_{d\!+\!1}}}[\Phi_{i}[J_{i}]]$}\!\!\!\!\!\!\!\!
  & QG Partition Function  \\
  with Sources $J_{i}(x)$%, Coupling $\l$ 	
  &&&&
  in \ads\ w/ $\Phi_{i}|_{\p}=J_{i}$ \\
\vspace{-0.2cm}& & & & \\
  QFT Partition Function
  &\multirow{2}{*}{$Z^{\l,N\gg1}_{\mathrm{QFT_{d}}}[J_{i}]$}
  &\multirow{2}{*}{$\longleftrightarrow$}
  &\multirow{2}{*}{$e^{-I_{\mathrm{GR_{d\!+\!1}}}[\Phi[J_{i}]]}$}\!\!\!\!\!\!\!\!
  & Classical GR Action\\
  at Strong Coupling%, $\l\gg1$ 	
  &&&
  & in \ads\ w/ $\Phi_{i}|_{\p}=J_{i}$\\
\vspace{-0.2cm}& & & & \\
  QFT $n$-Point    	&	
  \!\!\!\!\!\!\!\!\!\!\!\!\!\!\!\!\!\!\!\!\!\!\!\!\!
  \multirow{3}{*}{$ \vev{\CO_{1}(x_{1}) \dots \CO_{n}(x_{n})}$}	&&
  \multirow{3}{*}{${\delta^{n}   I_{\mathrm{GR_{d\!+\!1}}}[\Phi[J_{i}]]  \over \delta J_{1}(x_{1})\dots \delta J_{n}(x_{n})}{\Big |}_{J_{i}=0}$}
  \!\!\!\!\!\!\!\!\!\!\!\!\!\!\!\!\!\!\!\!\!\!\!\!\!
  & Classical Derivatives of \\
  Functions at  	& &$\longleftrightarrow$&& the On-Shell Classical \\
  Strong Coupling	& &			 	      && Gravitational Action \\
\vspace{-0.2cm}& & & & \\
\hline
\hline
\vspace{-0.2cm}& & & & \\
  Thermodynamic State 		&				&$\longleftrightarrow$	& 				& Black Hole  \\
  Temperature 				&$T$			&$\longleftrightarrow$	&$T_{H}$			& Hawking Temperature $\sim$ Mass \\
  Chemical Potential	 		&$\m$			&$\longleftrightarrow$	&$Q$			& Charge of Black Hole \\
  Free Energy 				&$F$			&$\longleftrightarrow$	&$I_{\rm GR}|_{\rm (on-shell)}$	& On-Shell Bulk Action \\
  Entropy 					&$S$			&$\longleftrightarrow$	&$A_{H}$ 		& Area of Horizon \\
& & & & \\
\hline
\end{tabular}
\end{center}
\caption{Elements of the holographic dictionary.}
%Each line gives a pair of quantities which translate into each other under the holographic dictionary.  Sometimes the arrow represents an equality upon suitable decoration (for example, the boundary value of a field must be renormalized before matching to the associated source in the boundary theory), while sometimes the relationship is more complicated (for example, the temperature depends on both the mass and charge of a black hole, but once we fix the chemical potential, only the mass is free to vary, so it is useful to think of the mass of the black hole as controlling the temperature of the QFT).}
\end{table}

\paragraph{Operators and Fields}
The first entry in the holographic dictionary relates the operators of the QFT to the matter fields in the bulk gravity: to every local operator $\CO_{i}(x)$  with quantum numbers (or \textit{charges}) $q_{i}$ and scaling dimension $\Delta_{i}$ in our QFT is associated a field $\Phi_{i}(x,r)$ in \ads\ which carries the Lorentz structure and quantum numbers required to act as a coupling for $\CO_{i}(x)$ in the Hamiltonian,
Note that the dimension of the operator, $\Delta_{i}$, is determined by the mass of the bulk field, as we saw above in (\ref{EQ:PhiAsymp}).\footnote{More generally, $\Delta$ depends on the Lorentz structure of the field $\Phi_{i}$ and on the full set of its bulk interactions. As a simple example, the relation for a scalar operator corresponding to a scalar field in the bulk is $m^{2}L^{2}=\Delta(\Delta-d)$.}   This means that the mass and other couplings of the bulk gravity theory generally do not correspond to interactions of the boundary theory in any simple way, but rather determine {which} boundary QFT we are studying.

\paragraph{An Equivalence of Partition Functions}
Given this map, the central claim of holographic duality~\cite{Maldacena:1997re,Gubser:1998bc,Witten:1998qj}\ can be succinctly expressed as
\be
\label{EQ:centralclaim}
\boxed{
\phantom{\int}Z_{\mathrm{QFT}}[J_{i}] ~=~ Z_{\mathrm{QG}}[\Phi[J_{i}]]\,,\phantom{\int}
}
\ee
where  $Z_{\mathrm{QFT}}[J_{i}]$ is the partition function of the QFT as a function of the sources $J_{i}$ for each operator $\CO_{i}$, while $Z_{\mathrm{QG}}[\Phi[J_{i}]]$ is the quantum partition function of the gravitational theory described in {Sec.}~(\ref{sssec:means}) computed in an \ads\ spacetime background.\footnote{In general, $Z_{\mathrm{QG}}$ means the partition function of closed string theory expanded around the \ads\ geometry.  In practice we can only compute this explicitly in special cases, and then only in regimes in which the string coupling and curvature are sufficiently weak.  Thus we may use this duality either to deduce the LHS {of Eq.~(\ref{EQ:centralclaim})} when the gravity is classical and weakly curved, or to define the RHS {of Eq.~(\ref{EQ:centralclaim})} when the QFT is computable.}   As discussed, the quantum gravity partition function must be evaluated on field configurations $\Phi_{i}$ which asymptote at the boundary to the sources $J_{i}$ of the QFT, hence the notation $\Phi[J]$.  This recalls the heuristic derivation above: the bulk fields are precisely the coupling constants of the QFT promoted to dynamical fields on the full RG-extended spacetime in which the RG scale becomes a physical coordinate.

An important ingredient in this recipe is a proper definition of the boundary value of the bulk field.  As we saw above for a bulk scalar $\Phi$ of mass $m$, dual to a scalar operator $\CO$ of dimension $\Delta$ given by $\Delta(d-\Delta)=m^{2}L^{2}$, the boundary value of the bulk field is in general divergent.  To specify the boundary condition we must rescale the bulk field by an appropriate wavefunction renormalization which picks off the leading divergence of the bulk field near the boundary,
\be
J(x)\equiv \phi_{d-\Delta}(x) =\lim_{r\to0}r^{\Delta-d}\Phi(x,r) \,.
\ee
We will return to this renormalization when we discuss the computation of one- and two-point functions.

\paragraph{Strong Coupling and Perturbative Gravity}
\label{sssec:strong}
In general, both sides of {Eq.~(\ref{EQ:centralclaim})} are complicated and computationally intractable objects.  {However, when one side or the other can be evaluated semi-classically,} we can use {Eq.~(\ref{EQ:centralclaim})} to generate a controlled strong-coupling expansion via the dual description.
Precisely whether, and when, such a limit is obtained is a delicate question that must be studied in detail on a case by case basis.
To get a sense for the general structure, let's consider the canonical example, the {large $N$} limit of the 4d $\CN=4$ gauge theory with gauge group $G=SU(N)$ and 't~Hooft coupling $\l$.    The dual theory turns out to be a closed string theory on \ads$_{5}$ with \ads-radius $L$.
The holographic dictionary then tells us that,
\be\label{EQ:Scalings}
N^{2}  = \({L\over \ell_{p}}\)^{d-1}\,,
~~~~~~~~
\l = \({L\over\ell_{s}}\)^{d}\,,
\ee
where $\ell_{p}$ is the Planck length determining the scale at which quantum effects occur in the bulk, and $\ell_{s}$ is the string length controlling the scale of stringy higher-curvature corrections to the bulk gravitational action, as discussed in Sec.~\ref{sssec:means}.   The first relation {in Eq.~(\ref{EQ:Scalings})} thus tells us that the bulk description becomes classical ($L\gg\ell_{p}$) only in the {large $N$} limit,\footnote{This provides an out from the Weinberg-Witten argument, which assumes that there are a finite number of local degrees of freedom.} and that the gravitational description is deeply quantum mechanical when $N$ is small.  In other words, the loop counting parameter in the bulk is $1\over N$.

With an eye toward the general class of holographic theories, it is worthwhile unpacking Eq. (\ref{EQ:Scalings}) in more detail.  In the case at hand, i.e. $SU(N)$ SYM, $N^{2}$ measures the number of degrees of freedom per point in the QFT.  In particular, all extensive thermodynamic quantities (entropy, energy, etc.) must scale as $N^{2}$.  In this light, what Eq. (\ref{EQ:Scalings}) shows is that the quantum corrections in the bulk are negligible when the number of degrees of freedom per point in the dual QFT is large, and vice-versa.  Indeed, in a host of controlled models in string theory, the appropriate version of the first equation in Eq. (\ref{EQ:Scalings}) takes the form, $N^{b}= \({L/ \ell_{p}}\)^{d-1}$, where $N$ is a conserved charge and $b$ is some real parameter~\cite{Silverstein:2003jp}.  Similarly, it is not always possible to interpret $\l$ as a \'t\,Hooft parameter in the dual QFT.  More generally, the role of $\l$ is played by the typical scale of anomalous dimensions in the QFT~\cite{Heemskerk:2009pn}.  We can understand this as follows: if the QFT is weakly-interacting, the effects of renormalization will generally be weak and most observables will be well-approximated by their classical cousins, i.e., quantum anomalous dimensions will be small; if interactions are strong, quantum effects will generally drive anomalous dimensions to be large.

While $N\gg1$ ensures that the Planck scale is negligible, so that the gravitational interactions can be treated with classical effective field theory, it does not tell us that the system is well approximated by pure {GR}. Indeed, in principle there will be a host of higher-curvature corrections to the classical GR action suppressed by powers of $\ell_{s}\over L$.  To ensure that these corrections are in fact negligible, {i.e.,} that $L\gg\ell_{s}$, we must furthermore take the 't~Hooft coupling to be large, $\l\gg1$.  Conversely, when the QFT is weakly coupled, $\l\ll1$, the bulk geometry is sufficiently strongly curved that GR is swamped by higher curvature corrections to the stringy effective action.

The main lesson here is that when the QFT is perturbative, the bulk gravity is out of control, and that when the QFT is strongly interacting and has a large entropy density, the dual gravitational description is simple semi-classical gravity.
To see why such a weakly-curved limit is interesting, let's consider our canonical example at large $N$,  $N\gg1$, and with strong 't~Hooft coupling, $\l\gg1$.  Let's also analytically continue to imaginary time, $\w=i\w_{E}$ with $\w_{E}>0$.  In this limit, the gravity sector reduces to classical Euclidean GR.  In terms of our fundamental relation, the RHS can thus be expressed as a sum over saddles.  Focusing on the dominant saddle, we get
\be
Z_{\mathrm{QFT}}[J] ~\simeq~ e^{-I_{\mathrm{GR}}[\Phi[J]]}\,,
\ee
where
\be
\label{EQ:IGR}
I_{\mathrm{GR}}[\Phi[J]] = {1\over 16\pi G_{N}} \int \!d^{d+1}\!x\, \sqrt{-g}\(R-2\LL +\CL_{m}\(\Phi\)\)
\ee
is the classical gravitational action expanded about the dominant classical saddle, subject again to the condition that the normalizable modes of all matter fields are determined by the sources of the dual QFT.  {Equation~(\ref{EQ:IGR})} thus defines a simple classical functional of the sources from which we can compute quantum correlation functions of the dual QFT via the expressions above as,
\be\label{EQ:ClassCorr}
\boxed{
\vev{\CO_{1}(x_{1}) \dots \CO_{n}(x_{n})}
=
{\delta^{n}   I_{\mathrm{GR}}[\Phi[J_{i}]]  \over \delta J_{1}(x_{1})\dots \delta J_{n}(x_{n})}{\Big |}_{J_{i}=0}
}\ee
The upshot is that, when such a dual semi-classical limit exists and is reliable, we can compute the quantum correlation functions of our QFT by finding solutions to the classical field equations of the dual gravitational system, evaluating the action on-shell as a functional of the boundary values of the fields, and taking appropriate derivatives.

\paragraph{Example: One- and Two-point Functions for a Scalar Operator}

As an example of this machinery at work, let's compute the one- and two-point functions for a boundary operator $\CO$ dual to a free bulk scalar field $\phi$.  For simplicity, we will work in Euclidean momentum space throughout.

For the one-point function, the basic form of the computation is straightforward.  Intuitively, our prescription (\ref{EQ:ClassCorr}) tells us that the one-point function corresponds to a variation of the classical {gravitational} action {with respect to} the boundary value of the bulk field.  But a variation of the action {with respect to} the boundary value of the field is precisely the conjugate momentum, $\Pi$, of the field $\Phi$ in the direction normal to the boundary,
\be\label{eq:momentum}
\Pi = -\sqrt{-g}g^{rr}\p_{r}\Phi\,.
\ee
Thus we should expect to find, with $k\equiv\(\w_{E},\vec{k}\)$,
\be
\vev{\CO(k)}
=
{\delta I_{\mathrm{GR}}[\Phi[J]]  \over \delta J(k)}\,
\sim \,\lim_{r\to0}\Pi(k,r)\,.
\ee
This is almost correct.  The trouble is that the on-shell classical action, $I_{\mathrm{GR}}[\phi]$, and the bulk field, $\Phi$, both generically diverge as we approach the boundary, {cf.} Eq (\ref{EQ:PhiAsymp}).  To absorb these divergences we must (a) regulate all quantities by evaluating them not at the boundary at $r=0$ but at $r=\eps\ll1$, (b) rescale the bulk field by an overall wavefunction normalization as in (\ref{EQ:NNRenorm}), and (c) add boundary counterterms to the action evaluated at $r=\eps$ so that the renormalized on-shell action remains finite as we remove the cutoff.  In retrospect, this should not be surprising: the radial coordinate $r$ is playing the role of a lattice cutoff, so we must work with properly renormalized quantities to avoid confusions. This \textit{holographic renormalization}~\cite{Balasubramanian:1999re,Bianchi:2001de,Bianchi:2001kw,Skenderis:2002wp,Heemskerk:2010hk,Faulkner:2010jy}\ is simply the bulk realization of the UV renormalization of the boundary QFT.

In fact, we have already seen this effect above.  Recall that the relationship between the amplitude of the non-normalizable mode $\phi_{d-\Delta}$ and the boundary value of the bulk field $\Phi$ also required a subtle {wavefunction renormalization} (\ref{EQ:NNRenorm}), so that the source is given by,
\be
J(k)=\phi_{d-\Delta}(k) =\lim_{r\to0}r^{\Delta-d}\Phi(k,r) \,.
\ee
Upon performing a similar renormalization of the bulk Euclidean action, one finds that the correct expression for the one-point function in terms of the renormalized radial momentum is\footnote{For a detailed discussion of this calculation see Appendix C of~\cite{CasalderreySolana:2011us}.  The basic strategy can be found in~\cite{Balasubramanian:1998de,Klebanov:1999tb,Papadimitriou:2004rz,Iqbal:2008by,Iqbal:2009fd}.}
\be
\vev{\CO(k)} = \lim_{r\to0}r^{d-\Delta}\Pi(k,r)\,.
\ee
Together with (\ref{eq:momentum}) and (\ref{EQ:PhiAsymp}), this gives
\be
\vev{\CO(k)} ={2\Delta-d\over L}\phi_{\Delta}(k)\,,
\ee
where $\phi_{\Delta}$ is the coefficient of the sub-leading term in  (\ref{EQ:PhiAsymp}).
Thus, just as the non-normalizable mode of the bulk field $\phi_{d-\Delta}$ determines the source, $J$, for the boundary operator $\CO_{\Delta}$, the subleading normalizable term $\phi_{\Delta}$ determines the response, $\vev{\CO}$.

This make computing the linear-response Green functions {easier.}  In linear response theory, an infinitesimal source $J(x)$ generates a response which is linearly proportional to the source, with the ratio defining the linear-response Green function,
\be
G_{E}(k)
\equiv
{\vev{\CO(k)} \over J(k)}
=
\lim_{r\to0}r^{2(\Delta-d)}{\Pi(k,r)\over\Phi(k,r)}\,.
\ee
Using our above expressions for the source, $J$, and response, $\vev{\CO}$, we thus find
\be
\label{EQ:limit}
G_{E}(k)
=
{2\Delta-d\over L}{\phi_{\Delta}(k)\over\phi_{d-\Delta}(k)}
\ee
Note that we could also have derived this result by taking two derivatives of the partition function.

Computing the 1- and 2-point functions thus {follows from} determining $\phi_{\Delta}$ in terms of $\phi_{d-\Delta}$.
As we have seen, in \ads\ these two modes are not independent: requiring the bulk solution to be regular at the Euclidean horizon imposes a relation which we can use to determine the response $\phi_{\Delta}$ in terms of the source $\phi_{d-\Delta}$.  In practice, finding the precise relation involves solving the bulk equation of motion subject to the boundary conditions, {i.e.,} to solving a set of $2^{\rm nd}$-order elliptic {partial differential equations.}  For simple examples this can be done analytically; more generally one is forced to use some form of matched asymptotic expansion or numerical integration to determine $\phi_{\Delta}$ in terms of $\phi_{d-\Delta}$.

The key point is that the computation of the quantum correlation function of the boundary QFT has {been} reduced to solving a set of classical {partial differential equations} with fixed boundary conditions.

An example where this can be done analytically is our example of a massive scalar in pure \ads, for which we have in fact already determined this relation in momentum {space, Eq.~(\ref{EQ:SourceResponse}).}  Plugging this in and again working in momentum space gives, for the Euclidean Green function,
\be\label{EQ:GE}
G_{E}(k) = {2\Delta-d\over L}{\Gamma({d\over2}-\Delta)\over2^{\Delta-{d\over2}}\Gamma(\Delta-{d\over2})}  \(k^{2}\)^{\Delta-{d\over2}}\,.
\ee
This is the exactly the form for the two point function of a scalar operator with scaling dimension $\Delta$ in a CFT.

\paragraph{Real-time Response and Retarded Green Functions}\label{sec:realtime}

So far we have focused on Euclidean correlation functions.  In fact, these holographic techniques can be readily extended to intrinsically Lorentzian computations needed for real-time response.

The simplest thing to do would be to simply compute the Euclidean Green function and analytically continue.  In practice this is often not tractable. For example, we will sometimes be interested in particles moving on light-like trajectories which are not easily described in Euclidean continuation.  Instead, one can construct an intrinsically real-time holographic prescription, as was first proposed by Son and Starinets~\cite{Son:2002sd}\ by essentially analytically continuing the Euclidean {prescription.} Their results have since been justified by developing a full Holographic Schwinger-Keldysh formalism\cite{Maldacena:2001kr,Herzog:2002pc,Skenderis:2008dh}.  Here we will forego the formalities.  We assume that such a justification can be made and proceed to the prescription.

The prescription requires repeating our Euclidean prescription step by step in the real-time Lorentzian geometry.  The key difference is that we must choose an appropriate boundary condition at the horizon.  The appropriate choice depends on which Green function we wish to compute.\footnote{Since we are only studying infinitesimal sources and responses, there is no reason to restrict to finite action modes, {i.e.,} to solutions which are regular at the horizon. This is a simplification afforded by linear response.  More generally, when considering non-trivial bulk field configurations, we should again impose regularity and in-falling boundary conditions.}  Intuitively, one might expect that the retarded Green function, $G_{R}$, should correspond to imposing causal in-falling boundary conditions at the horizon, while the advanced Green function, $G_{A}$, should involve acausal out-going boundary conditions.  This turns out to be precisely correct.  Once we have a solution satisfying the appropriate boundary conditions, we again compute the properly renormalized on-shell Lorentzian action, identify source and response from the asymptotic behavior of the solution near the boundary, and compute the {Green} function via linear response.  The final result takes {an analogous form to Eq.~(\ref{EQ:limit}),}
\be
G_{R}(\w,\vec{k})
=
\lim_{r\to0}r^{2(\Delta-d)}{\Pi_{r,\mathrm{in}}(\w,\vec{k},r)\over\Phi_{\mathrm{in}}(\w,\vec{k},r)}
=
{2\Delta-d\over L}{\phi_{\Delta,\mathrm{in}}(\w,\vec{k})\over\phi_{d-\Delta,\mathrm{in}}(\w,\vec{k})},
\ee
where $\w$ is the Lorentzian frequency and $\phi_{\Delta}$, $\phi_{d-\Delta}$ are found by solving the linearized {Lorentzian} equations in the bulk with in-falling boundary conditions at the horizon.  The advanced Green function is computed analogously with out-going boundary conditions enforced at the horizon.

\paragraph{Cautionary Notes on the Semi-Classical Limit}

While it is true that we have simplified our job by focusing on limits where the gravitational partition function $Z_{\mathrm{GR}}$ can be computed by saddle point approximation, this is not the same as expanding $Z_{\mathrm{QFT}}$ semi-classically; in the present saddle, the classical description is of classical gravity in one higher dimension.  Thus, rather than studying a semi-classical limit of the QFT governed by quasiparticle perturbation theory,  we are focusing on a very different emergent limit in which quasiparticles are replaced by geometry.

Note, too, a practical consistency condition: it cannot be the case that both $Z_{\mathrm{QFT}}$ and $Z_{\mathrm{QG}}$ are simultaneously perturbative,
for if that were the case we would learn that every classical field theory can be reorganized into classical gravity -- but this is explicitly forbidden by the Weinberg-Witten theorem. It is thus clear that this holographic duality had to be a strong-weak duality, with one side becoming strongly quantum mechanical whenever the dual becomes semi-classical.

Finally, we could just as easily have inverted our logic and studied a limit in which the QFT partition function became semi-classical.  While this is unlikely to teach us very much about the QFT, since perturbation theory is already well understood, it may teach us a great deal about the strongly quantum gravitational dual.  In particular, this approach provides the only non-perturbative definition of 4d quantum gravity currently known, and can be used to quickly and convincingly argue for the preservation of {unitarity} by the formation and evaporation of black holes.

\paragraph{Thermodynamics Encoded by Black Holes}

The final entry {in the holographic dictionary sketched in {Table~\ref{table:holdict}}} relates the thermodynamic state and ensemble in which we place our QFT to the precise geometry in which we study the bulk gravity~\cite{Witten:1998zw}.  When the QFT is exactly conformal, the bulk geometry is pure \ads.  Recalling that the UV physics of the boundary is associated with the asymptotic near-boundary region of \ads.  Deforming the theory away from this conformal UV fixed point to generate a QFT with non-trivial RG flow corresponds to studying a geometry which is asymptotically \ads\ near the boundary but flows away from pure \ads\ as we run into the bulk.  For example, turning on a finite temperature, $T$, and chemical potentials, $\mu_{i}$, in this CFT corresponds to studying asymptotically-\ads\ black hole spacetimes with Hawking temperature $T_{H}=T$ and charges $Q_{i}=\mu_{i}$.  More generally, the thermodynamic data of the QFT is entirely encoded in the thermodynamics of the black hole in the dual geometry (See Table~\ref{table:holdict} for the mapping between Thermodynamic parameters of the QFT and the quantum numbers of the dual black hole).

The classic example of the power of these thermodynamic relations is the computation of the thermodynamic properties of a {large $N$} $4d$ Yang-Mills {gauge} theory at strong 't~Hooft coupling.  The holographic dual of this system is a planar \ads\ black brane, {as discussed} in Sec. (\ref{sssec:adsbh}).  In the {large $N$,} large-$\l$  limit, the gravitational description reduces to GR and the black brane entropy becomes
\be
S_{\mathrm{BH}} = {A_{H}\over 4G_{N}}
\ee
where {$A_H$} is the area of the horizon.  Since the system is translationally invariant, this diverges, so it's convenient to consider instead the entropy per unit area in the field theory,
\be\label{EQ:EntropyDensity}
s = {S_{\mathrm{BH}}\over A_{\p}} = {A_{H}\over 4 G_{N}A_{\p}}\,,
\ee
where $A_{\p}$ is the area of the \ads\ boundary.  Plugging in factors, this gives a strong-coupling computation of the entropy density and energy density as,
\be
s_{\mathrm{strong}} = {\pi^{2}\over2}N^{2}T^{3}\,,
~~~~~~~~
\eps_{\mathrm{strong}} = {3\pi^{2}\over8}N^{2}T^{4}\,.
\ee
These strong coupling results are extremely close to the results in the free theory as computed in  perturbation theory:
\be
s_{\mathrm{free}} = {2\pi^{2}\over3}N^{2}T^{3}\,,
~~~~~~~~
\eps_{\mathrm{free}} = {\pi^{2}\over2}N^{2}T^{4}\,.
\ee
Remarkably, despite running from weak to strong coupling, $\l=0$ to $\l\to\infty$, all that happened to the thermodynamics is a mild renormalization by a factor of $3\over4$:
\be
{s_{\mathrm{strong}}\over s_{\mathrm{free}}} =
{\eps_{\mathrm{strong}}\over \eps_{\mathrm{free}}} = {3\over4}\,.
\ee
Studying the full $\l$-dependence of this ratio reveals that the energy and entropy densities flow smoothly and monotonically as we run from weak to strong coupling.  Analogous results {are obtained} for a wide variety of gauge groups, matter contents, and even dimensions, with the ratio ${s_{\mathrm{strong}}\over s_{\mathrm{free}}}$ typically within 10\% or so of the $\CN=4$ value, $3\over4$.

This holographic result already tells us an important fact: thermodynamic quantities like the free energy and the entropy density are not good probes of the strength of interaction of a quantum liquid.  Similar results apply to other thermodynamic quantities, for which the strong/weak ratio is typically $\CO(1)$ as well.\footnote{This can of course change if there is a quantum phase transition as parameters in the system are varied.}

%%%%%%%%%%%%%%%%
{\subsubsection{A Bit of History}}

Holography is far more general than the refined context in which it was discovered.  In particular, using holography as a tool does {not require supersymmetry, string theory, d-branes,} or 5-dimensional \ads, although all those ideas were central to its discovery and can provide useful guidance to the applied holographer.  It is useful to address this directly.

In its modern form, holography was first discovered in string theory
while studying very special toy systems with as much symmetry as possible, in particular, maximally supersymmetric 4d {$\CN$=4 $SU(N)$ Yang-Mills} theory at large $N$.  The reason supersymmetry was important in the discovery of holography is that SUSY turns off many quantum effects so that a semi-classical analysis can give, for at least some special quantities, exact results.  This allowed the computation of a host of correlation functions at weak coupling which could then be reliably extrapolated to strong coupling. For example, thermodynamic quantities like the free energy and entropy density, as well as transport properties such as conductivities, viscosities, susceptibilities and sound mode dispersion relations can all be calculated analytically, something one would never expect in a less constrained theory such as QCD.  Of particular interest were \textit{grey-body factors} of supersymmetric black holes, corresponding to the probability with which a particle would be absorbed by a black hole.  These could be computed at weak coupling via classical gravitational perturbation theory, then reliably extrapolated to strong coupling, giving a precise computation of a small set of very special absorption cross-sections of deeply quantum mechanical black holes.  {Thus, in the late 90's a number of researchers intensively computed strong-coupling results for} the string theorist's spherical cow.

The results were astonishing.  As first pointed out by Juan Maldacena in the fall of 1997~\cite{Maldacena:1997re}, while the weakly coupled $\CN=4$ field theory behaved much as familiar 4d gauge theories, at strong coupling the correlation functions of the theory reproduced, rather miraculously, classical scattering off black holes in one higher dimension!  More precisely, Maldacena~\cite{Maldacena:1997re} (and, shortly thereafter, Gubser-Klebanov-Polyakov~\cite{Gubser:1998bc}, and Witten~\cite{Witten:1998qj}) gave a precise relationship between the $4$-dimensional $\CN=4$ theory and a particular theory of gravity in 5-dimensional \ads.  Crucially, the 't Hooft coupling $\l$ and number of colors $N$ of the 4d {supersymmetric Yang-Mills} theory were related to the \ads curvature in units of the Planck and string lengths precisely as reviewed above in {Eq.~(\ref{EQ:Scalings}).}
Thus, when the 4d theory is weakly coupled, $\l\ll1$, the 5d theory is strongly curved and cannot be treated as a theory of perturbative gravity.  Conversely, when the 4d theory is strongly coupled, $\l\gg1$, and at large $N$, the 5d theory is weakly curved, so while the system cannot be treated as a weakly interacting gauge theory in 4d, it can be treated as a weakly coupled theory of gravity in 5d.

The point is this: while the original discovery of holographic duality required supersymmetry and conformal invariance, the duality itself does not depend on supersymmetry or conformality.  Indeed, we now have a long list of examples in which each of these constraints is weakened or removed, with every consistency check passed.  To date there is no example of a violation of a sharp holographic duality.  From the stringy point of view, holographic duality is simply a true, if as yet unproven, fact about quantum field theories and quantum gravity.

%%%%%%%%%%%%%%%%%%%%%%%%%%%%%%%%%%%%%%%%%%%%%%%%%%%%%%%%%%%%%%%%
%%%%%%%%%%%%%%%%%%%%%%%%%%%%%%%%%%%%%%%%%%%%%%%%%%%%%%%%%%%%%%%%
%%%%%%%%%%%%%%%%%%%%%%%%%%%%%%%%%%%%%%%%%%%%%%%%%%%%%%%%%%%%%%%%
%
%
%	Applied Holography
%
%
%%%%%%%%%%%%%%%%

\subsection{Applied Holography}
\label{sec_app_hol}

Applied holography is the application of holographic dualities to construct computationally tractable toy models of behavior which has proven difficult to model with traditional tools.  Key to this endeavor is the fact that various features of the physics which appear universal, or even trivial, on one side of the duality may be significant, or simply obscure, in the dual. Most current work in applied holography has used universal features of gravitational models, and in particular, the universality of black hole horizon physics, to make new discoveries about QFTs, though using universal features of weakly coupled field theories to learn about extreme phases of gravity is of interest as well.  Given the aim of this {Focus Issue,} we will emphasize the first direction in what follows.

The simplest application of holography involves studying QFTs for which the precise gravitational dual is explicitly known. In such cases we can use the perturbative QFT to compute any quantities of interest at weak coupling, $\l\to0$, then use the weakly coupled gravitational dual to compute the corresponding quantities at strong coupling, $\l\to\infty$.  This strategy has very limited applications because
the precise duals of most QFTs are not yet known. At the moment we only know the precise duality for a special set of non-generic quantum field theories, although they are fairly close cousins of the theories studied in particle physics.  For example, we know the precise gravitational dual of the $\CN=4$ theory discussed in detail above and have used it to learn a great deal about the structure and strong-coupling physics of this intricate and exotic theory (which we emphasize does not appear to be realized anywhere in nature).
By contrast, we do not know the dual of the Hubbard model.

This is not to say progress hasn't been made, just that it is difficult.  {An example} of this point is the {recent conjecture of a} duality between the 1d Ising model and pure Einstein gravity in \ads$_{3}$.
For this relation to hold the gravitational side must be treated not only quantum mechanically, but fully non-perturbatively, summing over an infinite number of topologically inequivalent locally-\ads$_{3}$ spacetimes, including a series of quantum corrections around each classical saddle, something that can only be done in closed form in the absolutely simplest possible gravitational systems, namely pure Einstein gravity in $2+1$ dimensions.  This conjecture underscores the difficulty of constructing the precise holographic dual of a garden-variety QFT.

If the precise dual is not known, a second approach to applied holography is to make use of a known exact duality with a class of QFTs that are structurally similar to the QFT of interest.  Universal properties of these sibling QFTs can then be inferred to apply to the original QFT, too.
An example of this strategy is the construction by Kachru {\textit{et al.}} of holographic dimer models~\cite{Kachru:2009xf,NJPfocusissue39_kachru}.  In this construction, the degrees of freedom of interest for the dimer problem form a small subset of the full set of holographic degrees of freedom. At low energy, this subset forms a relatively {generic} dimer model.  By exploiting the holographic description, much can be derived about the low-energy physics, including a novel phase transition between {Fermi-liquid and non-Fermi} liquid behavior.
The most powerful example of this approach so far has been the study of strongly-coupled quark-gluon liquids in large $N$ toy models of QCD, for which a menagerie of models have been considered, each emphasizing and reproducing various features of the observed phenomena.
As we shall see in detail in Sections (\ref{sec:eta}) and (\ref{sec:qgp}), this approach has been used to motivate the anomalously small viscosity of the RHIC fireball, to explain why the rapid thermalization of the RHIC droplet is not so surprising, and to estimate the jet quenching parameter.

{However, perhaps the most interesting use of holography} is to provide an entirely new way of defining consistent QFTs.  Traditionally, QFTs in dimensions greater than 2 are defined by specifying some Lagrangian or Hamiltonian governing the interactions of a set of well-defined quasiparticles.  Yet many systems, including some of the most theoretically and experimentally interesting systems, manifestly do not have any well-defined quasiparticles upon which to base a standard QFT.  From this point of view, holography looks interesting because it provides a recipe for computing quantum amplitudes that manifestly satisfy the conditions of a good QFT (local, causal, etc.) but which is defined without any appeal to a quasiparticle picture or any sort of conventional perturbation theory.  Holography thus provides an entirely new way to construct consistent models of many-body physics, replacing quasiparticles with geometry as the central organizing principle.

\subsubsection{ Viscous Hydrodynamics from Gravity}\label{sec:eta}

Suppose we are given a QFT at finite temperature and density and want to know whether it is weakly or strongly interacting.  A natural observable to query is the shear viscosity, $\eta$, which measures the efficiency of momentum transport across a velocity gradient.  Explicitly, the shear viscosity $\eta$ of a fluid determines the frictional force $F$ induced on plate of area $A$ moving at a velocity $v$ a distance $L$ over a fixed surface,
\be
F = \eta {A v \over L}\,.
\ee
This relation follows from the hydrodynamic stress tensor given in Eq.~(\ref{del_T_ij}).
As discussed in Sec.~\ref{sec_sQGP} shear viscosity has dimensions $\hbar n$, where
$n$ is a density. The natural thermodynamic quantity to which one can compare the viscosity
is thus the entropy density, $s$, with dimensions of $k_Bn$. $\eta/s$ thus gives a
simple dimensionless measure of the efficiency of transverse momentum transport.

>From kinetic theory, the transport of momentum proceeds by scattering quasiparticles between layers of the fluid, and is thus proportional to the mean free path between scattering {events between} fluid quasiparticles.  Since the mean free path decreases as the scattering cross section increases, the shear viscosity should be inversely related and thus strongly sensitive to the strength of interactions.
However, if we drive up the strength of interaction until the mean free path becomes of order the {Compton} wavelength of the would-be quasiparticle, the kinetic-theory relation between interaction and viscosity should no longer apply.  Meanwhile, in this post-quasiparticle setting,
the mean free path is much smaller than any typical velocity gradient, so a hydrodynamic description should set in.   Thus, as the interactions {become} strong, we expect the shear viscosity to become small, and furthermore expect coupling-dependence of the viscosity to qualitatively change, since the kinetic behavior must eventually cease.  But what values should $\eta/s$ take, and how will it vary as the coupling grows large?  \textit{A priori}, the answers to these questions would appear to be strongly system-dependent.

To address this question, let's compute $\eta/s$ holographically.\footnote{See~\cite{Cremonini:2011iq} for a summary of the current state of such computations, and~\cite{Iqbal:2008by,Iqbal:2009fd} for a more detailed exposition of the computation we outline, and for references to the original literature.}  This will give us a controlled strong-coupling estimate of $\eta/s$.  However, before proceeding, let's pause to consider, from the holographic perspective, why there must be any viscosity in the first place.  Recall that the holographic dual of a strongly-coupled relativistic large $N$ gauge theory at finite temperature is a black hole with non-zero mass.  But anytime we have a black hole, we will have dissipation: if we scatter a wave off the black hole, some fraction of its amplitude, and thus of the momentum it carried, will  be absorbed into the black hole.  So the fact that there is always a generic source of dissipation is very much built in to holographic duality.

A useful way to compute $\eta/s$ begins with the Kubo formula for $\eta$,
\be
\eta = -\lim_{\w,k\to0} {\Im G_{R}(\w,k)\over \w}\,,
\ee
which relates $\eta$ to the retarded Green function {$G_R$} for shear modes of the stress tensor,
\be
\label{EQ:Kubo}
G_{R}(t,x) = \vev{T_{xy}(t,x)T_{xy}(0,0)}\,\Theta(t)\,.
\ee
Note that we take $\w\to0$ only {after} sending $k\to0$. The Kubo formula follows from matching
linear response theory to the expected low frequency, low momentum limit of the correlation
function in hydrodynamics; see for example \cite{Son:2007vk}.
{To obtain $\eta$ we need} to compute $G_{R}(t,x)$.
{In general, this is difficult in a strongly coupled QFT. However, holographically this calculation turns} out to be relatively straightforward.  According to the holographic dictionary, the operator $T_{xy}$ is dual to the corresponding shear mode of the bulk metric, $g_{xy}$.  We thus need the response of the classical gravitational action, $I_{\mathrm{\ads}}[g]$, to variations of this mode, $\delta g_{xy}=h_{xy}$.
{So} long as we are expanding around pure \ads, $h_{xy}$ is governed by a simple massless scalar action of the form in Eq. (\ref{EQ:ScalarAction}), but with the normalization of the kinetic term determined by the normalization of the gravitational kinetic term, ${1\over 16\pi G_{N}}$,
\be
I_{\phi} = -  {1\over 16\pi G_{N}}\int \!d^{d+1}\!x\, \sqrt{g}\, \,\half\(\p\phi\)^{2}\,.
\ee
Since $m=0$, we have $\Delta(d-\Delta)=0$, so the dimension of our dual operator is $\Delta=d$.
Let's use this action and the formalism developed above to compute $G_{R}$.
With $\Delta=d$, the retarded Green function takes the form
\be
G_{R}(k)
=
\lim_{r\to0}{\Pi_{\mathrm{in}}(\w,\vec{k},r)\over\Phi_{\mathrm{in}}(\w,\vec{k},r)}\,.
\ee
It is readily checked ({cf.} \ref{EQ:GE}) that the real part of the ratio above scales, after sending $\vec{k}\to0$, as $\w^{3}$, and thus vanishes in our limit.  {We can thus replace the taking of the imaginary part in Eq.~(\ref{EQ:Kubo}) by $1\over i$ to get}
\be\label{EG:MyKubo}
\eta = -\lim_{\w,k\to0} \lim_{r\to0}{\Pi_{\mathrm{in}}(\w,\vec{k},r)\over i\w\Phi_{\mathrm{in}}(\w,\vec{k},r)}\,.
\ee

At this point something remarkable happens.  In the limit that $\w,\vec{k}\to0$, the radial evolution of $\Pi$ and $\w\Phi$ trivialize.  This can be seen by inspection of the definition of $\Pi$ and the Hamiltonian equation of motion, $\p_{r}\(\w\Phi\)\sim\w\Pi$ and $\p_{r}\Pi\sim k^{\m}k^{\n}\p_{\m}\p_{n}\Phi$.  Thus, in the limit   $\w,\vec{k}\to0$, both $\Pi$ and $\w\Phi$ become independent of $r$.

We can thus evaluate the ratio appearing in our Kubo formula at any radial coordinate we like.
A convenient choice is the horizon, where the solution must satisfy in-falling boundary conditions.  From our earlier discussion of the in-falling solutions (\ref{EQ:Ingoing}), which implied that $f\p_{r}\Phi_{\mathrm{in}} = i\w\Phi_{\mathrm{in}}$ at the horizon,
and using the definition
$\Pi_{\mathrm{in}} = -{1\over 16\pi G_{N}}\sqrt{-g}g^{rr}\p_{r}\Phi_{\mathrm{in}}$, we have that
\be
\lim_{r\to r_{H}} \Pi_{\mathrm{in}} = -{1\over 16\pi G_{N}r_{H}^{3}} i\w\Phi_{\mathrm{in}}(r_{H})\,.
\ee
The factor of $r_{H}^{-3}$ can be nicely interpreted as the area of the horizon in boundary units, ${A_{H}\over A_{\p}}$.  Plugging all of this into our Kubo formula, Eq.~(\ref{EG:MyKubo}) then becomes
\be
\eta = {A_{H}\over 16\pi G_{N} A_{\p}}\,,
\ee
where $A_{H}$ is the area of the horizon and $A_{\p}$ is the area of the boundary.  {From Eq.~(\ref{EQ:EntropyDensity}), ${A_{H}\over A_{\p}}=4 G_{N} s$.}  Putting these together, we see that the viscosity-to-entropy-density ratio takes the simple form now known as the KSS form after the authors of~\cite{Kovtun:2004de} (though see also e.g.~\cite{Policastro:2001yc,Buchel:2003tz,Kovtun:2003wp}),
\be \label{EQ:EtaS}
{\eta\over s} = {1\over 4\pi}\,,
\ee
which is orders of magnitude smaller than that of typical liquids.  In fact, the liquids which are believed to come close to this level of perfection are the QGP at RHIC and fermions at unitarity, as shown in Fig.~\ref{fig:ratio}.

~

Many features of this result are remarkable.
First, as motivated heuristically above,  $\eta/ s$ is indeed a good diagnostic of strong coupling in any many-body system.
Near weak coupling, QFT perturbation theory gives
\be
{\eta\over s}\sim {A\over\l^2\ln(\sqrt{\l})}\,,
\ee
where the constant of proportionality is system-dependent.
In particular, weakly coupled QFTs generally have ${\eta\over s}\gg1$.
Near strong coupling, by contrast, a more careful gravitational analysis gives
\be
{\eta\over s}={1\over4\pi} +O\({1\over\l^{3/2}}\)\,,
\ee
almost independently of the details of the system.
Thus, in any such system, the behavior of $\eta\over s$ changes dramatically as we run from weak to strong coupling.

Second, {Eq.~(\ref{EQ:EtaS})} is remarkably generic in holographic systems.  Indeed, in any QFT which is holographically dual to {GR}, whether the gauge group is $SO(N)$ or $SP(2N)$ or something more complicated, whether enjoying maximal supersymmetry or no supersymmetry, we {obtain} precisely ${\eta\over s} = {1\over 4\pi}$.  To see why this result is so general, note that  the radial evolution dropped out of our computation of $\eta$, so that all that we needed was the local physics near the horizon.  The rest of the holographic description (including the UV physics near the boundary) was moot.  Meanwhile, $s$ is explicitly proportional to the horizon area, so again is completely determined by the physics of the horizon.
The generality of this holographic result then follows from the exceptional universality of the physics of horizons in GR.

Observations such as these led Kovtun, Son and Starinets (KSS) to make the conjecture that $\eta\ge{1\over4\pi}$ is a universal lower bound on the shear viscosity of any quantum {many-body} system, with saturation occurring only for those {strongly coupled} theories which are holographically dual to pure Einstein gravity.
We now understand that this result is not universal.  Explicitly, Eq.~(\ref{EQ:EtaS}) holds when the spacetime geometry is \ads, the gravitational action is Einstein-Hilbert, and the fluid is time-independent, homogeneous and isotropic.  However, when higher-derivative terms are present in the gravitational action (corresponding to finite-$N$ and finite-$\l$ effects in the dual QFT), or when the geometry is deformed away from \ads\ such that the radial evolution does not factor out (as occurs for example when the geometry is time-dependent, spatially disordered or sufficiently anisotropic), $\eta\over s$ can depart from the KSS value, and indeed can be lower than $1\over4\pi$.

The cleanest example of such an effect involves~\cite{Brigante:2007nu,Brigante:2008gz,Kats:2007mq}\ adding the leading irrelevant operator to the bulk gravitational action, the $5d$ Gauss-Bonnet term, a quadratic scalar built out of the curvature tensors,
\be
I_{GB} = {1\over 16\pi G_{N}} \int d^{d+1}\!x\,\sqrt{-g}\(  -2\Lambda + R + \l_{GB}R^{2}_{GB} \)\,.
\ee
Such terms arise in known string theories where we know and have control over both sides of the duality.\footnote{In the QFT, the Gauss-Bonnet coupling $\l_{GB}$ is related to the Euler and Weyl anomaly coefficients.  While it cannot be tuned arbitrarily in these examples while remaining within the regime of validity of the classical gravitational description, it can certainly be taken away from zero in a controlled fashion~\cite{Kats:2007mq}.}  As it turns out, the holographic analysis including $\l_{GB}\neq0$ is {straightforward. All} that materially changes is the normalization of the action for the scalar mode $h_{xy}$, which picks up an extra factor of $(1-4\l_{GB})$, {yielding}
\be \label{EQ:EtaSGB}
{\eta\over s} = {1-4\l_{GB}\over 4\pi}
\ee
When $\l_{GB}>0$, the KSS conjecture is violated.\footnote{Note that $\l_{GB}<{1\over4}$ is a consistency condition for GB gravity.  When $\l_{GB}={1\over4}$, the kinetic term for the graviton degenerates and the theory becomes ill-defined.}

It is worth understanding why this happens. The value of $\eta/s$ is controlled by the normalization of the kinetic term in the scalar action for our tensor mode $h_{xy}$.  Typically, such coefficients are best understood as coupling constants, since scaling them away to make the kinetic term canonical shifts the coefficient into the interactions.\footnote{The universality of $\eta/s$ in GR thus physically derives from the equivalence principle: in Einstein gravity, the metric couples universally with a single overall coefficient to all forms of stress-energy.  When we muck with this universality, for example by breaking symmetries or by modifying the way the metric interacts with itself via higher-curvature corrections, we make our chosen tensor mode of the metric couple more, or less, strongly.}  When we make the coupling stronger, as in the case of Gauss-Bonnet with $\l_{GB}>0$, the viscosity goes down.  This is reminiscent of the dependence of viscosity on coupling in kinetic theory: {increasing the strength of quasiparticle scattering inhibits transport and decreases the shear viscosity.}  At strong coupling, there are no longer any well-defined quasiparticles to which to apply kinetic theory; instead, the effective degrees of freedom mediating low-energy momentum transport are weakly coupled gravitons in the holographic spacetime.

Interestingly, while this evidence rules out the KSS conjecture,
there exist compelling arguments for a weakened version of the KSS bound.  Physically, the arguments for such a refined bound involve demonstrating that violations of these bounds would lead, in a wide class of theories, either to violations of causality in the field theory~\cite{Brigante:2008gz,Buchel:2009tt} or to violations of the positivity of energy~\cite{Hofman:2009ug}; in fact these two pathologies are intimately related.  In the case of Gauss-Bonnet gravity which is holographically dual to 4$d$ QFTs with $\CN=1$ supersymmetry, both arguments lead to the same constraint, ${-7\over36}\le\l_{GB}\le{9\over100}$, leading to the refined lower bound,
\be
{\eta\over s}\ge {1\over4\pi}{16\over25}
\label{eq:GBbound}
\ee
In more general models, similar bounds obtain, but the precise values differ.
The meaning and reliability of such refined bounds remains a matter of active debate.
For example, it has been shown~\cite{deBoer:2009gx,Camanho:2009hu}\ that sufficiently contrived examples may be concocted which violate any such bound and in fact push ${\eta\over s}$ arbitrarily close to zero, and even negative!  However, it is not at all clear that these models are themselves well defined; indeed, one can immediately show~\cite{Camanho:2010ru} that all such models with ${\eta\over s}$ negative or vanishing are pathological, with causality and self-consistency enforcing a minimal value for ${\eta\over s}$.
Meanwhile, whether any such bound should be expected when we give up time-independence, homogeneity or isotropy is a topic of much current research.
To summarize, there are well-defined lower bounds on $\eta\over s$ in various classes of theories, but each such bound has been violated by considering a more general class of models. It remains possible that there is a simple, completely universal lower bound on $\eta\over s$, but the evidence for such a bound is increasingly tenuous.

~

Worrying about universality of a particular bound, however, misses the point.  The lasting import of Eq (\ref{EQ:EtaS}) is that it establishes a link between ultra-low shear viscosity and strongly interacting many body systems.  Said differently, if {one finds a fluid with an extremely small shear viscosity, one has} very strong evidence that it enjoys strong quantum correlations.  Conversely, extreme quantum fluids should be expected to have ultra-low viscosity of order $1\over4\pi$.  Whatever the final story about a hard lower limit, this connection is an important lesson from holography.

\subsubsection{ Diffusion, Jet Quenching and Dynamics in  the Holographic Quark-Gluon Plasma}\label{sec:qgp}

A related transport coefficient which can be fruitfully studied holographically is the diffusion constant for a heavy quark in a strongly-coupled holographic QGP.  It is an interesting example for us for several reasons.  First, the structure of the holographic computation is very different {from} the computations described above, and in fact owes a great deal to the string-theoretic origin of holographic duality.   Second, the result is very different from the simple universality of the viscosity computation.  {Specifically, the diffusion} constant for a heavy quark scales as
\be\label{eq:qdif}
D={2\over\pi T}{1\over\sqrt{\l}}
\ee
and is thus very strongly sensitive to the 't\,Hooft coupling, $\l$.  In particular, this suggests that even very heavy quarks should rapidly equilibrate in the QGP, in sharp contrast to perturbative estimates.  Note that this also  differs sharply from the perturbative YM result\footnote{This result corresponds to a specific numerical estimate of the logarithmic terms in Eq.~(\ref{D_QCD}); see Eq.~(6.1) of~\cite{Moore:2004tg}.}
\be
\label{eqn:diffusion}
D={6\over2\pi T} \({1\over 2\a_{s}}\)^2 \,,
\ee
for while Eqs.~(\ref{eq:qdif}) and~(\ref{eqn:diffusion}) look similar, the latter is only valid for $\a_{s}\ll1$, while the former is only valid for $\l\gg1$.

The basic structure of the holographic computation is described in~\cite{Herzog:2006gh,Gubser:2006bz}; see~\cite{Moore:2004tg}\ for a perturbative analysis.  Suppose we place a heavy quark in the QGP and treat it in a Born-Oppenheimer approximation as a slow, heavy, pointlike variable which sources the gauge field.  As we run the RG, this charge is conserved.  Holographically, this means that we should have a point-like defect not just on the UV boundary, but also on every slice of the holographic bulk.  The heavy quark on the boundary is thus the endpoint of a string in the bulk which hangs down from the boundary into the \ads\ horizon, hanging along a {straight} line according to the gravitational tug of \ads.  The string can be thought of as the holographic image of the color flux of the fundamental quark in the $3+1$ boundary.  Explicitly, the dynamics of the string are determined by a simple least action principle, where the action is the area of the surface in space-time swept out by the string in \ads.  For a static string, the extremal string is simply the string hanging straight down from the {boundary} quark.

To measure the diffusion constant, we need to give our quark a kick.  The easiest way to do this is to drag the quark through the QGP with a fixed velocity and then determine the drag force.  In equilibrium, we must have
\be
f=-\eta_{D} p\,,
\ee
where $\eta_{D}$ is the drag coefficient.
Holographically, the string hanging off the quark into the bulk will now dangle behind the moving quark, imparting a drag force to the quark.  Using the action principle we just defined, the drag force {is computed} to be
\be
f = -{\pi\over2}\sqrt{\l}T^{2}{v\over\sqrt{1-v^{2}}}\,.
\ee
For  heavy quarks, for which the thermal corrections to the kinetic mass are negligible, this gives
\be
\eta_{D} ={\pi\over2m}\sqrt{\l}T^{2} \,.
\ee
Using this in the relation $D=T/\eta_{D} m$ derived in Eq.~(\ref{Einstein:2})
yields Eq.~(\ref{eq:qdif}).

It is worth commenting on how this computation is similar to the viscosity calculation, and how it differs.  Perhaps the most important similarity is that in both cases we computed deeply quantum-mechanical transport quantities in our strongly-coupled QGP by solving simple, classical differential equations in the dual gravitational system.  This is a basic hallmark of holographic duality.  Secondly, the resulting computation was surprisingly independent of the details of the holographic setup, depending only on the gravitational dynamics of a string hanging into \ads\ from a moving point on the boundary.  This suggests that a heavy fundamental in any strongly coupled QGP should have similar qualitative behavior, namely, efficient diffusion and rapid equilibration.
By contrast, the final results differ in significant ways.  For example, while $\eta/s$ was remarkably insensitive to the coupling, $D$ depends sensitively, and is thus a sensitive probe of precisely how strongly coupled the theory is.

A  similar computation leads to a holographic estimate of the jet-quenching parameter, $\hat{q}$, introduced in Sec.~\ref{sec_jets}. Holographically, the role of the QGP fireball is played by the dual black hole, while the hard jet can be modeled as a heavy quark at very high energy, and thus moving on an effectively light-like trajectory.  Our job is thus to compute the transverse momentum gained by a light-like quark moving through the QGP.  This can be reduced~\cite{CasalderreySolana:2007zz,D'Eramo:2010xk,Liu:2006ug}\ to computing a light-like Wilson loop in the QGP.   Holographically, the expectation value of this Wilson loop can again be computed by studying the surface swept out by the trailing string, determined as before by extremizing the string action.  Since the Wilson loop is light-like, the computation must be performed in Lorentzian signature.  The resulting real-time holographic analysis, while somewhat intricate, leads to a simple prediction for the jet-quenching parameter in $\CN$=4 SYM~\cite{Liu:2006ug,Liu:2006he},
\be
\hat{q}_{\CN=4} = {\pi^{3/2}\Gamma({3\over4})\over \Gamma({5\over4})}\sqrt{\l}\,T^{3}.
\ee
Recalling that $\l=g^{2}N_{c}$ and $g^{2}=4\pi\a_{s}$, and plugging in $T\sim300MeV$ and $\a_{s}=\half$, which are reasonable estimates for the values at RHIC, this gives  $\hat{q}_{\CN=4}\sim 4.5{\mathrm{GeV}}^{2}/{\mathrm{fm}}$, which is comparable to results from RHIC; see Sec.~\ref{sec_jets}. Note that this holographic result leads to a semi-strongly coupled treatment of energy loss, treating
transverse momentum diffusion in strong coupling via a holographic estimate of $\hat{q}$, while the actual energy loss (longitudinal drag) is
done in perturbation theory, as discussed in Sec.~\ref{sec_jets}. For fully strongly-coupled treatments see {\em e.g.}~\cite{Chesler:2008uy,Arnold:2011qi}.

The important lesson of this holographic computation however is not the specific value of $\hat{q}$ we find but rather how $\hat{q}$ depends on the physical parameters of the system.  In particular, it turns out that in any holographic QGP dual to pure \ads, and thus governed by a dual CFT, one finds~\cite{Liu:2006he}
\be
{ \hat{q}_{\mathrm{CFT}} \over \hat{q}_{\CN=4} } = \sqrt{ s_{\mathrm{CFT}}  \over s_{\CN=4}  }\,,
\ee
where $s$ is the entropy density of the QGP.  The energy lost by a hard parton plowing through the QGP is thus not proportional to the entropy density, and so cannot be well modeled by scattering off a gas of persistent quasiparticles.

This holographic approach to studying the physics of a generic $SU(N)$ QGP to glean insight into the dynamics of e.g. the physical QGP at RHIC has been followed in many directions, including several of the papers in this {Focus Issue.}   For example,
Hubeny~\cite{NJPfocusissue37_hubeney}\ uses a simple but remarkably illuminating gravitational model for an accelerated heavy quark in {large $N$} QCD to study the collimation of synchrotron radiation emitted by a circling quark.  Like the diffusion and jet-quenching computations sketched above, Hubeny's calculation involves tracing the path of the bulk string which trails behind the accelerating quark into the bulk of \ads.  The result of the acceleration is that this bulk string becomes coiled, leading to the radiation emitted by the string remaining tightly collimated.  Curiously, this result at first appears to be in tension with standard holographic intuition.  Hubeny resolves this apparent conflict through a careful study of the gravitational backreaction induced by the coiled string.

\subsubsection{ Gravitational Engineering and Holographic Superconductors}

An interesting example of defining the field theory by its holographic dual is the holographic superconductor.  The idea, first outlined by Gubser~\cite{Gubser:2008px}\ and developed in detail by Hartnoll, Herzog and Horowitz~\cite{Hartnoll:2008vx,Hartnoll:2008kx}, goes roughly as follows.  Suppose we have a QFT which enters a superconducting phase at low temperatures by developing a non-zero condensate for some scalar operator, $\CO$, spontaneously breaking a global $U(1)$ symmetry.  What would that look like in the holographic context?  By appeal to the holographic dictionary in {Table~\ref{table:holdict},} the global current in the boundary is dual to a gauge field in the bulk, while the charged scalar order parameter in the boundary is dual to a scalar field in the bulk which is charged under the bulk gauge field.  Spontaneously breaking  (or \textit{Higgsing}) the global $U(1)$ {symmetry} on the boundary by giving a vacuum expectation value to $\CO$ is then dual to spontaneously breaking the bulk gauge $U(1)$ via a non-trivial profile for the bulk scalar field, $\Phi(r)$.

Our goal is thus to build a gravitational system which is dual to some QFT which exhibits this symmetry breaking pattern -- not a specific QFT, but any QFT with this structure.  By adjusting the gravitational theory we will then be able to explore a whole space of strongly-coupled QFTs with low-energy superfluid phases.  Procedurally, we will take the gravitational system as a definition of the dual QFT, computing the correlation functions etc of the QFT entirely by appeal to the holographic dual.

The holographic dictionary thus tells us that, whatever else exists in the gravitational dual of a {superconducting} QFT, we will need a vector field $A_{\mu}$ dual to the boundary current $\CJ_{m}$ and a scalar field $\Phi$ dual to the scalar order parameter $\CO$.
The dictionary, however, does not tell us the Lagrangian, {so we must choose it.} A minimal first guess is,
\be
I_{grav} = -{1\over 16\pi G_{N}}\int\sqrt{g}\(-2\Lambda +R -{1\over e^{2}}\[{1\over4}F^{2}+|D\Phi|^{2}+m^{2}|\Phi|^{2}\]\)
\ee
We can further simplify our lives by considering the \textit{probe limit}, $g\to\infty$, in which case the matter field stress tensor (which scales as $1\over e^{2}$) is negligible so that {the Einstein} equation, and thus the metric, is not modified {from} its original \ads\ form.

Our job is now to find non-trivial solutions of the bulk equations of motion which follow from our trial action expanded around an \ads\ black brane geometry, so as to put the system at finite temperature.  Non-trivial means such that the boundary operator $\CO$ acquiring a spontaneous {vacuum expectation value;} from our dictionary, this requires $\Phi$ to have non-trivial profile, Higgsing the vector, as expected.  Once we find such a solution, we can compute correlation functions in the resulting system by evaluating the bulk action on-shell as a functional of the non-normalizable modes and taking derivatives {with respect to} those modes, as discussed in Sec.~\ref{sssec:strong}.

For simplicity, let's focus on homogeneous configurations of the form $A=A_{t}(r)dt$ and $\Phi=\Phi(r)$.  The bulk equations of motion in our \ads\ black brane take the form
\be
\label{EQ:SC}
r^{2}f A_{t}'' - 2|\Phi|^{2}A_{t} = 0\,,
~~~~~~~~
r^{2}f\, \Phi'' +\(rf'\!-\!2f\)r\Phi'+ \({r^{2}A_{t}^{2}\over f}-m^{2}\)\Phi = 0\,.
\ee
Near the boundary, the general solution to the bulk field equations takes the form,
\begin{equation}
A_{t} \sim \m \,+ \r\,r + ...\,,
~~~~~~
\Phi \sim \phi_{d-\Delta} r^{d-\Delta} + \phi_{\Delta} r^{\Delta} + \dots\,,
\end{equation}
where $m^{2}L^{2}=\Delta(\Delta-d)$.  A convenient choice for the mass is $m^{2}L^{2}=-2$, which gives $\Delta=1$ or $\Delta=2$.  Let's focus on $\Delta=2$.
The holographic dictionary tells us to interpret $\mu$ as the source for the $\CJ_{t}$ component of the current, so $\mu$ plays the role of the chemical potential for the boundary charge density.  $\rho$, being the subleading piece of $A_{t}$ near the boundary, then represents the expectation value of the operator conjugate to $\mu$, $\rho=\vev{\CJ_{t}}$, so we can identify $\rho$ as the charge density
induced by $\mu$.  Similarly, $\phi_{d-\Delta}$, as the leading term in $\Phi$ near the boundary, is interpreted as the source for the boundary operator $\CO$, and $\phi_{\Delta}$ as the {vacuum expectation value,} $\phi_{\Delta}=\vev{\CO}$.

% figure 1: Lattice
\begin{figure}[!t]
\centerline{\epsfig{figure=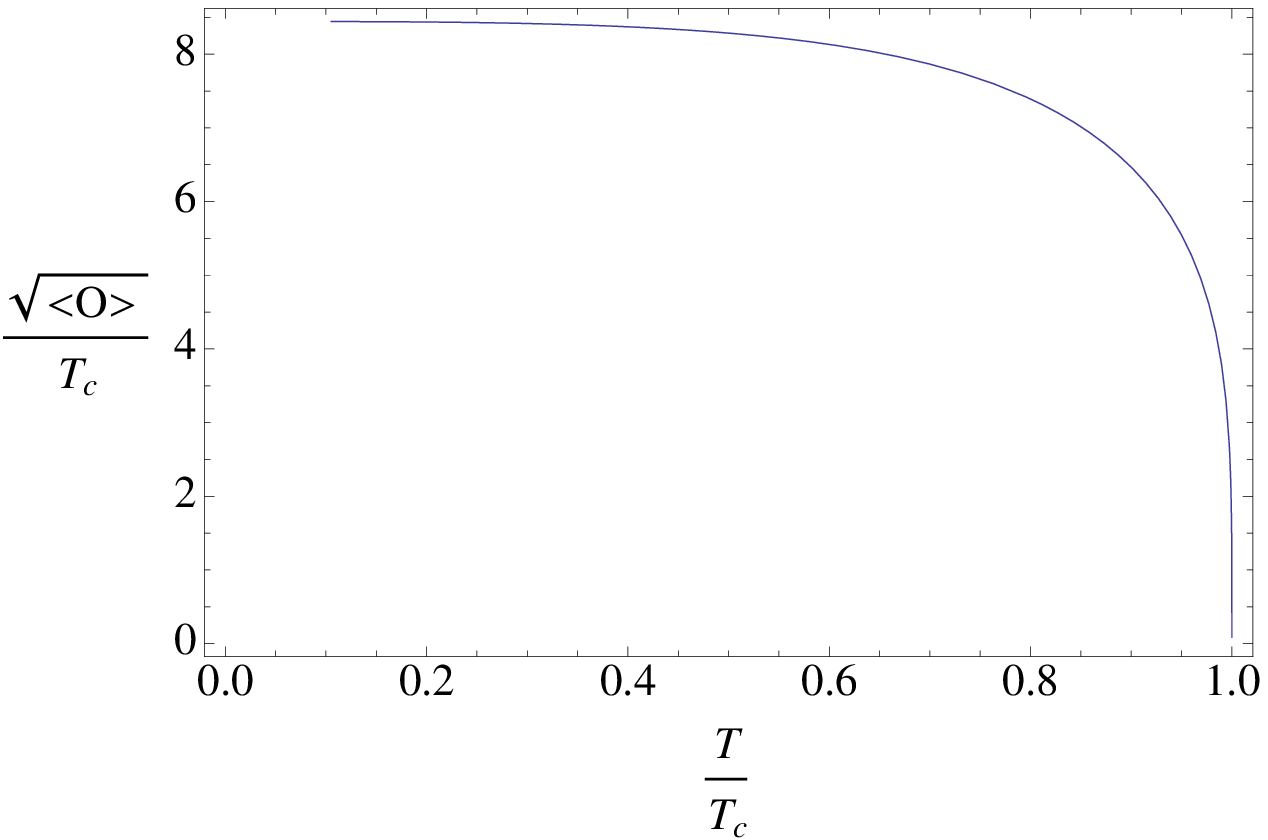, height=2.2 in}~~~~~~~~\epsfig{figure=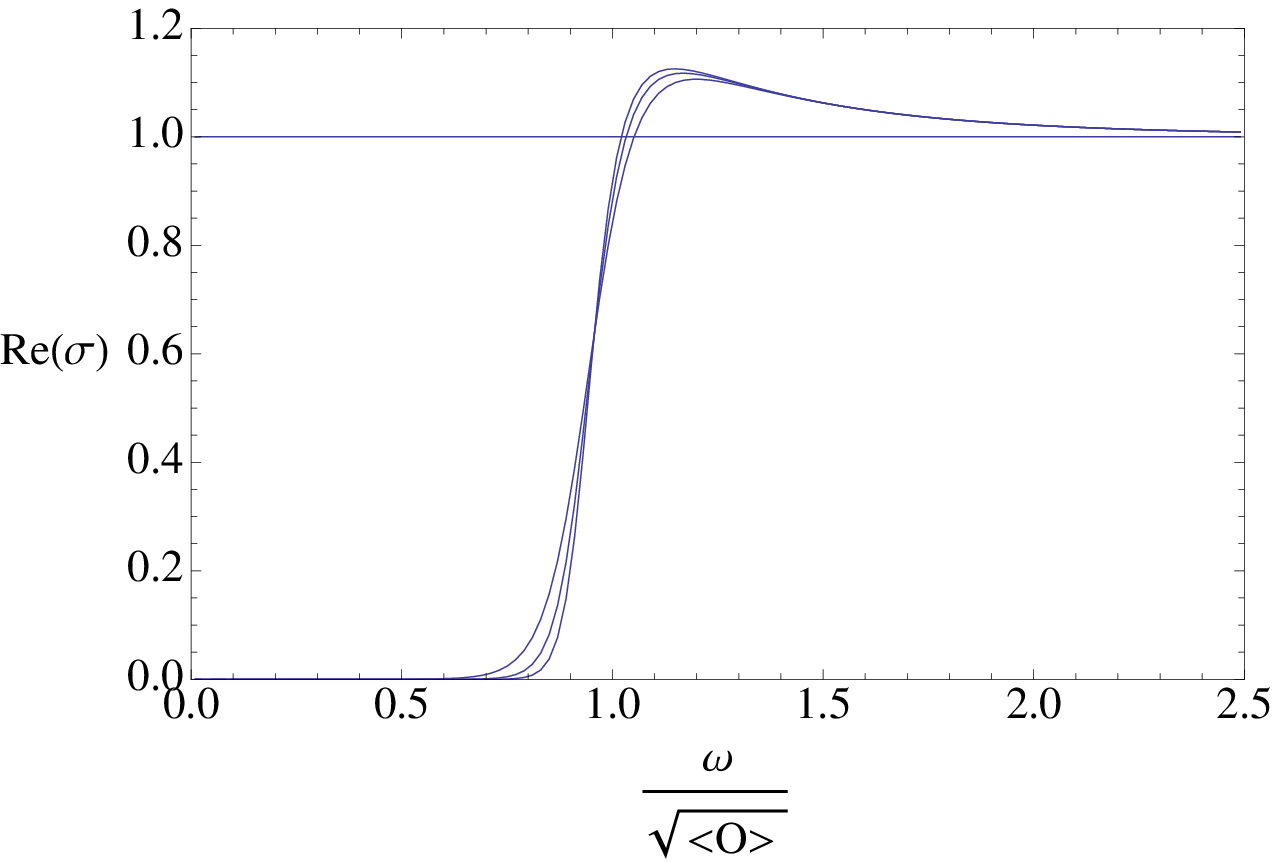, height=2.2 in}}
\caption{\em Condensates and conductivity of a holographic superconductor. Results adapted from code by C.Herzog~\cite{HerzogCode}\ presented the original work in~\cite{Hartnoll:2008vx}.
}
%\label{FIG:Lattice}
\end{figure}

To construct a spontaneous condensate, we should turn off the source, $\phi_{d-\Delta}=0$, fix a reference value for the chemical potential, $\mu=\mu_{o}$, and look for a solution of the bulk {equations of motion} which have, below some critical temperature $T_{c}$, a non-zero value for $\vev{\CO}=\phi_{\Delta}$.
This {requires that we solve Eqs.~(\ref{EQ:SC})} subject to the boundary conditions that $\Phi(0)=0$ (no source) and $A_{t}(0)=\mu_{o}$.  Generally this must be done numerically, {as Eqs.~(\ref{EQ:SC}) are nonlinearly coupled.}
By evaluating the on-shell action and using {the holographic dictionary in Table~\ref{table:holdict}, we can compute physical quantities.  For example, we find the critical temperature by searching for the minimum $T$ of the background black brane such that a non-trivial solution with these boundary conditions exists. We find the AC conductivities by looking at linear response for fluctuations of $A_{x}$ linearized around a non-trivial solution for $A_{t}$ and $\Phi$, etc.}
The precise behavior we find depends on the details of the gravitational model chosen.  By adding additional fields and interactions, we can use these techniques to engineer a wide range of dynamical phenomena in the dual QFT.

The point to emphasize here is that while we are explicitly computing correlation functions in a strongly-coupled QFT, and indeed studying the detailed physics of transport, the QFT under investigation is defined purely through its gravitational dual.  We thus call these superconductors \textit{holographic superconductors}, and more generally refer to such QFTs as \textit{holographic QFTs}.

Importantly, these models can be used to study much more than just the translationally invariant ground states of these strongly interacting superfluids.  For example, they provide a simple and powerful framework for studying spatially ordered phases and solitons.  For example, the paper by Ker\"anen \textit{et al.} in this Focus Issue~\cite{NJPfocusissue10_keranen} uses solitons in these holographic superconductors as precision probes of the superfluid, including in particular the excitation spectrum~\cite{keranen2009,keranen2009b} around a dark soliton at unitarity in the BCS-BEC crossover~\cite{NJPfocusissue33_carr,scottRG2010,liaoR2011,scottRG2012}.  As the solution of this problem reduces to a detailed study of coupled PDEs, considerable progress can be made with common computational techniques towards a concrete prediction observable in present experiments on ultracold quantum gases.

Notably, this holographic mechanism for spontaneously breaking a symmetry can be applied to any boundary QFT with symmetry breaking.  For example, the paper by Basu {\textit{et al.}} in this {Focus Issue~\cite{Basu:2011yg}} uses the holographic superconductor paradigm to study color superconductivity in holographic models of the QGP.  As good effective field theorists, Basu et al begin by writing down a gravitational effective theory with the minimal ingredients to mock up a color superconductor, then examine what constraints they must impose on the parameters of this effective description to reproduce the expected physics of a true color superconductor.

\subsubsection{ Non-Relativistic Holography and Cold Atoms}
\label{ssec:coldatoms}

Strongly correlated quantum liquids of cold atoms provide an excellent target for {applied holography.}  From a holographic point of view, cold atoms at unitarity are very similar to $\CN=4$ models of the RHIC fireball: they provide examples of strongly-correlated liquids at finite temperature and chemical potential that are governed by a conformal field theory.  The key difference is that these systems are non-relativistic, with dispersion relations scaling like\footnote{The scaling exponent $z$ is called the \textit{dynamical exponent}, and can be extracted either from the dispersion relation for low-lying modes or from the scaling of the characteristic timescale with the correlation length, $\tau\sim\xi^{z}$, as we approach a critical point~\cite{SachdevBook}.}
\be
\w\sim k^{z}\,,
\ee
so that the RG fixed points which govern their dynamics are non-relativistic conformal field theories (NRCFTs)~\cite{Mehen:1999nd,2006AnPhy.321..197S,Nishida:2007pj}.
Furthermore, in the case $z=2$ arising in cold fermions at unitarity, the symmetry algebra is enlarged~\cite{Nishida:2007pj}\ and includes a central \textit{Number operator}, $\hat{N}$, such that every operator has, in addition to a dimension, $\Delta$, a conserved \textit{particle number} eigenvalue, $N$.

Holographically, the non-relativistic scaling has a dramatic effect.  In the case of the $\CN=4$ theory, relativistic conformal invariance was enough to determine the dual space-time, with the conformal symmetry group of the QFT, $SO(4,2)$, mapping to the isometry group of the dual space-time; this fixes the geometry dual to the $\CN=4$ theory to be \ads$_{5}$.  More generally, the fact that relativistic QFTs flow to Lorentz-invariant RG fixed-points in the UV implies that their holographic duals should {asymptotically approach} pure \ads\ near the boundary.  If our QFT does not {possess Lorentz} invariance, the dual space-time cannot be simple \ads.  To build the holographic dual of an NRCFT, then, we must start from the beginning and identify the right space-time.

The simplest geometry with these properties is the \textit{Lifschitz metric}~\cite{Kachru:2008yh},
\be
ds^{2} = L^{2}\(-{dt^{2} \over r^{2z}} + {d\vec{x}^{2} +dr^{2} \over r^{2}}\)\,,
\ee
In this geometry, every slice at fixed $r$ is again a copy of flat space, but the scaling of the time-like coordinate and the spatial coordinates under a shift in $r$ is inhomogenous, with scaling $\w\sim k^{z}$.  At $z=1$, this reduces precisely to the \ads\ case, a good check.  However, at $z=2$ there is no enhancement of the isometry group of the geometry, so it's natural to look for another geometry which realizes the full enhanced $z=2$ non-relativistic conformal symmetry group.  Following this logic leads to the \textit{Schr\"odinger metric}~\cite{Son:2008ye,Balasubramanian:2008dm},
\be
ds^{2} = L^{2}\(-{dt^{2} \over r^{2z}} + {2dt\,d\xi + d\vec{x}^{2} +dr^{2} \over r^{2}}\)\,,
\ee
where $\xi$ is a new periodic variable.  Again we find that scaling implies $\w\sim k^{z}$. Now, however, the isometry group of the manifold is in fact enhanced at $z=2$.  Moreover, since $\xi$ is compact, we can expand  every field in eigenmodes of $\hat{N}=i\p_{\xi}$, so that  bulk fields, and thus all boundary operators to which they are dual, are labeled by both a dimension, $\Delta$, and also a $\xi$-momentum, {$N$}.  We thus have a geometry which enjoys both of the peculiar features of fermions at unitarity identified above.

Repeating the full construction in Sec.~\ref{SEC:EssentialHolography} in this non-relativistic setting remains an open problem.  However, considerable progress has been made in the context of both Lifschitz and Schr\"odinger systems, including the construction of charged black holes corresponding to NRCFTs at finite temperature and chemical potential~\cite{Herzog:2008wg,Maldacena:2008wh,Adams:2008wt,Balasubramanian:2009rx} and the construction of some toy non-relativistic superfluids (see e.g.~\cite{Adams:2011kb} in this Focus Issue).  Many of the basic universal results from the relativistic setting have been replicated.  For example, in the simplest models we again find ${\eta\over s}={1\over4\pi}$.  Note that, while we have suppressed factors of $c$ throughout, there are no such factors in this result for the viscosity ratio so we get a non-trivial prediction for non-relativistic systems.  Furthermore, while $\eta\over s$ is a natural observable in relativistic QFTs even if we
don't know about holography (it is the dimensionless momentum
diffusion constant in linearized hydrodynamics), the fact that $\eta\over s$ is relevant in non-relativistic systems is an interesting feature of holography; the momentum diffusion constant in non-relativistic hydrodynamics is $\eta\over n$.
In the context of Lifschitz scaling, considerably more progress has been made, including interesting results for models of strange metals~\cite{Hartnoll:2009ns}.\footnote{It is worth emphasizing that it is {straightforward} to build holographic models in which the dynamical exponent changes as the theory flows {from} the UV to IR.  A particularly illuminating example involves the fate of charged black holes with large ground-state degeneracies.  In at least some holographic models it has been shown that the ground state degeneracy is lifted by a dynamical instability which changes the near-horizon geometry from \ads\ to Lifschitz geometries which have no, or at least vastly diminished, ground state degeneracy.  Since the near-horizon region encodes the IR physics of the dual QFT, such holographic models describe QFTs where the dynamical exponent of the low-energy degrees of freedom arises due to strong interactions amongst constituents with no simple quasiparticle description, and is thus effectively invisible in the UV regime.  The dynamical scaling is in this sense \textit{emergent}.}

For models of cold atoms, on the other hand, our non-relativistic holographic models remain rather distant from the systems studied in actual experiments. Indeed, it seems likely that fermions at unitarity, like QCD, do not have weakly coupled holographic duals.
{However, unlike QCD, where we have already built holographic toy models which capture} a great deal of the strong-coupling phenomena, non-relativistic CFTs with dynamical exponent $z=2$ and Schr\"odinger symmetry have proven difficult to model holographically.
In particular, the thermodynamics and phase structure of current holographic models are not obviously related to any known systems, if not outright pathological~\cite{Herzog:2008wg,Maldacena:2008wh,Adams:2008wt}.  For example, we expect the finite-density ground state of these theories to form a superfluid which spontaneously breaks the $U(1)$ symmetry generated by the number operator, $\hat{N}$.  Holographically, $\hat{N}$ is mapped to $\p_{\xi}$, the generator of translations in the 2$^{nd}$ holographic direction.  Such a superfluid {ground state} must thus break translation invariance in the $\xi$-direction.  Unfortunately,  all of our current holographic models of $z=2$ Schr\"odinger CFTs feature manifest unbroken $\xi$-translation invariance.

This problem can be traced~\cite{Kovtun:2008qy,Barbon:2009az,Balasubramanian:2010uw} to the fact that all our current models were derived from {highly} symmetric stringy systems which enforce this symmetry.  As a result, these limitations do not appear to be intrinsic to the holographic framework but rather to the very specific models which have so far been studied.  It would be of great interest to develop more realistic models (see eg~\cite{Balasubramanian:2010uw} for recent work in this direction).
A step in this direction is described in the paper of Adams and Wang in this {Focus Issue~\cite{Adams:2011kb},} in which a superconducting ground state for such a non-relativistic holographic CFT is constructed by extending the ``holographic superconductor'' strategy from \ads\ to these non-relativistic geometries.  This leads to various interesting effects, including a surprising quantum phase transition as the density of the fluid is varied and a multicritical point where the transition to the superfluid state switches from 2$^{nd}$ to 1$^{st}$ order.  This does not solve the underlying problem, as the geometry remains unmodified, but is at least a proof of principle that such holographic {ground states} do exist.

%%%%%%%%%%%%%%%%%%%%%%%%%%%%%%%%%%%%%%%%%%%%%%%
%
%			New Sections 4.4, 4.5
%
\subsubsection{ Holographic Non-Fermi Liquids and Critical Phenomena}

Fermi surfaces are familiar beasts, and are generally well-described by standard quasiparticle perturbation theory.  This is true despite the fact that the electrons living on the Fermi surface may interact through the infinite-range Coulomb potential.  The key point is that the phase space available for two electrons to scatter is extremely limited, as both the initial electron states and the final electron states must all lie on the Fermi surface.  Folding in this phase space constraint, the effective interaction for (dressed) electrons scattering on the spherical Fermi surface mediated by the electromagnetic interaction is small, and hence the applicability of perturbation theory.  The corresponding state is called a fermi liquid.\footnote{We discuss the electron gas as an example of a Fermi liquid, but we emphasize that the arguments given here are only true for purely Coulombic forces. Current-current interactions, for example, are unscreened and destroy quasiparticle behavior~\cite{PhysRevB.8.2649}.}

Experimentally, many systems with a fermi surface are manifestly not well described by quasiparticle perturbation theory, and are thus referred to as non-fermi liquids.  Perhaps the most famous such examples are High-$T_{c}$ superconductors, whose spectral function (as measured e.g. by ARPES experiments) do not contain any sharp quasiparticle peaks.  While it is relatively straightforward to write down phenomenological models of such non-fermi-liquids, it has proven somewhat non-trivial to build controlled microscopic models which generate non-fermi-liquid behavior dynamically, and famously intractable to derive such non-fermi-liquid behavior from realistic models of the crystal lattices which lead, experimentally, to such behavior.

Since holographic QFTs neither require nor generally admit any quasiparticle description, it is natural to ask what happens when we put a holographic QFT at finite fermion number density~\cite{Lee:2008xf,Liu:2009dm,Cubrovic:2009ye,Faulkner:2009wj}.  The simplest way to do so is study a QFT with a fermion carrying a conserved charge under some global $U(1)$ for which we turn on a chemical potential.  Holographically, this corresponds to studying a fermionic field in \ads\ in the presence of a charged black hole.  The charge of the black hole induces a chemical potential in the QFT, while the boundary value of the bulk fermion specifies a source for the fermion in the QFT with which to probe the spectral function.  To compute the spectral function, which is equal to the imaginary part of the retarded Green function for the fermion in the QFT, we turn on a known source and measure the response.  Holographically, this maps to solving the Dirac equation in the background of a charged black hole in \ads\ subject to a Dirichlet boundary condition at the boundary and infalling boundary conditions at the horizon.  This turns out to be a straightforward numerical computation.  The resulting spectral functions show clear evidence of a sharp Fermi surface, but, as expected, enjoy no quasiparticle poles, and so the holographic QFT describes a non-fermi-liquid.

Much work has been done to clarify the physics of these holographic fermi liquids.  For example, it has been shown that incorporating gravitational effects can change the results dramatically at low energy, with the black hole in some cases vomiting forth its charge and disappearing, leaving behind an {\em electron star} which is extremely dense but  completely smooth, much like a neutron star but carrying a net charge~\cite{Hartnoll:2010gu,Hartnoll:2010xj}.  Meanwhile, completely different models of fermi-surfaces in holographic QFTs have been concocted~\cite{Hartnoll:2009ns}. Despite using very different ingredients, and having in some ways very different physics, all such models share the important property of having no simple quasiparticle poles in their spectral functions.

One of the key lessons of this study has been the role of emergent conformal symmetry in the near-horizon region~\cite{Faulkner:2009wj,Iqbal:2011in}.  Recall that the near-horizon region corresponds to the deep IR of the dual QFT.  Saying that this region has an emergent conformal symmetry then implies that the QFT has, at very low energy, an emergent scaling symmetry.  Importantly, the dynamical exponent of this conformal symmetry is generally larger than 1, and indeed often infinite, corresponding to ultra-local physics with a finite ground state degeneracy.  As usual, this enormous ground-state degeneracy is a marker of instability, and these systems are quick to decay into a host of other phases.  As such, these holographical critical points correspond to quantum critical points of the dual QFT.

All of these models of holographic Fermi surfaces and holographic critical phenomena share a generic two-component structure: first, a set of degrees of freedom, $\chi$, which correspond to the naive fermions living on the Fermi surface; second, operators $\psi$ in an emergent IR CFT which couple to the $\chi$ fields.  Integrating over the $\psi$-fields then generates new, strong interactions amongst the original $\chi$ fields, which is the origin of the strong interactions on the Fermi surface in these holographic models.  The natural question then becomes, if all we care about is the low energy physics, which is entirely encoded in the near-horizon region, why bother with the entire holographic framework?  Why not just build a truncated model which only includes the fermions of interest coupled to the emergent CFT which controls the interactions?  Doing so has come to be known as {\em semi-holography}~\cite{Faulkner:2010tq,Nickel:2010pr}, and has proven quite useful (see e.g.~\cite{Jensen:2011af,Iqbal:2011in,Iqbal:2011aj}).

{An example} of this semi-holographic approach is the paper by Nickel and Son in this {Focus Issue,}~\cite{Nickel:2010pr}.  In this paper, the authors argue that the transport of conserved charges is governed by an effective theory for a set of goldstone modes coupled to an emergent near-horizon CFT.  The fact that the physical modes are goldstone modes powerfully constrains the dynamics and enforces much of the structure of holographic transport.  The origin of the {Goldstone} bosons is quite beautiful.  From the holographic dictionary, we know that the conserved $U(1)$ of the boundary QFT maps into a gauged $U(1)$ in the bulk, and thus to a dynamical emergent vector field in the bulk.  Importantly, this vector has two boundary values, one at the boundary of \ads\ and one as we approach the horizon.  Now imagine we run the RG, integrating out the high energy modes (which live near the boundary) shell by shell.  If we integrate out the entire bulk, we are eventually left with a thin shell just outside the horizon, on of the boundaries of which we have a dynamical $U(1)$ gauge field, and thus a total symmetry group which is $U(1)\times U(1)$.  However, this is a fake -- there is only one $U(1)$ in the system -- so integrating out the degrees of freedom in the bulk between these two boundaries must spontaneously break  $U(1)\times U(1)$ to a single  $U(1)$.  As Nickel and Son show, this indeed happens, leaving behind a goldstone mode which remains coupled to the surviving emergent CFT.  A Wilsonian analysis of the most general action allowed by the symmetries of the system then efficiently reproduces many classic results about transport in holographic models.

These sorts of holographic approaches to quantum critical phenomena have also been applied to myriad problems in the physics of condensed matter broadly construed.  For example, the paper by Bayntun {\textit{et al.}} in this focus issue~\cite{Bayntun:2010nx} studies {plateau} transitions in {quantum-Hall-like} holographic models.  Such phenomenological models are becoming increasingly sophisticated and, as they improve, increasingly compelling.  In this case, the authors construct a model for which the scaling exponent for plateau transitions is $2\over5$, in rough agreement with experimental values of $\sim0.42$.

%%%%%%%%%%%%%%%%%%%%%%%%%%%%%%%%%%%%%%%%%%%%%%%
%
\subsubsection{Far-from-Equilibrium Physics in Strongly Coupled Quantum Field Theories}
\label{sssec:far}

Perhaps the most exciting application of holographic duality is to the study of far-from-equilibrium physics in strongly-interacting quantum many body systems.  The basic idea here is extremely simple.  As we have seen, the equilibrium thermal physics of a holographic QFT is encoded in the thermodynamics of a stationary\footnote{In GR, {\em stationary} means, roughly, time-independent, while {\em static} means, again roughly, non-rotating.} black hole horizon in the dual spacetime.  To study non-equilibrium physics in the QFT, then, we should study non-equilibrium physics in the dual gravity -- ie, time-dependent spacetimes whose black hole horizons evolve in time.

It is easy to visualize this connection.  Imagine the collision of two nuclei at RHIC or LHC.  At some initial time we have two approximately non-thermal objects in a head-on collision.  At some special moment they collide.  Quite rapidly -- in fact, amazingly rapidly according to experimental results -- the ensuing fireball thermalizes, then slowly evolves according to near-equilibrium hydrodynamics.  Now what would this process look like in a holographic dual?  Since the initial temperature is zero, the initial geometry has no black hole horizon.  Meanwhile, the initial non-thermal nuclei are basically just moving sources of stress-energy.  According to the holographic dictionary, the source for stress-energy in the boundary QFT is equal to the boundary value of the bulk metric.  Our moving packets of mass are thus represented holographically by traveling gravitational waves which extend from the boundary into the deep interior of \ads.  The collision of the nuclei then corresponds to a collision of the dual gravitational waves.  But we know what happens when you collide gravitational waves at sufficiently high energy -- they form a black hole.

However, since the waves are not perfectly thin and the process of black-hole formation is not perfectly efficient, the details of black-hole formation can be quite messy.  In particular, the horizon does not form instantaneously but grows and expands rapidly during the collision -- this black hole is far from equilibrium.  Meanwhile, while the initial collision, and thus the initial horizon, will generically be inhomogeneous, anisotropic and just generally messy, it is a theorem of black hole mechanics that the horizon always settles down to a completely uniform spherical surface with a uniform radius and a uniform temperature -- and indeed that it does so relatively rapidly, with fluctuations of the horizon damped by the viscosity we studied above.  Thus our collision of nuclei and thermalization of the QGP fireball at RHIC look, in the holographic dual, like a collision of gravitational waves which form an initially far-from-equilibrium black hole which then rapidly rings down to a simple uniform sphere.

Note the simplification afforded by this dual description: while the QFT description involves real-time quantum dynamics in a strongly-coupled many body system, which are extremely difficult to simulate, the holographic model involves solving a set of PDEs -- the Einstein equations, together with a well-defined set of initial data and boundary conditions -- which are considerably easier to simulate.

This dramatic simplification was exploited by Chesler and Yaffe~\cite{Chesler:2010bi}, who simulated the formation of a holographic QGP by numerically colliding gravitational shock waves to form a black hole in \ads.
An important lesson of the resulting numerical simulations is that these holographic QGPs thermalize exceedingly rapidly.  Rapid can be quantified in several ways.  One simple measure is that the thermalization time is of order the time required for light to cross the horizon of the final equilibrium black hole.  Alternatively, we can take the numerical results and rescale variables to reproduce the real scales of QCD.  Doing so~\cite{Chesler:2010bi} gives a very rough prediction of the RHIC thermalization time of $0.35$ fm/c, which is safely below the upper bound estimated via hydrodynamic simulations from the RHIC data; see Sec.~\ref{sec_qgp_open}. Another important lesson of this and related computations is that hydrodynamics become an excellent approximation very rapidly, with the pressure and energy tracking hydrodynamic predictions as fast as, and sometimes apparently even faster than allowed by thermalization.

Importantly, the holographic models studied thus far do not do nearly as good a job in other regards.  For example, the most simple models fail to reproduce the observed energy dependence of the total multiplicities.  This is perhaps not surprising given the simplicity of the models studied to this end, but is an important point to rectify.  Further, while these models give us useful insight into the phenomena, they do not give us any guidance on how to understand the relation between such rapid thermalization and asymptotic freedom or deep inelastic scattering.  Nonetheless, these are remarkable results.

%
%%%%%%%%%%%%%%%%%%%%%%%%%%%%%%%%%%%%%%%%%%%%%%%

%%%%%%%%%%%%%%%%%%%%%%%%%%%%%%%%%%%%%%%%%%%%%%%%%%%%%%%%%%%%%%%%%%%%
\section{Conclusions}
\label{sec:conclusions}
%%%%%%%%%%%%%%%%%%%%%%%%%%%%%%%%%%%%%%%%%%%%%%%%%%%%%%%%%%%%%%%%%%%%

 We conclude this review by providing a list of open questions in
the areas of ultracold quantum gases, QCD plasmas, and holographic
dualities. This list is far from complete, and the selection of
problems is guided by the topics discussed in the main body of
the review. Many of these problems will benefit from the connections
between the physics of ultracold quantum gases, the QGP, and holographic duality that are
being explored in this Focus Issue. For example, improved studies
of the hydrodynamic behavior of ultracold gases will benefit
from what has been learned about second order hydrodynamics
in the context of the QGP and the AdS/CFT
correspondence. Holographic dualities will also continue to
play an important role in studying the approach from the initial,
far-from-equilibrium, stage of a heavy ion collisions to the
hydrodynamic regime. Far-from-equilibrium states have also
been explored in ultracold atomic gases. Finally, new experimental
results on properties of the QGP obtained in the ongoing programs
at RHIC and LHC, as well as new results from experiments with
trapped gases may point to new, unexpected, connections between
these fascinating systems.

%%%%%%%%%%%%%%%%%%%%%%%%%%%%%%%%%%%%%%%%%%%%%%%%%%%%%%%%%%%%%%%%%%%%
\subsection{Open Problems and Questions in Ultracold Quantum Gases}
\label{ssec:ultracold}
%%%%%%%%%%%%%%%%%%%%%%%%%%%%%%%%%%%%%%%%%%%%%%%%%%%%%%%%%%%%%%%%%%%%

\begin{enumerate}

\item What is the structure of the unitary Fermi gas in both two and
three dimensions? In particular, is it possible to understand properties
of the system in terms of well defined quasiparticles?

At high temperature
the gas is always weakly correlated, independent of the strength of the
interaction. In the weak coupling regime we know that the quasiparticles
near $T_c$ are BCS-like gapped fermions, and that at very low temperature
the quasi-particles are phonons. At unitarity it has been argued that the
regime $T_c\lsim T \lsim 1.5 T_c$ can be understood in terms of a pseudogap
\cite{Janko:1997,Randeria:1998,Chen:2005}. This means that there is a gap in the
single particle spectrum, but no long range coherence. Pseudogap behavior
in three dimensions has been observed using radio frequency (RF)
spectroscopy~\cite{Chin:2004,Gaebler:2010}, and it has also been seen
in quantum Monte Carlo calculations~\cite{Magierski:2011}. Pseudogap
behavior was also reported in experiments with two-dimensional systems
\cite{Feld:2011}. However, it is not entirely clear to what extent
thermodynamic and transport properties can be understood in terms of
gapped quasiparticles.  One quasiparticle that may turn out to be relevant is
the polaron, a single spin
up particle immersed in a sea of spin down fermions~\cite{Schirotzek:2009}.
In highly polarized gases RF spectroscopy has
been used to measure the dispersion relation of polarons.
It was observed that the polaron description can be extended to more spin
balanced systems. In particular, a simple estimate based on a gas of
polarons provides a very good estimate for the critical polarization
at the Chandrasekhar-Clogston point, the first order transition from
an unpolarized superfluid to a polarized normal gas~\cite{Lobo:1999}.
Recently, polarons have also been observed in two-dimensional
Fermi gases~\cite{Koschorreck:2011,Zhang:2012}.

\item How do we make local measurements of transport coefficients
like shear and bulk viscosity, thermal conductivity and the spin
diffusion constant?

Measurements of elliptic flow and collective
mode damping provide trap averaged values of the shear viscosity
(see Sec.~\ref{ssec:transport}) but there is no direct measurement
of the local value of the shear viscosity $\eta(n,T)$. The status
is roughly analogous to the situation after the first experiments
aimed at measuring the equation of state. These experiments only
determined the total energy and entropy~\cite{ThermoLuo2009}. Local
measurements were made possible by new ideas like the Gibbs-Duhem
method employed by the ENS group~\cite{navonN2010} and the compressibility
thermometer introduced by the MIT group~\cite{kuJHMark2011}. It is
possible in principle to unfold the currently available data for
trap averaged viscosities, but this method requires a very careful
treatment of the dilute corona, and a more direct method would be
desirable. One possible direction is the development of new experimental
techniques that generate local shear flows and study their relaxation.
Another option is to measure the frequency dependent shear viscosity
via spectroscopic methods as suggested in~\cite{Taylor:2010ju}.
Similar questions apply to other transport properties, like
bulk viscosity, thermal conductivity and spin diffusion. Current
methods are mostly sensitive to average transport coefficients,
but more local measurements are needed.

\item Are there reliable approaches to transport theory beyond
the Boltzmann equation?

In the unitary limit, transport properties
can be computed reliably using the Boltzmann equation in the
limit $T\gg T_F$. For $T\sim T_F$ the quasiparticles are strongly
interacting, and the kinetic description in terms of binary scattering
between atoms breaks down. Enss \textit{et al.} have computed the shear
viscosity using a T-matrix resummation~\cite{Enss:2010qh}. A
similar approach is discussed by Guo \textit{et al.} in this Focus Issue
\cite{NJPfocusissue2_levin}. These methods are quite successful in
describing the thermodynamics, but there is no expansion parameter,
and it is not clear if non-equilibrium properties are predicted as
well as equilibrium features. In Sec.~\ref{ssec:uni} we briefly
discuss the application of $1/N$ and $\epsilon$ expansions to
equilibrium properties of the unitary gas, but except for the
recent work of \cite{Enss:2012} these methods
have not been extended to non-equilibrium properties. Similar
remarks apply to other methods as well: There are detailed
studies of equilibrium properties using the exact renormalization
group~\cite{Diehl:2009ma}, but no corresponding studies of transport
properties. Finally, there now exist very accurate quantum Monte
Carlo calculations of the equation of state, but there is only
a single, pioneering, study of transport properties
\cite{Wlazlowski:2012jb}.

\item Are there other universal quantum gases that can be
created by manipulating the dimensionality, the spin structure,
the type of interaction, or the dispersion relation?

The unitary gas exhibits a very high degree of universality:
in the limit of infinite scattering length and zero effective
range the many body system is completely characterized by
the fermion mass, the temperature, and the chemical potential.
The dependence on dimensionful combinations of these parameters
is completely governed by dimensional analysis. Universality
can be exploited to use the unitary gas as a model system
for dilute neutron matter; see Sec.~\ref{sec_phases}. The
question is whether this idea can be extended to more
complicated systems, for example three-species gases that
can serve as a model for the quark matter/nuclear matter
transition (see Sec.~\ref{ssec:newDirections}), or four-species
gases that correspond to self-bound nuclear matter, which is a
liquid of equal numbers of spin up and down protons and
neutrons. Universality in systems with more than two degrees
of freedom is more complicated, because the interaction also
depends on at least one three-body parameter~\cite{Braaten:2006vd},
and the system may require repulsive short range forces in order
to prevent collapse. New universal states can also be created
by combining Bose and Fermi gases in different dimensions
\cite{Nishida:2011ew}, or by manipulating the dispersion
relation of the atoms. Steps in this direction were recently
taken by Salger \textit{et al.}, who demonstrated linear dispersion
and Klein tunneling of a Bose-Einstein condensate in a one-dimensional
optical lattice \cite{Salger:2011}, and by Taruell \textit{et al.},
who created a two-dimensional optical lattice with the same honeycomb lattice
structure as graphene, and therefore obtained low energy fermions
with a linear dispersion relation~\cite{Tarruell:2012}.

\item What is the physics of the crossover in the discrete context of
optical lattices?

Will the predictions of holographic duality of a
lower limit on the ratio of viscosity of entropy hold true in the
lattice context, or not? What class of theories from holographic
duality might map onto the Fermi Resonance Hamiltonian or other
effective two-channel models?  How can we extend such models to treat the imbalanced Fermi gas?
How will a polaron gas be described in the lattice context?

\item What will be the interplay between disorder and interactions for the unitary gas?

The relative effects of disorder and interactions on both thermal and quantum phase transitions in quantum many body physics is an open question generally (see e.g.~\cite{pilati2010}).  An essential feature of disorder is Anderson localization, in which scattering off many small defects conspires to localize all states above a certain threshold in 3D.  This concept is in fact a wave physics or single particle quantum concept: it is so far unclear where and how interactions destroy or allow for Anderson localization in a generic sense.  Disorder was treated in this Focus Issue for ultracold Fermi gases specifically by Han and S\'a de Melo~\cite{NJPfocusissue14_sademelo}.  Here, too, we find potential connections to holographic duality, as explored in recent forays into disordered systems by Adams and Yaida~\cite{adamsA2011,adamsA2012}.  Although these explorations have not treated lattice physics particularly, nevertheless disorder is an essential feature of lattice systems arising in nature, and can be induced artificially in various ways for optical lattices~\cite{jendrzejewski2012}.  Can holographic duality capture the underlying lattice causes of disorder?  How will experiments in ultracold Fermi gases combine disordered optical lattices with unitary fermions at sufficiently low temperatures?  Might studies in this direction gain us some fundamental insight into the interplay between disorder and interactions in strongly correlated systems?

\end{enumerate}

%%%%%%%%%%%%%%%%%%%%%%%%%%%%%%%%%%%%%%%%%%%%%%%%%%%%%%%%%%%%%%%%%%%%
\subsection{Open Problems and Questions in Quantum Chromodynamic Plasmas}
\label{sec_qgp_open}
%%%%%%%%%%%%%%%%%%%%%%%%%%%%%%%%%%%%%%%%%%%%%%%%%%%%%%%%%%%%%%%%%%%%

%In this section we discuss the main challenges, questions
%and open problems that are being addressed by theorists and experimentalists
%at RHIC and the LHC.

\begin{enumerate}
\item Can we determine all the  transport properties of the QGP?

We would like
to determine, with fully quantified experimental and theoretical uncertainties, the value of the shear
viscosity to entropy density ratio $\eta/s$, the heavy quark diffusion
constant $D$, and the jet quenching parameter $\hat{q}$. This program
is probably closest to completion in the case of $\eta/s$. In this
case the basic method, second order relativistic viscous hydrodynamics,
is now fairly well understood. Final state corrections can be handled
by coupling to a hadronic cascade. There is a large set of data,
including the energy, impact parameter, and system size dependence
of elliptic flow, as well as the observation of higher harmonics
that points to the validity of the hydrodynamic description. The
internal consistency of hydrodynamics already implies a fairly
strong bound $\eta/s\lsim (0.5-1.0)$. The main problem is that we
have no direct method for establishing the initial conditions for
the hydrodynamic description. This problem may be addressed by
focusing on a larger set of observables, in particular higher
harmonics of the flow~\cite{Alver:2010dn,Schenke:2011bn}. There
also remain theoretical uncertainties regarding the coupling of
viscous fluid dynamics to a Boltzmann description, the role of
bulk viscosity, and the equation of state. A discussion of the
known uncertainties in $\eta/s$ can be found in~\cite{Song:2008hj}.

\hspace*{5mm}
The main strategy for determining the heavy quark diffusion constant
is to use a Langevin description of the diffusion process where
the background collective expansion is taken from hydrodynamics.
The most important problem at present is the fact that there are
only data for single leptons, and not for identified heavy quarks.
This problem will be solved by improved detection capabilities at
RHIC and LHC. The main theoretical issue is how to handle the
interplay between elastic and inelastic scattering. For very heavy
quarks elastic scattering is dominant, but for charm quarks
inelastic processes may be important. This also includes hadronization
effects such as coalescence.

\hspace*{5mm}
In the case of jet energy loss there are still a number of
theoretical issues that need to be resolved. The radiative energy
loss of an energetic parton ($E\gg T$) in a partonic medium can
be expressed in terms of a single medium parameter, the transverse
diffusion constant $\hat{q}$. It is not clear if this carries over
to the spectra of jets or leading particles produced in the medium.
Another source of uncertainty is that different implementations of
the in-medium radiation in perturbative QCD lead to significantly
different values of $\hat{q}$ when applied to the RHIC
data~\cite{Bass:2008rv}. There is a significant ongoing effort to
understand these differences~\cite{JET}. On the experimental side
the LHC represents a large step towards higher jet energies $E\sim
100$ GeV, and significantly improved statistics in the range $E\lsim
20$ GeV already explored at RHIC. Both at the LHC and RHIC
experimentalists are studying not only spectra of hard particles,
but also identified jets, jet shapes, and correlations between
hard particles and the associated soft emission.

\item Can we understand early thermalization?

The success of the hydrodynamic
description implies that the system must thermalize early. The
question is how to quantify this statement experimentally, and
how to understand early equilibration theoretically. The most
reliable constraint on the thermalization time arises from the
observation of elliptic flow. Elliptic flow is driven by the
anisotropy of the initial state. Since ballistic expansion
dilutes the initial anisotropy local equilibration must happen
early for elliptic flow to develop. A very conservative constraint,
discussed in~\cite{Heinz:2009xj}, is $\tau_0<2.5$ fm/c. Typical values
used in hydrodynamic fits are $\tau_0\lsim 1$ fm/c. Another hint of
early thermalization comes from the observation of thermal
photons with temperatures significantly larger than those
seen in hadronic spectra. The PHENIX collaboration has recently
reported a photon transverse momentum spectrum with a slope
parameter $T=221\pm 19\pm 19$ MeV~\cite{Adare:2008fqa}. The
spectra can be reproduced in hydrodynamic calculations
with thermalization times ranging from $\tau_0\lsim 0.15$ fm/c
to $\tau_0\lsim 0.6$ fm/c, corresponding to initial temperatures
$T_0\sim (300-600)$ MeV. Establishing experimental constraints on the equilibration
time is further complicated
by the fact that not all observables require full thermalization.
For example, for the development of radial and elliptic flow it
is not necessary that longitudinal momentum distributions are
fully equilibrated~\cite{Martinez:2012tu}.

\hspace*{5mm}
Understanding thermalization theoretically is difficult, even
if the early evolution is governed by weak coupling. The first
attempt to understand thermalization starting from an initial color-glass
state using $2\to 2$ and $2\to 3$ scattering is the  bottom-up
scenario described in~\cite{Baier:2000sb}. It was later realized
that this picture is modified by collective effects that lead
to plasma instabilities~\cite{Arnold:2003rq,Kurkela:2011ti,Mueller:2005un};  see
also the contribution by Dusling in this Focus Issue~\cite{Dusling:2011zw}.
In all of these approaches it is hard to understand how thermalization
at RHIC can occur on a time scale $\tau_0\lsim 1$ fm/c.

\hspace*{5mm}
Attempts to understand thermalization at strong coupling have focused
on holographic duality; see Sec.~\ref{sssec:far} and~\cite{Kovchegov:2009he,Gubser:2010nc}
for recent overviews. Holographic duality provides an interesting
geometric picture of thermalization: the initial state corresponds
to two colliding shock waves, and thermalization is signaled by
the formation of an event horizon. The hydrodynamic stage is described
by the relaxation, or \textit{ringdown} of the black hole. In holographic
duality one can achieve fast equilibration~\cite{Chesler:2010bi}, but
it is hard to make contact with asymptotic freedom and the well-established
theory and phenomenology of nuclear deep inelastic scattering, and
it is difficult to understand the observed energy and impact parameter
dependence of the total multiplicity~\cite{Gubser:2009sx}.

\item Does the quark-gluon plasma have a quasiparticle description?

We would like to understand how the properties of the hot and dense matter
produced at RHIC and the LHC are related to its structure, in particular
whether the matter can indeed be described as a plasma made of
quarks and gluons. Holographic duality would seem to indicate
that a strongly interacting quantum fluid near the viscosity bound
does not have a quasiparticle description; see Sec.~\ref{sec_sQGP}
and Sec.~\ref{sec:hd}. In Sec.~\ref{sec_sQGP} we discussed a number
of observables that have been suggested as probes of the quasiparticle
structure: fluctuations, the observed constituent quark scaling of
flow, and the relation between the heavy quark diffusion constant
and the shear viscosity. None of these observables provides a
conclusive test of the quasiparticle picture, and new ideas would
certainly be helpful. We also discussed the possibility to study
quasiparticles by extracting the spectral functions of the correlators
of conserved charges in lattice QCD, such as energy, momentum, flavor, etc.
These studies, too, are somewhat indirect because on the lattice we
only have direct access to imaginary time correlation functions. This
means that the spectral functions have to be extracted via analytic
continuation. Analytic continuation of numerical data is difficult,
but there has been significant progress over the last couple of
years~\cite{Meyer:2011gj}.

\item Can we experimentally locate the phase transition?

There is a continuing effort to
establish the presence of a critical endpoint in the QCD phase diagram.
The basic idea is to look for non-monotonic behavior of fluctuations or
correlations as a function of the energy of the colliding system.
The first data from the RHIC beam energy scan have recently been
released~\cite{Mohanty:2011nm}; see also the contribution by Mohanty
in this Focus Issue \cite{NJPfocusissue6_mohanty}.  The data contains new information
about the beam energy dependence of hydrodynamic flow which will help
to establish the onset of nearly perfect fluidity. Gupta \textit{et al.} have
argued that data on fluctuations can be directly compared to lattice
QCD, and that it can be used to set the scale for the phase transition
at zero chemical potential~\cite{Gupta:2011wh}. There is not yet
conclusive evidence for the existence of a critical point, but data
for additional beam energies and with better statistics will become
available in the near future.
In heavy ion collisions the growth of the correlation length is limited
by finite size and finite time effects, and it is important to identify
observables that provide the best sensitivity to critical behavior.
Observables that have been considered include fluctuations of
conserved charges, like baryon number, other fluctuation observables,
like the pion-to-kaon ratio, the mean $p_T$, or elliptic flow, and
higher moments like skewness and curtosis~\cite{Stephanov:2008qz}.
There is a parallel effort to locate the critical endpoint in lattice QCD.
This effort is complicated by the sign problem that affects lattice
calculations for fermions at finite chemical potential~\cite{deForcrand:2010ys}.

\hspace*{5mm}
The discovery of a critical point would establish the presence
of a phase transition between hadronic matter and the QGP.
At small chemical potential this transition is only a crossover. One
might hope that it would be possible to detect the crossover through
hydrodynamic effects associated with the minimum in the speed of
sound near the crossover temperature~\cite{VanHove:1982vk}. However, explicit
calculations seem to show that there is no direct link
between the presence of a minimum in the speed of sound and the
beam energy dependence of multiplicity and flow~\cite{Heinz:2009xj}.

\hspace*{5mm}
Finally, we would like to demonstrate the existence of quark matter
at high density and low temperature. It is not known whether the
cores of compact stars are sufficiently dense to contain quark matter
or other exotic phases. The presence of quark matter can potentially
be detected through its effect on the equation of state, which is
reflected in the mass-radius relationship. There is a very active
program to pin down  this relationship using observations of
compact stars~\cite{Steiner:2010fz,Ozel:2010fw}. Once the presence
of a high density phase has been established, more detailed information
regarding its properties can be derived using the cooling and spin-down
behavior of compact stars.

\end{enumerate}

%%%%%%%%%%%%%%%%%%%%%%%%%%%%%%%%%%%%%%%%%%%%%%%%%%%%%%%%%%%%%%%%%%%%
\subsection{Open Problems and Questions in Holographic Duality}
\label{ssec:HD}
%%%%%%%%%%%%%%%%%%%%%%%%%%%%%%%%%%%%%%%%%%%%%%%%%%%%%%%%%%%%%%%%%%%%

\begin{enumerate}

\item{What does it take to derive a holographic duality?}

We have no first-principles derivation of holographic duality.  This
is perhaps not surprising, since it is a strong-weak duality, but it
is frustrating.  We do have a number of heuristic arguments, such as
appeared in Sec.~\ref{ssec:heuristics} above.  We also know how to
derive a long but finite list of dual pairs as decoupling limits
of particular string theories, e.g., the original construction,
$\CN$=4 Yang-Mills in 4d.  But we do not know how to derive the
duality directly within the bulk or boundary theories.\footnote{This is
not for lack of trying; see\cite{Heemskerk:2009pn,Lee:2010ub,Douglas:2010rc}
for recent approaches.}

\hspace*{5mm}
This is important for two main reasons.  First, we do not have a
thorough understanding of the consistency conditions or regimes of
validity of our holographic dual descriptions.  We do have general
rules of thumb, such as the scalings discussed in Sec.~\ref{sssec:strong},
and we can check for self-consistency in specific examples.  But
particularly when we are defining a QFT via the holographic
prescription, it is not always  obvious what the precise regime
of validity is, and to what extent we should trust the chosen
gravitational truncation.

\hspace*{5mm}
Second, if we want to generalize holographic duality, it would be
very helpful to know why it is true in the more familiar examples.
For instance, we would like to generalize to intrinsically anisotropic
systems, to asymptotically flat or positive-curvature (dS) spacetimes,
or even to understand why this is not possible.

\item{Given a QFT, what is the gravitational dual?}

Even if a general derivation is not possible, it would be useful to
have a constructive map which allowed us to take an arbitrary QFT
and identify the appropriate gravitational dual.  For example, what
is the exact dual of QCD, not QCD plus a collection of exotic fields,
not of large $N$ QCD, not of some toy model, but pure QCD?  Similarly,
with an eye towards ultracold atoms, what is the gravitational dual
of the Hubbard model, or of the Kondo problem?

\hspace*{5mm}
Again, the fact that holography is a strong-weak duality should give
us pause.  As we have seen, when the gravitational system is weakly
coupled, the dual QFT generically does not have any simple quasiparticle
interpretation. If we want to define the QFT using conventional Lagrangian
methods, the best we will in general be able to do is say that it is the
strong-coupling limit of some explicit perturbative theory.  In the models
we understand best, the coupling that is getting strong is tightly
constrained by symmetries, e.g. gauge symmetries, flavor symmetries,
or supersymmetry, so we know how to follow the RG flow well beyond
the perturbative, quasiparticle regime. We do not know how to solve
the more general case without these extraordinary symmetries and the
constraints they imply.

\item{Can we derive gravity from field theory?}

The holographic graviton looks very much like an emergent vector
boson in more familiar condensed matter systems.  It arises as an
emergent dynamical gauge field which reorganizes the strongly-coupled
physics into a set of perturbative local fluctuations. However, in
holography the graviton (a) lives in one higher dimension and (b)
couples universally. Point (a) fits naturally with a basic feature
of gravity: since gauge transformations are coordinate transformations,
there can be no truly local degrees of freedom, so any emergent graviton
must come along with an emergent spacetime dimension with which to fix
its gauge redundancy, much as any emergent vector boson must come along
with an extra Hubbard-Stratanovich scalar to fix its gauge redundancy.
This was in fact the origin of the term holography in gravitational
physics.  Point (b) is more spectacular from the point of view of
emergence, but is in fact a necessary feature of any theory with a
fluctuating spin-2 gauge field.

\hspace*{5mm}
It is thus tempting to wonder whether we can in fact derive the
graviton in a general spacetime as an emergent gauge boson in a
strongly interacting quantum field theory.  Some progress has been
made on this front (\cf~\cite{Lee:2010ub}), but a detailed understanding
remains elusive.

\item{Do we always need the complete holographic description?}

In many of the examples discussed above, and indeed almost generically
in the holographic literature, most of the work is being done by a very
small fraction of the total degrees of freedom in the holographic
description.  For example, in the computation of $\eta/s$ in $
\CN$=4 SYM, whose holographic dual, string theory in \ads$_{5}$$\times$
$S^{5}$, includes an infinite number of dynamical massive fields, the
only fields that matter in the computation are the $5d$ metric and
a single scalar mode encoding one particular fluctuation of this
metric.  In practice, keeping track of the full holographic description
often seems to be overkill.\footnote{Note that this is
analogously true in many-body physics: to describe ground states
and low energy dynamics, one generally only needs a tiny corner
of Hilbert space. However, in the holographic description the
physics of a single bulk field represents an enormously complicated
and non-local set of interactions amongst the boundary degrees of
freedom.  So observing that the bulk description can be effectively
simplified to a small subset of bulk fields does not immediately
imply that the same is true of the boundary QFT.  For example, in
holographic models of non-Fermi liquids, the bulk physics can be
truncated rather dramatically, but the dual QFT has no quasiparticle
description and so has not proven amenable to any simple truncation;
concretely, the spectral function does not have quasiparticle poles.
With that said, it seems well worthwhile to search for relations between
known effective truncations in quantum many-body and holographic models.}

\hspace*{5mm}
This raises two natural questions.  First, when does it make sense
to truncate the full holographic description to a small subset of
modes?  Second, might this truncated holographic model have a more
straightforward QFT representation than the full holographic
description?  Said differently, might there be a ``semi-holographic''
formalism defined entirely within field theory which allows us to
build conventional QFTs which reproduce the interesting phenomenology
of fully holographic QFTs without using higher dimensions?

\hspace*{5mm}
Recent work on holographic Fermi liquids, holographic quantum critical
points and holographic quantum liquids (see~\cite{Faulkner:2010tq}
and especially the work of Nickel and Son in this Focus Issue
\cite{Nickel:2010pr}) demonstrates that at least some of the
interesting holographic effects can indeed be realized in a purely
quantum field theoretic formalism.  On the other hand, for questions
where the non-linear dynamics of the geometry such as far-from
equilibrium physics, as opposed to kinematics or linear response,
are important, it seems unlikely than any semi-holographic
description can usefully capture the salient physics.  Just how
far a semi-holographic approach can be pushed, then, is an extremely
interesting question.

\item{What predictions does holography make for generic strongly
coupled systems?}

Once we let go of the idea that our models must be defined by
traditional quasiparticle effective QFTs, we can start asking
general questions about the strongly coupled QFTs illuminated
by our holographic examples.  What common properties do these
theories share?  What general lessons  can we draw about their
kinematics, dynamics, phase structures, critical behavior,
equilibration, etc?  For example, we already know to expect
nearly inviscid hydrodynamic behavior at finite density and
low temperature.  What else should we expect in general?
Perhaps most importantly, what lessons can we draw for systems
we really care about?

\item{What is the holographic dual of a non-relativistic QFT?}

As discussed in Sec.~\ref{ssec:coldatoms}, non-trivial progress has
been made in studying non-relativistic QFTs and many body systems.
This has led to some simple heuristic lessons, for example that we
should expect the viscosity to entropy ratio to again be extremely
low, and that the \textit{Bertsch parameter}\footnote{We defined the
Bertsch parameter of the unitary Fermi gas in Sec.~\ref{ssec:uni}.
Here we use the term to refer to ratio of the ground state
energy at strong coupling to that at weak coupling in a general
non-relativistic CFT.}
should be expected to be of order unity and remain relatively
insensitive to the coupling constant near strong coupling. However,
the state of current models is rather poor, with awkward thermodynamics
and bizarre phase structure.  What are we missing?  Can we build a
holographic dual in the same universality class as fermions at unitarity?

\item{How (and why) is quantum information encoded in the bulk geometry?}

Holographically, information about the thermodynamic state of the
boundary quantum field theory is encoded in the detailed geometry
and matter field backgrounds of the bulk theory.  For example, the
temperature and chemical potential are encoded in the mass and
charge of the bulk black hole. Similarly, when the QFT is in a
pure state, the bulk geometry should encode the quantum correlations
of the boundary QFT at zero temperature.  But how?

\hspace*{5mm}
One approach to this question is the Ryu-Takanayagi conjecture
\cite{Ryu:2006bv,Nishioka:2009un}, which states that the entanglement
entropy associated to a region $R$ in a holographic QFT is given by
the area of the minimal-area surface in the bulk which ends on the
boundary of the region $R$. Holographically, this conjecture matches the relation
between the thermodynamic
entropy of a black hole and the area of its horizon.  Interestingly, it suggests
an intimate connection between quantum and thermal entropy -- and more
generally, between quantum and thermal fluctuations -- all of which are
encoded by the physics of horizons in the holographic dual.

\hspace*{5mm}
While a general proof
remains lacking, considerable evidence has been found for this conjecture,
and conversely numerous predictions have been made for the
entanglement entropy of strongly-coupled many-body systems by exploiting
this conjecture (see e.g. \cite{Casini:2012rn, Callan:2012ip, Headrick:2007km, Hubeny:2007xt, Hayden:2011ag, Casini:2011kv}\ and references therein).  For example, in this focus issue, Albash and Johnson
\cite{Albash:2010mv}\ study the evolution of the Ryu-Takanayagi entropy
under a thermal quench, leading both to evidence supporting the conjecture
and to predictions for the evolution of the entanglement entropy.

\item{How does spatial inhomogeneity change the story?}

As we have seen, holography translates quantum computations in the
boundary QFT (i.e. Feynman diagrams and path integrals) into classical
GR calculations in the bulk (i.e. solving nonlinear partial differential
equation boundary value problems).  As such, it is clear why most research
to date has focused on homogeneous, isotropic problems. However, from
the quantum many-body side we know that inhomogeneity can radically
alter the dynamics, and that the precise way they do so provides a
detailed probe of the physics driving those dynamics.  The same is
true for anisotropy. Moreover, much of what we know about inhomogeneity
and anisotropy in QFT comes from some form of perturbation theory.
It would thus be particularly interesting to study the dynamics of
holographic systems in the presence of inhomogeneity and anisotropy.
Will we find evidence of many-body localization?  Do (2+1)d QFTs at
strong coupling and in the hydrodynamic regime display a canonical
turbulent cascade, or an inverse cascade as is usually though not
universally expected in two spatial dimensions?

\end{enumerate}

Many of the issues discussed in Secs.~\ref{ssec:ultracold}-\ref{ssec:HD}
are addressed in the contributions to this Focus Issue,
and we hope that this summary, as well as the insights
contained in these contributions, will spur further
advances in the intriguing subject of strongly
correlated quantum fluids.

\section{Acknowledgments}

This work was supported in part by the U.S. Department of Energy under cooperative research agreement
DE-FG02-03ER41260 (T.S.) and DE-FG02-05ER-41360 (A.A.), and by the Division of Materials Science and
Engineering,  the Office of Basic Energy
Sciences, Office of Science grant  DE-SC-0002712 (J.E.T.); by the U.S. National Science Foundation under grants PHY-0547845, PHY-0903457,
and PHY-1067973 (L.D.C.) and PHY-1067873 (J.E.T.); by the Army Research Office grant  W911NF-11-1-0420 (J.E.T.); and by the U.S. Air Force
Office of Scientific Research under grants FA9550-10-1-0072 (J.E.T.) and FA9550-11-1-0224 (L.D.C.).  We developed and wrote this Review in part at the Aspen Center for
Physics workshops on Quantum Simulation/Computation with Cold Atoms and Molecules (L.D.C.), String Duals of Finite
Temperature and Low-Dimensional Systems (A.A.), and Critical Behavior of Lattice Models in Condensed Matter and
Particle Physics (L.D.C.); at the Kavli Institute for Theoretical Physics workshops on Beyond Standard Optical
Lattices (L.D.C.), Disentangling Quantum Many-body Systems: Computational and Conceptual Approaches (L.D.C.),
and Holographic Duality and Condensed Matter Physics (A.A.); and at the Heidelberg Center for Quantum
Dynamics (L.D.C.).  We thank Joaquin Drut for providing some of the data in Fig.~\ref{fig:ratio}.
We thank Aurel Bulgac, Chenglin Cao, Oliver DeWolfe, Ethan Dyer, Ethan Elliot, James Joseph, Kathy Levin, John McGreevy,
Jessie Petricka, Michael Wall, Haibin Wu, and Martin Zwierlein for useful discussions.

%\bibliographystyle{unsrt}
%\bibliography{refs}

\end{document}

%% file: allan_macros.tex
\def\({\left(}
\def\){\right)}
\def\[{\left[}
\def\]{\right]}

\def\frac#1#2{{#1 \over #2}}

\def\vev#1{\langle{#1}\rangle}

\newcommand{\half}{\frac{1}{2}}

\renewcommand{\a}{\alpha}
\renewcommand{\b}{\beta}

\renewcommand{\r}{\rho}

\renewcommand{\l}{\lambda}

\newcommand{\w}{\omega}
\newcommand{\m}{\mu}
\newcommand{\n}{\nu}
\newcommand{\eps}{\epsilon}

\newcommand{\LL}{\Lambda}

\renewcommand{\S}{\Sigma}
\newcommand{\CA}{{\cal A}}

\newcommand{\CC}{{\cal C}}

\newcommand{\CF}{{\cal F}}

\newcommand{\CJ}{{\cal J}}

\newcommand{\CL}{{\cal L}}

\newcommand{\CN}{{\cal N}}
\newcommand{\CO}{{\cal O}}

\def\Im{{\rm Im\hskip0.1em}}

\def\cf{{\it c.f.}}

\def\p{\partial}

\def\ads{AdS}

\def\II{\relax{I\kern-.10em I}}

\def\ms{m_s}

\newcommand{\rH}{r_{\mathrm H}}

\newcommand{\scale}{{u}}